\documentclass[12pt]{article}
\usepackage[a4paper]{geometry}
\usepackage{amsfonts,amssymb,amsmath,epsfig,enumerate,cases,lscape}
\usepackage[all]{xy}
\DeclareMathAlphabet{\mathpzc}{OT1}{pzc}{m}{it}
\usepackage{ulem}
\usepackage{multirow}
\DeclareMathAlphabet{\mathpzc}{OT1}{pzc}{m}{it} 
\advance\oddsidemargin -1cm
\renewcommand{\thesection}{\Roman{section}}
\renewcommand{\theequation}{\thesection.\arabic{equation}}
\numberwithin{equation}{section}
\begin{document}
\newcommand{\HRule}{\rule{\linewidth}{0.1mm}}
\newcommand{\as}{\alpha_{\textrm{s}}\,}
\newcommand{\ds}{\displaystyle}
\newcommand{\aB}{a_{\textrm{B}}\,}
\newcommand{\Bstar}{ \overset{*\!\!}{B} }
\newcommand{\Gstar}{ \overset{*\!\!}{G} }
\newcommand{\lP}{l_{\wp}}
\newcommand{\nP}{n_{\wp}}
\newcommand{\SP}{S_\wp}
\newcommand{\GG}{\textnormal I \! \Gamma}
\newcommand{\A}{\mathcal{A}}
\newcommand{\B}{\mathcal{B}}
\newcommand{\D}{\mathcal{D}}
\newcommand{\F}{\mathcal{F}}
\newcommand{\M}{\mathcal{M}}
\newcommand{\vr}{\vec{r}}
\newcommand{\sdot}{\,{\scriptscriptstyle{}^{\bullet}}\,}
\newcommand{\R}{\mathcal{R}}
\newcommand{\J}{\mathcal{J}}
\renewcommand{\S}{\mathcal{S}}
\newcommand{\Z}{\mathcal{Z}}
\newcommand{\ZP}{\mathcal{Z}_{\wp}}

\newcommand{\eklo}[2]{\, {}^{[\mathrm{#1}]}{\!#2}}

\newcommand{\rklo}[2]{\, {}^{(\mathrm{#1})}{ {#2} }}

%
\newcommand{\ekloi}[3]{\, {}^{[\mathrm{#1}]}{#2}_\mathrm{#3}}

%
\newcommand{\rkloi}[3]{\, {}^{(\mathrm{#1})}{#2}_\mathrm{#3}}

%
\newcommand{\ekru}[2]{#1_\mathrm{[#2]}}

%
\newcommand{\tekru}[2]{\tilde{#1}_\mathrm{[#2]}}

%
\newcommand{\orrk}[2]{ \mathbb{#1}^{(\textrm{#2})}  }

%
\newcommand{\vri}[2]{ \vec{#1}_{\textrm{#2}}   }

%
\newcommand{\nrft}[2]{ \tilde{\mathbb{#1}}_{[\mathrm{#2}]} }

%
\newcommand{\nrf}[2]{ {\mathbb{{#1}}}_{\mathrm{ {#2}}} }

%
\newcommand{\en}[1]{\varepsilon_{\textrm{#1}}}

%
\newcommand{\Ea}[3]{ #1^{(\mathrm{#2})}_{\mathrm{#3}}}

%
\newcommand{\Eb}[3]{ #1^{(\mathrm{#2})}_{\mathrm{[#3]}}}

%
\newcommand{\ru}[2]{#1_\mathrm{#2}}

%
\newcommand{\tru}[2]{\tilde{#1}_\mathrm{#2}}

%
\newcommand{\roek}[2]{ \mathbb{#1}^{[ \mathrm{#2} ] }  }

%
\newcommand{\rork}[2]{ #1^{( \mathrm{#2} ) }  }

%
\newcommand{\rorkt}[2]{ \tilde{#1}^{( \mathrm{#2} ) }  }

%
\newcommand{\gkloit}[3]{\, {}^{\{\mathrm{#1}\}}{ { \tilde{#2}  }  }_\mathrm{#3}}

\newcommand{\nsn}{\nu_*^{\{n\}}}

%
%
\newcommand{\rugk}[2]{#1_{\{\rm #2\}} }

\newcommand{\rogk}[2]{#1^{\{\rm #2\}} }

\newenvironment{mysubsection}[1]
{  
\addcontentsline{toc}{subsection}{\emph{\textbf{#1}} } { \emph{\textbf{#1}} }\nopagebreak
}

\newenvironment{mysubsubsection}[1]
{  
\addcontentsline{toc}{subsubsection} {\emph{#1}}  { \emph{#1} }\nopagebreak
}

\newenvironment{mytable}[1]
{  
\addcontentsline{toc}{subsubsection}{#1} { }\nopagebreak
}

\newenvironment{myfigure}[1]
{  
\addcontentsline{toc}{subsubsection}{\emph{#1}} { }\nopagebreak
}

\newenvironment{myappendix}[1]
{\renewcommand{\thesection}{\Alph{section}}
\addcontentsline{toc}{subsection}{\textbf{\emph{\appendixname}}\
  \emph{\textbf{\thesection :}}\   \textbf{\emph{{#1}}}}  {}
}


\newcommand{\slo}[1]{\,'{\!#1}}

%
\newcommand{\gkloi}[3]{\, {}^{\{\mathrm{#1}\}}{\!{#2}}_\mathrm{#3}}

%
\newcommand{\gklo}[2]{\, {}^{\{\mathrm{#1}\}}{\!{#2}}}

%
\newcommand{\rrklo}[2]{\, {}^{(\mathrm{#1})}{\! {#2} }}
%
\newcommand{\srklo}[2]{\, {}^{({\sf #1})}{\! {#2} }}


\newcommand{\e}{\operatorname{e}}
\newcommand{\crm}{\mathrm{c}}
\newcommand{\mustbe}{\stackrel{!}{=}}
\newcommand{\rt}{(r,\vartheta)}
\newcommand{\tablesize}{\scriptstyle}
\newcommand{\elp}{\ell_\mathcal{P}}
\newcommand{\elz}{\ell_z}
\newcommand{\hLz}{\hat{L}_z}
\newcommand{\hvL}{\hat{\vec{L}}}
\newcommand{\spins}{s}
\newcommand{\ver}{\vec{\mathrm{e}}_r}
\newcommand{\vet}{\vec{\mathrm{e}}_\vartheta}
\newcommand{\vep}{\vec{\mathrm{e}}_\phi}

\newcommand{\dn}{\delta_{0}} 
\newcommand{\dO}{\delta_{\Omega}}

\newcommand{\tanu}{\tilde{a}_\nu}
\newcommand{\tanuw}{\tilde{a}_{\nu w}}

\newcommand{\bjn}{{}^{(b)}\!j_0(\vec{r})} 

\newcommand{\bko}{{}^{(b)}k_0}
\newcommand{\pko}{{}^{(p)}k_0}
\newcommand{\pgko}{{}^{\{p\}}k_0}
\newcommand{\pkr}{{}^{\{p\}}k_r}
\newcommand{\pkt}{{}^{\{p\}}k_\vartheta}
\newcommand{\pkp}{{}^{\{p\}}k_\phi}
\newcommand{\bkp}{{}^{\{b\}}k_\phi}
\newcommand{\gpko}{{}^{\{p\}}k_0}
\newcommand{\bbko}{{}^{[b]}k_0}
\newcommand{\bkn}{{}^{(b)}\!k_0\left(\vec{r}\right)}
\newcommand{\akn}{{}^{(a)}\!k_0\left(\vec{r}\right)}
\newcommand{\ak}{\vec{k}_a}
\newcommand{\akr}{{}^{(a)}\!k_r}
\newcommand{\akt}{{}^{(a)}\!k_\vartheta}
\newcommand{\akp}{{}^{(a)}\!k_\phi}
\newcommand{\bkv}{\vec{k}_b(\vec{r})} 
\newcommand{\bkphi}{{}^{(b)\!}k_\phi}
\newcommand{\bk}{\vec{k}_b} 
\newcommand{\bgkn}{{}^{\{b\}}\!k_0} 

\newcommand{\bpp}{{}^{(b)}\varphi_{+}(\vec{r})}
\newcommand{\bpm}{{}^{(b)}\varphi_{-}(\vec{r})}
\newcommand{\bppm}{{}^{(b)}\varphi_{\pm}(\vec{r})}
\newcommand{\bppk}{{}^{(b)}\varphi_{+}^\dagger(\vec{r})}
\newcommand{\bpmk}{{}^{(b)}\varphi_{-}^\dagger(\vec{r})}

\newcommand{\tNGe}{{\tilde{N}_\textrm{G}^\textrm{(e)}}}
\newcommand{\tNGee}{{\tilde{N}_\textrm{G}^\textrm{[e]}}}
\newcommand{\tNPhi}{{\tilde{N}_{\Phi}}}
\newcommand{\tNNO}{\tilde{\mathbb{N}}_{\Omega}}
\newcommand{\tNNPhi}{\tilde{\mathbb{N}}_{\Phi}}
\newcommand{\tNNGe}{{\tilde{\mathbb{N}}_\textrm{G}^\textrm{e}}} 
\newcommand{\tNNGee}{{\tilde{\mathbb{N}}_\textrm{G}^\textrm{[e]}}}
\newcommand{\tNNGer}{{\tilde{\mathbb{N}}_\textrm{G}^\textrm{(e)}}} 
\newcommand{\tNNGeg}{{\tilde{\mathbb{N}}_\textrm{G}^\textrm{\{e\}}}} 
\newcommand{\tNNegan}{{\tilde{\mathbb{N}}^\textrm{\{e\}}_{\,\text{an}}}} 
\newcommand{\tNNegiii}{{\tilde{\mathbb{N}}^\textrm{\{e\}}_{\mathsf{III}}}}
\newcommand{\tND}{\tilde{N}_\textrm{D}}

\newcommand{\Mee}{M^\textrm{(e)}}
\newcommand{\tMe}{\tilde{M}^\textrm{(e)}}
\newcommand{\tMee}{\tilde{M}^\textrm{[e]}}
\newcommand{\tMMee}{\tilde{\mathbb{M}}^\textrm{[e]}}
\newcommand{\tMMeg}{\tilde{\mathbb{M}}^\textrm{\{e\}}}
\newcommand{\tMMegiii}{\tMMeg_\mathsf{III}}
\newcommand{\tMMegv}{\tMMeg_\mathsf{V}}

\newcommand{\MRpm}{\mathcal{R}_\pm}
\newcommand{\MSpm}{\mathcal{S}_\pm}
\newcommand{\bMR}{{}^{(b)}\!\mathcal{R}}
\newcommand{\bMRpm}{{}^{(b)}\!\mathcal{R}_\pm}
\newcommand{\bMRp}{{}^{(b)}\!\mathcal{R}_+}
\newcommand{\bMRm}{{}^{(b)}\!\mathcal{R}_-}
\newcommand{\bMRpS}{{}^{(b)}\!{\overset{*}{\mathcal{R}}{} }_+}
\newcommand{\bMRmS}{{}^{(b)}\!{\overset{*}{\mathcal{R}}{} }_-}
\newcommand{\bMS}{{}^{(b)}\!\mathcal{S}}
\newcommand{\bMSpm}{{}^{(b)}\!\mathcal{S}_\pm}
\newcommand{\bMSp}{{}^{(b)}\!\mathcal{S}_+}
\newcommand{\bMSm}{{}^{(b)}\!\mathcal{S}_-}
\newcommand{\bMSpS}{{}^{(b)}\!{\overset{*}{\mathcal{S}}{} }_+}
\newcommand{\bMSmS}{{}^{(b)}\!{\overset{*}{\mathcal{S}}{} }_-}

\newcommand{\aMRpm}{{}^{(a)}\!\mathcal{R}_\pm}
\newcommand{\aMRp}{{}^{(a)}\!\mathcal{R}_+}
\newcommand{\aMRm}{{}^{(a)}\!\mathcal{R}_-}
\newcommand{\aMRpS}{{}^{(a)}\!{\overset{*}{\mathcal{R}}{} }_+}
\newcommand{\aMRmS}{{}^{(a)}\!{\overset{*}{\mathcal{R}}{} }_-}
\newcommand{\aMSpm}{{}^{(a)}\!\mathcal{S}_\pm}
\newcommand{\aMSp}{{}^{(a)}\!\mathcal{S}_+}
\newcommand{\aMSm}{{}^{(a)}\!\mathcal{S}_-}
\newcommand{\aMSpS}{{}^{(a)}\!{\overset{*}{\mathcal{S}}{} }_+}
\newcommand{\aMSmS}{{}^{(a)}\!{\overset{*}{\mathcal{S}}{} }_-}

\newcommand{\pMR}{{}^{(p)}\!\R}
\newcommand{\pMRpm}{{}^{(p)}\!\R_\pm}
\newcommand{\pMRp}{{}^{(p)}\!\R_+}
\newcommand{\pMRm}{{}^{(p)}\!\R_-}
\newcommand{\pMS}{{}^{(p)}\!\mathcal{S}}
\newcommand{\pMSpm}{{}^{(p)}\!\mathcal{S}_\pm}
\newcommand{\pMSp}{{}^{(p)}\!\mathcal{S}_+}
\newcommand{\pMSm}{{}^{(p)}\!\mathcal{S}_-}
\newcommand{\pSRpm}{{}^{(p)}\!\tilde{R}_\pm}
\newcommand{\pSRp}{{}^{(p)}\!\tilde{R}_+}
\newcommand{\pSRm}{{}^{(p)}\!\tilde{R}_-}
\newcommand{\pSSpm}{{}^{(p)}\!\tilde{S}_\pm}
\newcommand{\pSSp}{{}^{(p)}\!\tilde{S}_+}
\newcommand{\pSSm}{{}^{(p)}\!\tilde{S}_-}
\newcommand{\pMRpS}{{}^{(p)}\!{\overset{*}{\mathcal{R}}{} }_+}
\newcommand{\pMRmS}{{}^{(p)}\!{\overset{*}{\mathcal{R}}{} }_-}
\newcommand{\pMSpS}{{}^{(p)}\!{\overset{*}{\mathcal{S}}{} }_+}
\newcommand{\pMSmS}{{}^{(p)}\!{\overset{*}{\mathcal{S}}{} }_-}
\newcommand{\ptRpm}{{}^{(p)}\!\tilde{R}_\pm}
\newcommand{\ptMRpm}{{}^{(p)}\!\tilde{\mathcal{R}}_\pm}
\newcommand{\ptRp}{{}^{(p)}\!\tilde{R}_+}
\newcommand{\ptMRp}{{}^{(p)}\!\tilde{\mathcal{R}}_+}
\newcommand{\ptRm}{{}^{(p)}\!\tilde{R}_-}
\newcommand{\ptMRm}{{}^{(p)}\!\tilde{\mathcal{R}}_-}
\newcommand{\ptSpm}{{}^{(p)}\!\tilde{S}_\pm}
\newcommand{\ptMSpm}{{}^{(p)}\!\tilde{\mathcal{S}}_\pm}
\newcommand{\ptSp}{{}^{(p)}\!\tilde{S}_+}
\newcommand{\ptMSp}{{}^{(p)}\!\tilde{\mathcal{S}}_+}
\newcommand{\ptSm}{{}^{(p)}\!\tilde{S}_-}
\newcommand{\ptMSm}{{}^{(p)}\!\tilde{\mathcal{S}}_-}
\newcommand{\hRpm}{\hat{R}_\pm}
\newcommand{\hSpm}{\hat{S}_\pm}
\newcommand{\hRp}{\hat{R}_+}
\newcommand{\hSp}{\hat{S}_+}
\newcommand{\hRm}{\hat{R}_-}
\newcommand{\hSm}{\hat{S}_-}
\newcommand{\hMTpm}{\hat{\mathcal{T}}_\pm}
\newcommand{\hMTp}{\hat{\mathcal{T}}_+}
\newcommand{\hMTm}{\hat{\mathcal{T}}_-}
\newcommand{\hMLz}{\hat{\mathcal{L}}_z}

\newcommand{\SPp}{\tilde{\Phi}_+}
\newcommand{\SPm}{\tilde{\Phi}_-}

\newcommand{\bpMR}{{}^{(b/p)}\!\R}
\newcommand{\bpMRpm}{{}^{(b/p)}\!\R_\pm}
\newcommand{\bpMS}{{}^{(b/p)}\!\mathcal{S}}
\newcommand{\bpMSpm}{{}^{(b/p)}\!\mathcal{S}_\pm}

\newcommand{\MCa}{\mathcal{C}_{(a)}}
\newcommand{\MCaS}{\overset{*}{\mathcal{C}}_{(a)}}

\newcommand{\Kb}{K_{\{b\}}} 
\newcommand{\Kpg}{K_{\{p\}}}
\newcommand{\Kpgiii}{K_{\{p\}}^\mathsf{III}}
\newcommand{\Aan}{A^{\text{an}}}
\newcommand{\peAo}{{}^{[p]}\!A_0} 
\newcommand{\peAw}{{}^{[p]}\!A_w} 
\newcommand{\bAe}{{}^{[b]}\!A_0}
\newcommand{\bAn}{{}^{(b)}\!A_0}
\newcommand{\pAn}{{}^{(p)}\!A_0}
\newcommand{\pgA}{{}^{\{p\}}\!A}
\newcommand{\pgAiii}{\pgA^{\textsf{III}}}
\newcommand{\aiii}{a_{\textsf{III}}}
\newcommand{\pgAn}{{}^{\{p\}}\!A_0}
\newcommand{\pgAan}{{}^{\{p\}}\!A^{\text{an}}}
\newcommand{\pAvn}{{}^{(p)}\!\vec{A}_0}
\newcommand{\peAn}{{}^{[p]}\!A_0}
\newcommand{\bgAe}{{}^{\{b\}}\!A_0} 
\newcommand{\bgAan}{{}^{\{b\}}\!A^{\text{an}}} 
\newcommand{\bgA}{{}^{\{b\}}\!A} 
\newcommand{\bgAiii}{\bgA^{\textsf{III}}} 
\newcommand{\bgAv}{\bgA^{\textsf{V}}} 
\newcommand{\bAmu}{{}^{(b)}\!A_\mu} 
\newcommand{\Acnuiii}{\mathcal{A}_\nu^\mathsf{III}} 
\newcommand{\Aciii}{\mathcal{A}^\mathsf{III}} 
\newcommand{\bgAcnuiii}{{}^{\{b\}}\mathcal{A}_\nu^\mathsf{III}} 
\newcommand{\bgAciii}{{}^{\{b\}}\mathcal{A}^\mathsf{III}} 
\newcommand{\MMe}{\mathbb{M}^\textrm{(e)}}
\newcommand{\MMeg}{\mathbb{M}^\textrm{\{e\}}} 
\newcommand{\tMMegan}{\tilde{\mathbb{M}}^\textrm{\{e\}}_{\ \text{an}}} 
\newcommand{\muegan}{\mu^\textrm{\{e\}}_{\ \text{an}}} 
\newcommand{\muegiii}{\mu^\textrm{\{e\}}_{\mathsf{III}}} 
\newcommand{\LLnu}{\mathbb{L}_\nu} 
\newcommand{\KKnu}{\mathbb{K}_\nu} 

\newcommand{\EE}{\mathbb{E}}
\newcommand{\eER}{E_\textrm{R}^\textrm{(e)}}
\newcommand{\ERe}{{E_\textrm{R}^\textrm{(e)}}}
\newcommand{\ERee}{{E_\textrm{R}^\textrm{[e]}}}
\newcommand{\eeER}{E_\textrm{R}^\textrm{[e]}} 
\newcommand{\egER}{E_\textrm{R}^\textrm{\{e\}}} 
\newcommand{\ET}{E_{\textrm{T}}}
\newcommand{\tETT}{\tilde{E}_{\textrm{[T]}}}
\newcommand{\tEePhi}{\tilde{E}_{\textrm{[$\Phi$]}}}
\newcommand{\tEEePhi}{\tilde{\mathbb{E}}_{\textrm{[$\Phi$]}}}
\newcommand{\EePhif}{E_{\textrm{[$\Phi$]}}^f}
\newcommand{\tEephif}{\tilde{E}_{\textrm{[$\phi$]}}^f}
\newcommand{\tEgT}{\tilde{E}_{\textrm{\{T\}}}}
\newcommand{\tEeT}{\tilde{E}_{\textrm{[T]}}}
\newcommand{\tEEeT}{\tilde{\mathbb{E}}_{\textrm{[T]}}}
\newcommand{\tEEgT}{\tilde{\mathbb{E}}_{\textrm{\{T\}}}}
\newcommand{\ETiv}{E^\textrm{(IV)}_{\textrm{[T]}}}
\newcommand{\EETiv}{\mathbb{E}^\textrm{(IV)}_{\textrm{[T]}}}
\newcommand{\EgTiv}{\mathbb{E}^\textrm{(IV)}_{\textrm{\{T\}}}}
\newcommand{\EET}{\mathbb{E}_\textrm{[T]}} 
\newcommand{\EETT}{\mathbb{E}_\textrm{[T]}} 
\newcommand{\EEivbn}{\mathbb{E}^\textrm{(IV)}(\beta,\nu)}
\newcommand{\EEgiv}{\mathbb{E}^{\{\textsf{IV}\}}} 
\newcommand{\tEEgiv}{\tilde{\mathbb{E}}^{\{\textsf{IV}\}}} 
\newcommand{\EEgivbn}{\mathbb{E}^{\{\textsf{IV}\}}(\beta,\nu)} 
\newcommand{\Egivbn}{E^{\{\textsf{IV}\}}(\beta,\nu)} 
\newcommand{\EEeiv}{\mathbb{E}^{[\textsf{IV}]}} 
\newcommand{\EEeivbn}{\mathbb{E}^{[\textsf{IV}]}(\beta,\nu)} 
\newcommand{\EEO}{\mathbb{E}_\Omega} 
\newcommand{\tEEO}{\tilde{\mathbb{E}}_{[\Omega]}}
\newcommand{\EEOe}{\mathbb{E}_{[\Omega]}} 
\newcommand{\GtEEO}{{}^{(G)}\!\tEEO} 
\newcommand{\GetEO}{{}^{[G]}\!\tEO} 
\newcommand{\DtEEO}{{}^{(D)}\!\tEEO} 
\newcommand{\etEEePhi}{{}^{(e)}\!\tilde{\mathbb{E}}_{[\Phi]}}
\newcommand{\etEEPhi}{{}^{(e)}\!\tilde{\mathbb{E}}_{\Phi}}
\newcommand{\DtEEPhi}{{}^{(D)}\!\tilde{\mathbb{E}}_{\Phi}}
\newcommand{\DtEEePhi}{{}^{(D)}\!\tilde{\mathbb{E}}_{[\Phi]}}
\newcommand{\GtEEePhi}{{}^{(G)}\!\tilde{\mathbb{E}}_{[\Phi]}}
\newcommand{\DEEO}{{}^{(D)}\!\EEO} 
\newcommand{\DEEOe}{{}^{(D)}\!\EEOe} 
\newcommand{\etEEO}{{}^{(e)}\!\tEEO} 
\newcommand{\eetEEO}{{}^{[e]}\!\tEEO} 
\newcommand{\antEEO}{{}^{\{\text{an}\}}\!\tEEO} 
\newcommand{\anrtEEO}{{}^{(\text{an})}\!\tEEO} 
\newcommand{\eetEO}{{}^{[e]}\!\tEO} 
\newcommand{\Egiv}{E_\textrm{\{T\}}^{\textrm{(IV)}}} 
\newcommand{\EEP}{\mathbb{E}_\mathcal{P}} 

\newcommand{\Ekin}{{E_\textrm{kin}}}
\newcommand{\EKIN}{{E_\textrm{KIN}}} 
\newcommand{\ekin}{ {\varepsilon}_{\textrm{kin}} } 
\newcommand{\eKIN}{ {\varepsilon}_{\textrm{KIN}} } 
\newcommand{\pEkin}{{}^{(p)}\!\Ekin}
\newcommand{\rEkin}{{}^{(r)}\!\Ekin} 
\newcommand{\rEKIN}{{}^{(r)}\!\EKIN} 
\newcommand{\rekin}{{}^{(r)}\!\ekin} 
\newcommand{\reKIN}{{}^{(r)}\!\eKIN} 
\newcommand{\thEKIN}{{}^{(\vartheta)}\!\EKIN} 
\newcommand{\thekin}{{}^{(\vartheta)}\!\ekin} 
\newcommand{\theKIN}{{}^{(\vartheta)}\!\eKIN} 

\newcommand{\epot}{ {\varepsilon}_{\textrm{pot}} }
\newcommand{\ePOT}{ {\varepsilon}_{\textrm{POT}} }

\newcommand{\etot}{ {\varepsilon}_{\textrm{tot}} }
\newcommand{\eegtot}{ {\varepsilon}^{\{e\}}_{\textrm{tot}} }
\newcommand{\eagtot}{ {\varepsilon}^{\{a\}}_{\textrm{tot}} }

\newcommand{\tEEOo}{\tilde{\mathbb{E}}_\Omega} 
\newcommand{\EReg}{{E_\textrm{R}^\textrm{\{e\}}}} 
\newcommand{\Eegan}{{E^\textrm{\{e\}}_\text{an}}} 
\newcommand{\Eeg}{{E^\textrm{\{e\}}}}
\newcommand{\Eegiii}{{E_\mathsf{III}^\textrm{\{e\}}}}
\newcommand{\EEegiii}{{\mathbb{E}_\mathsf{III}^\textrm{\{e\}}}}
\newcommand{\Ew}{{E_\textrm{w}}}
\newcommand{\Ewee}{{\Ew^\textrm{[e]}}} 
\newcommand{\cEk}{{\cal{E}_\textrm{kin}}}

\newcommand{\tPhi}{\tilde{\Phi}}
\newcommand{\tPhipm}{\tilde{\Phi}_\pm}
\newcommand{\tPhip}{\tilde{\Phi}_+}
\newcommand{\tPhim}{\tilde{\Phi}_-}
\newcommand{\tPhib}{\tPhi_b} 
\newcommand{\tPhinuw}{\tPhi_{\nu w}} 

\newcommand{\tO}{\tilde{\Omega}}
\newcommand{\tOpm}{\tilde{\Omega}_\pm}
\newcommand{\tOp}{\tilde{\Omega}_+}
\newcommand{\tOm}{\tilde{\Omega}_-}

\newcommand{\lO}{\ell_\mathcal{O}}
\newcommand{\dlO}{\dot{\ell}_\mathcal{O}}
\newcommand{\ddlO}{\ddot{\ell}_\mathcal{O}}

\newcommand{\Sag}{S^{\{a\}}}
\newcommand{\SOag}{\Sag_\mathcal{O}}

\newcommand{\lGe}{ {\lambda_\textrm{G}^{(\textrm{e})}}\!}

\newcommand{\Du}{\mathcal{D}_\mu}

\newcommand{\Tmunu}{T_{\mu\nu}}
\newcommand{\Too}{T_{00}}
\newcommand{\TTmunu}{{}^{(T)}\!\Tmunu}
\newcommand{\DTmunu}{{}^{(D)}\!\Tmunu}
\newcommand{\GTmunu}{{}^{(G)}\!\Tmunu}
\newcommand{\TToo}{{}^{(T)}\!\Too}
\newcommand{\DToo}{{}^{(D)}\!\Too}
\newcommand{\GToo}{{}^{(G)}\!\Too}

\newcommand{\ap}{{}^{(a)}\!\varphi_{\pm}(\vec{r})}
\newcommand{\app}{{}^{(a)}\!\varphi_{+}(\vec{r})}
\newcommand{\apm}{{}^{(a)}\!\varphi_{-}(\vec{r})}
\newcommand{\appm}{{}^{(a)}\!\varphi_{\pm}(\vec{r})}
\newcommand{\appmd}{{}^{(a)}\!\varphi_{\pm}^{\dagger}(\vec{r})}
\newcommand{\apmd}{{}^{(a)}\!\varphi_{-}^{\dagger}(\vec{r})}
\newcommand{\appd}{{}^{(a)}\!\varphi_{+}^{\dagger}(\vec{r})}
\newcommand{\bpx}{{}^{(b/p)}\!}
\newcommand{\bpxpp}{{}^{(b/p)}\!\varphi_{+}(\vec{r})}
\newcommand{\bpxpm}{{}^{(b/p)}\!\varphi_{-}(\vec{r})}
\newcommand{\pxppm}{{}^{(p)}\!\varphi_{\pm}(\vec{r})}
\newcommand{\pxpp}{{}^{(p)}\!\varphi_{+}(\vec{r})}
\newcommand{\pxpm}{{}^{(p)}\!\varphi_{-}(\vec{r})}
\newcommand{\pxppmd}{{}^{(p)}\!\varphi_{\pm}^{\dagger}(\vec{r})}
\newcommand{\pxpmd}{{}^{(p)}\!\varphi_{-}^{\dagger}(\vec{r})}
\newcommand{\pxppd}{{}^{(p)}\!\varphi_{+}^{\dagger}(\vec{r})}

\newcommand{\PsiPO}{\Psi_{\mathcal{P},\mathcal{O}}}
\newcommand{\PsiP}{\Psi_\mathcal{P}}
\newcommand{\PsiO}{\Psi_\mathcal{O}}

\newcommand{\zetaejm}{\zeta_{\ e}^{j,\;m}}
\newcommand{\zetapp}{\zeta^{\frac{1}{2},\frac{1}{2}}}
\newcommand{\zetapm}{\zeta^{\frac{1}{2},-\frac{1}{2}}}
\newcommand{\zetappm}{\zeta^{\frac{1}{2},\pm\frac{1}{2}}}
\newcommand{\xip}{\xi^{(+)}}
\newcommand{\xim}{\xi^{(-)}}
\newcommand{\etap}{\eta^{(+)}}
\newcommand{\etam}{\eta^{(-)}}
\newcommand{\omp}{\omega^{(+)}}
\newcommand{\omm}{\omega^{(-)}}

\newcommand{\Jhz}{\hat{\J}_z}
\newcommand{\Jhzpm}{\hat{\J}_z^{(\pm)}}
\newcommand{\Jhzp}{\hat{\J}_z^{(+)}}
\newcommand{\Jhzm}{\hat{\J}_z^{(-)}}
\newcommand{\Jlhz}{\hat{J}_z}
\newcommand{\Jlhzpm}{\hat{J}_z^{(\pm)}}
\newcommand{\Jlhzp}{\hat{J}_z^{(+)}}
\newcommand{\Jlhzm}{\hat{J}_z^{(-)}}
\newcommand{\Lhz}{\hat{\mathcal{L}}_z}
\newcommand{\Lhzp}{\hat{\mathcal{L}}_z^{(+)}}
\newcommand{\Lhzm}{\hat{\mathcal{L}}_z^{(-)}}
\newcommand{\Llhz}{\hat{L}_z}
\newcommand{\Llhzp}{\hat{L}_z^{(+)}}
\newcommand{\Llhzm}{\hat{L}_z^{(-)}}
\newcommand{\Shz}{\hat{\mathcal{S}}_z}

\newcommand{\ajz}{{}^{(a)}\!j_z}
\newcommand{\pjz}{{}^{(p)}\!j_z}
\newcommand{\bjz}{{}^{(b)}\!j_z}

\newcommand{\ptTkin}{{}^{(p)}\!\tilde{T}_\text{kin}}
\newcommand{\pTkin}{{}^{(p)}\!T_\text{kin}}
\newcommand{\ptTkinePhi}{{}^{(p)}\!\tilde{T}_{\text{kin}\,[\Phi]}}
\newcommand{\ptTr}{{}^{(p)}\!\tilde{T}_r}
\newcommand{\pTr}{{}^{(p)}\!T_r}
\newcommand{\ptTth}{{}^{(p)}\!\tilde{T}_\vartheta}
\newcommand{\pTth}{{}^{(p)}\!T_\vartheta}
\newcommand{\ptTphi}{{}^{(p)}\!\tilde{T}_\phi}
\newcommand{\pTphi}{{}^{(p)}\!T_\phi}

\newcommand{\Zp}{\mathcal{Z}_\mathcal{P}}
\newcommand{\tZp}{\tilde{\mathcal{Z}}_\mathcal{P}}

\newcommand{\tNf}{\tilde{N}_f}

\newcommand{\hQ}{\hat{Q}}
\newcommand{\hQQ}{\hat{\mathpzc{Q}}}

\newcommand{\pr}{{}^{(p)}\!}
\newcommand{\lpr}{{}^{(\elp)}\!}
\newcommand{\threer}{{}^{(3)}\!}

\newcommand{\Pnuiii}{\mathcal{P}_\nu^\mathsf{III}}

\newcommand{\bbar}{{\mathchoice
{{\vcenter{\offinterlineskip\vskip.1ex\hbox{$\,\tilde{}$}\vskip-1.85ex\hbox{$b$}\vskip.4ex}}}
{{\vcenter{\offinterlineskip\vskip.1ex\hbox{$\,\tilde{}$}\vskip-1.75ex\hbox{$b$}\vskip.4ex}}}
{{\vcenter{\offinterlineskip\vskip.1ex\hbox{$\scriptstyle\,\tilde{}$}\vskip-1.2ex\hbox{$\scriptstyle b$}\vskip.2ex}}}
{{\vcenter{\offinterlineskip\vskip.1ex\hbox{$\scriptscriptstyle\,\tilde{}$}\vskip-.9ex\hbox{$\scriptscriptstyle b$}\vskip.4ex}}}
}}

\title{\bf Quadrupole Approximation for Para-Positronium\\ in\\ Relativistic Schr\"odinger Theory}
\author{M.\ Mattes and M.\ Sorg} 
\date{ }
\maketitle
\begin{abstract}
The non-relativistic energy levels of para-positronium are calculated in the \\quadrupole
approximation of the interaction potential. This approximation technique takes into
account the anisotropy of the electrostatic electron-positron interaction in the lowest
order. The states due to different values of the quantum number~$(l_z)$ of angular
momentum are found to be no longer degenerate as is the case in the conventional
theory. The physical origin of this elimination of the conventional degeneracy may
intuitively be attributed to the state-dependent inertial \emph{broadening} of the rotating charge
clouds; the corresponding \emph{anisotropic} deformation (in the quadrupole approximation)
lowers then the negative electrostatic interaction energy. The result of this influence of
anisotropy is that the states with~$l_z=0$ adopt smaller binding energy whereas the
states with maximal value of~$|l_z|$ (for fixed principal quantum number~$n$) have the
largest binding energy within the angular momentum multiplet~$(-|l_{z,\mathrm{max}}| \le
l_z \le |l_{z,\mathrm{max}}|)$. This yields a certain kind of electric fine-structure
splitting with the splitted RST levels being placed in a relatively narrow band around the
(highly degenerated) conventional levels.
 \vspace{2.5cm}
 \noindent


\end{abstract}


\begin{center}
  {\Large\textbf{Contents}}\\[3em]
\end{center}

\begin{itemize}
\item[\textbf{I}] {\large \textbf{Introduction and Survey of Results \dotfill \textbf{6} }}
 \begin{itemize}
   \item[] {\large \emph{Fermionic and Bosonic States \dotfill  10 }}
   \item[] {\large \emph{Angular Momentum Quantization in RST \dotfill 11}}
   \item[] {\large \emph{Quadrupole Approximation \dotfill 12 }}
	\item[] {\large \emph{Para-Positronium Spectrum \dotfill 13 }}
    \end{itemize}
\item[\textbf{II}] {\large \textbf{Positronium Eigenvalue Problem \dotfill \textbf{15} }}
  \begin{itemize}
  \item[\textbf{1.}] \emph{\textbf{Relativistic Schr\"{o}dinger Equations} \dotfill 15 }
  \item[\textbf{2.}] \emph{\textbf{Dirac Equations} \dotfill 16 }
  \item[\textbf{3.}] \emph{\textbf{Maxwell Equations} \dotfill 18 }
  \item[\textbf{4.}] \emph{\textbf{Conservation Laws} \dotfill 20}
  \item[\textbf{5.}] \emph{\textbf{Stationary Field Configurations} \dotfill 23 }
    \begin{itemize}
    \item[] {\large \emph{Gauge Field Subsystem \dotfill 23}}
    \item[] {\large \emph{Matter Subsystem \dotfill 25}}
    \end{itemize}
  \end{itemize}
\item[\textbf{III}] {\large \textbf{Ortho/Para Dichotomy \dotfill \textbf{28} }}
  \begin{itemize}
  \item[\textbf{1.}] \emph{\textbf{Mass Eigenvalue Equations} \dotfill 29 }
  \item[\textbf{2.}] \emph{\textbf{Poisson Equations} \dotfill 33 }
  \item[\textbf{3.}] \emph{\textbf{Non-Unique Spinor Fields} \dotfill 34 }
    \begin{itemize}
    \item[] {\large \emph{Fermionic States in RST \dotfill 36 }}
    \item[] {\large \emph{Bosonic States in RST \dotfill 38}}
    \item[] {\large \emph{Uniqueness of the Physical Densities \dotfill 42 }}
    \end{itemize}
  \end{itemize}
\newpage
\item[\textbf{IV}] {\large \textbf{Para-Positronium \dotfill \textbf{45} }}
  \begin{itemize}
  \item[\textbf{1.}] \emph{\textbf{Mass Eigenvalue Equations in Terms of Amplitude Fields}
      \dotfill 45 }
  \item[\textbf{2.}] \emph{\textbf{Angular Momentum Quantization in RST} \dotfill 50 }
    \begin{itemize}
    \item[] {\large \emph{Figure IV.A: Azimuthal Current Density $\pkp(\vartheta)$
          \dotfill 59 }}
    \end{itemize}
  \item[\textbf{3.}] \emph{\textbf{Energy Functional and Principle of Minimal Energy}
      \dotfill 63 }
    \begin{itemize}
    \item[] {\large \emph{Relativistic Mass Renormalization \dotfill 64 }}
    \item[] {\large \emph{Kinetic Energy \dotfill 66 }}
    \item[] {\large \emph{Electric Interaction Energy \dotfill 69 }}
    \item[] {\large \emph{Relativistic Principle of Minimal Energy \dotfill 70 }}
    \item[] {\large \emph{Non-Relativistic Energy Functional \dotfill 74 }}
    \item[] {\large \emph{Spherically Symmetric Eigenvalue Equations \dotfill 77 }}
    \item[] {\large \emph{Anisotropic Interaction Pontential \dotfill 79 }}
    \item[] {\large \emph{Energy of the Anisotropic Gauge Fields \dotfill 83 }}
    \item[] {\large \emph{Non-Relativistic Gauge Field Equations \dotfill 87 }}
    \item[] {\large \emph{Generalized Eigenvalue Equation for $\tPhi(r)$ \dotfill 89 }}
    \item[] {\large \emph{Selecting a Plausible Trial Amplitude $\tPhi(r)$ \dotfill 90 }}
    \item[] {\large \emph{Partial Extremalization \dotfill 93 }}
    \end{itemize}
  \item[\textbf{4.}] \emph{\textbf{Para-Spectrum} \dotfill 95 }
    \begin{itemize}
    \item[] {\large \emph{RST Elimination of the $\elz$-Degeneracy \dotfill 95 }}
    \item[] {\large \emph{Fig.IV.B: Elimination of the Degeneracy for~$\lP=3$ \dotfill 97 }}
    \item[] {\large \emph{Spectrum due to $\elz = \pm \elp$ \dotfill 98}}
    \item[] {\large \emph{Fig.IV.C: Lowest-Lying RST Levels \dotfill 102}}
    \item[] {\large \emph{Magnitude of Fine-Structure Splitting \dotfill 103 }}
    \item[] {\large \emph{Fig.IV.D: RST Levels~$\EEP^{( n)}\Big|_{l_z=\lP-1}$ \dotfill 107 }}
    \item[] {\large \emph{Fig.IV.E: Band Structure of the RST Levels \dotfill 109 }}
    \end{itemize}
  \end{itemize}
\item[] {\large \textbf{Appendix A: \quad Quadrupole Energy $\Eegiii$ \dotfill \textbf{110}}}
  \begin{itemize}
  \item[]
  \begin{itemize}
  \item[] {\large \emph{Exact Quadrupole Solution \dotfill 111 }}
  \item[] {\large \emph{Quadrupole Energy \dotfill 114 }}\vspace{5mm}
  \end{itemize}
  \end{itemize}
\item[] {\large \textbf{Appendix B: \quad Special Case~$l_z=0,\forall \lP$\dotfill \textbf{117}}}
  \begin{itemize}
  \item[] {\large \emph{Fig.B.I: RST Levels due to~$l_z=0,\,\forall \lP $ \dotfill 121 }}
  \end{itemize}\vspace{5mm}
\item[] {\large \textbf{Appendix C: \quad Universality of the Quadrupole Ratios\\*
      \phantom{Appendix C}\quad\quad\boldmath $\frac{\rklo{p}{f}_3}{\rklo{p}{e}_3}$
      and~$\frac{\rklo{p}{m}_3}{\rklo{p}{e}_3}$ \dotfill \textbf{122} }}
\item[] {\large \textbf{Appendix D: \quad Angular-Momentum Algebra \\*
      \phantom{Appendix D: }\quad for Para-Positronium \dotfill \textbf{124}}}
\begin{itemize}
\item[]
\begin{itemize}
  \item[] {\large \emph{Eigenvalue Problem for Angular Momentum \dotfill 124 }}
  \item[] {\large \emph{Second-Order Form of the Eigenvalue Problem \dotfill 127 }}
  \item[] {\large \emph{Ladder Operators \dotfill 129}}
  \item[] {\large \emph{Ladder Formalism for $\lP=3 $\dotfill 136 }}
  \item[] {\large \emph{Fig.D.I: Ladder Operation for
        $\rkloi{\lP}{f}{R,l_z}(\vartheta,\phi)$\dotfill 137 }}
  \item[] {\large \emph{Fig.D.II: Ladder Operation for
        $\rkloi{\lP}{f}{S,l_z}(\vartheta,\phi)$\dotfill 142 }}
  \item[] {\large \emph{Determination of the ladder coefficients \dotfill 143 }}
  \item[] {\large \emph{Compact Representation \dotfill 148 }}
\end{itemize}
\end{itemize}
\vspace{5mm}
\item[] {\large \textbf{References \dotfill \textbf{151} }}
\end{itemize}

\section{Introduction and Survey of Results}
\indent

The present paper is intended to increase the accuracy of the RST predictions concerning
the (non-relativistic) energy levels of para-positronium. In the lowest order of
approximation (\emph{spherically symmetric approximation}), the corresponding RST
predictions deviated from their conventional counterparts by some 5--10\% for principal
quantum numbers up to $n \simeq 100$ [1--4]. This provided us with sufficient motivation
in order to consider now also the next higher order of approximation where the anisotropy
of the interaction between positron and electron is taken into account, albeit only in the
\emph{quadrupole approximation}. But amazingly enough, this low degree of approximation is
already sufficient in order to eliminate the energetic degeneracy which is present in the
spherically symmetric approximation (and in the conventional theory).

The latter kind of RST approximation is due to the assumption that the interaction
potential between the positronium constituents (i.\,e. electron and positron) is
spherically symmetric, i.\,e. it depends exclusively on the radial variable $r$ of the
spherical polar coordinates $(r, \vartheta, \phi)$. Such a spherical symmetry of the
interaction potential is adopted also in the conventional theory, namely in form of the
Coulomb potential entering the conventional Hamiltonian $\hat{H}$ in the usual way:
\begin{equation}
\label{eq:I.1}
\hat{H} = \frac{1}{2\,M}\,\left[ \hat{p}_1^2 + \hat{p}_2^2 \right] - \frac{e^2}{\left|\left| \vr_1 - \vr_2 \right|\right|} \;.
\end{equation}
The corresponding energy eigenvalue problem can be solved exactly [5], and the conventional eigenvalues $E_C^{(n)}$ $(n=1,2,3,\ldots)$ turn out as
\begin{gather}
\label{eq:I.2}
E_C^{(n)} = -\frac{e^2}{4\,a_B} \cdot \frac{1}{n^2} = -\frac{6{,}8029\ldots}{n^2}\;\text{[eV]} \\
\left( a_B = \frac{\hbar^2}{Me^2}\,\ldots\,\text{Bohr radius} \right) \nonumber \;.
\end{gather}
Since the conventional quantum number of angular momentum $\ell$ ranges here from zero to $n-1$, and its $z$-component $\elz$ from $-\ell$ up to $+\ell$ $(-\ell \leq \elz \leq +\ell)$, the conventional degree of degeneracy is thus given by
\begin{equation}
\label{eq:I.3}
\sum_{\ell=0}^{n-1} \left( 2\,\ell + 1 \right) = n^2 \;.
\end{equation}
This result is a mathematical consequence of the fact that the original three-dimensional
eigenvalue problem can be separated by a product ansatz into two eigenvalue problems of
their own, namely into the eigenvalue problem for the angular momentum operator and into a
residual one-dimensional problem (in radial direction) for the energy.

Such a separation of the original three-dimensional eigenvalue problem into a
two-dimensional angular and a one-dimensional radial problem is also possible in RST, but
only if one resorts to the \emph{spherically symmetric approximation} for the interaction
potential. In this case, the angular momentum problem can also be solved exactly quite
similarly as in the conventional theory (see \textbf{Subsect.IV.2} ``\emph{Angular
  Momentum Quantization in RST}''), but the residual one-dimensional eigenvalue problem
(in radial direction) for determining the energy eigenvalues is in RST much more
complicated than in the conventional theory. And furthermore the conventional
$\elz$-degeneracy is eliminated in RST. The reason is here that the interaction potential
(even in the spherically symmetric approximation) depends on the considered quantum state
and thus is not fixed once and forever as is the Coulomb potential in the conventional
Hamiltonian $\hat{H}$ (\ref{eq:I.1}). Indeed, the RST interaction potential must be
determined from a Poisson equation which has the charge density of the considered quantum
state as its source, even in the spherically symmetric approximation (see
``\emph{Spherically Symmetric Eigenvalue Equations}`` in \textbf{Subsect.IV.3}).

The physical consequence of this state-dependence of the interaction potential is now that
the states with different values of angular momentum $\elz$ are subjected to different
interaction potentials and this entails of course the elimination of the otherwise present
$\elz$-degeneracy. Such a result seems not to be supported by the conventional theory [6];
but one must not forget the fact that the present elimination of the $\elz$-degeneracy in
RST occurs in the \emph{quadrupole} approximation; and therefore it may well be the case
that the higher orders of approximation do weaken that effect of degeneracy elimination so
that it might escape the experimental verification by means of the present-day techniques.

Indeed, there is some plausible reason why such an elimination should actually
occur. Namely, the states due to different values of $\elz$ may intuitively be associated
with different magnitudes of rotation (around the $z$-axis) which then would entail the
emergence of different strengths of the centrifugal forces. As a consequene, the original
spherically symmetric charge distributions would be flattened (or rolled out) in a
different way by these different centrifugal forces, cf. \textbf{Fig.s
  IV.A.0--IV.A.3}. But such deformed electric charge distributions should then generate
(via the Poisson equations) an electric interaction potential with a certain deviation
from the purely spherical shape; i.\,e. the electric interaction energy (and therefore
also the binding energy) should vary with the values of $\elz$ ($\leadsto$ elimination of
the $\elz$-degeneracy). The fact, that this plausible effect is seemingly not observed
experimentally, may be understood as a further hint at the ``unreasonable'' logic of the
quantum world.

Subsequently, these results are elaborated through the following arrangement:

In \textbf{Sect.II} the fundamentals of RST are briefly put forward, with an emphasis on
the two-particle Dirac equation for the \emph{fluid-dynamic quantum matter} and on the
(non-Abelian) Maxwell equations as the field equations for the matter interactions,
cf. equations (\ref{eq:II.13}) and (\ref{eq:II.15}). This coupled system of Dirac and
Maxwell equations is of such a logical structure that the standard conservation laws of
charge $(j_\mu)$ and energy-momentum ($\Tmunu$) do emerge as an immediate and natural
consequence, see equations (\ref{eq:II.23}) and (\ref{eq:II.32}) below. The important
point with such a logical structure is that a plausible definition of the total energy
$\ET$, due to an RST field configuration, suggests itself: namely as the spatial integral
of the time component $\Too$ of the energy-momentum density $\Tmunu$ over whole
three-space, cf. equation (\ref{eq:II.38}) below. Such a definition of energy seems
adequate for a \emph{fluid-dynamic} theory, whereas in the conventional quantum theory (as
a \emph{probabistic point-particle} theory) the concept of energy is based rather upon the
eigenvalues of some energy operator ($\hat{H}$).

In any case, a definite energy can be expected only for a stationary system, such as
positronium which is the object to be studied in the present paper. Here it is assumed
that the corresponding gauge fields, mediating the interactions between the electron and
the positron, are time-independent and the matter fields appear as the product of the
usual exponential time factor ($\sim \e^{-iE\frac{t}{\hbar}}$) and a time-independent
pre-factor $\psi(\vr)$, cf. (\ref{eq:II.50}). As usual, these assumptions convert the
general RST field equations to the Poisson equations for the gauge potentials,
cf. (\ref{eq:II.43})--(\ref{eq:II.44}), and to the ``mass eigenvalue equations'' for the
matter fields which we write down for the present purpose in Pauli form,
cf. (\ref{eq:II.52a})--(\ref{eq:II.52b}), rather than in Dirac form. This coupled set of
Pauli and gauge field equations is then our point of departure for working out the
non-relativistic energy spectrum of \emph{para-positronium}.

Indeed, the RST treatment of positronium (i.\,e. the bound system of electron and positron) leads one to a certain observation which one encounters also in the conventional quantum theory: namely, this specific two-particle system occurs in two forms, i.\,e. ortho- and para-positronium, which we hereafter briefly term \emph{ortho/para dichotomy} (\textbf{Sect.III}). However, this phenomenon of dichotomy is characterized in the conventional theory and in RST by means of quite different concepts, although both approaches do refer here to the particle spin. More concretely, the conventional theory resorts to the addition theorem of angular momenta for the spins of electron ($s_e$) and positron ($s_p$) which is realized by taking the \emph{tensor product} of the single-particle spin spaces, such that the total spin $S = s_e \pm s_p$ is zero ($S=0$) for para-positronium and $S=1$ for ortho-positronium ($s_e = s_p = \frac{1}{2}$). In contrast to this, RST is based upon the \emph{Whitney sum} of single-particle vector (or spinor) bundles; and this means that each single particle does occupy a definite one-particle wave function which then carries \emph{individually} the angular momentum of the \emph{composite} system! Consequently, each positronium constituent (i.\,e. electron and positron) is described by a one-particle wave function due to either vanishing spin ($s_\mathcal{P} = s_e - s_p = 0$ $\leadsto$ ``\emph{para-positronium}'') or due to spin $s_\mathcal{O} = 1\, (= s_e + s_p)$, i.\,e. ''\emph{ortho-positronium}''. In this way, both manifestations of positronium appear to be equipped with integer spin (i.\,e. $s_\mathcal{P} = 0, s_\mathcal{O} = 1$) and thus are to be considered \emph{bosons}.

On the other hand, any single Dirac particle must be considered a \emph{fermion}; and this entails of course the necessity to provide now in RST a clear definition of the bosonic and fermionic states for few-particle systems being composed of Dirac particles.

\begin{center}
  \large{\textit{Fermionic and Bosonic States in RST}}
\end{center}

The resolution of this composition problem works as follows: the fermionic or bosonic character of a one-particle state is defined by the angular momentum eigenvalue of the selected spinor basis. Consequently, a fermionic basis (such as (\ref{eq:III.30a})--(\ref{eq:III.30d})) carries a \emph{half-integer} value of angular momentum and a bosonic basis (such as (\ref{eq:III.35}) or (\ref{eq:III.36a})--(\ref{eq:III.36b})) carries an integer-valued angular momentum. Any Dirac spinor $\psi$ must now be decomposed with respect to such a spinor basis and thereby inherits its fermionic or bosonic character from just that selected spinor basis. By such a spinor decomposition there do arise as usual the corresponding spinor components (``\emph{amplitude fields}'') which are required to be always single-valued fields and therefore do always carry integer angular momentum, cf. (\ref{eq:III.43a})--(\ref{eq:III.43b}), whereas the spinor basis will in general be multi-valued, cf. (\ref{eq:III.50a})--(\ref{eq:III.50b}).

This multi-valuedness of the RST spinor fields does, however, not induce phathological elements in the logical structure because the observable objects of the theory (such as charge and energy-momentum densities) are always \emph{unique} compositions of the non-unique spinor fields. For instance, the uniqueness of the charge density $\akn$ as a bilinear construction of the spinor fields $\psi_a(\vr)$ $(a=1,2)$ ensures the uniqueness of the corresponding gauge potentials which feel the charge densities as their sources, see the Poisson equations (\ref{eq:III.21})--(\ref{eq:III.22}). In this way, one can arrive at a unique eigenvalue problem for the determination of the positronium binding energy, although the underlying RST spinor fields are multi-valued, see the eigenvalue system (\ref{eq:IV.2a})--(\ref{eq:IV.2d}) below.

By means of all these rather technical preparations, one can now in \textbf{Sect.IV}
tackle the proper problem of the positronium energy spectrum, where we are satisfied for
the moment with the restriction to the para-case ($\leadsto s_p = 0$). This problem is of
course not exactly solvable so that we have to resort to various approximation
assumptions. The first one of these refers to the \emph{non-relativistic} situation; the
next one concerns the \emph{electrostatic} approximation which neglects the magnetic
effects altogether. Finally we take the \emph{spherically symmetric approximation} as our
point of departure and furthermore treat the anisotropy effect only in the lowest order of
approximation, i.\,e. the \emph{quadrupole approximation}. Here, the assumption of
``spherical symmetry'' refers exclusively to the electrostatic gauge potential
$\pgAn(\vr)$ (putting $\pgAn(\vr) \Rightarrow \peAn(r)$; $r = \left|\left| \vr
  \right|\right|$), not to the matter fields which are admitted to be anisotropic. Indeed,
it is just this approximation assumption of spherical symmetry $\pgAn(\vr) \Rightarrow
\peAn(r)$ which allows us to separate exactly the original mass eigenvalue equation into
an angular and a radial problem, where as usual the radial problem ultimately provides the
wanted energy levels; but the angular problem is independent of the special interaction
potential and thus may be solved exactly for all spherically symmetric interaction
potentials.

\begin{center}
  \large{\textit{Angular Momentum Quantization in RST}}
\end{center}

The angular problem is defined by the two coupled equations
(\ref{eq:IV.8a})--(\ref{eq:IV.8b}) below; and its solution is just what one may term
``\emph{angular momentum quantization in RST}'' because it corresponds to the analogous
phenomenon in the conventional quantum theory: there are two quantum numbers of angular
momentum (termed $\elp$ and $\elz$) where $\elp$ takes its values in the set of natural
numbers ($\elp = 0,1,2,3,\ldots$) and $\elz$ is an integer whose range is $-\elp \leq \elz
\leq +\elp$. It is only the quantum number $\elp$ which explicitly enters the radial
eigenvalue problem, cf. equations (\ref{eq:IV.9a})--(\ref{eq:IV.9b}) below, so that the
second quantum number $\elz$ cannot \emph{immediately} influence the energy spectrum! This
entails the effect of degeneracy of the energy levels with respect to $\elz$ which occurs
also \emph{in the conventional theory} (where the spherically symmetric Coulomb potential
is adopted as the interaction potential between both positronium constituents). The
various RST angular distributions of the current density, due to the possible values of
$\elz$, are displayed for $\boldsymbol{\elp = 3}$ in the \textbf{Fig.s IV.A} on p.s
\pageref{figIV.A.0}-\pageref{figIV.A.3}. A preliminary first analysis of this RST algebra
of angular momentum is presented in \textbf{App.D}.

\begin{center}
  \large{\textit{Quadrupole Approximation}}
\end{center}

This kind of~$l_z$-degeneracy would be also present in RST, but \emph{only} if one is satisfied
with the spherically symmetric approximation of the interaction potential! The
corresponding result for the energy spectrum has been studied in a precedent paper [4],
where it has been found that those ``spherically symmetric'' RST predictions undergo
certain deviations from the conventional spectrum (\ref{eq:I.2}) ranging up to some 10\%
for principal quantum numbers $n$ in the range $1 \leq n \leq 100$. This appears to us as
an acceptable result in view of the very rough approximation techniques being applied up
to now within the RST framework. Especially, one becomes now curious whether perhaps the
RST predictions for the positronium spectrum will come even closer to the corresponding
conventional predictions (\ref{eq:I.2} ) if those anisotropy effects are taken into
account which are to be neglected in the very spirit of the spherically symmetric
approximation.

In order to clarify this question, one must first consider the anisotropy of the
interaction potential and then calculate the effect of this anisotropy on the binding
energy. Since the gauge potential $\pgAn\rt$ is tied up to the charge density $\pgko\rt$
via a Poisson-type equation, cf. (\ref{eq:IV.10}), and since furthermore the charge
density $\pgko$ is itself built up by the anisotropic wave functions, the latter objects
do transfer their anisotropy directly to the interaction potential $\pgAn\rt$. Therefore
it must be possible to express the anisotropy of the gauge potential $\pgAn\rt$ in terms
of the anisotropy of the wave functions (or amplitude fields, resp.); and indeed, this can
be realized by expanding the anisotropic part of the gauge potential in terms of the
angular pre-factors of the wave functions, cf. equation (\ref{eq:IV.103}) below. The first
non-trivial term of this perturbation expansion is the \emph{quadrupole} term which
appears as the product of an angular-dependent function $\pgA^\mathsf{III}(\vartheta)$ and
a radial function $\pgA^\mathsf{III}(r)$, cf. (\ref{eq:IV.107}).

The pleasant property of such type of perturbation expansion is now that the angular
functions (such as $\pgA^\mathsf{III}(\vartheta)$ for the quadrupole case) can be exactly
traced back to the angular-dependent pre-factors of the wave amplitudes; and since the
latter pre-factors are uniquely associated with the (discrete) values $\elz$ of angular
momentum one thus obtains a unique link of the angular dependence of the gauge potentials
to the various values of $\elz$, see equation (\ref{eq:IV.104c}) below. And it is now just
this effect which brings about the \emph{elimination} of the ``spherically symmetric''
$\elz$-degeneracy of the energy spectrum! Clearly, any specific angular dependence of the
interaction potential (due to the various values of $\elz$) must necessarily entail a
different interaction energy and therefore also a different binding energy, see the
estimate of quadrupole energies (\ref{eq:IV.129}) due to the various values of $\elz$ for
$\boldsymbol{\elp = 3}$ (where the possible values of $\elz$ are $\elz = 0, \pm 1, \pm 2, \pm 3$). A
certain peculiarity emerges here for $\elp = 3$, where it turns out that the quadrupole
correction term for $\elz = \pm 2$ is zero so that an improvement of the corresponding
``spherically symmetric'' energy level can arise not until the octupole approximation is
considered, see equation (\ref{eq:IV.104c}).

\begin{center}
  \large{\textit{Para-Positronium Spectrum}}
\end{center}

But once the total energy (as sum of the kinetic and the electrostatic interaction energy)
is determined, one can (at least approximately) calculate the non-relativistic energy
spectrum of para-positronium. As a nearby approximation procedure, one chooses some
plausible trial configuration with included variational parameters for the concerned RST
fields, cf. (\ref{eq:IV.131}), and then one extremalizes the energy functional on that
selected set of trial configurations. As an example, we discuss here explicitly the
situation of eliminating the $\elz$-degeneracy in the quadrupole approximation for
$\boldsymbol{\elp = 3}$, see \textbf{Fig.IV.B},~p.~\pageref{figIV.B}. The result is as
expected; namely, the splitted RST levels become simultaneously shifted towards the
conventional prediction (\ref{eq:I.2}). The RST level closest to the conventional
prediction is due to $\elz = \pm 3$. Its deviation from its conventional counterpart
(\ref{eq:I.2}) is 0,005~[eV] corresponding to a relative deviation of 1,2\%.

This motivates one to consider the whole spectrum for $\elz = \pm \elp$ in the presently
discussed quadrupole approximation. The results are collected in \textbf{Fig.IV.E} on
p.~\pageref{figIV.E}. The most striking effect of these results refers to the fact that
the RST-splitted levels do form some kind of energy band around the (highly degenerated)
conventional levels~$\Ea{E}{n}{C}$ (\ref{eq:I.2}). For low principal quantum numbers~$n$,
the bands are well-separated; but it seems that for sufficiently large~$n$ the bands will
overlap and thus constitute a rather intricate spectrum, rather different from the highly
regular conventional spectrum (\ref{eq:I.2}). It remains to be clarified whether such a
result is merely an artefact being caused by all the applied approximations (especially
quadrupole approximation) or whether the emerging band structure is also confirmed by the
exact (hitherto unknown) solution of the RST eigenvalue problem.

\section{Positronium Eigenvalue Problem}
\indent

In order that the paper be sufficiently self-contained, it may appear useful to mention
briefly some fundamental facts about RST. As its very notation says, the central idea is
the Relativistic Schr\"odinger Equation (\ref{eq:II.1}) which leads one in a rather
straight-forward way to the Dirac equation (\ref{eq:II.13}) for few-particle (or
many-particle) systems. In order to ultimately end up with a closed dynamical system for
the fluid-dynamic quantum matter, one adds the (generally non-Abelian) Maxwell equations
(\ref{eq:II.15}) where this coupled system of matter and gauge field dynamics
automatically entails certain conservation laws, such as those for charge (\ref{eq:II.25})
or energy-momentum (\ref{eq:II.32}). This fundamental structure of RST is then
subsequently specified down to the non-relativistic positronium system, especially to its
para-form, with the main interest aiming at its energy spectrum.

\begin{center}
  \emph{\textbf{1.\ Relativistic Schr\"odinger Equations}}
\end{center}

A subset of problems within the general framework of RST concerns the (stationary) bound
systems. The simplest of those systems is positronium which consists of two oppositely
charged particles of the same rest mass ($M$). The physical behaviour of its matter
subsystem is assumed here to obey the \emph{Relativistic Schr\"odinger Equation}
\begin{equation}
\label{eq:II.1}
i\hbar\crm\;\Du\,\Psi = \mathcal{H}_\mu\,\Psi
\end{equation}
where the two-particle wave function $\Psi(x)$ is the direct sum of the two one-particle wave functions $\psi_a(x)\ (a=1,2)$
\begin{equation}
\label{eq:II.2}
\Psi(x) = \psi_1(x) \oplus \psi_2(x) \;.
\end{equation}
The gauge-covariant derivative $\D$ on the left-hand side of the basic wave equation (\ref{eq:II.1}) is defined in terms of the $\mathfrak{u}(2)$-valued gauge potential $\A_\mu$ as usual
\begin{equation}
\label{eq:II.3}
\Du\,\Psi = \partial_\mu\,\Psi + \A_\mu\,\Psi \;,
\end{equation}
or rewritten in component form
\begin{subequations}
\begin{align}
\label{eq:II.4a}
D_\mu\,\psi_1 &= \partial_\mu\,\psi_1 - i\,A^2_\mu\,\psi_1 \\
\label{eq:II.4b}
D_\mu\,\psi_2 &= \partial_\mu\,\psi_2 - i\,A^1_\mu\,\psi_2 \;.
\end{align}
\end{subequations}
Here, the electromagnetic four-potentials $A^a_\mu\ (a=1,2)$ are the components of the
original gauge potential $\A_\mu$ with respect to some suitable basis $\tau_\alpha\
(\alpha=1, \ldots 4)$ of the $\mathfrak{u}(2)$-algebra
\begin{equation}
\label{eq:II.5}
\A_\mu (x) = A^\alpha_\mu (x)\,\tau_\alpha = A^a_\mu (x)\,\tau_a + B_\mu(x)\,\chi - B^*_\mu(x)\,\bar{\chi} \;.
\end{equation}
Here, the electromagnetic generators $\tau_a\ (a=1,2)$ do commute
\begin{equation}
\label{eq:II.6}
\left[ \tau_1, \tau_2 \right] = 0
\end{equation}
and the exchange potential $B_\mu$ is put to zero ($\leadsto B_\mu(x) \equiv 0$) because the two positronium constituents (i.\,e. electron and positron) do count as \emph{non-identical} particles. Recall that the exchange effects, being mediated by the exchange potential $B_\mu(x)$, do occur exclusively for \emph{identical} particles so that $B_\mu(x)$ is inactive for the positronium constituents ($\leadsto B_\mu(x) \equiv 0$). Thus the bundle connection $\A_\mu(x)$ (\ref{eq:II.5}) becomes reduced to its $\mathfrak{u}(1) \oplus \mathfrak{u}(1)$ projection
\begin{equation}
\label{eq:II.7}
\A_\mu(x)\ \Rightarrow\ A^a_\mu(x)\,\tau_a \;.
\end{equation}

\begin{center}
  \emph{\textbf{2.\ Dirac Equations}}
\end{center}

For Dirac particles, which are to be described by four-spinors $\psi_a(x)$, the Hamiltonian $\mathcal{H}_\mu$ in the Relativistic Schr\"odinger Equation (\ref{eq:II.1}) obeys the relation
\begin{equation}
\label{eq:II.8}
\GG^\mu\,\mathcal{H}_\mu = \M\crm^2 \;,
\end{equation}
where $\GG^\mu$ is the total velocity operator and thus is the direct sum of the Dirac matrices $\gamma^\mu$
\begin{equation}
\label{eq:II.9}
\GG^\mu = \left( -\gamma^\mu \right) \oplus \gamma^\mu \;.
\end{equation}
The mass operator $\M$ specifies the two particle masses $M^a\ (a=1,2)$
\begin{equation}
\label{eq:II.10}
\M = i\,M^a\,\tau_a
\end{equation}
and is required to be Hermitian ($\bar{\M} = \M$) and covariantly constant
\begin{equation}
\label{eq:II.11}
\Du\,\M \equiv 0 \;.
\end{equation}
This requirement is trivially satisfied for particles of identical rest masses ($M^1 = M^2 \doteqdot M$) since for such a situation the mass operator becomes proportional to the identity operator
\begin{equation}
\label{eq:II.12}
\M = M\ \mathbf{1} \;.
\end{equation}
Thus the result is that, by virtue of the relation (\ref{eq:II.8}), the Relativistic Schr\"odinger Equation (\ref{eq:II.1}) becomes the two-particle Dirac equation
\begin{equation}
\label{eq:II.13}
i\hbar\mathrm{c}\,\GG^\mu\,\D_\mu\,\Psi = \M\crm^2\,\Psi \;,
\end{equation}
or in component form
\begin{subequations}
\begin{align}
\label{eq:II.14a}
i\hbar\mathrm{c}\,\gamma^\mu\,D_\mu\,\psi_1 &= -M\crm^2\,\psi_1 \\
\label{eq:II.14b}
i\hbar\mathrm{c}\,\gamma^\mu\,D_\mu\,\psi_2 &= M\crm^2\,\psi_2 \;,
\end{align}
\end{subequations}
where the gauge-covariant derivatives ($D$) of the single-particle wave functions $\psi_a(x)$ are given by equations (\ref{eq:II.4a})--(\ref{eq:II.4b}).

For \emph{identical} particles, the Dirac equations (\ref{eq:II.14a})--(\ref{eq:II.14b}) would couple both particles much more directly since the exchange potential $B_\mu$ is generated cooperatively by both particles and simultaneously does act back on any individual particle which then entails the phenomenon of self-coupling. However, for the present situation of \emph{non-identical} particles the coupling is more indirect: any particle does generate a Dirac four-current $k_{a\,\mu}(x)\ (a=1,2)$ which is the source of the four-potential $A^a_\mu(x)$ (see below). And then this four-potential $A^a_\mu$ of the $a$-th particle acts on the wave-function $\psi_b(x)$ of the other particle ($b \neq a$) as shown by equations (\ref{eq:II.14a})--(\ref{eq:II.14b}) in connection with the gauge-covariant derivatives $D$ (\ref{eq:II.4a})--(\ref{eq:II.4b}).

\begin{center}
  \emph{\textbf{3.\ Maxwell Equations}}
\end{center}

\begin{sloppypar}
The bundle connection $\A_\mu(x)$ (\ref{eq:II.5}) is itself a dynamical object of the theory (just as is the wave function $\Psi(x)$) and therefore must be required to obey some field equation. This is the (generally non-Abelian) Maxwell equation
\begin{gather}
\label{eq:II.15}
\D^\mu\F_{\mu\nu} = -4\pi i\as\J_\nu \\
\left(\as \doteqdot \frac{e^2}{\hbar \crm} \right) \;. \nonumber
\end{gather}
Here, the bundle curvature $\F_{\mu\nu}$ is defined in terms of the bundle connection $\A_\mu$ as usual, i.\,e.
\begin{equation}
\label{eq:II.16}
\F_{\mu\nu} = \nabla_\mu\A_\nu - \nabla_\nu\A_\mu + \left[\A_\mu,\A_\nu \right] \;.
\end{equation}
For the present situation of non-identical particles, the connection $\A_\mu$ becomes reduced to its (Abelian) $\mathfrak{u}(1) \oplus \mathfrak{u}(1)$ projection, cf. (\ref{eq:II.7}), which then also holds for its curvature $\F_{\mu\nu}$
\begin{equation}
\label{eq:II.17}
\F_{\mu\nu}\ \Rightarrow\ \nabla_\mu\A_\nu - \nabla_\nu\A_\mu \;.
\end{equation}
Decomposing here both the curvature $\F_{\mu\nu}$ and current operator $\J_\mu$ with respect to the chosen basis of commuting generators $\tau_a\ (a=1,2)$
\begin{subequations}
\begin{align}
\label{eq:II.18a}
\F_{\mu\nu}\ &\Rightarrow\ F^a_{\mu\nu} \tau_a \\
\label{eq:II.18b}
\J_\nu\ &\Rightarrow\ i {j^a}_\nu \tau_a \;,
\end{align}
\end{subequations}
one obtains the Maxwell equations (\ref{eq:II.15}) in component form as
\begin{equation}
\label{eq:II.19}
\nabla^\mu F^a_{\mu\nu} = 4\pi \as j^a_\nu \;.
\end{equation}
Since for the present Abelian situation the curvature components $F^a_{\mu\nu}$ (\ref{eq:II.18a}) are linked to the connection components $A^a_\mu$ (\ref{eq:II.7}) as usual in Maxwellian electrodynamics (in its Abelian form)
\begin{equation}
\label{eq:II.20}
F^a_{\mu\nu} = \nabla_\mu A^a_\nu - \nabla_\nu A^a_\mu \;,
\end{equation}
the Maxwell equations (\ref{eq:II.19}) for the field strengths $F^a_{\mu\nu}$ become converted to the d'Alembert equations for the four-potentials $A^a_\mu$
\begin{equation}
\label{eq:II.21}
\square\,A^a_\mu = 4\pi \as j^a_\mu \;,
\end{equation}
provided the gauge potentials $A^a_\mu$ do obey the Lorentz gauge condition
\begin{equation}
\label{eq:II.22}
\nabla^\mu A^a_\mu \equiv 0 \;.
\end{equation}
\end{sloppypar}

\newpage
\begin{center}
  \emph{\textbf{4.\ Conservation Laws}}
\end{center}

One of the most striking features in the description of physical systems is that both classical and quantum matter do obey certain conservation laws. For the presently considered Relativistic Schr\"odinger Theory, as a \emph{fluid-dynamic} theory, this means that there should exist certain \emph{local} conservation laws, preferably concerning charge and energy-momentum. Moreover, these local laws should turn out as an immediate consequence of the basic dynamical equations, i.\,e. the Relativistic Schr\"odinger Equation (\ref{eq:II.1}) and the Maxwell equations (\ref{eq:II.15}).

In this regard, a very satisfying feature of the Relativistic Schr\"odinger Theory is now that such conservation laws are automatically implied by the dynamical equations themselves. In order to elaborate this briefly, consider first the conservation of total charge which as a local law reads
\begin{equation}
\label{eq:II.23}
\nabla^\mu j_\mu \equiv 0 \;.
\end{equation}
But such a continuity equation for the total four-current $j_\mu$ can easily be deduced from \emph{both} the matter equation (\ref{eq:II.1}) and the gauge field equations (\ref{eq:II.15}); and this fact signals the internal consistency of the RST dynamics. First, consider the gauge field dynamics (\ref{eq:II.15}) and observe here the identity
\begin{equation}
\label{eq:II.24}
\D^\mu\D^\nu\F_{\mu\nu} \equiv 0
\end{equation}
which holds in any flat space-time. Obviously, the combination of this identity with the Maxwell equations (\ref{eq:II.15}) yields the following continuity equation in operator form
\begin{equation}
\label{eq:II.25}
\D^\mu\J_\mu \equiv 0 \;.
\end{equation}
Decomposing here the current operator in component form yields
\begin{equation}
\label{eq:II.26}
D^\mu j^\alpha_\mu \equiv 0
\end{equation}
which furthermore simplifies to
\begin{gather}
\label{eq:II.27}
\nabla^\mu\;j^a_\mu \equiv 0 \\
(a = 1,2) \nonumber
\end{gather}
under the Abelian reduction (\ref{eq:II.18a})--(\ref{eq:II.18b}). But when the individual Maxwell currents $j^a_\mu$ do obey such a continuity equation (\ref{eq:II.27}), then the total current $j_\mu$
\begin{equation}
\label{eq:II.28}
j_\mu \doteqdot \sum_{a=1}^{2}\, j^a_\mu
\end{equation}
must also obey a continuity equation which is just the requirement (\ref{eq:II.23}).

On the other hand, we can start also from the matter dynamics (\ref{eq:II.1}) and can define the total current $j_\mu$ by
\begin{equation}
\label{eq:II.29}
j_\mu \doteqdot \bar{\Psi} \GG_\mu \Psi \;.
\end{equation}
The divergence of this current is
\begin{equation}
\label{eq:II.30}
\nabla^\mu j_\mu = \left( \D^\mu \bar{\Psi} \right) \GG_\mu + \bar{\Psi} \GG_\mu \left( \D^\mu \Psi \right) + \bar{\Psi} \left( \D^\mu \GG_\mu \right) \Psi \;.
\end{equation}
Here, one requires that the gauge-covariant derivative of the total velocity operator $\GG_\mu$ is covariantly constant
\begin{equation}
\label{eq:II.31}
\D^\mu \GG_\mu \equiv 0 \;,
\end{equation}
and furthermore one evokes the Relativistic Schr\"odinger equation together with the Hamiltonian condition (\ref{eq:II.8}) which then ultimately yields again the desired continuity equation (\ref{eq:II.23}). Thus, the conservation of total charge is actually deducible from both subdynamics of the whole RST system; and this fact supports the mutual compatibility of both subdynamics (i.\,e. the matter dynamics (\ref{eq:II.1}) and the gauge field dynamics (\ref{eq:II.15})).

A further important conservation law does refer to the energy-momentum content of the considered physical system. Aiming again at a \emph{local} law, one may think of a continuity equation of the following form
\begin{equation}
\label{eq:II.32}
\nabla^\mu\, \TTmunu \equiv 0 \;,
\end{equation}
where $\TTmunu$ is the total energy-momentum density, i.\,e. the sum of the Dirac matter part $\left( \DTmunu \right)$ and the gauge field part $\left( \GTmunu \right)$:
\begin{equation}
\label{eq:II.33}
\TTmunu = \DTmunu + \GTmunu \;.
\end{equation}
Clearly, the validity of the total law (\ref{eq:II.32}) does not require an analogous law for the subdensities but merely requires the right balance of the energy-momentum exchange between the subsystems, i.\,e.
\begin{equation}
\label{eq:II.34}
\nabla^\mu \, \DTmunu = - \nabla^\mu \, \GTmunu \;.
\end{equation}

Indeed, the matter part has been identified as
\begin{equation}
\label{eq:II.35}
\DTmunu = \frac{i\hbar\crm}{4} \left[ \bar{\Psi} \GG_\mu (\D_\nu \Psi) - (\D_\nu \bar{\Psi}) \GG_\mu \Psi + \bar{\Psi}\GG_\nu (\D_\mu \Psi ) - (\D_\mu \bar{\Psi}) \GG_\nu \Psi \right]
\end{equation}
and the gauge field part by
\begin{equation}
\label{eq:II.36}
\GTmunu = \frac{\hbar\crm}{4\pi\as} K_{\alpha\beta} \left( F^\alpha_{\ \mu\lambda} F^{\beta\ \lambda}_{\ \nu} - \frac{1}{4} g_{\mu\nu} F^{\alpha}_{\ \sigma\lambda} F^{\beta\sigma\lambda} \right) \;,
\end{equation}
where $K_{\alpha\beta}$ is the fibre metric in the associated Lie algebra bundle. The (local) conservation law (\ref{eq:II.32}) comes now actually about through the mutual annihilation (\ref{eq:II.34}) of the sources of both energy-momentum densities, i.\,e.
\begin{equation}
\label{eq:II.37}
\nabla^\mu\,\DTmunu = -\nabla^\mu\,\GTmunu = \hbar\crm\, F^{\alpha}_{\ \mu\nu}\, j^{\ \mu}_{\alpha} \;.
\end{equation}
Obviously the sources of the partial densities $\DTmunu$ and $\GTmunu$ are just the well-known \emph{Lorentz forces} in non-Abelian form.

It should now appear self-suggesting that the definition of the total energy ($\ET$) of an RST field configuration is to be based upon the time component $\TToo$ of the energy-momentum density $\TTmunu$, i.\,e.
\begin{equation}
\label{eq:II.38}
E_T \doteqdot \int d^3\vr\;\TToo(\vr) \;.
\end{equation}
But since the total density $\TTmunu$ is the sum of a matter part and a gauge field part, cf. (\ref{eq:II.33}), the total energy $\ET$ (\ref{eq:II.38}) naturally breaks up in an analogous way
\begin{equation}
\label{eq:II.39}
\ET = E_\mathrm{D} + E_\mathrm{G} \;,
\end{equation}
with the self-evident definitions
\begin{subequations}
\begin{align}
\label{eq:II.40a}
E_\mathrm{D} &\doteqdot \int d^3\vr\;\DToo(\vr) \\
\label{eq:II.40b}
E_\mathrm{G} &\doteqdot \int d^3\vr\;\GToo(\vr) \;.
\end{align}
\end{subequations}
But clearly, such a preference of the time component $\Too$ among all the other components $\Tmunu$ entails the selection of a special time axis for the space-time manifold. This then induces a similar space-time splitting of all the other objects in the theory, i.\,e. we have to consider now \emph{stationary} field configurations which are generally thought to represent the basis of the energy spectra of the bound systems.

\begin{center}
  \emph{\textbf{5.\ Stationary Field Configurations}}
\end{center}

In the present context, the notion of stationarity is coined with regard to the time-independence of the physical observables of the theory, i.\,e. the physical densities and the electromagnetic fields generated by them. In contrast to this, the wave functions do not count as observables and therefore are not required to be time-independent. But their time-dependence must be in such a way that the associated densities become truly time-independent.

\begin{center}
  \large{\textit{Gauge-Field Subsystem}}
\end{center}

The simplest space-time splitting refers to the four-potentials $A^a_{\ \mu}$, which for the stationary states become time-independent and thus appear in the following form:
\begin{gather}
\label{eq:II.41}
A^a_{\ \mu}(x)\ \Rightarrow\ \left\{ {}^{(a)}\!A_0(\vr)\;;\ -\vec{A}_a(\vr) \right\} \\
(a = 1,2) \;. \nonumber
\end{gather}
A similar arrangement does apply also to the Maxwell four-currents $j^a_{\ \mu}$
\begin{equation}
\label{eq:II.42}
j^a_{\ \mu}\ \Rightarrow\ \left\{ {}^{(a)}\!j_0(\vr)\;;\ -\vec{j}_a(\vr) \right\} \;,
\end{equation}
so that the d'Alambert equations (\ref{eq:II.21}) become split up into the Poisson equations, for both the scalar potentials ${}^{(a)}\!A_0(\vr)$
\begin{equation}
\label{eq:II.43}
\Delta {}^{(a)}\!A_0(\vr) = -4\pi\as {}^{(a)}\!j_0(\vr)
\end{equation}
and the three-vector potentials $\vec{A}_a(\vr)$
\begin{equation}
\label{eq:II.44}
\Delta \vec{A}_a(\vr) = -4\pi\as \vec{j}_a(\vr) \;.
\end{equation}
Recall here that the standard solutions of these equations are formally given by
\begin{subequations}
\begin{align}
\label{eq:II.45a}
{}^{(a)}\!A_0(\vr) &= \as \int d^3\vr\,'\;\frac{{}^{(a)}\!j_0(\vr\,')}{||\vr-\vr\,'||} \\
\label{eq:II.45b}
\vec{A}_a(\vr) &= \as \int d^3\vr\,'\;\frac{\vec{j}_a(\vr\,')}{||\vr-\vr\,'||} \;.
\end{align}
\end{subequations}
Similar arguments would apply also to the exchange potential $B_\mu = \left\{ B_0, -\vec{B} \right\}$ but since we are dealing with non-identical particles the exchange potential must be put to zero $(B_\mu(x) \equiv 0)$.

It is true, the particle interactions are organized here via the (electromagnetic and exchange) \emph{potentials} which, according to the \emph{principle of minimal coupling}, are entering the covariant derivatives $D_\mu \psi_a$ of the wave functions $\psi_a$ as shown by equations (\ref{eq:II.4a})--(\ref{eq:II.4b}). But nevertheless it is very instructive to glimpse also at the \emph{field strengths} $F^a_{\ \mu\nu}$. Their space-time splitting is given by
\begin{subequations}
\begin{align}
\label{eq:II.46a}
\vec{E}_a = \left\{ {}^{(a)}\!E^j \right\} &\doteqdot \left\{ F^a_{\ 0j} \right\} \\
\label{eq:II.46b}
\vec{H}_a = \left\{ {}^{(a)}\!H^j \right\} &\doteqdot \left\{ \frac{1}{2}\, \varepsilon^{jk}_{\ \ l}\, F^{a\ l}_{\ k} \right\}
\end{align}
\end{subequations}
and thus the \emph{linear} Maxwell equations (\ref{eq:II.19}) do split up in three-vector form $(a=1,2)$ to the scalar equations for the electric fields
\begin{equation}
\label{eq:II.47}
\vec{\nabla} \sdot \vec{E}_a = 4\pi\as {}^{(a)}\!j_0
\end{equation}
and to the curl equations for the magnetic fields
\begin{equation}
\label{eq:II.48}
\vec{\nabla} \times \vec{H}_a = 4\pi\as \vec{j}_a \;.
\end{equation}
There is a pleasant consistency check for these linearized (but still relativistic) field equations in three-vector form; namely one may first link the field-strengths to the potentials in three-vector notation (cf. (\ref{eq:II.20}) for the corresponding relativistic link):
\begin{subequations}
\begin{align}
\label{eq:II.49a}
\vec{E}_a(\vr) &= - \vec{\nabla} {}^{(a)}\!A_0(\vr) \\
\label{eq:II.49b}
\vec{H}_a(\vr) &= \vec{\nabla} \times \vec{A}_a(\vr)
\end{align}
\end{subequations}
and then one substitutes these three-vector field strengths into their source and curl equations (\ref{eq:II.47})--(\ref{eq:II.48}). In this way one actually recovers the Poisson equations (\ref{eq:II.43})--(\ref{eq:II.44}) for the electromagnetic potentials ${}^{(a)}\!A_0, \vec{A}_a$.

\begin{center}
  \large{\textit{Matter Subsystem}}
\end{center}

Concerning now the stationary form of the matter dynamics, one resorts of course to the generally used factorization of the wave functions $\psi_a(\vr, t)$ into a time and a space factor
\begin{equation}
\label{eq:II.50}
\psi_a(\vr, t) = \exp\left[ -i\, \frac{M_a\crm^2}{\hbar}t \right] \cdot \psi_a(\vr) \;.
\end{equation}
Here, the mass eigenvalues $M_a\ (a=1,2)$ are the proper objects to be determined from the mass eigenvalue equations which we readily put forward now. For this purpose, observe first that the Dirac four-spinors $\psi_a$ may be conceived as the direct sum of Pauli two-spinors ${}^{(a)}\!\varphi_\pm$:
\begin{equation}
\label{eq:II.51}
\psi_a(\vr) = {}^{(a)}\!\varphi_+(\vr) \oplus {}^{(a)}\!\varphi_-(\vr) \;.
\end{equation}
Consequently, the Dirac mass eigenvalue equations (to be deduced from the general Dirac equations (\ref{eq:II.14a})--(\ref{eq:II.14b}) by means of the factorization ansatz (\ref{eq:II.50})) are recast to their equivalent Pauli form for the two-spinors ${}^{(a)}\varphi_\pm(\vr)$
\begin{subequations}
\begin{align}
\label{eq:II.52a}
i\, \vec{\sigma} \sdot \vec{\nabla}\;{}^{(1)}\!\varphi_\pm(\vr) + {}^{(2)}\!A_0(\vr) \cdot
{}^{(1)}\!\varphi_\mp(\vr) &= \frac{\pm M_p - M_1}{\hbar}\,\crm \cdot {}^{(1)}\!\varphi_\mp(\vr) \\
\label{eq:II.52b}
i\, \vec{\sigma} \sdot \vec{\nabla}\;{}^{(2)}\!\varphi_\pm(\vr) + {}^{(1)}\!A_0(\vr) \cdot
{}^{(2)}\!\varphi_\mp(\vr) &= -\frac{M_2 \pm M_e}{\hbar}\,\crm \cdot {}^{(2)}\!\varphi_\mp(\vr) \;.
\end{align}
\end{subequations}
(Observe here that we do neglect for the moment the magnetic effects by putting the three-vector potentials $\vec{A}_a(\vr)$ (\ref{eq:II.41}) to zero: $\vec{A}_a(\vr) \Rightarrow 0$). The \emph{mass eigenvalue} for the positron (with rest mass $M_p$) is denoted by $M_1$ and for the electron (with rest mass $M_e$) by $M_2$.

Summarizing, the RST eigenvalue system consists of the \emph{mass eigenvalue equations}
(\ref{eq:II.52a})--(\ref{eq:II.52b}) for the Pauli spinors ${}^{(a)}\!\varphi_\pm(\vr)$ in
combination with the Poisson equations (\ref{eq:II.43}). Since the magnetic effects are
neglected, the Poisson equations (\ref{eq:II.44}) need not be considered here. But what is
necessary in order to close the whole eigenvalue problem is the prescription for the link
of the Pauli spinors ${}^{(a)}\!\varphi_\pm(\vr)$ to the Maxwell charge densities
${}^{(a)}\!j_0(\vr)$, or more generally to the Maxwellian four-currents $j^a_{\ \mu}$
(\ref{eq:II.42}) as the sources of the four-potentials $A^a_{\ \mu}$, cf. the d'Alembert
equations (\ref{eq:II.21}). Surely, such a link between the wave functions $\psi_a$ and
the currents $j^a_{\ \mu}$ will have something to do with the Dirac four-currents
$k_{a\mu}$ which are usually defined by $(a=1,2)$
\begin{equation}
\label{eq:II.53}
k_{a\mu} \doteqdot \bar{\psi_a}\, \gamma_u\, \psi_a \;.
\end{equation}
Indeed, a more profound scrutiny reveals the following link~\cite{7}
\begin{subequations}
\begin{align}
\label{eq:II.54a}
j^1_{\ \mu} &\equiv k_{1\mu} = \bar{\psi_1}\, \gamma_u\, \psi_1 \\
\label{eq:II.54b}
j^2_{\ \mu} &\equiv -k_{2\mu} = -\bar{\psi_2}\, \gamma_u\, \psi_2 \;.
\end{align}
\end{subequations}
The change in sign of both Dirac currents reflects the positive and negative charge of both particles. In terms of the Pauli spinors $\appm\ (a=1,2)$ the space and time components of the Dirac currents read
\begin{subequations}
\begin{align}
\label{eq:II.55a}
\akn &= \appd \, \app + \apmd \, \apm \\
\label{eq:II.55b}
\vec{k}_a(\vr) &= \appd \, \vec{\sigma} \, \apm + \apmd \, \vec{\sigma} \, \apm \;.
\end{align}
\end{subequations}
Although the magnetic effects, which originate from the three-currents $\vec{k}_a(\vr)$ via the magnetic Poisson equations (\ref{eq:II.44}), are neglected in the present paper these current densities nevertheless play now an important part for identifying two essentially different kinds of positronium.

\section{Ortho/Para Dichotomy}
\indent

In the conventional theory, the manifestation of two principally different kinds of
positronium is traced back to the two possibilities of combining the spins of the electron
(e) and positron (p)~\cite{8}: if both spins $\spins_e$ and $\spins_p$ add up to the
total Spin $S = 1$
\begin{gather}
\label{eq:III.1}
s_O = \spins_p + \spins_e = 1 \\
\left( \spins_p = \spins_e = \frac{1}{2} \right) \;, \nonumber
\end{gather}
one has \emph{ortho-positronium}; and zero spin
\begin{equation}
\label{eq:III.2}
s_\wp = \spins_p - \spins_e = 0
\end{equation}
yields \emph{para-positronium}. This is the well known ortho/para dichotomy mentioned in
any textbook on relativistic quantum mechanics. In contrast to this (generally valid)
composition rule for angular momenta, the ortho/para dichotomy \emph{in RST} is based upon
the (anti) parallelity of the Maxwellian three-currents $\vec{j}_a(\vr)$. Here, it is
assumed that both particles do occupy physically equivalent one-particle states
$\psi_a(\vr)\ (a=1,2)$ in the sense that the Dirac currents (and therefore also the
Maxwell currents) are either parallel or antiparallel. Thus we propose the following
characterization of \textbf{ortho-positronium}~\cite{9}:
\begin{subequations}
\begin{align}
\label{eq:III.3a}
\vec{k}_1(\vr) &\equiv -\vec{k}_2(\vr) \doteqdot \vec{k}_b(\vr) \\
\label{eq:III.3b}
\vec{j}_1(\vr) &\equiv \vec{j}_2(\vr) \doteqdot \vec{j}_b(\vr) \equiv \vec{k}_b(\vr) \\
\label{eq:III.3c}
\vec{A}_1(\vr) &\equiv \vec{A}_2(\vr) \doteqdot \vec{A}_b(\vr) \\
\label{eq:III.3d}
\vec{H}_1(\vr) &\equiv \vec{H}_2(\vr) \doteqdot \vec{H}_b(\vr) \;,
\end{align}
\end{subequations}
and analogously for \textbf{para-positronium}:
\begin{subequations}
\begin{align}
\label{eq:III.4a}
\vec{k}_1(\vr) &\equiv \vec{k}_2(\vr) \doteqdot \vec{k}_p(\vr) \\
\label{eq:III.4b}
\vec{j}_1(\vr) &\equiv -\vec{j}_2(\vr) \doteqdot \vec{j}_p(\vr) \equiv \vec{k}_p(\vr) \\
\label{eq:III.4c}
\vec{A}_1(\vr) &\equiv -\vec{A}_2(\vr) \doteqdot \vec{A}_p(\vr) \\
\label{eq:III.4d}
\vec{H}_1(\vr) &\equiv -\vec{H}_2(\vr) \doteqdot \vec{H}_p(\vr) \;.
\end{align}
\end{subequations}

The interesting point with such a subdivision of positronium into two classes is the fact
that this subdivision is based upon the \emph{magnetic} effects which however are
\emph{neglected} for the present paper; but despite this neglection the subdivision is of
great relevance also for the presently considered \emph{electrostatic} approximation!
Namely, even within the framework of the latter approximation scheme, there do emerge
different distributions of electrostatic charge $\left( {}^{(a)}\!k_0(\vr) \right)$ for
ortho- and para-positronium with the corresponding quantum numbers ($\leadsto$
\emph{ortho/para dichotomy}); and moreover there does arise also a certain ambiguity of
the electric charge distribution even within the subclass of the ortho-configurations due
to the same quantum number ($\leadsto$ \emph{ortho-dimorphism}). In order to elaborate
these effects it is necessary to first specify the general eigenvalue problem down to the
subcases and then to look for the corresponding solutions.

\begin{center}
  \emph{\textbf{1.\ Mass Eigenvalue Equations}}
\end{center}

The hypothesis of physically equivalent states for both positronium constituents entails that both mass eigenvalues $M_a$ are actually identical, i.\,e.
\begin{equation}
\label{eq:III.5}
M_1 = -M_2 \doteqdot -M_* \;.
\end{equation}
Next, one considers the Maxwellian charge densities ${}^{(a)}\!j_0(\vr)$ which must of course differ in sign for oppositely charged particles
\begin{equation}
\label{eq:III.6}
{}^{(1)}\!j_0(\vr) = -{}^{(2)}\!j_0(\vr)
\end{equation}
and this must be true for both ortho- and para-positronium. On the other hand, the link (\ref{eq:II.54a})--(\ref{eq:II.54b}) of the Maxwellian currents $j^a_{\ \mu}$ to the Dirac currents $k_{a\mu}$ shows that the requirement (\ref{eq:III.6}) demands the identity of the Dirac densities ${}^{(a)}\!k_0(\vr)$, i.\,e.
\begin{equation}
\label{eq:III.7}
{}^{(1)}\!k_0(\vr) \equiv {}^{(2)}\!k_0(\vr) \;,
\end{equation}
or rewritten in terms of the Pauli spinors (\ref{eq:II.55a})
\begin{equation}
\label{eq:III.8}
{}^{(1)}\!\varphi_+^\dagger(\vr) \, {}^{(1)}\!\varphi_+(\vr) + {}^{(1)}\!\varphi_-^\dagger(\vr) \, {}^{(1)}\!\varphi_-(\vr) = {}^{(2)}\!\varphi_+^\dagger(\vr) \, {}^{(2)}\!\varphi_+(\vr) + {}^{(2)}\!\varphi_-^\dagger(\vr) \, {}^{(2)}\!\varphi_-(\vr) \;.
\end{equation}

However, the situation is different for the Dirac three-currents $\vec{k}_a(\vr)$, since
they may differ in sign, cf. (\ref{eq:III.3a}) vs. (\ref{eq:III.4a}). Expressing the
(anti) parallelity of both Dirac currents in terms of the Pauli spinors,
cf. (\ref{eq:II.55b}), one requires
\begin{align}
\label{eq:III.9}
&{}^{(1)}\!\varphi_+^\dagger(\vr)\,\vec{\sigma}\,{}^{(1)}\!\varphi_-(\vr) + {}^{(1)}\!\varphi_-^\dagger(\vr)\,\vec{\sigma}\,{}^{(1)}\!\varphi_+(\vr)  \\
&= \mp \left\{ {}^{(2)}\!\varphi_+^\dagger(\vr)\,\vec{\sigma}\,{}^{(2)}\!\varphi_-(\vr) + {}^{(2)}\!\varphi_-^\dagger(\vr)\,\vec{\sigma}\,{}^{(2)}\!\varphi_+(\vr) \right\} \;, \nonumber
\end{align}
where the upper/lower sign refers to the ortho/para case, resp. Both conditions (\ref{eq:III.8}) and (\ref{eq:III.9}) can be satisfied by putting
\begin{subequations}
\begin{align}
\label{eq:III.10a}
{}^{(1)}\!\varphi_+(\vr) &= \mp i\,{}^{(2)}\!\varphi_+(\vr) \doteqdot {}^{(b/p)}\!\varphi_+(\vr) \\
\label{eq:III.10b}
{}^{(1)}\!\varphi_-(\vr) &= i\,{}^{(2)}\!\varphi_-(\vr) \doteqdot {}^{(b/p)}\!\varphi_-(\vr)
\end{align}
\end{subequations}
where the upper/lower sign refers again to ortho/para-positronium, resp.

It is true, the disposal (\ref{eq:III.10a})--(\ref{eq:III.10b}) satisfies both algebraic
requirements (\ref{eq:III.8}) and (\ref{eq:III.9}), but additionally there must be
satisfied also a differential requirement: actually, the spinor identifications
(\ref{eq:III.10a})--(\ref{eq:III.10b}) leave us with just one spinor field
(i.\,e. ${}^{(b)}\!\varphi_\pm(\vr)$ for ortho-positronium and
${}^{(p)}\!\varphi_\pm(\vr)$ for para-positronium); and therefore both spinor equations
(\ref{eq:II.52a})--(\ref{eq:II.52b}) must collapse without contradiction to only one
spinor equation (either for ${}^{(b)}\!\varphi_\pm(\vr)$ or
${}^{(p)}\!\varphi_\pm(\vr)$). In order to validate this requirement, we have to make a
disposal also for the electrostatic gauge potentials ${}^{(a)}\!A_0(\vr)$. But this can
easily be done by observing the link between the Dirac densities ${}^{(a)}\!k_0(\vr)$ and
${}^{(a)}\!A_0(\vr)$ as it is implemented by the Poisson equations
(\ref{eq:II.43}). Indeed, this link entails that the potentials ${}^{(a)}\!A_0(\vr)$ must
differ (or not) in sign when this is (or is not) the case also for the Maxwell densities
${}^{(a)}\!j_0(\vr)$. Therefore one concludes that for the positronium situation both
electrostatic potentials must \emph{always} differ in sign
\begin{equation}
\label{eq:III.11}
{}^{(1)}\!A_0(\vr) = -{}^{(2)}\!A_0(\vr) \doteqdot {}^{(b/p)}\!A_0(\vr) \;.
\end{equation}
But when this circumstance is duly respected, both mass eigenvalue equations (\ref{eq:II.52a})--(\ref{eq:II.52b}) actually do collapse to a single one for \textbf{ortho-positronium}:
\begin{equation}
\label{eq:III.12}
i\vec{\sigma}\sdot  \vec{\nabla} \, {}^{(b)}\!\varphi_\pm(\vr) - {}^{(b)}\!A_0(\vr) \cdot {}^{(b)}\!\varphi_\mp(\vr) = \frac{M_* \pm M}{\hbar}\crm \cdot {}^{(b)}\!\varphi_\mp(\vr) \;.
\end{equation}
The existence of one and the same eigenvalue equation (\ref{eq:III.12}) for both ortho-positronium constituents thus validates our original hypothesis that both the electron and the positron should occupy physically equivalent states.

The relativistic pair (\ref{eq:III.12}) of Pauli equations has a single Schr\"odinger-like equation as its non-relativistic limit. Indeed, assuming that the ``negative'' Pauli-spinor ${}^{(p)}\!\varphi_-(\vr)$ is always considerably smaller than its ``positive'' companion ${}^{(b)}\!\varphi_+(\vr)$ one can solve the upper one of the equations (\ref{eq:III.12}) for ${}^{(b)}\!\varphi_-(\vr)$ approximately in the following form:
\begin{equation}
\label{eq:III.13}
{}^{(b)}\!\varphi_-(\vr) \simeq \frac{i\hbar}{2M\crm}\,\vec{\sigma} \sdot \vec{\nabla}\,{}^{(b)}\!\varphi_+(\vr) \;,
\end{equation}
and if this is substituted into the lower equation (\ref{eq:III.12}) one finally ends up with the following non-relativistic eigenvalue equation of the Pauli form
\begin{equation}
\label{eq:III.14}
-\frac{\hbar^2}{2M} \, \Delta \, {}^{(b)}\!\varphi_+(\vr) - {}^{(b)}\!A_0(\vr) \cdot {}^{(b)}\!\varphi_+(\vr) = E_* \cdot {}^{(b)}\!\varphi_+(\vr) \;.
\end{equation}
Here, the non-relativistic eigenvalue $E_*$ emerges as the difference of the rest mass $M$ and the relativistic mass eigenvalue $M_*$, i.\,e.
\begin{equation}
\label{eq:III.15}
E_* \doteqdot \left( M_* - M \right) \crm^2 \;.
\end{equation}
Subsequently, we will be satisfied with clarifying the phenomenon of the ortho-dimorphism in the non-relativistic version (\ref{eq:III.14}) of the original relativistic eigenvalue equation (\ref{eq:III.12}).

But the case of para-positronium is a little bit more complicated. To begin with the
\emph{positron} equation (\ref{eq:II.52a}), this becomes transcribed by the
identifications (\ref{eq:III.10a})--(\ref{eq:III.10b}), lower case, to the following form
(\textbf{para-positronium}):
\begin{equation}
\label{eq:III.16}
i\,\vec{\sigma} \sdot \vec{\nabla}\,{}^{(p)}\!\varphi_\pm(\vr) - {}^{(p)}\!A_0(\vr) \cdot {}^{(p)}\!\varphi_\mp(\vr) = \frac{M_* \pm M}{\hbar}\crm \cdot {}^{(p)}\!\varphi_\mp(\vr) \;.
\end{equation}
Obviously, this positron equation is just of the same form as the joint positron/electron
equation (\ref{eq:III.12}) for ortho-positronium. However, the \emph{electron} equation
(\ref{eq:II.52b}) of para-positronium becomes transcribed by the identifications
(\ref{eq:III.10a})--(\ref{eq:III.10b}) to a somewhat different form:
\begin{equation}
\label{eq:III.17}
i\,\vec{\sigma} \sdot \vec{\nabla} \, {}^{(p)}\!\varphi_\pm(\vr) + {}^{(p)}\!A_0(\vr) \cdot {}^{(p)}\!\varphi_\mp(\vr) = -\frac{M_* \pm M}{\hbar}\crm \cdot {}^{(p)}\!\varphi_\mp(\vr) \;.
\end{equation}
Since the sign of the potential term $\left( \sim {}^{(p)}\!A_0(\vr) \right)$ is reversed
here in comparison to the positron equation (\ref{eq:III.16}), the latter electron
equation (\ref{eq:III.17}) is the ''\emph{charge conjugated}'' form of the first equation
(\ref{eq:III.16}).

The \emph{charge conjugation} is defined here by the following replacements:
\begin{subequations}
\begin{align}
\label{eq:III.18a}
{}^{(p)}\!\varphi_+(\vr)\ &\Rightarrow\ {}^{(p)}\!\varphi_-(\vr)\quad,\quad {}^{(p)}\!\varphi_-(\vr)\ \Rightarrow\ {}^{(p)}\!\varphi_+(\vr) \\
\label{eq:III.18b}
{}^{(p)}\!A_0(\vr)\ &\Rightarrow\ -{}^{(p)}\!A_0(\vr) \\
\label{eq:III.18c}
M_*\ &\Rightarrow\ -M_* \;.
\end{align}
\end{subequations}
Indeed, one is easily convinced that the two forms of eigenvalue equations
(\ref{eq:III.16}) and (\ref{eq:III.17}) for para-positronium are transcribed to one
another by these replacements (\ref{eq:III.18a})--(\ref{eq:III.18c}). This means that any
solution of the positron equation (\ref{eq:III.16}) can be interpreted also to be a
solution of the electron equation (\ref{eq:III.17}); however, the charge conjugation does
not leave invariant the Poisson equations, cf. (\ref{eq:III.21})-(\ref{eq:III.22})
below. Therefore, we prefer here the use of solutions with the same non-relativistic
limit! Indeed, it follows from both equations (\ref{eq:III.16}) and (\ref{eq:III.17}) that
the ``negative'' Pauli spinor ${}^{(p)}\!\varphi_-(\vr)$ can be approximately expressed in
terms of the ``positive'' spinor ${}^{(p)}\!\varphi_+(\vr)$ through
\begin{equation}
\label{eq:III.19}
{}^{(p)}\!\varphi_-(\vr) \simeq \pm\frac{i\hbar}{2M\crm}\,\vec{\sigma} \sdot \vec{\nabla}\,{}^{(p)}\!\varphi_+(\vr) \;,
\end{equation}
where the upper case refers to (\ref{eq:III.16}) and the lower case to (\ref{eq:III.17}). This result may then be substituted in either residual equation (\ref{eq:III.16}) and (\ref{eq:III.17}) which in both cases yields the \emph{same} Schr\"odinger-like equation for the ``positive'' spinor ${}^{(p)}\!\varphi_+(\vr)$:
\begin{equation}
\label{eq:III.20}
-\frac{\hbar^2}{2M}\,\Delta\,{}^{(p)}\!\varphi_+(\vr) - \hbar\crm {}^{(p)}\!A_0(\vr) \cdot {}^{(p)}\!\varphi_+(\vr) = E_* \cdot {}^{(p)}\!\varphi_+(\vr) \;.
\end{equation}

Thus it becomes again evident that the corresponding solutions of (\ref{eq:III.16}) and
(\ref{eq:III.17}) do actually describe physically equivalent states. The notion of
``physical equivalence'' is meant here to refer in first line to the numerical identity of
the energy being carried by anyone of the constituents of para-positronium, see below.

\begin{center}
  \emph{\textbf{2.\ Poisson Equations}}
\end{center}

\begin{sloppypar}
  The mass eigenvalue equations do not yet represent a closed system and therefore cannot
  be solved before an equation for the interaction potential ${}^{(b/p)}\!A_0(\vr)$ has
  been specified. On principle, this has already been done in form of equation
  (\ref{eq:II.43}) so that we merely have to further specify that equation in agreement
  with the ortho/para dichotomy. Observing here the circumstance that the Maxwellian
  current of the first particle ($a=1$, positron) agrees with the Dirac current,
  cf. (\ref{eq:II.54a}) and (\ref{eq:II.55a}), the Poisson equation (\ref{eq:II.43}) reads
  in terms of the Pauli spinors
\begin{align}
\label{eq:III.21}
\Delta\,{}^{(b/p)}\!A_0(\vr) &= -4\pi\as {}^{(b/p)}\!k_0(\vr) \\
&= - 4\pi\as \left\{ {}^{(b/p)}\!\varphi_+^\dagger(\vr)\,{}^{(b/p)}\!\varphi_+(\vr) + {}^{(b/p)}\!\varphi_-^\dagger(\vr)\,{}^{(b/p)}\!\varphi_-(\vr) \right\} \;. \nonumber
\end{align}
This Poisson equation closes the relativistic eigenvalue systems, both for
ortho-positronium (\ref{eq:III.12}) and for para-positronium
(\ref{eq:III.16})--(\ref{eq:III.17}). For the non-relativistic limit, one merely
suppresses the ``negative'' Pauli spinors ${}^{(b/p)}\!\varphi_-(\vr)$ so that the
relativistic Poisson equations (\ref{eq:III.21}) simplify to
\begin{equation}
\label{eq:III.22}
\Delta\,{}^{(b/p)}\!A_0(\vr) = -4\pi\as\,{}^{(b/p)}\!\varphi_+^\dagger(\vr)\,{}^{(b/p)}\!\varphi_+(\vr)
\end{equation}
which then closes both the non-relativistic eigenvalue equations (\ref{eq:III.14}) for ortho-positronium and (\ref{eq:III.20}) for para-positronium.
\end{sloppypar}

Clearly, these coupled systems of eigenvalue and Poisson equations cannot be solved exactly (though exact solutions do surely exist), and consequently we have to resort to some adequate approximation procedure. But this suggests itself when we subsequently will establish the variational \emph{principle of minimal energy}.
\pagebreak

\begin{center}
  \emph{\textbf{3.\ Non-Unique Spinor Fields}}
\end{center}

Despite the fact that we originally subdivided the whole set of positronium configurations
into two subclasses, i.\,e. ortho- and para-positronium
(\ref{eq:III.3a})--(\ref{eq:III.4d}), it may seem now that by the neglection of the
magnetic forces we ended up with an eigenvalue problem, which does no longer offer any
handle for sticking to that original subdivision into two peculiar subsets. Indeed, the
adopted \emph{electrostatic approximation} does admit exclusively an interaction force of
the purely electric type (being described by the electric potential
${}^{(b/p)}\!A_0(\vr)$), whereas the original ortho/para dichotomy was based upon the
three-currents $\ak(\vr)$ as the curls of the magnetic fields, cf. (\ref{eq:II.48}). As a
result of this neglection of magnetism, the Poisson equation (\ref{eq:III.21}), or
(\ref{eq:III.22}), resp., holds equally well for both the ortho-configurations (b) and the
para-configurations (p). But also the mass eigenvalue equations, especially in their
non-relativistic forms (\ref{eq:III.14}) and (\ref{eq:III.20}) are formally the same for
ortho- and para-positronium. Does this mean that, through passing over to the
electrostatic approximation, the difference between ortho- and para-positronium has gone
lost? This is actually not the case because the different angular momentum (in combination
with the hypothesis of the physical equivalence of both constituent states) leaves its
footprint also on the electrostatic approximation.

The crucial point here refers to the fluid-dynamic character of RST, as opposed to the
probabilistic character of the conventional quantum theory. This entails that in RST the
angular momenta of the subsystems cannot be combined (to the total angular momentum of the
whole system) in such a way as it is the case in the conventional theory ($\leadsto$
addition theorem for angular momenta~\cite{10}). More concretely: if we wish to insist on
the viewpoint that the (observable) angular momentum of the considered two-particle system
should emerge as the eigenvalue $j_z$ of the angular momentum operator $\Jhz = \Lhz +
\Shz$, i.\,e.
\begin{equation}
\label{eq:III.23}
\Jhz\,\Psi_{b/p}(\vr) = {}^{(b/p)}\!j_z\,\hbar \cdot \Psi_{b/p}(\vr) \;,
\end{equation}
then the quantum number ${}^{(b/p)}\!j_z$ due to the whole two-particle system must be
carried already by any individual particle! Namely, the wave function $\Psi_{b/p}(\vr)$
refers here to the two-particle system as a whole and thus, according to the RST
philosophy, is to be conceived as the direct (Whitney) sum of the one-particle constituent
wave functions $\psi_a(\vr)\ (a=1,2)$
\begin{equation}
\label{eq:III.24}
\Psi_{b/p}(\vr) = {}^{(b/p)}\!\psi_1(\vr) \oplus {}^{(b/p)}\!\psi_2(\vr) \;.
\end{equation}

According to this sum structure, the total angular momentum $\Jhz$ is also the sum of the individual angular momenta
\begin{equation}
\label{eq:III.25}
\Jhz = {}^{(1)}\!\hat{J}_z \oplus {}^{(2)}\!\hat{J}_z \;.
\end{equation}
Therefore the eigenvalue equations (\ref{eq:III.23}) for the total angular momentum $\hat{\J}_z$ can be decomposed as follows
\begin{equation}
\label{eq:III.26}
\Jhz\,\Psi_{b/p}(\vr) = \left( {}^{(1)}\!\hat{J}_z\,\psi_1(\vr) \right) \oplus \left( {}^{(2)}\!\hat{J}_z\,\psi_2(\vr) \right) \;.
\end{equation}
Consequently both one-particle spinors $\psi_1(\vr)$ and $\psi_2(\vr)$ must obey the same eigenvalue equation as the total wave function (\ref{eq:III.24}), i.\,e.
\begin{subequations}
\begin{align}
\label{eq:III.27a}
{}^{(1)}\!\hat{J}_z\,{}^{(b/p)}\!\psi_1(\vr) &= {}^{(b/p)}\!j_z\,\hbar \cdot {}^{(b/p)}\!\psi_1(\vr) \\
\label{eq:III.27b}
{}^{(2)}\!\hat{J}_z\,{}^{(b/p)}\!\psi_2(\vr) &= {}^{(b/p)}\!j_z\,\hbar \cdot {}^{(b/p)}\!\psi_2(\vr) \;.
\end{align}
\end{subequations}

Furthermore, the Dirac four-spinors $\psi_a(\vr)$ can also be conceived as the direct sum of Pauli two-spinors $\app$ and $\apm$, i.\,e.
\begin{equation}
\label{eq:III.28}
\psi_a(\vr) = \app \oplus \apm \;.
\end{equation}
Therefore the same eigenvalue equation must also hold for the individual Pauli spinors, especially after those identifications (\ref{eq:III.10a})--(\ref{eq:III.10b}):
\begin{subequations}
\begin{align}
\label{eq:III.29a}
\Jlhzp\;\bpxpp &= \bpx j_z\, \hbar\; \bpxpp \\
\label{eq:III.29b}
\Jlhzm\;\bpxpm &= \bpx j_z\, \hbar\; \bpxpm \;.
\end{align}
\end{subequations}
This means mathematically that the eigenvalue $\bpx j_z$ of the two-particle state
$\Psi_{b/p}$ (\ref{eq:III.23}) becomes transferred to any \emph{individual} Pauli
component of this two-particle state! In physical terms, the bosonic or fermionic character
of the two-particle state $\Psi$ becomes thus incorporated in any individual constituent
of the two-particle system.

But now it is clear that positronium as a whole carries bosonic properties ($\leadsto$
$j_z$ is integer-valued); and this must therefore hold also for any Pauli constituent
$\appm$ of both particles $(a = 1,2)$, cf. (\ref{eq:III.29a})--(\ref{eq:III.29b}). On the
other hand, it is well known that the Pauli spinors do form a half-integer representation
of the rotation group SO(3). This means that one can select in any two-dimensional spinor
space a certain spinor basis $\left\{ \zetaejm \right\}$ with the following eigenvalue
properties:
\begin{subequations}
\begin{align}
\label{eq:III.30a}
\hat{\vec{J}}\,^2\;\zetaejm &= j(j+1) \hbar^2\,\zetaejm \\
\label{eq:III.30b}
\hat{J}_z\;\zetaejm &= m \hbar\,\zetaejm \\
\label{eq:III.30c}
\hat{\vec{L}}\,^2\;\zetaejm &= \ell(\ell+1) \hbar^2\,\zetaejm \\
\label{eq:III.30d}
\hat{\vec{S}}\,^2\;\zetaejm &= s(s+1) \hbar^2\,\zetaejm \;.
\end{align}
\end{subequations}
Here the electron/positron spin is $s=\frac{1}{2}$; the orbital angular momentum is $\ell = 0,1,2,3,\ldots$ and thus the lowest possible value of $j(=\ell \pm s)$ is $j=\frac{1}{2}$ with $\ell=0$ or $\ell=1$. Therefore we have two basis systems for $j=\frac{1}{2}$, namely $\left\{ \zetapp_0; \zetapm_0 \right\}$ and $\left\{ \zetapp_1; \zetapm_1 \right\}$.

\begin{center}
  \large{\textit{Fermionic States}}
\end{center}

The existence of these two basis systems admits us to decompose now a \emph{fermionic} state in the following way
\begin{subequations}
\begin{align}
\label{eq:III.31a}
\app &= \aMRp(\vr) \cdot \zetapp_0 + \aMSp(\vr) \cdot \zetapm_0 \\
\label{eq:III.31b}
\apm &= -i \left\{ \aMRm(\vr) \cdot \zetapp_1 + \aMSm(\vr) \cdot \zetapm_1 \right\} \;.
\end{align}
\end{subequations}
The action of the angular momentum operator $\Jlhz$ on such a fermionic state is obviously
\begin{subequations}
\begin{align}
\label{eq:III.32a}
\Jlhzp\,\app &= \left( \Llhz\,\aMRp(\vr) \right) \cdot \zetapp_0 + \left( \Llhz\,\aMSp(\vr) \right) \cdot \zetapm_0 \\
&\quad + \aMRp(\vr) \cdot \left( \Jlhzp\,\zetapp_0 \right) + \aMSp(\vr) \cdot \left( \Jlhzp\,\zetapm_0 \right) \nonumber \\
&= \left[ \left( \Llhz + \frac{\hbar}{2} \right) \aMRp(\vr) \right] \cdot \zetapp_0 + \left[ \left( \Llhz - \frac{\hbar}{2} \right) \aMSp(\vr) \right] \cdot \zetapm_0 \nonumber \quad\quad \\[1em]
\label{eq:III.32b}
\Jlhzm\,\apm &= -i \left( \Llhz\,\aMRm(\vr) \right) \cdot \zetapp_1 - i \left( \Llhz\,\aMSm(\vr) \right) \cdot \zeta_1^{\frac{1}{2},-\frac{1}{2}} \\
&\quad - i \, \aMRm(\vr) \cdot \left( \Jlhzm\,\zetapp_1 \right) - i \, \aMSm(\vr) \cdot \left( \Jlhzm\,\zetapm_1 \right) \nonumber \\
&= -i \left[ \left( \Llhz + \frac{\hbar}{2} \right) \aMRm(\vr) \right] \cdot \zetapp_1 - i \left[ \left( \Llhz - \frac{\hbar}{2} \right) \aMSm(\vr) \right] \cdot \zetapm_1 \;. \nonumber \quad\quad
\end{align}
\end{subequations}
The required results (\ref{eq:III.29a})--(\ref{eq:III.29b}) of the action of the angular momentum operator $\Jlhz$ on the Pauli spinors $\appm$ (with $a=1,2$ or $a=b/p$) are now deducible from the present equations (\ref{eq:III.32a})--(\ref{eq:III.32b}) by making the following arrangements:
\begin{subequations}
\begin{align}
\label{eq:III.33a}
\Llhz\,\aMRpm(\vr) &= \elz \hbar \cdot \aMRpm(\vr) \\
\label{eq:III.33b}
\Llhz\,\aMSpm(\vr) &= (\elz + 1) \hbar \cdot \aMSpm(\vr) \;.
\end{align}
\end{subequations}
Indeed, with these disposals the equations (\ref{eq:III.32a})--(\ref{eq:III.32b}) adopt the required form of the eigenvalue equations (\ref{eq:III.29a})--(\ref{eq:III.29b}) with the eigenvalue of angular momentum being found as
\begin{equation}
\label{eq:III.34}
\ajz = \elz + \frac{1}{2} \;.
\end{equation}
Since the quantum number of orbital angular momentum is adopted as integer $(\elz = 0, \pm 1, \pm 2, \pm 3, \ldots)$, we actually end up with half-integer quantum numbers $\ajz$ (\ref{eq:III.34}) for \emph{fermionic} states!

These fermionic states can obviously be realized by use of \emph{unique} amplitude fields $\aMRpm(\vr), \aMSpm(\vr)$ and also \emph{unique} spinor basis fields $\zetappm_0, \zetappm_1$. Evidently, the latter fields work as the carriers of the spin, whereas the amplitude fields do contribute the orbital angular momentum. If this philosophy is tried also for the bosonic states we are forced to give up the uniqueness of the spinor basis!

\begin{center}
  \large{\textit{Bosonic States}}
\end{center}

\begin{sloppypar}
  Joining here the general conviction that bosonic states should have integer quantum
  numbers $\ajz$ (\ref{eq:III.29a})--(\ref{eq:III.29b}), i.\,e. $\ajz = 0, \pm 1, \pm 2,
  \pm 3, \ldots$, it suggests itself to think that the amplitude fields should furthermore
  carry integer quantum numbers $(\elz)$ of orbital angular momentum; i.\,e. such
  equations as (\ref{eq:III.33a})--(\ref{eq:III.33b}) should persist also for the bosonic
  states. Thus the necessary modification must refer to the basis spinor fields $\zetaejm$
  (\ref{eq:III.30a})--(\ref{eq:III.30d}). More concretely, we think of four basis spinor
  fields $\xip_0, \xim_0, \xip_1, \xim_1$ which for \textbf{para-positronium} obey the
  following eigenvalue equations
\begin{equation}
\label{eq:III.35}
\Jlhzp\,\xip_0 = \Jlhzp\,\xim_0 = \Jlhzm\,\xip_1 = \Jlhzm\,\xim_1 = 0 \;;
\end{equation}
and similarly for \textbf{ortho-positronium} one wishes to work with a spinor basis $\etap_0, \etam_0, \etap_1, \etam_1$ of the following kind:
\begin{subequations}
\begin{align}
\label{eq:III.36a}
\Jlhzp\,\etap_0 &= \hbar\,\etap_0 \\
\label{eq:III.36b}
\Jlhzp\,\etam_0 &= - \hbar\,\etam_0 \\
\label{eq:III.36c}
\Jlhzm\,\etap_1 &= \hbar\,\etap_1 \\
\label{eq:III.36d}
\Jlhzm\,\etam_1 &= - \hbar\,\etam_1 \;.
\end{align}
\end{subequations}
This says that for the para-case (\ref{eq:III.35}) the spins $\spins_a\
\left(=\frac{1}{2}\right)$ of both constituent particles ($a=1,2$) do combine to zero spin
quantum number $\spins_\mathcal{P}$ of the para-type $(\spins_\mathcal{P} \doteqdot
\spins_1 - \spins_2 = 0,$ $\spins_1 = \spins_2 = \frac{1}{2})$; whereas for the ortho-case
(\ref{eq:III.36a})--(\ref{eq:III.36d}) the individual spins combine to unity
$(\spins_\mathcal{O} \doteqdot \spins_1 + \spins_2 = 1)$. Thus both basis systems
(\ref{eq:III.35}) and (\ref{eq:III.36a})--(\ref{eq:III.36d}) carry integer spin and
therefore may be used for the corresponding decomposition of the Pauli spinors $\appm$ due
to a bound two-particle system.
\end{sloppypar}

For para-positronium $(\spins_\mathcal{P} = 0)$ one has now in place of the fermionic
situation (\ref{eq:III.31a})--(\ref{eq:III.31b}) the following decomposition
\begin{subequations}
\begin{align}
\label{eq:III.37a}
\pxpp &= \pMRp(\vr) \cdot \xip_0 + \pMSp(\vr) \cdot \xim_0 \\
\label{eq:III.37b}
\pxpm &= -i \left\{ \pMRm(\vr) \cdot \xip_1 + \pMSm(\vr) \cdot \xim_1 \right\} \;.
\end{align}
\end{subequations}
Here, the action of the angular momentum operator $\Jlhz$ looks now as follows
\begin{subequations}
\begin{align}
\label{eq:III.38a}
\Jlhzp\,\pxpp &= \left( \Llhz\,\pMRp(\vr) \right) \cdot \xip_0 + \left( \Llhz\,\pMSp(\vr) \right) \cdot \xim_0 \\
\label{eq:III.38b}
\Jlhzm\,\pxpm &= -i \left( \Llhz\,\pMRm(\vr) \right) \cdot \xip_1 - i \left( \Llhz\,\pMSm(\vr) \right) \cdot \xim_1 \;.
\end{align}
\end{subequations}
Consequently in order to satisfy again the para-form ($p$) of the eigenvalue equations (\ref{eq:III.29a})--(\ref{eq:III.29b}) one puts here
\begin{subequations}
\begin{align}
\label{eq:III.39a}
\Llhz\,\pMRpm(\vr) &= \elz\,\hbar \cdot \pMRpm(\vr) \\
\label{eq:III.39b}
\Llhz\,\pMSpm(\vr) &= \elz\,\hbar \cdot \pMSpm(\vr) \;,
\end{align}
\end{subequations}
so that the quantum number $\pjz$ (\ref{eq:III.29a})--(\ref{eq:III.29b}) is solely due to \emph{orbital} angular momentum:
\begin{gather}
\label{eq:III.40}
\pjz \equiv \elz \\
(\elz = 0, \pm 1, \pm 2, \pm 3, \ldots) \nonumber
\end{gather}

For ortho-positronium $(\spins_\mathcal{O} = 1)$, the situation is somewhat different. First, the decomposition of the ortho-spinors $\bppm$ looks quite similar to the para-case (\ref{eq:III.37a})--(\ref{eq:III.37b}):
\begin{subequations}
\begin{align}
\label{eq:III.41a}
\bpp &= \bMRp(\vr) \cdot \etap_0 + \bMSp(\vr) \cdot \etam_0 \\
\label{eq:III.41b}
\bpm &= -i \left\{ \bMRm(\vr) \cdot \etap_1 + \bMSm(\vr) \cdot \etam_1 \right\} \;.
\end{align}
\end{subequations}
But the action of the angular momentum operator $\Jlhzpm$ on these ortho-states looks now as follows
\begin{subequations}
\begin{align}
\label{eq:III.42a}
\Jlhzp\,\bpp &= \left[ \left( \Llhz + \hbar \right) \bMRp(\vr) \right] \cdot \etap_0 + \left[ \left( \Llhz - \hbar \right) \bMSp(\vr) \right] \cdot \etam_0 \\
\label{eq:III.42b}
\Jlhzm\,\bpm &= -i \left\{ \left[ \left( \Llhz + \hbar \right) \bMRm(\vr) \right] \cdot \etap_1 + \left[ \left( \Llhz - \hbar \right) \bMSm(\vr) \right] \cdot \etam_1 \right\} \;.
\end{align}
\end{subequations}
For satisfying here again the eigenvalue requirement (\ref{eq:III.29a})--(\ref{eq:III.29b}) in its ortho-form ($b$) one puts in a self-evident way
\begin{subequations}
\begin{align}
\label{eq:III.43a}
\Llhz\,\bMRpm &= \elz\,\hbar \cdot \bMRpm \\
\label{eq:III.43b}
\Llhz\,\bMSpm &= (\elz+2)\,\hbar \cdot \bMSpm \;.
\end{align}
\end{subequations}
This arrangement yields then the following eigenvalue equations for the ortho-spinors
\begin{equation}
\label{eq:III.44}
\hat{J}_z^{\pm} \,\bppm = (\elz + 1)\,\hbar \cdot \bppm \;.
\end{equation}
Thus one finds the quantum numbers of the ortho-system:
\begin{gather}
\label{eq:III.45}
\bjz = \elz + 1 \\
(\bjz = 0, \pm 1, \pm 2, \pm 3, \ldots) \;. \nonumber
\end{gather}

\begin{sloppypar}
For a realization of the required basis spinors one takes the fermionic basis $\left\{ \zetappm_0, \zetappm_1 \right\}$ as the point of departure and introduces a general spinor basis $\left\{ \omp_0, \omm_0, \omp_1, \omm_1 \right\}$ through
\begin{subequations}
\begin{align}
\label{eq:III.46a}
\omp_0 &= \e^{-i \bbar \phi} \cdot \zetapp_0 \\
\label{eq:III.46b}
\omm_0 &= \e^{i \bbar \phi} \cdot \zetapm_0 \\
\label{eq:III.46c}
\omp_1 &= \e^{-i \bbar \phi} \cdot \zetapp_1 \\
\label{eq:III.46d}
\omm_1 &= \e^{i \bbar \phi} \cdot \zetapm_1 \;.
\end{align}
\end{subequations}
Here it is easy to see emerging the following relations concerning angular momentum
\begin{subequations}
\begin{align}
\label{eq:III.47a}
\Jlhzp\,\omp_0 &= - \left( \bbar - \frac{1}{2} \right)\,\hbar \cdot \omp_0 \\
\label{eq:III.47b}
\Jlhzp\,\omm_0 &= \left( \bbar - \frac{1}{2} \right)\,\hbar \cdot \omm_0 \\
\label{eq:III.47c}
\Jlhzm\,\omp_1 &= - \left( \bbar - \frac{1}{2} \right)\,\hbar \cdot \omp_1 \\
\label{eq:III.47d}
\Jlhzm\,\omm_1 &= \left( \bbar - \frac{1}{2} \right)\,\hbar \cdot \omm_1 \;.
\end{align}
\end{subequations}
Obviously, this $\omega$-basis contains some free parameter $\bbar$ (i.\,e. the
\emph{boson number}) and if this is chosen as $\bbar = \frac{1}{2}$ we obtain the
desired $\xi$-basis (\ref{eq:III.35}) for para-positronium; and if $\bbar$ is chosen as
$\bbar = -\frac{1}{2}$ we obtain the $\eta$-basis (\ref{eq:III.36a})--(\ref{eq:III.36d})
for ortho-positronium. For $\bbar = 0$ we get back the purely fermionic $\zeta$-basis
(\ref{eq:III.30a})--(\ref{eq:III.30d}).
\end{sloppypar}

With the choice of the $\omega$-basis (\ref{eq:III.46a})--(\ref{eq:III.46d}) the loss of
uniqueness becomes now evident: since the original $\zeta$-basis
(\ref{eq:III.30a})--(\ref{eq:III.30d}) is \emph{unique} (i.\,e. the $\zetaejm(\vartheta,
\phi)$ constitute a ``unique'' spinor field on the 2-sphere $S^2$), the other two basis
systems $\xi$ (\ref{eq:III.35}) and $\eta$ (\ref{eq:III.36a})--(\ref{eq:III.36d}) are
double-valued. More concretely, for both the $\xi$- and the $\eta$-basis $\left( \leadsto
  \bbar = \pm \frac{1}{2} \right)$ one finds by performing one revolution around the
$z$-axis $\left( 0 \leq \phi \leq 2\pi \right)$:
\begin{subequations}
\begin{align}
\label{eq:III.48a}
\xi^{(\pm)}_{0,1}(\phi + 2\pi) &= \e^{\pm i \pi} \cdot \xi^{(\pm)}_{0,1}(\phi) = -\xi^{(\pm)}_{0,1} \\
\label{eq:III.48b}
\eta^{(\pm)}_{0,1}(\phi + 2\pi) &= \e^{\pm i \pi} \cdot \eta^{(\pm)}_{0,1}(\phi) = -\eta^{(\pm)}_{0,1} \;,
\end{align}
\end{subequations}
and this says that we need two revolutions around the $z$-axis $\left( 0 \leq \phi \leq
  4\pi \right)$ in order to return to the original basis configurations. Since we adopt
all the amplitude fields $\bpMRpm(\vr), \bpMSpm(\vr)$ to be unique scalar fields, the
double-valuedness of the para- and ortho-basis becomes transferred to the para-spinors
$\pxppm$ (\ref{eq:III.37a})--(\ref{eq:III.37b}) and ortho-spinors $\bppm$
(\ref{eq:III.41a})--(\ref{eq:III.41b}) and from here ultimately to the Dirac spinors
$\PsiPO$
\begin{subequations}
\begin{align}
\label{eq:III.49a}
\PsiP(\vr) &= \pxpp \oplus \pxpm \\
\label{eq:III.49b}
\PsiO(\vr) &= \bpp \oplus \bpm \;.
\end{align}
\end{subequations}
Summarizing, the Dirac wave functions $\PsiP(\vr)$ and $\PsiO(\vr)$ for ortho- and
para-positronium must in RST be double-valued in the following sense:
\begin{subequations}
\begin{align}
\label{eq:III.50a}
\PsiP(r, \vartheta, \phi+2\pi) &= - \PsiP(r, \vartheta, \phi) \\
\label{eq:III.50b}
\PsiO(r, \vartheta, \phi+2\pi) &= - \PsiO(r, \vartheta, \phi) \;.
\end{align}
\end{subequations}
\pagebreak
\begin{center}
  \large{\textit{Uniqueness of the Physical Densities}}
\end{center}

Naturally, in a fluid-dynamic theory (such as the present RST) the proper observable
objects are the physical densities, such as those of charge, current, energy, linear and
angular momentum, etc. A plausible condition on these densities is surely given by the
demand that these physical densities should be \emph{single-valued} tensor fields. But, as
we will readily demonstrate, this condition can be satisfied also by \emph{non-unique}
wave functions; and this fact allows us to actually deal with such non-unique wave
functions, as given for example by the double-valued positronium states
(\ref{eq:III.50a})--(\ref{eq:III.50b}). Therefore one wishes to have some condition on the
wave functions which on the one hand admits their non-uniqueness but on the other hand
ensures the uniqueness of the associated physical densities!

Now according to the present RST philosophy, the non-uniqueness of the (Dirac) wave
functions is to be traced back to the spinor basis, whereas the amplitude fields are
furthermore required to be unique. Therefore the non-uniqueness of the wave functions is
measured by the boson number $\bbar$, cf. (\ref{eq:III.46a})--(\ref{eq:III.46d}); and thus
the condition of uniqueness of the densities is to be retraced to some condition for the
fixation of the boson number $\bbar$. Such a fixation may be attained now by considering
specifically the Dirac density $\akn$ and the three-current $\ak(\vr)$ which read in terms
of the Pauli spinors $\appm$ as shown by equations
(\ref{eq:II.55a})--(\ref{eq:II.55b}). Decomposing these Pauli spinors with respect to the
$\omega$-basis (\ref{eq:III.46a})--(\ref{eq:III.46d}) lets then appear the (Dirac) charge
densities (\ref{eq:II.55a}) in the following form:
\vspace{0.5em}
\begin{equation}
\label{eq:III.51}
\akn = \frac{\aMRpS \cdot \aMRp + \aMSpS \cdot \aMSp + \aMRmS \cdot \aMRm + \aMSmS \cdot \aMSm}{4\pi} \;.
\vspace{0.5em}
\end{equation}
Evidently, these charge densities are unique in any case and therefore do not yet provide an immediate handle for fixing the parameter $\bbar$.

This situation changes now when one considers also the Dirac currents $\ak$ (\ref{eq:II.55b}), which by their very definitions are always real-valued objects:
\vspace{0.5em}
\begin{subequations}
\begin{align}
\label{eq:III.52a}
\akr &= \frac{i}{4\pi} \left\{ \aMRpS \cdot \aMRm + \aMSpS \cdot \aMSm - \aMRmS \cdot \aMRp - \aMSmS \cdot \aMSp \right\} \\[0.5em]
\label{eq:III.52b}
\akt &= - \frac{i}{4\pi} \left\{ \e^{2i(\bbar - \frac{1}{2})\phi} \cdot \;\MCa - \e^{-2i(\bbar - \frac{1}{2})\phi} \cdot \;\MCaS \right\} \\
&( \MCa \doteqdot \aMRpS \cdot \aMSm + \aMRmS \cdot \aMSp ) \nonumber \\[0.5em]
\label{eq:III.52c}
\akp &= \frac{\sin \vartheta}{4\pi} \left\{ \aMRpS \cdot \aMRm + \aMRmS \cdot \aMRp - \aMSpS \cdot \aMSm - \aMSmS \cdot \aMSp \right\} \\
&\quad - \frac{\cos \vartheta}{4\pi} \left\{ \e^{2i(\bbar - \frac{1}{2})\phi} \cdot \;\MCa + \e^{-2i(\bbar - \frac{1}{2})\phi} \cdot \;\MCaS \right\} \;. \nonumber
\end{align}
\end{subequations}
But here a nearby restriction upon the parameter $\bbar$ suggests itself, namely through the plausible demand that the Dirac currents $\ak$ (\ref{eq:II.55b}), with their components being specified by (\ref{eq:III.52a})--(\ref{eq:III.52c}), must be \emph{unique} (!) vector fields over three-space (albeit only apart from the origin $r = 0$ and the $z$ axis $\vartheta = 0,\pi$). Evidently this demand of uniqueness reads in terms of the spherical polar coordinates $\{ r, \vartheta, \phi \}$
\begin{equation}
\label{eq:III.53}
\ak(r, \vartheta, \phi + 2\pi) = \ak(r, \vartheta, \phi) \;,
\end{equation}
and thus the values of $\bbar$ become restricted to the range
\begin{gather}
\label{eq:III.54}
\bbar = \frac{1}{2} \left( n + 1 \right) \\
(n = 0, \pm 1, \pm 2, \pm 3, \ldots) \nonumber
\end{gather}
which then entails also \emph{(half-)integer} quantum numbers for the $z$ component of the angular momentum (\ref{eq:III.47a})--(\ref{eq:III.47d}) of the spinor basis:
\begin{equation}
\label{eq:III.55}
s_z = \pm \left( \bbar - \frac{1}{2} \right) = \pm \frac{n}{2} \;.
\end{equation}
Here it will suffice to admit for the spinor basis (of the two-particle systems)
exclusively the values $\bbar=\pm\frac{1}{2}$; other values of~$\bbar$ come into play for
bound systems of more than two fermions.

Notice here that this \emph{(half-)integrity} arises as a consequence of the demand of
uniqueness with respect to certain physical \emph{densities} (i.\,e. Dirac current),
whereas the corresponding \emph{integral} quantum numbers of conventional non-relativistic
quantum mechanics are mostly traced back in the textbooks to the uniqueness requirement
for the \emph{wave functions} themselves (not the \emph{densities}). The lowest values of
$s_z$ (\ref{eq:III.55}) are $s_z = \pm\frac{1}{2}$ for $\bbar = 0$ and $s_z = 0, \pm 1$
for $\bbar = \pm \frac{1}{2}$. Thus for the first case $(\bbar = 0)$ we have a
\emph{fermionic basis} and for the second case $\left( \bbar = \pm\frac{1}{2} \right)$ one
deals with a \emph{bosonic basis}. In this sense, a Dirac particle is said to occupy a
fermionic quantum state $\psi$ if the ``\emph{boson number}'' $\bbar$ of its spinor basis
is zero ($\bbar = 0$), and a bosonic quantum state if the boson number $\bbar$ equals
$\pm\frac{1}{2}$. Observe that through this arrangement the fermionic or bosonic character
of the quantum state of a Dirac particle is defined by reference to the corresponding
spinor \emph{basis}.

\section{Para-Positronium ($\bbar = \frac{1}{2}$)}
\indent

Though the best perspective on the ortho/para-dichotomy would surely result from a
detailed synopsis of both configurations, it is nevertheless very instructive to first
consider the case of para-positronium separately. The interesting point with the latter
kind of positronium is here that the situation is much simpler than for the
ortho-counterpart, so that a treatment of the para-case appears as an ideal preparation to
the later discussion of ortho-positronium which itself decays into two subcases
(``\emph{dimorphism}''). Furthermore, the simpler case of para-positronium is also
well-suited in order to present the quantization of angular momentum in RST from a more
general viewpoint.

\begin{center}
  \emph{\textbf{1.\ Mass Eigenvalue Equations in Terms of Amplitude Fields}}
\end{center}

Once the para-spinors $\pxppm$ are decomposed with respect to a bosonic basis,
cf. (\ref{eq:III.37a})--(\ref{eq:III.37b}), one can substitute this decomposition into the
two original eigenvalue equations in spinor form,
cf. (\ref{eq:III.16})--(\ref{eq:III.17}), and will then obtain the corresponding
eigenvalue equations in terms of the amplitude fields $\pMRpm(\vr)$ and $\pMSpm(\vr)$. But
since both eigenvalue equations are assumed to yield physically equivalent states, there
must exist a peculiar relationship between both eigenvalue equations. It is
this relationship which first must be worked out.

The identification (\ref{eq:III.10a})--(\ref{eq:III.10b}) of the positron spinors
${}^{(1)}\!\varphi_\pm(\vr)$ with the para-spinors $\pxppm$ reads in terms of the
amplitude fields
\begin{subequations}
\begin{align}
\label{eq:IV.1a}
{}^{(1)}\!\R_\pm(\vr) &\doteqdot \pMRpm(\vr) \\
\label{eq:IV.1b}
{}^{(1)}\!\mathcal{S}_\pm(\vr) &\doteqdot \pMSpm(\vr)
\end{align}
\end{subequations}
so that the spinor form (\ref{eq:III.16}) of the eigenvalue equation reappears in terms of the amplitude fields as
\begin{subequations}
\begin{align}
\label{eq:IV.2a}
&\frac{\partial\,\pMRp}{\partial r} + \frac{i}{r} \cdot \frac{\partial\,\pMRp}{\partial \phi} + \frac{1}{2r} \cdot \pMRp - {}^{(p)}\!A_0 \cdot \pMRm \\
&+ \frac{1}{r} \cdot \frac{\partial\,\pMSp}{\partial \vartheta} + \frac{\cot \vartheta}{r} \left[ \frac{1}{2}\,\pMSp - i\,\frac{\partial\,\pMSp}{\partial \phi} \right] = \frac{M+M_*}{\hbar}\,\crm \cdot \pMRm \nonumber \\[1em]
\label{eq:IV.2b}
&\frac{\partial\,\pMSp}{\partial r} - \frac{i}{r} \cdot \frac{\partial\,\pMSp}{\partial \phi} + \frac{1}{2r} \cdot \pMSp - {}^{(p)}\!A_0 \cdot \pMSm \\
&- \frac{1}{r} \cdot \frac{\partial\,\pMRp}{\partial \vartheta} - \frac{\cot \vartheta}{r} \left[ \frac{1}{2}\,\pMRp + i\,\frac{\partial\,\pMRp}{\partial \phi} \right] = \frac{M+M_*}{\hbar}\,\crm \cdot \pMSm \nonumber \\[1em]
\label{eq:IV.2c}
&\frac{\partial\,\pMRm}{\partial r} - \frac{i}{r} \cdot \frac{\partial\,\pMRm}{\partial \phi} + \frac{\frac{3}{2}}{r} \cdot \pMRm + {}^{(p)}\!A_0 \cdot \pMRp \\
&- \frac{1}{r} \cdot \frac{\partial\,\pMSm}{\partial \vartheta} - \frac{\cot \vartheta}{r} \left[ \frac{1}{2}\,\pMSm - i\,\frac{\partial\,\pMSm}{\partial \phi} \right] = \frac{M-M_*}{\hbar}\,\crm \cdot \pMRp \nonumber \\[1em]
\label{eq:IV.2d}
&\frac{\partial\,\pMSm}{\partial r} + \frac{i}{r} \cdot \frac{\partial\,\pMSm}{\partial \phi} + \frac{\frac{3}{2}}{r} \cdot \pMSm + {}^{(p)}\!A_0 \cdot \pMSp \\
&+ \frac{1}{r} \cdot \frac{\partial\,\pMRm}{\partial \vartheta} + \frac{\cot \vartheta}{r} \left[ \frac{1}{2}\,\pMRm + i\,\frac{\partial\,\pMRm}{\partial \phi} \right] = \frac{M-M_*}{\hbar}\,\crm \cdot \pMSp \nonumber \;.
\end{align}
\end{subequations}
On the other hand, those identifications (\ref{eq:III.10a})--(\ref{eq:III.10b}) of the
electronic spinors ${}^{(2)}\!\varphi_\pm(\vr)$ with the para-spinors $\pxppm$ read in
terms of the amplitude fields
\begin{subequations}
\begin{align}
\label{eq:IV.3a}
{}^{(2)}\!\R_\pm(\vr) &= -i\,\pMRpm(\vr) \\
\label{eq:IV.3b}
{}^{(2)}\!\mathcal{S}_\pm(\vr) &= -i\,\pMSpm(\vr)
\end{align}
\end{subequations}
so that the spinor form (\ref{eq:III.17}) of the electron equation reappears now from
(\ref{eq:IV.2a})--(\ref{eq:IV.2d}) by means of the replacements
(\ref{eq:III.18a})--(\ref{eq:III.18c}).

In order to get some information about the peculiarities of this para-system
(\ref{eq:IV.2a})--(\ref{eq:IV.2d}) one first recalls the hypothesis that the spin
$\spins_p$ of the constituents should be zero, cf. (\ref{eq:III.35}), so that their total
angular momentum is of purely orbital nature, cf. (\ref{eq:III.40}). Since, according to
our hypothesis, the orbital part of total angular momentum is carried by the amplitude
fields, we may fix their general form as follows:
\begin{subequations}
\begin{align}
\label{eq:IV.4a}
\pMRpm(\vr) &= \frac{\e^{i\elz\phi}}{\sqrt{r \sin \vartheta}} \cdot \pSRpm(r, \vartheta) \\
\label{eq:IV.4b}
\pMSpm(\vr) &= \frac{\e^{i\elz\phi}}{\sqrt{r \sin \vartheta}} \cdot \pSSpm(r, \vartheta) \;.
\end{align}
\end{subequations}

If this form is substituted into the present eigenvalue equations (\ref{eq:IV.2a})--(\ref{eq:IV.2d}), this system reappears in terms of the real-valued amplitudes $\pSRpm(r, \vartheta)$ and $\pSSpm(r, \vartheta)$
\begin{subequations}
\begin{align}
\label{eq:IV.5a}
\frac{\partial}{\partial r} \pSRp - \frac{\elz}{r} \cdot \pSRp - {}^{(p)}\!A_0 \cdot \pSRm + \frac{1}{r} \left[ \frac{\partial}{\partial \vartheta} \pSSp + \elz \cot \vartheta \cdot \pSSp \right] \\
= \frac{M+M_*}{\hbar}\,\crm \cdot \pSRm \nonumber \\[1em]
\label{eq:IV.5b}
\frac{\partial}{\partial r} \pSSp + \frac{\elz}{r} \cdot \pSSp - {}^{(p)}\!A_0 \cdot \pSSm - \frac{1}{r} \left[ \frac{\partial}{\partial \vartheta} \pSRp - \elz \cot \vartheta \cdot \pSRp \right] \\
= \frac{M+M_*}{\hbar}\,\crm \cdot \pSSm \nonumber \\[1em]
\label{eq:IV.5c}
\frac{\partial}{\partial r} \pSRm + \frac{1+\elz}{r} \cdot \pSRm + {}^{(p)}\!A_0 \cdot \pSRp - \frac{1}{r} \left[ \frac{\partial}{\partial \vartheta} \pSSm + \elz \cot \vartheta \cdot \pSSm \right] \\
= \frac{M-M_*}{\hbar}\,\crm \cdot \pSRp \nonumber \\[1em]
\label{eq:IV.5d}
\frac{\partial}{\partial r} \pSSm + \frac{1-\elz}{r} \cdot \pSSm + {}^{(p)}\!A_0 \cdot \pSSp + \frac{1}{r} \left[ \frac{\partial}{\partial \vartheta} \pSRm - \elz \cot \vartheta \cdot \pSRm \right] \\
= \frac{M-M_*}{\hbar}\,\crm \cdot \pSSp \nonumber \;.
\end{align}
\end{subequations}

This eigenvalue problem may perhaps look somewhat complicated, but the assumption of a spherically symmetric potential ${}^{(p)}\!A_0(r)$ ($\leadsto$ \emph{spherically symmetric approximation}) admits us to separate it into an angular and a radial problem. For such a separation, one first resorts to the following product ansatz
\begin{subequations}
\begin{align}
\label{eq:IV.6a}
\pSRpm(r, \vartheta) &= f_R(\vartheta) \cdot \,'\!R_\pm(r) \\
\label{eq:IV.6b}
\pSSpm(r, \vartheta) &= f_S(\vartheta) \cdot \,'\!S_\pm(r) \;,
\end{align}
\end{subequations}
and furthermore one makes the following identifications
\begin{subequations}
\begin{align}
\label{eq:IV.7a}
'\!R_+(r) &\equiv\;'\!S_+(r) \doteqdot \SPp(r) \\
\label{eq:IV.7b}
'\!R_-(r) &\equiv\;'\!S_-(r) \doteqdot \SPm(r) \;,
\end{align}
\end{subequations}
which then ultimately separates the mass eigenvalue system (\ref{eq:IV.5a})--(\ref{eq:IV.5d}) into an angular problem
\begin{subequations}
\begin{align}
\label{eq:IV.8a}
\frac{d\,f_R(\vartheta)}{d \vartheta} - \elz \cot \vartheta \cdot f_R(\vartheta) = \left( \lP + \elz  \right) \cdot f_S(\vartheta) \\
\label{eq:IV.8b}
\frac{d\,f_S(\vartheta)}{d \vartheta} + \elz \cot \vartheta \cdot f_S(\vartheta) = \left( \elz - \lP  \right) \cdot f_R(\vartheta)
\end{align}
\end{subequations}
and a purely radial problem:
\begin{subequations}
\begin{align}
\label{eq:IV.9a}
\frac{d\,\SPp(r)}{dr} - \frac{\lP}{r} \cdot \SPp(r) - {}^{(p)}\!A_0(r) \cdot \SPm(r) &= \frac{M+M_*}{\hbar}\,\crm \cdot \SPm(r) \\
\label{eq:IV.9b}
\frac{d\,\SPm(r)}{dr} + \frac{1+\lP}{r} \cdot \SPm(r) + {}^{(p)}\!A_0(r) \cdot \SPp(r) &= \frac{M-M_*}{\hbar}\,\crm \cdot \SPp(r) \;.
\end{align}
\end{subequations}

Observe here that the eigenvalue $\elz$ of orbital angular momentum does not enter the
radial problem (\ref{eq:IV.9a})--(\ref{eq:IV.9b}) which ensures the occurence of energy
degeneracy within the set of solutions belonging all to the same quantum number
$\ell_\mathcal{P}$ but to different values of $\elz$, \emph{provided the interaction
  potential ${}^{(p)}\!A_0(r)$ is fixed from the outside.} However, for our present
two-body problem the interaction potential must obey the Poisson equation
(\ref{eq:III.21}), which for the present case of para-positronium adopts the following
form:
\begin{align}
\label{eq:IV.10}
\Delta\,{}^{(p)}\!A_0(\vr) &= -4\pi\as\,{}^{(p)}\!k_0(\vr) \\
&= -\as \frac{f_R^2(\vartheta) + f_S^2(\vartheta)}{\sin \vartheta} \cdot \frac{\SPp^2(r) + \SPm^2(r)}{r} \;. \nonumber
\end{align}
Therefore the interaction potential $\pAn(\vr)$ depends upon the quantum state of
positronium, i.\,e. more concretely: $\pAn(\vr)$ will in general depend also upon the
quantum number $\elz$ of angular momentum; and this dependence will then be transferred
also to the binding energy of the quantum state. Thus, it is just this effect which lets
us suppose that the angular momentum degeneracy of the conventional prediction
(\ref{eq:I.2}) will be lifted in RST. Clearly, it remains to be clarified whether this
lifting is (or is not) a negligibly small effect.

But in any case we have arrived now at the definite form of the energy eigenvalue problem
for para-positronium which (in the electrostatic and spherically symmetric approximation)
evidently consists of the angular equations (\ref{eq:IV.8a})--(\ref{eq:IV.8b}), the mass
eigenvalue equations (\ref{eq:IV.9a})--(\ref{eq:IV.9b}) and the Poisson equation
(\ref{eq:IV.10}). Here it should be clear that \emph{exact} solutions of this eigenvalue
problem cannot be found so that we will have to be content with the elaboration of
approximate solutions which, however, are expected to display some of the qualitative
features of the unknown exact solutions. The central point of our approximation procedure
refers now to the solution $\pAn(\vr)$ of the Poisson equation (\ref{eq:IV.10}) which may
formally be written as
\begin{align}
\label{eq:IV.11}
\pAn(\vr) &= \as \int d^3 \vr\,'\,\frac{\pko(\vr\,')}{||\vr-\vr\,'||} \\
&= \frac{\as}{4\pi} \int d\Omega' \, \frac{f_R^2(\vartheta') + f_S^2(\vartheta')}{\sin \vartheta'} \int dr'\;r' \, \frac{\SPp^2(r') + \SPm^2(r')}{||\vr - \vr\,'||} \;. \nonumber \quad\quad
\end{align}
From here it is easily seen that the interaction potential $\pAn(\vr)$ adopts the Coulomb form in the asymptotic region ($r \rightarrow \infty$)
\begin{equation}
\label{eq:IV.12}
\underset{r \rightarrow \infty}{\lim} \pAn(\vr) = \frac{\as}{r} \;,
\end{equation}
provided the Dirac charge density $\pko(\vr)$ is normalized to unity
\begin{equation}
\label{eq:IV.13}
\int d^3\vr\;\pko(\vr) = 1 \;.
\end{equation}
Since the density $\pko(\vr)$ is factorized, cf. (\ref{eq:IV.10})
\begin{equation}
\label{eq:IV.14}
\pko(\vr) = \gpko(\vartheta) \cdot \gpko(r) \;,
\end{equation}
the normalization condition (\ref{eq:IV.13}) can be passed on separately to the longitudinal and radial factors:
\begin{subequations}
\begin{align}
\label{eq:IV.15a}
\int d\Omega \, \gpko(\vartheta) &= \int \frac{d\Omega}{4\pi}\, \frac{f_R^2(\vartheta) + f_S^2{\vartheta}}{\sin \vartheta} = 1 \\
\label{eq:IV.15b}
\int\limits_0^\infty dr\;r^2 \, \gpko(r) &= \int\limits_0^\infty dr\;r\,\left[ \SPp^2(r) + \SPm^2(r) \right] = 1 \;.
\end{align}
\end{subequations}

\begin{center}
  \emph{\textbf{2.\ Angular Momentum Quantization in RST}}
\end{center}

Obviously, both angular equations (\ref{eq:IV.8a})--(\ref{eq:IV.8b}) represent a closed
eigenvalue problem for the quantum numbers $\ell_\mathcal{P}$ and $\elz$, i.\,e. the
quantization problem of angular momentum in RST. The solution of this problem must yield
the possible values of $\ell_\mathcal{P}$ and $\elz$ which then does apply quite generally
for all bound~$O(3)$-symmetric systems, since the special form of the interaction
potential $\pAn(r)$ does not enter the angular problem at all! Clearly, this is the RST
counterpart of the corresponding eigenvalue problem in conventional quantum
mechanics~\cite{10}
\begin{subequations}
\begin{align}
\label{eq:IV.16a}
\hat{\vec{L}}^2\ \left|\ell,m\right> &= \hbar^2 \ell(\ell+1)\ \left|\ell,m\right> \\
\label{eq:IV.16b}
\hat{L}_z\ \left|\ell,m\right> &= m\,\hbar\ \left|\ell,m\right> \;.
\end{align}
\end{subequations}

For a closer inspection of the coupled angular problem (\ref{eq:IV.8a})--(\ref{eq:IV.8b}) it is very instructive to decouple those equations by differentiating once more which then yields the following decoupled second-order equations
\begin{subequations}
\begin{align}
\label{eq:IV.17a}
\frac{d^2\,f_R(\vartheta)}{d\vartheta^2} + \elp^2 \cdot f_R(\vartheta) - \elz(\elz - 1) \cdot \frac{f_R(\vartheta)}{\sin^2 \vartheta} &= 0 \\
\label{eq:IV.17b}
\frac{d^2\,f_S(\vartheta)}{d\vartheta^2} + \elp^2 \cdot f_S(\vartheta) - \elz(1 + \elz) \cdot \frac{f_S(\vartheta)}{\sin^2 \vartheta} &= 0 \;.
\end{align}
\end{subequations}
Here, the first striking item refers to the fact that the substitution $\elz\ \Rightarrow\ -\elz$ entails the replacements $f_R(\vartheta)\ \Rightarrow\ f_S(\vartheta)$, $f_S(\vartheta)\ \Rightarrow\ -f_R(\vartheta)$. Of course, this pleasant property of the solutions $f_R(\vartheta), f_S(\vartheta)$ could have been deduced also from the original equations (\ref{eq:IV.8a})--(\ref{eq:IV.8b}). We will take advantage of this symmetry by solving the angular system only for a restricted number of possible values of $\elz$ (e.\,g. $\elz \geq 0$) and will deliver subsequently the solutions for the other $\elz$ by the above mentioned replacements.

Concerning now the \emph{non-singular} solutions of the decoupled system (\ref{eq:IV.17a})--(\ref{eq:IV.17b}), it suggests itself to try an ansatz in form of a \emph{finite} power series expansion with respect to $\sin\vartheta$, i.\,e. we put ($x \doteqdot \sin \vartheta$)
\begin{subequations}
\begin{align}
\label{eq:IV.18a}
f_R(\vartheta)\ &\Rightarrow\ F_R(x) \\
\label{eq:IV.18b}
f_S(\vartheta)\ &\Rightarrow\ F_S(x)
\end{align}
\end{subequations}
with
\begin{subequations}
\begin{align}
\label{eq:IV.19a}
F_R(x) &= \sum_n \rho_n\,x^n \\
\label{eq:IV.19b}
F_S(x) &= \sum_n \sigma_n\,x^n \;.
\end{align}
\end{subequations}
The range of the powers ($n$) remains to be determined by substituting this ansatz back into the second-order system (\ref{eq:IV.17a})--(\ref{eq:IV.17b}) which reappears now through the alterations (\ref{eq:IV.18a})--(\ref{eq:IV.18b}) in the following form
\begin{subequations}
\begin{align}
\label{eq:IV.20a}
(1-x^2)\,\frac{d^2\,F_R(x)}{dx^2} - x\,\frac{d\,F_R(x)}{dx} + \elp^2 \cdot F_R(x) - \elz(\elz - 1) \cdot \frac{F_R(x)}{x^2} &= 0 \\[0.7em]
\label{eq:IV.20b}
(1-x^2)\,\frac{d^2\,F_S(x)}{dx^2} - x\,\frac{d\,F_S(x)}{dx} + \elp^2 \cdot F_S(x) - \frac{\elz(1 + \elz)}{x^2} \cdot F_S(x) &= 0 \;.
\end{align}
\end{subequations}
The solutions hereof are then obtained via the recurrence formulae
\begin{subequations}
\begin{align}
\label{eq:IV.21a}
\rho_{n+2} &= \frac{n^2 - \elp^2}{(n+2)(n+1) - \elz(\elz - 1)} \cdot \rho_n \\
\label{eq:IV.21b}
\sigma_{n+2} &= \frac{n^2 - \elp^2}{(n+2)(n+1) - \elz(\elz + 1)} \cdot \sigma_n \;.
\end{align}
\end{subequations}
In order that both series (\ref{eq:IV.19a})--(\ref{eq:IV.19b}) remain finite one obviously has to impose the halt condition
\begin{gather}
\label{eq:IV.22}
n_\text{max} = \elp \\
(\elp > 0) \nonumber
\end{gather}
which fixes $\elp$ as an integer ($\elp = 0,1,2,3,4,\ldots$). Furthermore, the starting power $n_\text{min}$ is found as
\begin{equation}
\label{eq:IV.23}
n_\text{min} = \begin{cases}
\;\elz \\
\;1+\elz
\end{cases}
\text{for (\ref{eq:IV.21a})}
\end{equation}
and
\begin{equation}
\label{eq:IV.24}
n_\text{min} = \begin{cases}
\;1+\elz \\
\;-\elz
\end{cases}
\text{for (\ref{eq:IV.21b})}
\end{equation}
Therefore $\elz$ is also found to be an integer $(-\elp \leq \elz \leq \elp)$. Thus the desired angular functions $f_R(\vartheta)$ and $f_S(\vartheta)$ are ultimately found in terms of powers of trigonometric functions.

Perhaps it is most instructive to consider a simple example ($\elp = 3$, say). First, taking $\elz = 0$ simplifies both equations (\ref{eq:IV.17a})--(\ref{eq:IV.17b}) to
\begin{subequations}
\begin{align}
\label{eq:IV.25a}
\frac{d^2\,f_R(\vartheta)}{d\vartheta^2} + 9\,f_R(\vartheta) &= 0 \\
\label{eq:IV.25b}
\frac{d^2\,f_S(\vartheta)}{d\vartheta^2} + 9\,f_S(\vartheta) &= 0
\end{align}
\end{subequations}
with the obvious solutions
\begin{subequations}
\begin{align}
\label{eq:IV.26a}
f_R(\vartheta) &= \sqrt{\frac{2}{\pi}}\;\cos(3\vartheta) \\
\label{eq:IV.26b}
f_S(\vartheta) &= - \sqrt{\frac{2}{\pi}}\;\sin(3\vartheta) \;.
\end{align}
\end{subequations}
This solution satisfies the first-order system (\ref{eq:IV.8a})--(\ref{eq:IV.8b}) for $\elp = 3, \elz = 0$
\begin{subequations}
\begin{align}
\label{eq:IV.27a}
\frac{d\,f_R(\vartheta)}{d\vartheta} &= 3\,f_S(\vartheta) \\
\label{eq:IV.27b}
\frac{d\,f_S(\vartheta)}{d\vartheta} &= -3\,f_R(\vartheta) \;,
\end{align}
\end{subequations}
and moreover it satisfies also the angular normalization condition (\ref{eq:IV.15a}).

Because of its untypical simplicity, the present solution
(\ref{eq:IV.26a})--(\ref{eq:IV.26b}) for $\elz = 0, \elp = 3$ could be guessed directly
from the second-order system (\ref{eq:IV.25a})--(\ref{eq:IV.25b}) or also from the
first-order system (\ref{eq:IV.27a})--(\ref{eq:IV.27b}). But of course it must be possible
to find it also by following our systematic procedure being based upon the formal
solutions (\ref{eq:IV.19a})--(\ref{eq:IV.19b}). These may be written down beyond the present
example more generally as

\begin{subnumcases}{F_R(x) =}
\label{eq:IV.28a}
\;\sum_{n=\elz}^{\elp} \rho_n\,x^n \ \ \quad ;\quad(\elp-\elz) \Rightarrow \text{even}; \quad\elz \geq 0\\
\label{eq:IV.28b}
\;\sum_{n=1-\elz}^{\elp} \rho_n\,x^n \quad ;\quad(\elp-\elz) \Rightarrow \text{odd}; \quad\elz \leq 1 \;\ \ \ 
\end{subnumcases}
\begin{subnumcases}{F_S(x) =}
\label{eq:IV.29a}
\;\sum_{n=-\elz}^{\elp} \sigma_n\,x^n \ \quad ;\quad(\elp-\elz) \Rightarrow \text{even}; \quad\elz \leq 0\\
\label{eq:IV.29b}
\;\sum_{n=\elz+1}^{\elp} \sigma_n\,x^n \quad ;\quad(\elp-\elz) \Rightarrow \text{odd}; \quad\elz \geq -1 \;.
\end{subnumcases}
Because of $\elp = 3$ and $\elz = 0$, we have to start here with either the second-order solution $F_R(x)$ (\ref{eq:IV.28b}) and then have to construct $f_S(\vartheta)$ from the first-order equation (\ref{eq:IV.8a}), or one can start also from $F_S(x)$ (\ref{eq:IV.29b}) and then one determines $f_R(\vartheta)$ from (\ref{eq:IV.8b}).

For this situation ($\boldsymbol{\elp = 3}, \boldsymbol{\elz = 0}$) one finds from the recurrence formula (\ref{eq:IV.21b}) by starting, say, with $F_S(x)$ (\ref{eq:IV.29b})
\begin{equation}
\label{eq:IV.30}
\sigma_3 = -\frac{4}{3}\, \sigma_1
\end{equation}
so that the desired second-order solution $F_S(x)$ (\ref{eq:IV.29b}) becomes specified to
\begin{gather}
\label{eq:IV.31}
F_S(x)\ \Rightarrow\ \sigma_1 \left( x-\frac{4}{3}\,x^3 \right) \\
(\boldsymbol{\elp = 3}, \boldsymbol{\mathbf{\elz} = 0}) \;. \nonumber
\end{gather}
This then yields for the corresponding angular function $f_S(\vartheta)$ (\ref{eq:IV.18b})
\begin{align}
\label{eq:IV.32}
f_S(\vartheta)\ &\Rightarrow\ \sigma_1 \left( \sin \vartheta - \frac{4}{3}\,\sin^3 \vartheta \right) \\
&\equiv \frac{1}{3}\, \sigma_1 \sin (3\vartheta) \;. \nonumber 
\end{align}
Of course, this function $f_S(\vartheta)$ must satisfy both the second-order equation (\ref{eq:IV.17b}) and the first-order equation (\ref{eq:IV.8b}) from which we deduce the first angular function $f_R(\vartheta)$ as
\begin{align}
\label{eq:IV.33}
f_R(\vartheta)\ &\Rightarrow\ \sigma_1 \left\{ \cos \vartheta - \frac{4}{3}\,\cos^3 \vartheta \right\} \\
&\equiv -\frac{1}{3}\,\sigma_1 \cos (3\vartheta) \;. \nonumber
\end{align}

Finally, by means of the angular normalization condition (\ref{eq:IV.15a}) the constant $\sigma_1$ becomes (up to sign) fixed to
\begin{equation}
\label{eq:IV.34}
\sigma_1 = -3 \sqrt{\frac{2}{\pi}} \;,
\end{equation}
so that our method of decoupled second-order equations with the results (\ref{eq:IV.32})--(\ref{eq:IV.33}) actually reproduces the immediate first-order result (\ref{eq:IV.26a})--(\ref{eq:IV.26b})! (Would we have started with $F_R(x)$ (\ref{eq:IV.28b})) in place of $F_S(x)$ (\ref{eq:IV.29b}), we had found both functions $f_R(\vartheta), f_S(\vartheta)$ being interchanged in that result (\ref{eq:IV.26a})--(\ref{eq:IV.26b})). Clearly, this equips us with sufficient confidence into the second-order method (\ref{eq:IV.20a})--(\ref{eq:IV.20b}) in order to construct by means of it the higher angular momentum states due to $\elz = \pm 1, \pm 2, \pm 3$.

For instance, for $\boldsymbol{\mathbf{\elz} = 1}$ (and still $\boldsymbol{\elp = 3}$) we have to start now from the solution (\ref{eq:IV.28a}) of the second-order equation (\ref{eq:IV.20a}), i.\,e. more concretely
\begin{equation}
\label{eq:IV.35}
F_R(x) = \rho_1 x + \rho_3 x^3
\end{equation}
with the coefficient $\rho_3$ to be taken from the recurrence formula (\ref{eq:IV.21a}) as
\begin{equation}
\label{eq:IV.36}
\rho_3 = -\frac{4}{3}\,\rho_1 \;.
\end{equation}
Thus the second-order solution $F_R(x)$ becomes explicitly
\begin{gather}
\label{eq:IV.37}
F_R(x) = \rho_1 \left( x - \frac{4}{3}\,x^3 \right) \\
(\boldsymbol{\elp = 3}, \boldsymbol{\mathbf{\elz} = 1}) \;, \nonumber
\end{gather}
and consequently the associated angular function $f_R(\vartheta)$ (\ref{eq:IV.18a}) must look as follows
\begin{equation}
\label{eq:IV.38}
f_R(\vartheta) = \rho_1 \left( \sin \vartheta - \frac{4}{3}\,\sin^3 \vartheta \right) \;.
\end{equation}
As a check, one is easily convinced that this function actually obeys the second-order equation (\ref{eq:IV.17a}). Furthermore, one substitutes this angular function $f_R(\vartheta)$ (\ref{eq:IV.38}) in the first-order equation (\ref{eq:IV.8a}) and thus finds from here the associated angular function $f_S(\vartheta)$
\begin{equation}
\label{eq:IV.39}
f_S(\vartheta) = -\frac{2}{3}\,\rho_1 \cos\vartheta \sin^2 \vartheta
\end{equation}
which, of course, must then obey the second-order equation (\ref{eq:IV.17b}). Finally, one determines again the residual coefficient $\rho_1$ by means of the angular normalization condition (\ref{eq:IV.15a})
\begin{equation}
\label{eq:IV.40}
\rho_1 = \sqrt{\frac{24}{\pi}} \;,
\end{equation}
which puts the desired solution into its final form
\begin{subequations}
\begin{align}
\label{eq:IV.41a}
f_R(\vartheta) &= \sqrt{\frac{24}{\pi}} \sin \vartheta \left( 1 - \frac{4}{3}\,\sin^2 \vartheta \right) \\
\label{eq:IV.41b}
f_S(\vartheta) &= -\frac{2}{3} \sqrt{\frac{24}{\pi}} \sin^2 \vartheta \cos \vartheta \\
(&\boldsymbol{\elp = 3}, \boldsymbol{\mathbf{\elz} = 1}) \;. \nonumber
\end{align}
\end{subequations}
The residual cases for $\elz = -1, \pm 2, \pm 3$ and $\elp = 3$ can be dealt with in the
same way and the results are collected in the following table. Observe here also the
realization of the symmetries $f_R(\vartheta)\ \Rightarrow\ f_S(\vartheta),\
f_S(\vartheta)\ \Rightarrow\ -f_R(\vartheta)$ being included by $\elz\ \Rightarrow\
-\elz$, as was already remarked below equation (\ref{eq:IV.17b})! (For a more systematic
computation of the angular eigenfunctions see \textbf{App.D}).
\clearpage
\vskip 0.5cm
\begin{center}
\label{tablefR}
\begin{tabular}{|c||c|c|}
\hline
$\elz$ & $f_R(\vartheta)$ & $f_S(\vartheta)$ \\
\hline\hline
$\parbox[0pt][3em][c]{0cm}{}0$ & $\sqrt{\frac{2}{\pi}} \cos (3\vartheta) = \sqrt{\frac{2}{\pi}} \left( -3\cos \vartheta + 4\cos^3 \vartheta \right)$ & $-\sqrt{\frac{2}{\pi}} \sin (3\vartheta) = -\sqrt{\frac{2}{\pi}} \left( 3\sin \vartheta - 4\sin^3 \vartheta \right)$ \\
\hline\hline
$\parbox[0pt][3em][c]{0cm}{}+1$ & $\sqrt{\frac{24}{\pi}} \cdot \left( \sin \vartheta -
  \frac{4}{3}\,\sin^3\vartheta \right)$ & $ -\sqrt{\frac{32}{3\pi}} \cdot \sin^2 \vartheta \cos \vartheta$ \\
\hline
$\parbox[0pt][3em][c]{0cm}{}-1$ & $-\sqrt{\frac{32}{3\pi}} \cdot \sin^2 \vartheta \cos \vartheta$ & $-\sqrt{\frac{24}{\pi}} \cdot \left( \sin \vartheta - \frac{4}{3}\,\sin^3\vartheta \right)$ \\
\hline\hline
$\parbox[0pt][3em][c]{0cm}{}+2$ & $\sqrt{\frac{80}{3\pi}} \cdot \sin^2 \vartheta \cos \vartheta$ & $-\sqrt{\frac{16}{15\pi}} \cdot \sin^3 \vartheta$ \\
\hline
$\parbox[0pt][3em][c]{0cm}{}-2$ & $\sqrt{\frac{16}{15\pi}} \cdot \sin^3 \vartheta$ & $\sqrt{\frac{80}{3\pi}} \cdot \sin^2 \vartheta \cos \vartheta$ \\
\hline\hline
$\parbox[0pt][3em][c]{0cm}{}+3$ & $\sqrt{\frac{32}{5\pi}} \cdot \sin^3 \vartheta$ & $0$ \\
\hline
$\parbox[0pt][3em][c]{0cm}{}-3$ & $0$ & $-\sqrt{\frac{32}{5\pi}} \cdot \sin^3 \vartheta$ \\
\hline
\end{tabular}
\end{center}
\vskip 0.5cm

As pleasant as this RST method of angular momentum quantization may appear, there is also a somewhat critical point which refers to the case $\boldsymbol{\mathbf{\elz} = 0}$. Indeed for this case of vanishing $\elz$, there arises an additional singularity on the $z$-axis! This is readily realized by considering the Dirac current $\vec{k}_p(\vr)$ (\ref{eq:III.4b}) which may be decomposed with respect to the spherical polar coordinates $\left\{ r, \vartheta, \phi \right\}$ as usual
\begin{equation}
\vec{k}_p = \pkr \cdot \ver + \pkt \cdot \vet + \pkp \cdot \vep \;, \nonumber
\end{equation}
where the components $\pkr, \pkt, \pkp$ have already been specified in terms of the amplitude fields by equations (\ref{eq:III.52a})--(\ref{eq:III.52c}). Substituting therein the present parametrization of the amplitude fields $\MRpm, \MSpm$ by means of the angular functions $f_R(\vartheta), f_S(\vartheta)$ (\ref{eq:IV.6a})--(\ref{eq:IV.6b}) and the radial fields $\tPhipm(r)$ (\ref{eq:IV.7a})--(\ref{eq:IV.7b}), one ultimately finds that only the azimuthal component is non-trivial and factorizes to a product form with respect to the spherical polar coordinates, i.\,e. one finds
\begin{subequations}
\begin{align}
\label{eq:IV.42a}
\pkr(r, \vartheta) &\equiv 0 \\
\label{eq:IV.42b}
\pkt(r, \vartheta) &\equiv 0 \\
\label{eq:IV.42c}
\pkp(r, \vartheta) &= \pkp(\vartheta) \cdot \pkp(r) \;, 
\end{align}
\end{subequations}
with the azimuthal factors being given by
\begin{subequations}
\begin{align}
\label{eq:IV.43a}
\pkp(\vartheta) &= \frac{\sin \vartheta \cdot \left[ f_R^2(\vartheta) - f_S^2(\vartheta) \right] - 2\cos \vartheta \cdot f_R(\vartheta) f_S(\vartheta)}{2\pi \sin \vartheta} \\
\label{eq:IV.43b}
\pkp(r) &= \frac{\tPhip(r) \cdot \tPhim(r)}{r} \;.
\end{align}
\end{subequations}
Now substituting here in the angular part (\ref{eq:IV.43a}) the solution (\ref{eq:IV.26a})-(\ref{eq:IV.26b}) for $\elz = 0$ yields
\begin{equation}
\label{eq:IV.44}
\pkp(\vartheta)\big|_{\elz = 0} = \pm \frac{1}{\pi^2}\,\frac{\sin(7\vartheta)}{\sin \vartheta} \;,
\end{equation}
and this result remains finite on the $z$-axis ($\vartheta = 0, \pi$). Clearly, this
signals the presence of a magnetic singularity on the $z$-axis, cf. the curl equation
(\ref{eq:II.48}). Thus the conclusion is that for the states with $\elz = 0$ the presently
adopted \emph{electrostatic approximation} is possibly of limited use, and magnetism
should be taken into account.

From this reason, it seems worthwile to consider also the other states with non-vanishing
$\elz$. Intuitively, one expects that the centrifugal forces will press matter off the
axis of rotation (i.\,e. the $z$-axis; $\vartheta = 0,\pi$) so that the (azimuthal)
current density $\pkp(\vartheta)$ becomes zero on the $z$-axis and adopts its maximal
value the farer away from this axis the larger is the value $|\elz|$ of angular
momentum. For instance, consider the largest possible angular momentum
(i.\,e. $\boldsymbol{\mathbf{\elz} = \pm 3}$ for $\boldsymbol{\elp = 3}$) and deduce from
the above table that the corresponding azimuthal part of the current density
$\pkp(\vartheta)$ (\ref{eq:IV.43a}) becomes
\begin{equation}
\label{eq:IV.45}
\bkp(\vartheta)\big|_{\elz = \pm 3} = \pm \frac{16}{5\pi^2}\,\sin^6 \vartheta \;.
\end{equation}
Obviously, this current density vanishes like $\sim \vartheta^6$ when approaching the $z$-axis ($\vartheta \rightarrow 0$).

Next, turn to the case $\boldsymbol{\mathbf{\elz} = \pm 2}$ and find again from the corresponding results in the above table for the (angular) current strength $\pkp(\vartheta)$ (\ref{eq:IV.43a})
\begin{equation}
\label{eq:IV.46}
\pkp(\vartheta) \big|_{\elz = \pm 2} = \pm \frac{56}{3\pi^2}\,\sin^4 \vartheta \left\{ 1 - \frac{36}{35}\,\sin^2 \vartheta \right\} \;.
\end{equation}
Comparing this to the precedent result (\ref{eq:IV.45}) for $\boldsymbol{\mathbf{\elz} = \pm 3}$ we see that now for $\boldsymbol{\mathbf{\elz} = \pm 2}$ the current density vanishes somewhat slower ($\sim \vartheta^4$) for approaching the $z$-axis ($\vartheta \rightarrow 0$). Clearly, this rule is continued also to the case $\boldsymbol{\mathbf{\elz} = \pm 1}$:
\begin{equation}
\label{eq:IV.47}
\pkp(\vartheta)\big|_{\elz = \pm 1} = \pm \frac{4}{\pi^2}\,\sin^2 \vartheta \left\{ 7 - \frac{56}{3}\,\sin^2 \vartheta + 12 \sin^4 \vartheta \right\} \;,
\end{equation}
which meets with the expectation that for $\boldsymbol{\mathbf{\elz} = \pm 1}$ the current density vanishes now slowest (i.\,e. $\sim \vartheta^2$) for approaching the $z$-axis ($\vartheta \rightarrow 0$), see \textbf{Fig. IV.A}.
\pagebreak
\begin{center}
\epsfig{file=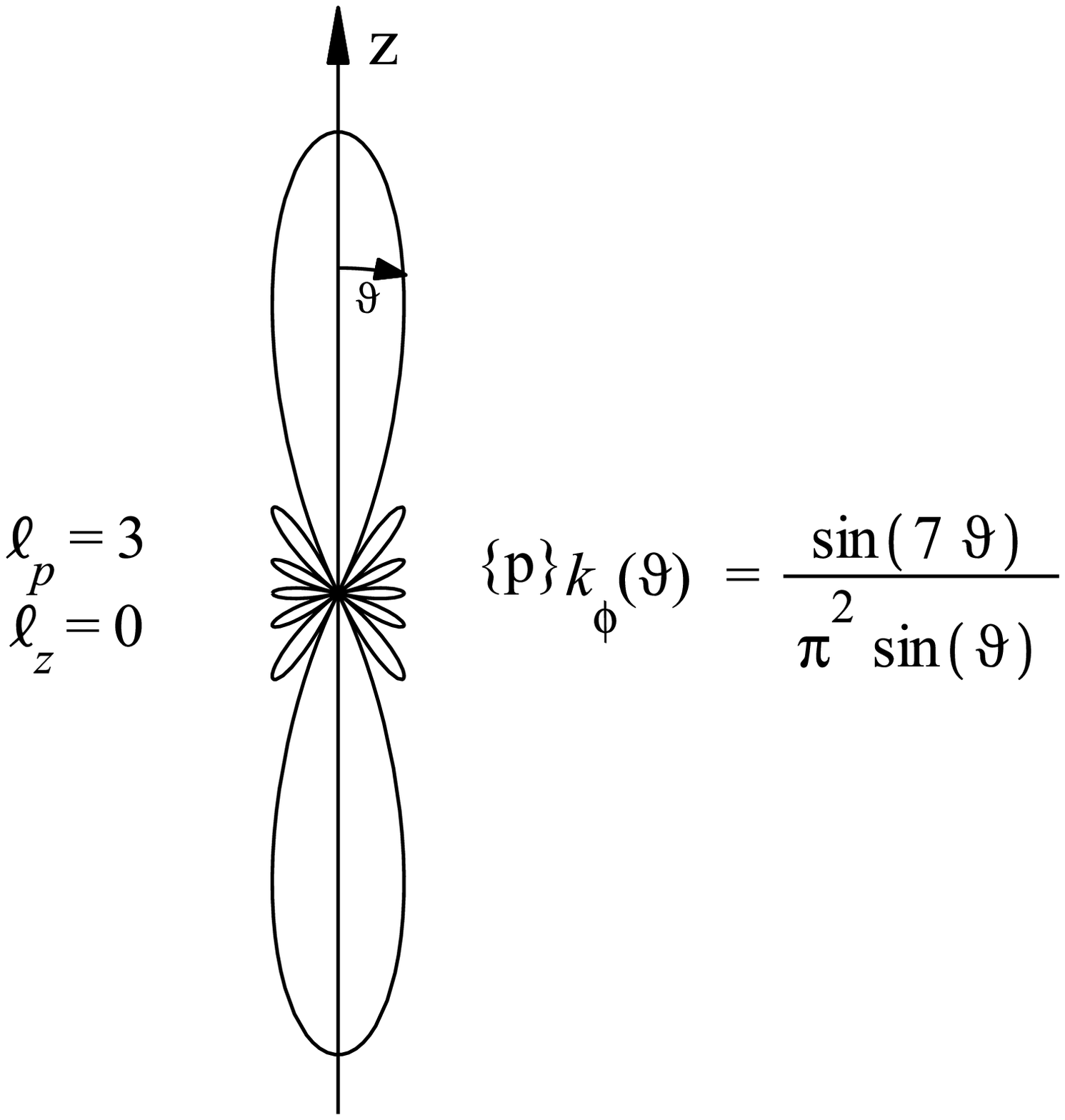,height=12cm}
\end{center}
\vspace{-1cm}
{\textbf{Fig.~IV.A.0}\hspace{5mm} \emph{\large\textbf{Azimuthal Current Density
\boldmath$\gkloi{p}{k}{\phi}(\vartheta)$  (\ref{eq:IV.44})      }   }  }
\myfigure{Fig.~IV.A.0: {Azimuthal Current Density
\boldmath$\gkloi{p}{k}{\phi}(\vartheta) $}}
\indent\\*
\enlargethispage{1cm}
\label{figIV.A.0}
For~$l_z=0$, the current density~$\gkloi{p}{k}{\phi}$ is non-zero on the
z-axis~$(\vartheta=0,\pi)$ which signals the existence of a \emph{magnetic} singularity
just on the axis of rotation. For the groundstate~\mbox{$(\lP=0)$}, there occurs an
additional singularity of the \emph{electric} type (see Fig.4b in ref~\cite{4}). This fact
might eventually be the origin of the relatively large deviation of the RST prediction for
the groundstate energy (-7,2305\,[eV])) from the conventional prediction of
$-6,8029\ldots\,$[eV] see equation (\ref{eq:I.2}) for~$n=1$. Thus the expectation is that
the RST groundstate prediction will be shifted towards its conventional counterpart by a
more adequate treatment of both groundstate singularities. Observe also the enlargement of
the current density close to the axis of rotation (z-axis) which keeps the angular
momentum as small as possible~$(\leadsto l_z=0)$.

\begin{center}
\epsfig{file=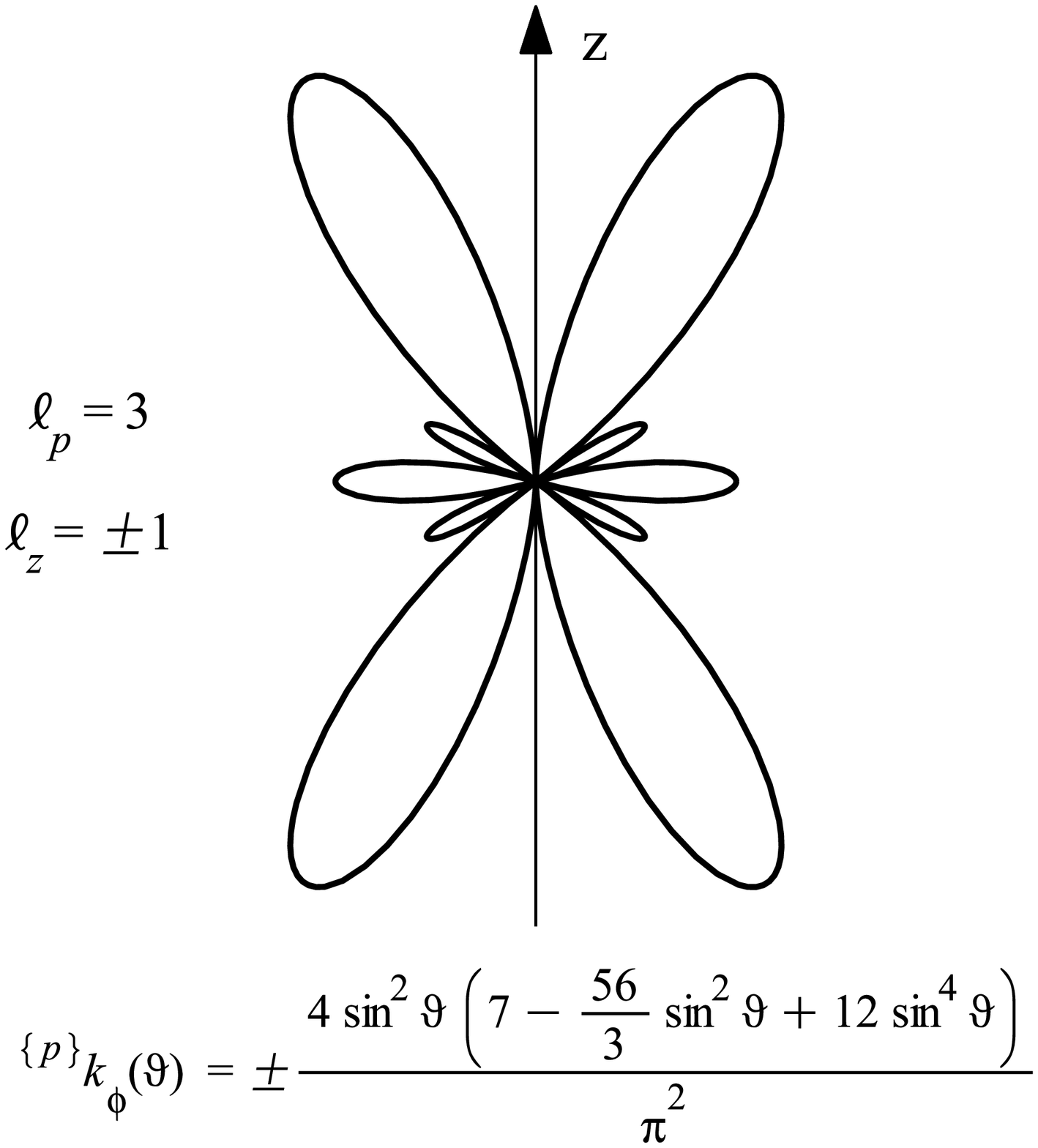,height=12cm}
\end{center}
{\textbf{Fig.~IV.A.1}\hspace{5mm} \emph{\large\textbf{Azimuthal Current Density
\boldmath$\gkloi{p}{k}{\phi}(\vartheta)$ (\ref{eq:IV.47}) }  }  }
\myfigure{Fig.~IV.A.1: {Azimuthal Current Density
\boldmath$\gkloi{p}{k}{\phi}(\vartheta) $}}
\indent\\*

For~$l_z=\pm 1$, the current density~$\gkloi{p}{k}{\phi}(\vartheta)$ takes its maximum
aside of the axis of rotation (in contrast to~$l_z=0$, precedent figure). This may be
intuitively associated with the larger amount of angular momentum (i.e.~$l_z=\pm 1$ in
place of~$l_z=0$). Since the current density vanishes now on the
z-axis~$(\vartheta=0,\pi)$ one expects that no magnetic singularity can occur here.

\begin{center}
\epsfig{file=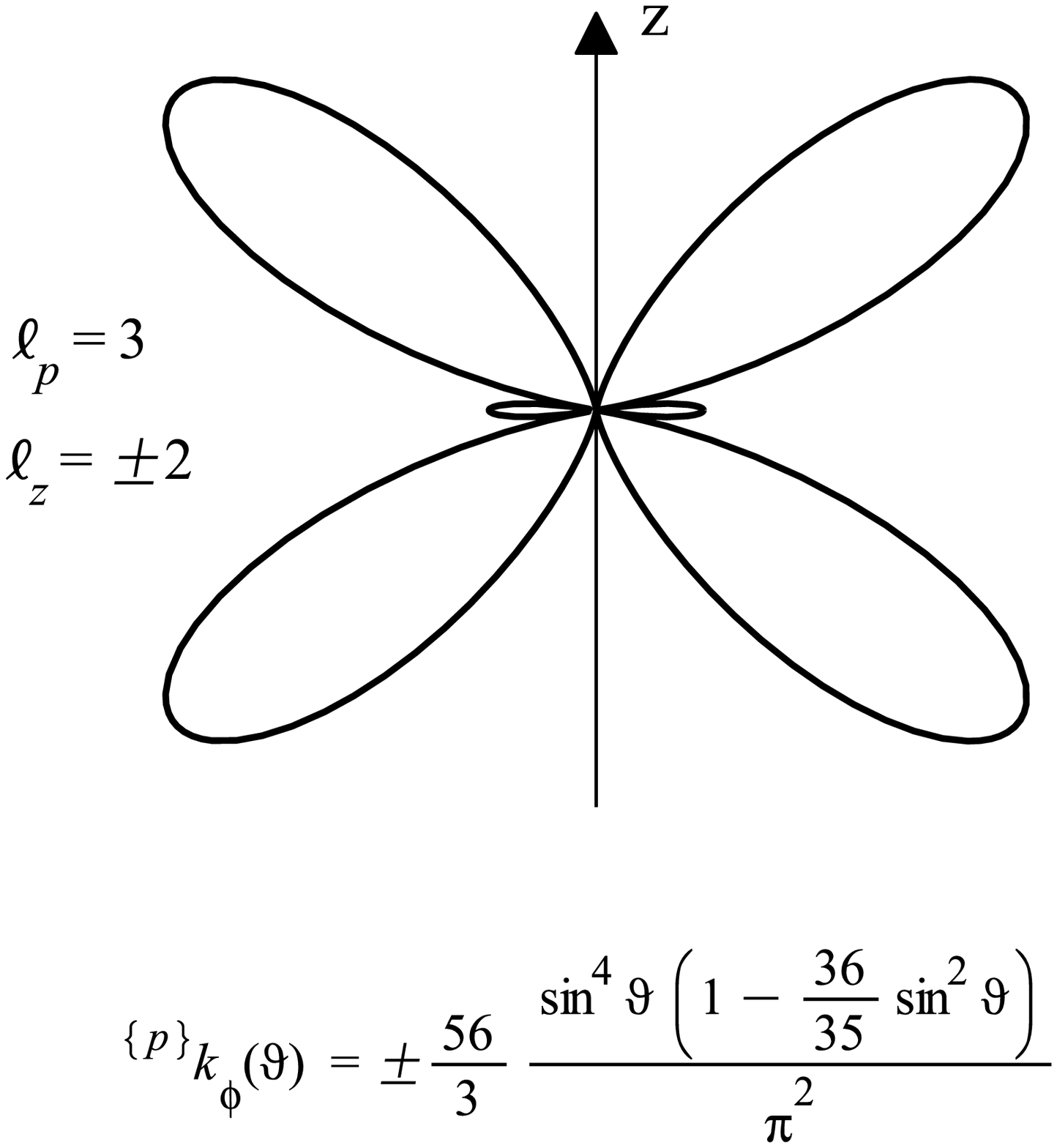,height=12cm}
\end{center}
{\textbf{Fig.~IV.A.2}\hspace{5mm} \emph{\large\textbf{Azimuthal Current Density
\boldmath$\gkloi{p}{k}{\phi}(\vartheta)$ (\ref{eq:IV.46}) }}  }
\myfigure{Fig.~IV.A.2: {Azimuthal Current Density
\boldmath$\gkloi{p}{k}{\phi}(\vartheta) $}}
\indent\\*

For~$l_z=\pm 2$, the maximum of the current density~$\gkloi{p}{k}{\phi}(\vartheta)$
becomes shifted farther away from the axis of rotation~$(\vartheta=0,\pi)$ which may
intuitively considered again to be due to the even larger amount of angular momentum
(i.e.~$|l_z|=2$ in place of~$|l_z|=1$, precedent figure). The peculiarity of this specific
angular distribution of charge is that the associated quadrupole correction
(\ref{eq:IV.104c}) of the spherically symmetric potential~$\eklo{p}{A}_0(r)$ is zero; see
the discussion below equation (\ref{eq:IV.105}). The small region around~$\vartheta=\pi/2$
has opposite direction of rotation.

\begin{center}
\epsfig{file=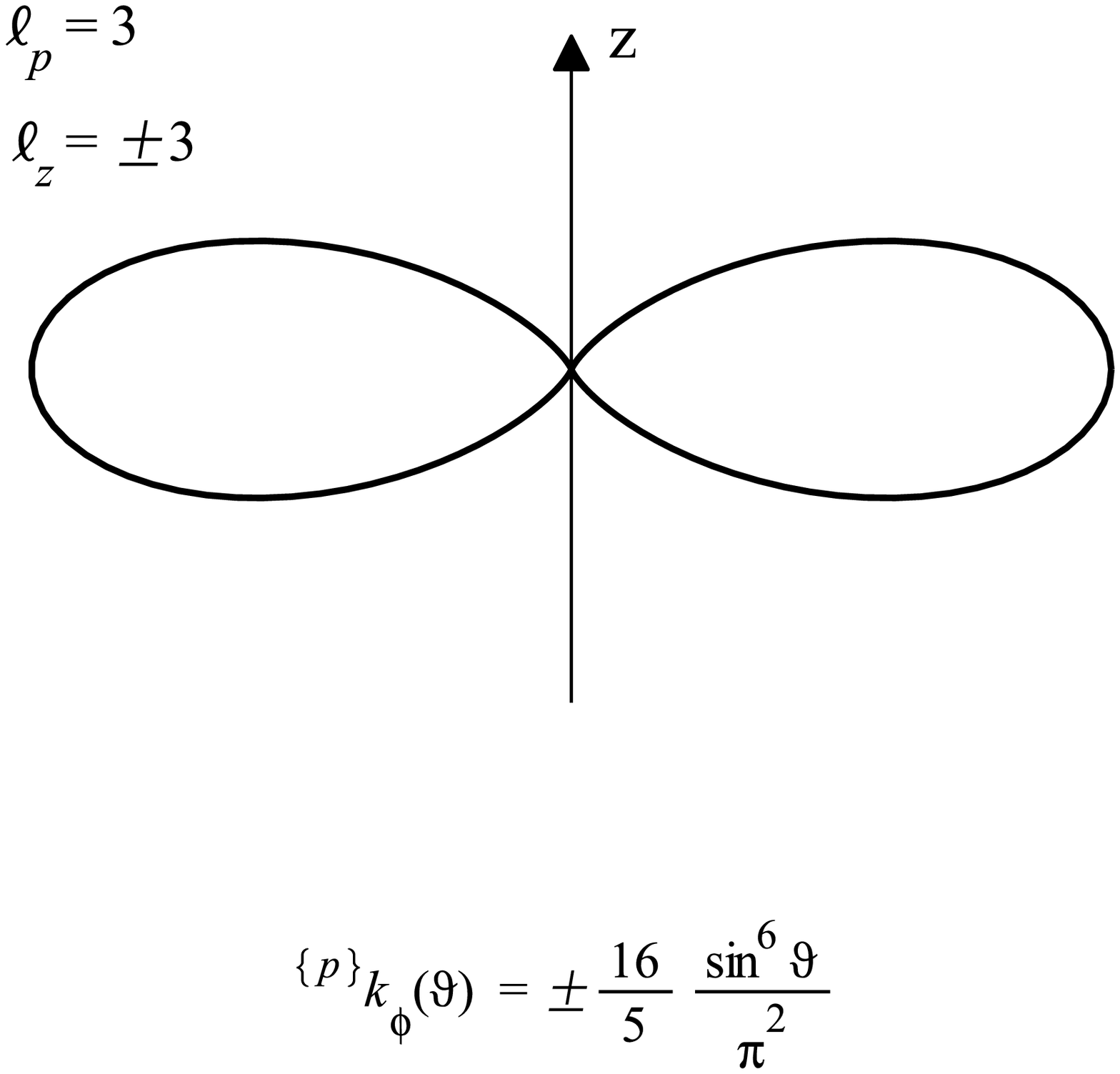,height=12cm}
\end{center}
{\textbf{Fig.~IV.A.3}\hspace{5mm} \emph{\large\textbf{Azimuthal Current Density
\boldmath$\gkloi{p}{k}{\phi}(\vartheta)$ (\ref{eq:IV.45}) }}  }
\myfigure{Fig.~IV.A.3: {Azimuthal Current Density
\boldmath$\gkloi{p}{k}{\phi}(\vartheta) $}}
\indent\\*
\label{figIV.A.3}
\enlargethispage{1cm} For~$l_z=\pm 3$, the maximum of the current
density~$\gkloi{p}{k}{\phi}(\vartheta)$ is now farthest away from the axis of rotation
which meets with the intuitive expectation that such a configuration should possess the
largest value~$l_z$ of angular momentum due to a given value of~$l_\wp$ (i.e.~$l_z=\pm
l_\wp$). Simultaneously, this specific geometry of the flow pattern generates also the
largest amount of electrostatic(!) anisotropy energy~$E_{|||}^{\{e\}}$ (\ref{eq:IV.128}) within
an~$\lP$-multiplet, see the corresponding estimate (\ref{eq:IV.129}). Thus, this effect of
dislocation of charge and current to the outside regions with increasing angular
momentum~$l_z$ turns out to be the origin of the elimination of the~$l_z$-degeneracy
occurring in the spherically symmetric approximation. The whole energy spectrum due to the
present subset defined by~$l_z=\pm l_\wp$ is studied in \textbf{Subsect.IV.4}.

\pagebreak
\begin{center}
  \emph{\textbf{3.\ Energy Functional and Principle of Minimal Energy}}
\end{center}

In order to decide whether the states due to different values of $\elz$ (but the same
value of $\elp$) do carry the same energy, one needs an energy functional ($\tEgT$,
say). This object will then be equipped with two intimately related meanings: First, its
value on any solution of the positronium eigenvalue problem (precedent subsections) yields
the binding energy of the corresponding positronium state; and second, it provides us with
the basis of the variational \emph{principle of minimal energy}. This fact may then be
exploited in order to construct approximate variational solutions so that the values of
the energy functional $\tEgT$ on these solutions gives to us a first rough estimate of the
energy spectrum of para-positronium.

Of course, in view of its importance the construction of such an energy functional $\tEgT$
has been the subject of several precedent investigations [Zitate] so that we can restrict
ourselves here to merely quoting the relevant results. In principle, the desired
functional is composed of two different kinds of contributions, i.\,e. the physical part
$\EgTiv$ and the constraint terms which are neccessary in order to reproduce the right
extremal equations. Of course, the latter must then be identical with the system of
eigenvalue equations. Thus, the choice looks as follows~\cite{4}
\begin{equation}
\label{eq:IV.48}
\tETT = \EETiv + 2 \lambda_D \cdot \tND + \lGe \cdot \tNGe
\end{equation}
with the physical part being given by
\begin{equation}
\label{eq:IV.49}
\ETiv = 2 M\crm^2 \cdot \mathcal{Z}_\mathcal{P}^2 + 4\; \pTkin + \eER \;.
\end{equation}
No doubt, the more interesting part is the latter one, because for the practical purpose
of extremalizing the total functional (\ref{eq:IV.48}) one will select those trial
configurations which \emph{per se} annihilate the constraints $\tND$ and $\tNGe$ (see
below) so that one can deal exclusively with the physical part (\ref{eq:IV.49}), which is
the sum of renormalized rest mass energy, kinetic energy and gauge field
energy. Naturally, all these different contributions are to be understood as functionals
of the basis RST fields (i.\,e. matter field $\Psi$ and bundle connection $\mathcal{A}$);
but in the course of our investigations we did reparametrize these fields ultimately in
terms of the angular functions $f_R(\vartheta), f_S(\vartheta)$
(\ref{eq:IV.6a})--(\ref{eq:IV.6b}), spherically symmetric amplitude fields $\tPhipm(r)$
(\ref{eq:IV.7a})--(\ref{eq:IV.7b}), and electric part $\pAn(\vr)$ of the gauge field where
we first may resort to its spherically symmetric approximation $\peAn(r)\ (r \doteqdot
|\vr|)$. Consequently, we first have to specify all the energy contributions in terms of
these latter fields before we can determine the energy spectrum of para-positronium,
namely through deriving great benefit from that energy functional (\ref{eq:IV.48}).

\begin{center}
  \large{\textit{Relativistic Mass Renormalization}}
\end{center}

The first term of the physical part $\EETiv$ (\ref{eq:IV.49}) is not simply the rest mass energy ($2 M \crm^2$) of both particles (i.\,e. electron plus positron) but contains also the renormalization factor $\Zp^2$ which is defined by
\begin{equation}
\label{eq:IV.50}
\Zp^2 \doteqdot \int d^3\vr\;\bar{\psi}_\mathcal{P}(\vr) \psi_\mathcal{P}(\vr) \;,
\end{equation}
or in terms of the Pauli spinors $\pxppm$ in place of the Dirac spinors $\psi_\mathcal{P}(\vr)$ (\ref{eq:III.49a})
\begin{equation}
\label{eq:IV.51}
\Zp^2 = \int d^3\vr\;\left\{ \pxppd \pxpp - \pxpmd \pxpm \right\} \;.
\end{equation}
Inserting here the former decomposition (\ref{eq:III.37a})--(\ref{eq:III.37b}) of the
Pauli spinors with respect to the bosonic spinor basis yields the renormalization factor
in terms of the amplitude fields $\pMRpm, \pMSpm$ as
\begin{align}
\label{eq:IV.52}
\Zp^2 = \int \frac{d^3\vr}{4\pi}\;\bigg\{ &\pMRpS \cdot \pMRp + \pMSpS \cdot \pMSp  \\
 - &\pMRmS \cdot \pMRm - \pMSmS \cdot \pMSm \bigg\} \;, \nonumber
\end{align}
and since furthermore the original amplitude fields $\pMRpm, \pMSpm$ have been ultimately transformed to the angular functions $f_R(\vartheta), f_S(\vartheta)$ and radial amplitudes $\tPhipm(r)$ via the equations (\ref{eq:IV.4a})--(\ref{eq:IV.4b}), (\ref{eq:IV.6a})--(\ref{eq:IV.6b}), and (\ref{eq:IV.7a})--(\ref{eq:IV.7b}) one ends up with the following form of the relativistic normalization factor
\begin{equation}
\label{eq:IV.53}
\Zp^2\ \Rightarrow\ \tZp^2 = \int \frac{d\Omega}{4\pi \sin\vartheta} \left( f_R^2(\vartheta) + f_S^2(\vartheta) \right) \int\limits_{r=0}^{\infty} dr\;r \left[ \tPhip^2(r) - \tPhim^2(r) \right] \;.
\end{equation}
Thus the final form of the (renormalized) rest mass energy of the two particles is found to appear as
\begin{equation}
\label{eq:IV.54}
2M\crm^2 \cdot \tZp^2 = 2M\crm^2 \cdot \int\limits_0^\infty dr\;r \left[ \tPhip^2(r) - \tPhim^2(r) \right] \;,
\end{equation}
provided one agrees to apply here the separate angular normalization condition (\ref{eq:IV.15a}).

Since (for the present prelimary estimate of the positronium spectrum) we will be
satisfied with the non-relativistic approximation, we have to reveal now the
non-relativistic limit of the renormalization factor $\tZp^2$ (\ref{eq:IV.54}). For this
purpose, one might perhaps be tempted to simply neglect completely the ``negative''
constituent $\tPhim(r)$ of the amplitude field and consider the ``positive'' constituent
$\tPhip(r)$ as the true non-relativistic approximation. However, such an argument would be
somewhat too superficial. Actually, we have to observe here that for the relativistic
treatment the ``negative'' constituent $\tPhim(r)$ is needed in order to obey the radial
normalization condition (\ref{eq:IV.15b}). Even if we take the viewpoint that $\left|
  \tPhim(r) \right|$ is always much smaller than $\left| \tPhip(r) \right|$, it cannot be
neglected for the transition to the non-relativistic limit! Thus we have to keep in view
$\tPhim(r)$ at least approximately; and we do this by solving (approximately) the first
eigenvalue equation (\ref{eq:IV.9a}) for $\tPhim(r)$ in the following form
\begin{equation}
\label{eq:IV.55}
\tPhim(r) \simeq \frac{\hbar}{2M\crm} \cdot \left( \frac{d}{dr}\; {-\frac{\elp}{r}} \right) \tPhip(r)
\end{equation}
where the mass eigenvalue $M_*$ is (approximately) identified with the rest mass $M$ of the particles. Next, substituting this result for $\tPhim(r)$ in the radial normalization condition (\ref{eq:IV.15b}) yields
\begin{equation}
\label{eq:IV.56}
\int\limits_{r=0}^{\infty} dr\;r\,\tPhip^2(r) = 1 - \left( \frac{\hbar}{2M\crm} \right)^2 \int\limits_0^\infty dr \left\{ \left( \frac{d\tPhip(r)}{dr} \right)^2 + \frac{\elp^2}{r^2}\cdot \tPhip^2(r)  \right\} \;,
\end{equation}
and thus the radial part of the renormalization factor $\tZp$ (\ref{eq:IV.53}) becomes
\begin{equation}
\label{eq:IV.57}
\int\limits_0^\infty dr\;r \left[ \tPhip^2(r) - \tPhim^2(r) \right] \simeq 1 - 2\left(
  \frac{\hbar}{2M\crm} \right)^2 \int\limits_0^\infty dr\;r \left\{ \left(
    \frac{d\tPhip(r)}{dr} \right)^2 + \frac{\elp^2}{r^2} \cdot \tPhip^2(r) \right\}\ .
\end{equation}
So we see that the non-relativistic approximation of the mass renormalization term
(\ref{eq:IV.53}) is not built up only of the particle rest mass energy ($M\crm^2$) but
also of the non-relativistic kinetic energy $\rkloi{p}{E}{kin}$, i.\,e. we find
\begin{subequations}
\begin{align}
\label{eq:IV.58a}
M\crm^2\,\tZp^2\ \Rightarrow\ M\crm^2 - \int \frac{d\Omega}{4\pi\sin\vartheta} \left(
  f_R^2(\vartheta) + f_S^2(\vartheta) \right) \cdot \pEkin_{[\Phi]} = Mc^2 -
\pEkin_{[\Phi]} \\
\label{eq:IV.58b}
\pEkin_{[\Phi]} \doteqdot \frac{\hbar^2}{2M} \int\limits_0^\infty dr\;r \left\{ \left( \frac{d\tPhip(r)}{dr} \right)^2 + \frac{\elp^2}{r^2} \cdot \tPhip^2(r) \right\} \;.
\end{align}
\end{subequations}

This result is interesting in so far as it says that the rest mass contribution (first
term) to the physical energy $\ETiv$ (\ref{eq:IV.49}) actually is the proper rest mass
energy \emph{minus} the non-relativistic kinetic energy $\pEkin$. This is the reason why
the physical energy $\ETiv$ (\ref{eq:IV.49}) (for \emph{two} particles!) contains the
\emph{four-fold} kinetic energy $\ptTkin$ of a single particle, namely in order that the
\emph{two-fold} negative kinetic energy in the rest mass term $2M\crm^2 \cdot \tZp^2$
(\ref{eq:IV.58a}) can be compensated for. Clearly, for the non-relativistic form $\tEEeT$
of the relativistic $\tEeT$ (\ref{eq:IV.48}) we can omit that ordinary rest mass term
($2M\crm^2$) because it does not take part in the dynamics at all.

\begin{center}
  \large{\textit{Kinetic Energy}}
\end{center}

The second physical contribution to the energy functional $\tETT$
(\ref{eq:IV.48})--(\ref{eq:IV.49}) is four times the one-particle kinetic energy
$\ptTkin$. Analogously to the precedent case of the rest mass energy, this contribution
must now also be expressed in terms of the fields $\tPhipm$ and $f_R(\vartheta),
f_S(\vartheta)$. To this end, it is very convenient and also instructive to subdivide the
total kinetic energy $\ptTkin$ into three parts which are associated with the motion along
the three spatial directions being specified by the spherical polar coordinates $r,
\vartheta, \phi$:
\begin{equation}
\label{eq:IV.59}
\pTkin = \pTr + \pTth + \pTphi \;.
\end{equation}
Turning here first to the radial part $\ptTr$ one is easily convinced that this energy contribution is itself the sum of two terms
\begin{equation}
\label{eq:IV.60}
\pTr = \pTr [\R] + \pTr [\mathcal{S}] \;,
\end{equation}
where the first one is built up solely by the  amplitudes $\ptRpm\rt$ (\ref{eq:IV.4a})
\begin{align}
\label{eq:IV.61}
\pTr [\R]\ \Rightarrow\ \ptTr [\R] = &\frac{\hbar\crm}{4} \int d^2 \vr \big\{ \ptRm\rt \cdot \frac{\partial}{\partial r} \ptRp\rt \\
- &\ptRp\rt \cdot \frac{\partial}{\partial r} \ptRm\rt - \frac{\ptRp\rt \cdot \ptRm\rt}{r} \big\} \;, \nonumber \\
&\quad\quad\quad(d^2\vr \doteqdot r\,dr\,d\vartheta) \nonumber
\end{align}
and the second term $\ptTr[\mathcal{S}]$ looks identical apart from the replacement
$\ptRpm\rt\ \Rightarrow\ \ptSpm\rt$. Thus we are merely left here with the task to apply
the last transformations (\ref{eq:IV.6a})--(\ref{eq:IV.7b}) which then ultimately brings
the total kinetic energy $\ptTr$ of the radial type (\ref{eq:IV.60}) to the following form
\begin{align}
\label{eq:IV.62}
\ptTr = \frac{\hbar\crm}{4} \int\limits_0^\pi d\vartheta \left[ f_R^2(\vartheta) + f_S^2(\vartheta) \right] \int\limits_0^\infty dr\;r \bigg\{ &\tPhim(r) \cdot \frac{d\tPhip(r)}{dr} \\
- &\tPhip(r) \cdot \frac{d\tPhim(r)}{dr} - \frac{\tPhip(r) \cdot \tPhim(r)}{r} \bigg\} \;. \nonumber
\end{align}

\begin{sloppypar}
Properly speaking, the product ansatz (\ref{eq:IV.6a})--(\ref{eq:IV.7b}) represents a kind of pre-selection among all the possible matter field configurations so that the radial kinetic energy (\ref{eq:IV.62}) becomes a functional of only four very restricted functions of $r$ and $\vartheta$, i.\,e. $\tPhipm(r)$ and $f_R(\vartheta), f_S(\vartheta)$. But it is also possible to go one step further by demanding that the angular functions obey the normalization condition (\ref{eq:IV.15a}) which may also be written as
\begin{equation}
\label{eq:IV.63}
\tNf \doteqdot \frac{1}{2} \int\limits_0^\pi d\vartheta \left( f_R^2(\vartheta) + f_S^2(\vartheta) \right) - 1 = 0 \;.
\end{equation}
In this case, the radial kinetic energy $\ptTr$ becomes reduced to a function of $\tPhipm(r)$, i.\,e.
\begin{align}
\label{eq:IV.64}
\ptTr\ \Rightarrow\ \ptTr[\Phi] = \frac{\hbar\crm}{2} \int\limits_0^\infty dr\;r \bigg\{ \tPhip(r) \cdot \frac{d\tPhip(r)}{dr} - &\tPhip(r) \cdot \frac{d\tPhim(r)}{dr} \\
- &\frac{\tPhip(r) \cdot \tPhim(r)}{r} \bigg\} \;, \nonumber
\end{align}
so that the extremalization procedure (due to the \emph{principle of minimal energy}) can
refer only to the amplitudes $\tPhipm(r)$. On the other hand, one may prefer to subject
the energy functional also to the extremalization with respect to the angular functions
$f_R(\vartheta), f_S(\vartheta)$ which then has the consequence that we have to work with
the original functional $\ptTr$ (\ref{eq:IV.62}) but have to add in this case the angular
normalization condition (\ref{eq:IV.63}) as a constraint. Observe also that the radial
normalization condition (\ref{eq:IV.15b}) has to be satisfied in any case.
\end{sloppypar}

Next, the energy $\ptTth$ due to the longitudinal motion is found to appear in terms of
the amplitude fields $\ptRpm, \ptSpm$ in the following form:
\begin{align}
\label{eq:eq:IV.65}
\ptTth = \frac{\hbar\crm}{4} \int \frac{d^2\vr}{r} \bigg\{ &\ptRm\rt \cdot \frac{\partial}{\partial\vartheta} \ptSp\rt - \ptSp\rt \cdot \frac{\partial}{\partial\vartheta} \ptRm\rt \\
+ &\ptRp\rt \cdot \frac{\partial}{\partial \vartheta} \ptSm\rt - \ptSm\rt \cdot \frac{\partial}{\partial\vartheta} \ptRp\rt \bigg\} \;. \nonumber
\end{align}
If the former product ansatz (\ref{eq:IV.6a})--(\ref{eq:IV.7b}) is again introduced
herein, one finally ends up with the longitudinal analogue $\ptTth$ of the radial energy
$\ptTr$ (\ref{eq:IV.62}), i.\,e.
\begin{equation}
\label{eq:IV.66}
\ptTth = \frac{\hbar\crm}{2} \int\limits_0^\pi d\vartheta \left[ f_R(\vartheta) \cdot \frac{d\,f_S(\vartheta}{d\vartheta} - f_S(\vartheta) \cdot \frac{d\,f_R(\vartheta)}{d\vartheta} \right] \; \int\limits_0^\infty dr\;\tPhip(r) \cdot \tPhim(r) \;.
\end{equation}

Finally, the azimuthal part $\ptTphi$ of the kinetic energy $\ptTkin$ (\ref{eq:IV.59}) is found as
\begin{align}
\label{eq:IV.67}
\ptTphi = &\frac{\hbar\crm}{2} \elz \int \frac{d^2\vr}{r} \, \bigg\{ \ptSp\rt \cdot \ptSm\rt - \ptRp\rt \cdot \ptRm\rt \\
+ &\cot\vartheta \left[ \ptRp\rt \cdot \ptSm\rt + \ptRm\rt \cdot \ptSp\rt \right] \bigg\} \;, \nonumber
\end{align}
or by applying again those transformations (\ref{eq:IV.6a})--(\ref{eq:IV.7b})
\begin{equation}
\label{eq:IV.68}
\ptTphi = \frac{\hbar\crm}{2} \elz \int\limits_0^\pi d\vartheta \, \bigg\{f_S^2(\vartheta) - f_R^2(\vartheta) + 2\cot\vartheta f_R(\vartheta) f_S(\vartheta) \bigg\}\,\int\limits_0^\infty dr\;\tPhip(r) \cdot \tPhim(r) \;.
\end{equation}

\begin{center}
  \large{\textit{Electric Interaction Energy}}
\end{center}

According to the \emph{principle of minimal coupling}, the interaction of both particles
occurs via the gauge potential $\A_\mu$, cf. (\ref{eq:II.3})--(\ref{eq:II.5}); but since
we are dealing with non-identical particles and do also neglect the magnetic forces we
have to observe merely the electric part $\pAn(\vr)$ of the non-Abelian bundle connection
$\A_\mu$. Moreover, recall that we are presently satisfied with the \emph{spherically
  symmetric approximation} where the interaction potential $\pAn(\vr)$ is assumed to
depend only upon the radial variable $r$ of the spherical polar coordinates (i.\,e. we put
$\pAn(\vr)\ \Rightarrow\ \peAn(r)$). The interaction energy can now be expressed by the
interaction potential $\peAn(r)$ in two different ways. The first possibility consists in
identifying the interaction energy with the gauge field energy ($\ERee$, say) due to the
electro-static potential~\cite{2}
\begin{equation}
\label{eq:IV.69}
\ERe = -\frac{\hbar\crm}{4\pi\as} \int d^3\vr\;\left|\left|\vec{\nabla} \pAn(\vr)\right|\right|^2
\end{equation}
which becomes simplified in our electrostatic approximation to
\begin{equation}
\label{eq:IV.70}
\ERe\ \Rightarrow\ \ERee = -\frac{\hbar\crm}{\as} \int\limits_0^\infty dr\;r^2\,\left( \frac{d\,\peAn(r)}{dr} \right)^2 \;.
\end{equation}

The second possibility refers to the ``\emph{mass equivalent}'' of the gauge field energy
$\ERe$ which is defined in the following general way
\begin{equation}
\label{eq:IV.71}
\Mee\crm^2 = -\hbar\crm \int d^3\vr\;\pAn(\vr) \cdot \pko(\vr) \;,
\end{equation}
i.\,e. in the special case of our spherically symmetric approximation
\begin{equation}
\label{eq:IV.72}
\Mee\crm^2\ \Rightarrow\ \tMee\crm^2 = -\hbar\crm \int d\Omega\;\pgko(\vartheta) \int\limits_0^\infty dr\;r^2\,\peAn(r) \cdot \pgko(r) \;,
\end{equation}
where we have already made use of the factorization (\ref{eq:IV.14}) of the charge density
$\pgko(\vr)$. Here it is also reasonable to apply the angular normalization condition
(\ref{eq:IV.15a}) so that the mass equivalent $\tMee\crm^2$ (\ref{eq:IV.72}) appears in
its final form as
\begin{equation}
\label{eq:IV.73}
\tMee\crm^2 = -\hbar\crm \int\limits_0^\infty dr\;r\,\peAn(r) \cdot \left[ \tPhip^2(r) + \tPhim^2(r) \right] \;.
\end{equation}
By this procedure, the angular functions $f_R(\vartheta), f_S(\vartheta)$ disappear from the mass equivalent quite similarly to the situation with the kinetic energy $\ptTr$. Therefore we have to include the angular normalization condition (\ref{eq:IV.63})  as an extra constraint for the subsequent \emph{principle of minimal energy}. But the crucial point with the mass equivalent $\tMee\crm^2$ is now that it must be identical to the gauge field energy $\ERee$ (\ref{eq:IV.70}) which may be reformulated as a further constraint for the variational principle:
\begin{equation}
\label{eq:IV.74}
\tNGee \doteqdot \ERee - \tMee\crm^2 = 0 \;.
\end{equation}

\begin{center}
  \large{\textit{Relativistic Principle of Minimal Energy}}
\end{center}

Collecting now all the results, we find that our original energy functional $\tETT$ (\ref{eq:IV.48}) reappears in its final form (but restricted to the electrostatic and spherically symmetric approximation) as follows:
\begin{align}
\label{eq:IV.75}
\tETT\ \Rightarrow\ \tEephif = &2M\crm^2 \cdot \int\limits_0^\infty dr\;r\,\left[\tPhip^2(r) - \tPhim^2(r) \right] \\
+ &2\hbar\crm \int\limits_0^\infty dr\,r \left\{ \tPhim(r) \cdot \frac{d\,\tPhip(r)}{dr} - \tPhip(r) \cdot \frac{d\,\tPhim(r)}{dr} - \frac{\tPhip(r) \cdot \tPhim(r)}{r} \right\} \nonumber \\
+ &2\hbar\crm \int\limits_0^\pi d\vartheta\;\left[ f_R(\vartheta) \cdot \frac{d\,f_S(\vartheta)}{d\vartheta} - f_S(\vartheta) \cdot \frac{d\,f_R(\vartheta)}{d\vartheta} \right] \int\limits_0^\infty dr\;\tPhip(r) \cdot \tPhim(r) \nonumber \\
+ &2 \elz \hbar \crm \int\limits_0^\pi d\vartheta\;\left\{ f_S^2(\vartheta) - f_R^2(\vartheta) + 2\cot\vartheta\,f_R(\vartheta) f_S(\vartheta) \right\} \int\limits_0^\infty dr\;\tPhip(r) \cdot \tPhim(r) \nonumber \\
- &\frac{\hbar\crm}{\as} \int\limits_0^\infty dr\;r^2\,\left( \frac{d\,\peAn(r)}{dr} \right)^2 + \lGe \cdot \tNGee + 2\lambda_f \cdot \tNf + 2 \lambda_D \cdot \tNPhi \;. \nonumber
\end{align}
In this result, one recognizes the original form (\ref{eq:IV.48})--(\ref{eq:IV.49}) of the energy functional $\tETT$ in a more elaborate form: the last three terms are the constraints, where the normalization (\ref{eq:IV.13}) of the density $\pko(\vr)$ in the constraint form
\begin{equation}
\label{eq:IV.76}
N_D \doteqdot \int d^3\vr\;\pko(\vr) - 1 = 0 \;,
\end{equation}
with the Langrangean multiplier $2\lambda_D$, is split up into the separate angular
normalization constraint (\ref{eq:IV.63}) and its radial analogue
\begin{equation}
\label{eq:IV.77}
\tNPhi \doteqdot \int dr\;r\,\left[ \tPhip^2(r) + \tPhim^2(r) \right] - 1 = 0 \;.
\end{equation}
Of course, the latter constraint is nothing else than the radial normalization condition (\ref{eq:IV.15b}). The associated Langrangean multipliers are denoted by $\lambda_f$ and $\lambda_D$, resp. All other terms in the final form (\ref{eq:IV.75}) of the energy functional $\tETT$ do refer to its physical part $\ETiv$ (\ref{eq:IV.49}), i.\,e. the sum of the (renormalized) rest mass energy (\ref{eq:IV.54}), three kinds of kinetic energy (\ref{eq:IV.64}), (\ref{eq:IV.66}), (\ref{eq:IV.68}), and the gauge field energy (\ref{eq:IV.70}).

Once the right energy functional is written down, one can look now for the extremalizing configurations in order to check whether the system of eigenvalue equations is really reproduced. First, turn to the extremal configurations of $\tEePhi$ (\ref{eq:IV.75}) with respect to the angular functions $f_R(\vartheta)$ and $f_S(\vartheta)$ and find just those first-order equations (\ref{eq:IV.8a})--(\ref{eq:IV.8b}), provided the Langrangean multiplier $\lambda_f$ is identified with the quantum number $\elp$ in the following way
\begin{equation}
\label{eq:IV.78}
\frac{\lambda_f}{2 \hbar\crm \int\limits_0^\infty dr\;\tPhip(r) \cdot \tPhim(r)} = \elp \;.
\end{equation}

Next, one wishes to look for the extremal equations for the amplitudes $\tPhipm(r)$ and for the potential $\peAo(r)$. But for this purpose it is very convenient to first eliminate the angular functions $f_R(\vartheta), f_S(\vartheta)$ from the energy functional $\EePhif$ (\ref{eq:IV.75}), namely just by use of their extremal equations (\ref{eq:IV.8a})--(\ref{eq:IV.8b}). In this way, one is left with the reduced functional $\tEePhi$ which then acts only over the configuration space of the amplitudes $\tPhipm(r)$ and the potential $\pAn(r)$:
\begin{align}
\label{eq:IV.79}
\tEePhi = &2M\crm^2 \int\limits_0^\infty dr\;r\,\left[ \tPhip^2(r) - \tPhim^2(r) \right] - \frac{\hbar\crm}{\as} \int\limits_0^\infty dr\;r^2\,\left( \frac{d\,\peAn(r)}{dr} \right)^2 \\
+ &2\hbar\crm \int\limits_0^\infty dr\;r\,\left\{ \tPhim(r) \cdot \frac{d\,\tPhip(r)}{dr} - \tPhip(r) \cdot \frac{\tPhim(r)}{dr} - \left( 1+2\elp \right) \cdot \frac{\tPhip(r) \cdot \tPhim(r)}{r} \right\} \nonumber \quad\\
+ &\lGe \cdot \tNGee + 2 \lambda_D \cdot \tNPhi \;. \nonumber
\end{align}

Observe here again, that the constraints appear in the third line; all other terms are of
truly physical nature, i.\,e. the rest mass energy and the electrostatic interaction
energy (first line) and the kinetic energy (second line). Together with the angular
functions $f_R(\vartheta), f_S(\vartheta)$ there has also disappeared the quantum number
$\elz$ of angular momentum which says that the energy does \emph{not explicitly} depend on
$\elz$; however, there is of course an \emph{indirect} dependence on $\elz$ via the
interaction potential $\peAn(r)$. The reason for this is that the potential $\peAn(r)$
depends on the electrostatic charge distribution $\pko\rt$, see the Poisson equation
(\ref{eq:IV.10}), which takes account of the angular functions $f_R(\vartheta),
f_S(\vartheta)$ and therefore also of the quantum number $\elz$. Consequently we have to
expect a certain lifting of the energy degeneracy with respect to angular momentum $\elz$!

But concerning our reduced functional $\tEePhi$ (\ref{eq:IV.79}), it is now a standard variational procedure to reproduce from it the system of eigenvalue equations: the extremalization with respect to the amplitude fields $\tPhipm(r)$ generates the mass eigenvalue equations (\ref{eq:IV.9a})--(\ref{eq:IV.9b}), with the Langrangean multiplier $\lambda_D$ being given by
\begin{equation}
\label{eq:IV.80}
\lambda_D = -M_* \crm^2 \;.
\end{equation}
However, since we are satisfied for the moment with the spherically symmetric approximation, the extremalization of $\tEePhi$ (\ref{eq:IV.79}) with respect to the gauge potential $\peAn(r)$ does of course not yield the Poisson equation (\ref{eq:IV.10}) but rather the spherically symmetric approximation thereof:
\begin{equation}
\label{eq:IV.81}
\Delta_r\,\peAn(r) = -\as \cdot \frac{\tPhip^2(r) + \tPhim^2(r)}{r} \;.
\end{equation}
This may be understood as a kind of angular average (over the two-sphere) of the original Poisson equation (\ref{eq:IV.10}).

The spherically symmetric Poisson equation (\ref{eq:IV.81}) closes now the present RST eigenvalue problem in the spherically symmetric and electrostatic approximation. Its formal solution is given by
\begin{equation}
\label{eq:IV.82}
\peAn(r) = \frac{\as}{4\pi} \int \frac{d^3\vr\,'}{r'} \; \frac{\tPhip^2(r') + \tPhim^2(r')}{\left|\left| \vr - \vr\,' \right|\right|}
\end{equation}
which may be used in order to calculate the potential $\peAn(r)$ in terms of the amplitude
fields $\tPhipm(r)$. For the latter fields one may try some ansatz, as realistic as
possible, and may then substitute the result of integration $\peAn(r)$ in the energy
functional $\tEePhi$ (\ref{eq:IV.79}) in order to determine the ansatz parameters in the
trial amplitudes through extremalization of $\tEePhi$. The result of this procedure would
then be the desired energy spectrum (see below). However, a somewhat better picture of
this spectrum will surely be attained if we succeed to go (at least partially) beyond the
spherically symmetric approximation $\peAn(r)$. For this purpose, one may reconsider the
original Poisson equation (\ref{eq:IV.10}) whose formal solution is given by equation
(\ref{eq:IV.11}). Clearly, one expects that if we will be able to use that
\emph{anisotropic} potential $\pAn(\vr)$ (\ref{eq:IV.11}) in place of its \emph{isotropic}
aproximation $\peAn(r)$ (\ref{eq:IV.82}), then we will get an improved energy spectrum
(see below). The improvement must especially help to clarify the existence of an energetic
degeneracy of the states due to different quantum numbers $\elz$ but the same $\elp$.

\begin{center}
  \large{\textit{Non-Relativistic Energy Functional}}
\end{center}

Up to now, our line of arguments referred always to the relativistic situation; but the
main purpose of the present paper aims at the deviations of the RST predictions from the
conventional \emph{non-relativistic} spectrum (\ref{eq:I.2}) (a comparison of the
\emph{relativistic} corrections between both theoretical frameworks must be deferred to a
later study). Therefore one is now first interested in the non-relativistic approximation
to both the energy functional $\tEePhi$ (\ref{eq:IV.79}) and its extremal equations
(i.\,e. the non-relativistic form of the eigenvalue equations
(\ref{eq:IV.9a})--(\ref{eq:IV.9b}) in combination with the Poisson equation
(\ref{eq:IV.10})). Of course, one will demand from reasons of internal consistency that
the wanted non-relativistic field equations should represent just the extremal equations
due to the non-relativistic form of the energy functional. Thus we will first look now for
the non-relativistic energy functional ($\tEEePhi$, say) by working out the
non-relativistic form of its individual energy contributions.

The non-relativistic form of the first contribution to the relativistic functional
$\tEePhi$ (\ref{eq:IV.79}), i.\,e. the renormalized rest mass energy, has already been
determined by equation (\ref{eq:IV.58a}) and ultimately appeared in the following form:
\begin{equation}
\label{eq:IV.83}
2\,M\crm^2 \Zp^2\ \Rightarrow\ 2\,M\crm^2 - 2\; \pEkin_{[\Phi]} \;,
\end{equation}
namely by making use of the angular normalization condition (\ref{eq:IV.15a}). Here, the
non-relativistic kinetic one-particle energy $\pEkin$ is specified by equation
(\ref{eq:IV.58b}) for which the ``negative'' Pauli amplitude $\tPhim(r)$ has simply been
neglected. Indeed, this amplitude can be eliminated completely for passing over to the
non-relativistic approximation so that we can deal here exclusively with the ``positive''
Pauli component $\tPhip(r)$, being henceforth simply rewritten as $\tPhi(r)$. Thus the
non-relativistic one-particle energy $\pEkin$ appears simply as
\begin{equation}
\label{eq:IV.84}
\pEkin_{[\Phi]} = \frac{\hbar^2}{2M} \int\limits_0^\infty dr\;r\,\left\{ \left( \frac{d\,\tPhi(r)}{dr} \right)^2 + \elp^2 \cdot \frac{\tPhi^2(r)}{r^2} \right\} \;.
\end{equation}

The second energy contribution to the functional $\tEePhi$ (\ref{eq:IV.79}) (first line)
is the electrostatic gauge field energy $\ERee$ (\ref{eq:IV.70}), which is built up
exclusively by the electrostatic potential $\peAn(r)$. Therefore this term does not at all
change its formal appearance for the transcription to the non-relativistic situation. But,
of course, the relativistic $\pAn(r)$ cannot preserve its functional form
for the non-relativistic approach. Indeed, it is easy to deduce this desired
non-relativistic form from its relativistic predecessor (\ref{eq:IV.82}); namely by simply
omitting again the ``negative'' Pauli component $\tPhim(r)\ \left( \tPhip(r)\ \Rightarrow\
  \tPhi(r) \right)$
\begin{equation}
\label{eq:IV.85}
\peAn(r) = \frac{\as}{4\pi} \int \frac{d^3\,\vr\,'}{r'}\,\frac{\tPhi^2(r')}{\left|\left| \vr - \vr\,' \right|\right|} \;,
\end{equation}
see the corresponding remarks above equation (\ref{eq:IV.84}). However, one must not forget here the fact that both potentials $\peAn(r)$ (\ref{eq:IV.82}) and (\ref{eq:IV.85}) are a consequence of the spherically symmetric assumption $\left( \pAn(\vr)\ \Rightarrow\ \peAn(r) \right)$. Subsequently we will go one step further beyond this spherically symmetric approximation, namely be expanding the original \emph{anisotropic} potential $\pAn(\vr)$ (\ref{eq:IV.11}) around the \emph{isotropic} potential $\peAn(r)$ (\ref{eq:IV.85}). But for the moment we are satisfied with merely deducing the non-relativistic approximation of the relativistic functional $\tEePhi$ (\ref{eq:IV.79}) within the framework of the spherically symmetric approximation.

The last contribution of physical nature to the energy $\tEePhi$ (\ref{eq:IV.79}) is the kinetic energy $\ptTkinePhi$ (second line)
\begin{align}
\label{eq:IV.86}
\ptTkinePhi = 2\,\hbar\crm \int\limits_0^\infty dr\;r\,\left\{ \tPhim(r) \cdot \frac{d\,\tPhip(r)}{dr} - \tPhip(r) \cdot \frac{d\,\tPhim(r)}{dr} \right. \\
\left. - \left( 1 + 2\,\elp \right)\,\frac{\tPhip(r) \cdot \tPhim(r)}{r} \right\} \;. \nonumber
\end{align}
The non-relativistic version hereof can easily be obtained by simply solving (approximately) the first eigenvalue equation (\ref{eq:IV.9a}) with respect to $\tPhim(r)$ as
\begin{equation}
\label{eq:IV.87}
\tPhim(r) \simeq \frac{\hbar}{2M\crm}\,\left\{ \frac{d\,\tPhip(r)}{dr} - \frac{\elp}{r} \cdot \tPhip(r) \right\}
\end{equation}
and then substituting this into the second eigenvalue equation (\ref{eq:IV.9b}) which
yields the four-fold non-relativistic kinetic energy ${}^{(p)}\Ekin_{[\Phi]}$ (\ref{eq:IV.58b}):
\begin{align}
\label{eq:IV.88}
\ptTkinePhi\ &\Rightarrow\ 4\,{}^{(p)}\Ekin_{[\Phi]} \\
&= \frac{2\,\hbar^2}{M} \int\limits_0^\infty dr\;r\,\left\{ \left( \frac{d\,\tPhi(r)}{dr} \right)^2 + \frac{\elp^2}{r^2} \cdot \tPhi^2(r) \right\} \nonumber \;.
\end{align}

Finally, it remains to consider also the non-relativistic forms of the constraints for the \emph{principle of minimal energy} (see last line of equation (\ref{eq:IV.79})). The relativistic Poisson constraint $\tNGee$ (\ref{eq:IV.74}) is nothing else than the difference of the electrostatic field energy $\ERee$ (\ref{eq:IV.70}) and its mass equivalent $\tMee\crm^2$ (\ref{eq:IV.73}). The non-relativistic version ($\tMMee\crm^2$, say) of the mass equivalent is obtained again by simply omitting the ``negative'' Pauli amplitude $\tPhim(r)$ and putting $\tPhip(r)\Rightarrow\tPhi(r)$:
\begin{equation}
\label{eq:IV.89}
\tMee\crm^2\ \Rightarrow\ \tMMee\crm^2 \doteqdot -\hbar\crm \int\limits_0^\infty dr\;r\,\peAn(r)\,\tPhi^2(r) \;,
\end{equation}
so that the non-relativistic Poisson constraint $(\tNNGee)$ reads
\begin{equation}
\label{eq:IV.90}
\tNGee\ \Rightarrow\ \tNNGee \doteqdot \ERee - \tMMee\crm^2 = 0 \;.
\end{equation}
And finally, the non-relativistic normalization term $(\tNNPhi)$ due to its relativistic predecessor $\tNPhi$ (\ref{eq:IV.77}) must evidently look as follows
\begin{equation}
\label{eq:IV.91}
\tNPhi\ \Rightarrow\ \tNNPhi \doteqdot \int\limits_0^\infty dr\;r\,\tPhi^2(r) - 1 = 0 \;.
\end{equation}

Thus, collecting all the non-relativistic contributions, one ultimately finds for the non-relativistic approximation $\tEEePhi$ of the relativistic case $\tEePhi$ (\ref{eq:IV.79}):
\begin{equation}
\label{eq:IV.92}
\tEePhi\ \Rightarrow\ \tEEePhi = 2\,{}^{(p)}\Ekin_{[\Phi]} + \ERee + \lGe \cdot \tNNGee + 2\,\lambda_s \cdot \tNNPhi
\end{equation}
(here the Langrangean multiplier $\lambda_s$ is to be understood as the non-relativistic
version of $\lambda_D$). Observe also again that the proper rest mass energy $2\,M\crm^2$
is omitted (because of its dynamical ineffectiveness) and that the relativistic
renormalized rest energy $2\,M\Zp^2$ combines with the four-fold relativistic kinetic
energy $\ptTkin$, cf. (\ref{eq:IV.86}), just to the two-fold non-relativistic
single-particle energy ${}^{(p)}\Ekin_{[\Phi]}$, as it must be expected for a two-particle
system. Clearly, this supports the intrinsic consistency of the relativistic and
non-relativistic versions of our energy functional.

\begin{center}
  \large{\textit{Spherically Symmetric Eigenvalue Equations}}
\end{center}

A further consistency test refers to the extremal equations of the non-relativistic functional $\tEEePhi$ (\ref{eq:IV.92}). Indeed, one expects here that those extremal equations will coincide with the non-relativistic approximation of the relativistic eigenvalue system (\ref{eq:IV.9a})--(\ref{eq:IV.9b}) to be complemented by the non-relativistic version of the relativistic Poisson equation (\ref{eq:IV.81}), i.\,e.
\begin{equation}
\label{eq:IV.93}
\Delta\,\peAn(r) = -\as \frac{\tPhi^2(r)}{r} \;,
\end{equation}
which has the formal solution displayed by equation (\ref{eq:IV.85}). But indeed, it is easy to see that this consistency test is positive: the extremalization of the present non-relativistic functional $\tEEePhi$ (\ref{eq:IV.92}) with respect to the amplitude field $\tPhi(r)$ generates the following Schr\"{o}dinger-like equation
\begin{align}
\label{eq:IV.94}
-\frac{\hbar^2}{2M} \left( \frac{d^2}{dr^2} + \frac{1}{r} \frac{d}{dr} \right) \tPhi(r) + \frac{\hbar^2}{2Mr^2} \, \elp^2 \cdot \tPhi(r) 
-\hbar\crm\ \peAn(r) \cdot \tPhi(r) = -\lambda_s \cdot \tPhi(r) \;,
\end{align}
and furthermore the extremalization with respect to the gauge potential $\peAn(r)$ lets
emerge just the spherically symmetric Poisson equation (\ref{eq:IV.93}). On the other
hand, we can deduce also this Schr\"{o}dinger-equation (\ref{eq:IV.94}) as well directly
from the relativistic eigenvalue system (\ref{eq:IV.9a})--(\ref{eq:IV.9b}): for this
purpose, resort again to the former equation (\ref{eq:IV.87}) and substitute this into the
second equation (\ref{eq:IV.9b}) in order to find $\left( \tPhip(r)\Rightarrow\tPhi(r)
\right)$
\begin{align}
\label{eq:IV.95}
-\frac{\hbar^2}{2M} \left( \frac{d^2}{dr^2} + \frac{1}{r} \frac{d}{dr} \right) \tPhi(r) &+ \frac{\hbar^2}{2Mr^2}\,\elp^2 \cdot \tPhi(r) - \hbar\crm\,\peAn(r) \cdot \tPhi(r) \\
&= \left( M_* - M \right) \crm^2 \cdot \tPhi(r) \;. \nonumber
\end{align}
Comparing this result to the former one (\ref{eq:IV.94}) yields the identification of the non-relativistic Lagrangean multiplier $\lambda_s$ with the difference of mass energies
\begin{equation}
\label{eq:IV.96}
\lambda_s = \left( M- M_* \right) \crm^2 \doteqdot - E_* \;,
\end{equation}
i.\,e. the non-relativistic multiplier $\lambda_s$ plays the part of the Schr\"{o}dinger eigenvalue $E_*$ in the Schr\"{o}dinger-like equation (\ref{eq:IV.95}).

The present spherically symmetric approach to the positronium eigenvalue problem,
consisting of the equations (\ref{eq:IV.92})--(\ref{eq:IV.96}) has been used in order to
get a first rough estimate of the energy spectrum~\cite{3}; and it has been found that the
corresponding predictions deviate from the conventional predictions (\ref{eq:I.2}) by some
5\,--\,10\,\%. This result is not too bad, especially when one recalls that it came about
by neglecting the anisotropy of the interaction potential $\pAn(\vr)$. But in the strict
sense, this potential cannot be spherically symmetric, not even for the product ansatz
described by the equations (\ref{eq:IV.4a})--(\ref{eq:IV.4b}),
(\ref{eq:IV.6a})--(\ref{eq:IV.6b}) and (\ref{eq:IV.7a})--(\ref{eq:IV.7b}), see the
integral representation (\ref{eq:IV.11}) in terms of the charge density $\pko(\vr)$. This
anisotropy of $\pAn(\vr)$ does exist also for all the excited states even if they have
vanishing $z$-component of angular momentum $\left( \leadsto \elz = 0, \elp \geq 0
\right)$. Therefore, if we wish to get a more accurate RST prediction of the positronium
spectrum (especially for $\elz \neq 0$), we have to regard also the anisotropy of that
interaction potential $\pAn(\vr)$, at least in some approximative sense.

\begin{center}
  \large{\textit{Anisotropic Interaction Potential}}
\end{center}

The point of departure for studying the anisotropy in question is, of course, the integral
representation (\ref{eq:IV.11}) of $\pAn(\vr)$, which in the non-relativistic
approximation becomes simplified to
\begin{equation}
\label{eq:IV.97}
\pgAn\rt = \as \int \frac{d\,\Omega'}{4\pi}\,\frac{f_R^2(\vartheta') + f_S^2(\vartheta')}{\sin\vartheta'} \int\limits_0^\infty dr'\;r'\,\frac{\tPhi^2(r')}{\left|\left| \vr - \vr\,' \right|\right|} \;.
\end{equation}

For the subsequent expansion of this anisotropic potential with respect to the magnitude
of anisotropy it is very helpful to first clarify its relation to the spherically
symmetric approximation $\peAn(r)$ (\ref{eq:IV.85}). To this end, consider the average of
the anisotropic potential $\pgAn\rt$ over the 2-sphere; indeed, for this one finds by
explicitly integrating over the solid angle $\Omega$:
\begin{equation}
\label{eq:IV.98}
\int\frac{d\,\Omega}{4\pi}\,\pgAn\rt = \int\frac{d\,\Omega'}{4\pi}\,\frac{f_R^2(\vartheta') + f_S^2(\vartheta')}{\sin\vartheta'} \cdot \peAn(r) = \peAn(r)
\end{equation}
where one has made use also of the angular normalization condition (\ref{eq:IV.15a}). Thus, restating this result because of its importance for the subsequent perturbation expansion, we note down for our spherically symmetric approximation
\begin{equation}
\label{eq:IV.99}
\peAn(r) \equiv \int \frac{d\,\Omega}{4\pi}\,\pgAn\rt \;.
\end{equation}

This namely suggests to define the anisotropic part $\Aan\rt$ of $\pgAn\rt$ as just its
deviation from the spherically symmetric approximation, i.\,e. we put~\cite{1}
\begin{equation}
\label{eq:IV.100}
\pgAn\rt = \peAn(r) + \Aan\rt \;.
\end{equation}
Or conversely, combining both integral representations (\ref{eq:IV.85}) and
(\ref{eq:IV.97}) one gets the following integral representation for the anisotropic part:
\begin{align}
\label{eq:IV.101}
\Aan\rt &= \pgAn\rt - \peAn(r) \\
&= \as \int \frac{d\,\Omega'}{4\pi}\,\left[ \frac{f_R^2(\vartheta') + f_S^2(\vartheta')}{\sin\vartheta'} - 1 \right]\,\int\limits_0^\infty dr'\;r'\,\frac{\tPhi^2(r')}{\left|\left| \vr - \vr\,' \right|\right|} \nonumber \;.
\end{align}
Would we neglect here the angular dependence of the radial (i.\,e. the last) integral, then we would find the anisotropic part vanishing $\left( \Aan\rt \Rightarrow 0 \right)$ and thus the potential $\pgAn\rt$ (\ref{eq:IV.97}) would coincide with its spherically symmetric approximation $\peAn(r)$. Therefore the angular dependence of $\pgAn\rt$ (\ref{eq:IV.97}) comes about through the regard of the angular dependence of the radial integral! This fact may now be exploited in order to establish a perturbation expansion of the anisotropic part $\Aan\rt$ (\ref{eq:IV.101}) with respect to the magnitude of the anisotropy, namely just by constructing such an expansion for that radial integral.

The point of departure for the wanted perturbation series is the following expansion of the denominator in the radial part of (\ref{eq:IV.101})
\begin{align}
\label{eq:IV.102}
\frac{1}{\left|\left| \vr - \vr\,' \right|\right|} &= \frac{1}{\left( r^2 + {r'}^2 \right)^\frac{1}{2}} + \frac{r \cdot r'}{\left( r^2 + {r'}^2 \right)^\frac{3}{2}}\,\left( \hat{\vr} \sdot \hat{\vr}\,' \right) + \frac{3}{2}\,\frac{r^2 \cdot {r'}^2}{\left( r^2 + {r'}^2 \right)^\frac{5}{2}}\,\left( \hat{\vr} \sdot \hat{\vr}\,' \right)^2 \\
&+ \frac{5}{2}\,\frac{r^3 \cdot {r'}^3}{\left( r^2 + {r'}^2 \right)^\frac{7}{2}}\,\left( \hat{\vr} \sdot \hat{\vr}\,' \right)^3 + \frac{35}{8}\,\frac{r^4 \cdot {r'}^4}{\left( r^2 + {r'}^2 \right)^\frac{9}{2}}\,\left( \hat{\vr} \sdot \hat{\vr}\,' \right)^4 + \ldots \nonumber
\end{align}
Substituting this series expansion into the integral representation (\ref{eq:IV.101}) for
the anisotropic part $\Aan\rt$ yields an expansion of the following kind:
\begin{align}
\label{eq:IV.103}
\Aan\rt &= \pgA^\mathsf{I}(\vartheta) \cdot \pgA^\mathsf{I}(r) + \pgA^\mathsf{II}(\vartheta) \cdot \pgA^\mathsf{II}(r) + \pgA^\mathsf{III}(\vartheta) \cdot \pgA^\mathsf{III}(r) \\
&+ \pgA^\mathsf{IV}(\vartheta) \cdot \pgA^\mathsf{IV}(r) + \pgA^\mathsf{V}(\vartheta) \cdot \pgA^\mathsf{V}(r) + \ldots \nonumber
\end{align}
Here, the angular pre-factors in front of the radial auxiliary potentials
$\pgA^\mathsf{I}(r)$, $\pgA^\mathsf{II}(r)$, $\pgA^\mathsf{III}(r)$ depend on the angular
functions $f_R(\vartheta)$, $f_S(\vartheta)$ in the following way
\begin{subequations}
\begin{align}
\label{eq:IV.104a}
\pgA^\mathsf{I}(\vartheta) &\doteqdot \int \frac{d\,\Omega'}{4\pi}\,\left[ \frac{f_R^2(\vartheta') + f_S^2(\vartheta')}{\sin\vartheta'} - 1 \right] \equiv 0 \\
\label{eq:IV.104b}
\pgA^\mathsf{II}(\vartheta) &\doteqdot \int \frac{d\,\Omega'}{4\pi}\,\left( \hat{\vr} \sdot \hat{\vr}\,' \right)\,\left[ \frac{f_R^2(\vartheta') + f_S^2(\vartheta')}{\sin\vartheta'} - 1 \right] \equiv 0 \\
\label{eq:IV.104c}
\pgA^\mathsf{III}(\vartheta) &\doteqdot \frac{3}{2} \int \frac{d\,\Omega'}{4\pi}\,\left(
  \hat{\vr} \sdot \hat{\vr}\,' \right)^2\,\left[ \frac{f_R^2(\vartheta') +
    f_S^2(\vartheta')}{\sin\vartheta'} - 1 \right]  \\
\text{i.e.\ for~\boldmath$\lP=3$:}\nonumber \\*
&\pgA^\mathsf{III}(\vartheta)= \begin{cases}
\ \frac{12}{32} \left[ \cos^2\vartheta - \frac{1}{3} \right]\ \ ,\ \ \elz = 0 \\
\ \frac{9}{32} \left[ \cos^2\vartheta - \frac{1}{3} \right]\ \ ,\ \ \elz = \pm 1 \\
\ \quad\quad\quad 0 \quad\quad\quad\ \ ,\ \ \elz = \pm 2 \\
\ -\frac{15}{32} \left[ \cos^2\vartheta - \frac{1}{3} \right] ,\ \ \elz = \pm 3
\end{cases}
\nonumber
\end{align}
\end{subequations}
(the result for~$l_z=0$ does hold for any~$\lP$, see \textbf{App.B}).

As a check of the results for these angular pre-factors, one deduces from the combination
of equations (\ref{eq:IV.99}) and (\ref{eq:IV.100}) that their angular average must
vanish, i.\,e.
\begin{equation}
\label{eq:IV.105}
\int d\Omega\;\pgA^\mathsf{III}(\vartheta) = \int d\Omega\;\pgA^\mathsf{V}(\vartheta) = \ldots = 0 \;.
\end{equation}
This is trivially satisfied for $\pgA^\mathsf{III}(\vartheta)$ (\ref{eq:IV.104c}) with
$\elz = \pm 2$; indeed the vanishing of $\pgA^\mathsf{III}(\vartheta)$ for $\elz = \pm 2$
says that in this case the spherically symmetric approximation $\peAn(r)$ is a very good
approximation and corrections can arise only in the higher perturbation orders. This
circumstance equips the spherically symmetric approximation with additional relevance,
since its predictions remain partially true even in the first order of
perturbation. Observe also that the first non-trivial order of anisotropic perturbation is
of the third kind (\ref{eq:IV.104c}); the first two correction terms
$\pgA^\mathsf{I}(\vartheta)$ (\ref{eq:IV.104a}) and $\pgA^\mathsf{II}(\vartheta)$
(\ref{eq:IV.104b}) must \emph{always} vanish.

For the radial correction factors $\pgA^\mathsf{I}(r)$, $\pgA^\mathsf{II}(r)$, $\ldots$ of the anisotropic part $\Aan\rt$ (\ref{eq:IV.103}) one obtains the following results~\cite{1}
\begin{subequations}
\begin{align}
\label{eq:IV.106a}
\pgA^\mathsf{I}(r) &= \as \cdot \int\limits_0^\infty dr'\;\frac{r' \cdot \tPhi^2(r')}{\left( r^2 + {r'}^2 \right)^\frac{1}{2}} \\
\label{eq:IV.106b}
\pgA^\mathsf{II}(r) &= \as \cdot r \int\limits_0^\infty dr'\;\frac{{r'}^2 \cdot \tPhi^2(r')}{\left( r^2 + {r'}^2 \right)^\frac{3}{2}} \\
\label{eq:IV.106c}
\pgA^\mathsf{III}(r) &= \as \cdot r^2 \int\limits_0^\infty dr'\;\frac{{r'}^3 \cdot \tPhi^2(r')}{\left( r^2 + {r'}^2 \right)^\frac{5}{2}} \\
\label{eq:IV.106d}
\pgA^\mathsf{IV}(r) &= \as \cdot r^3 \int\limits_0^\infty dr'\;\frac{{r'}^4 \cdot \tPhi^2(r')}{\left( r^2 + {r'}^2 \right)^\frac{7}{2}} \\
\label{eq:IV.106e}
\pgA^\mathsf{V}(r) &= \as \cdot r^4 \int\limits_0^\infty dr'\;\frac{{r'}^5 \cdot \tPhi^2(r')}{\left( r^2 + {r'}^2 \right)^\frac{9}{2}} \\
&\text{\Huge{\vdots}} \nonumber
\end{align}
\end{subequations}

After having selected some trial amplitude $\tPhi(r)$, one could in principle explicitly
calculate these radial potentials in terms of the ansatz parameters contained in the
selected trial amplitude. However, our method of approximation just consists in leaving
unspecified these (non-trivial) radial correction potentials $\pgA^\mathsf{III}(r)$,
$\pgA^\mathsf{V}(r)$ etc. and determining them through solving their extremal equations
due to the \emph{principle of minimal energy}. Thus, the next step must concern the
reformulation of the energy functional $\tEEePhi$ (\ref{eq:IV.92}) in terms of the chosen
set of field variables $\tPhi(r)$, $\peAn(r)$, $\pgA^\mathsf{III}(r)$. For a first rough
estimate of the para-spectrum we restrict ourselves to the first non-trivial anisotropic
correction $\left( \sim \pgA^\mathsf{III}(r) \right)$, i.\,e. we let the expansion
(\ref{eq:IV.103}) of $\pgAan\rt$ stop at the third term:
\begin{equation}
\label{eq:IV.107}
\pgAan\rt \simeq \pgA^\mathsf{III}(\vartheta) \cdot \pgA^\mathsf{III}(r) \;.
\end{equation}

\newpage
\begin{center}
  \large{\textit{Energy of the Anisotropic Gauge Fields}}
\end{center}

\begin{sloppypar}
  In order to get a first impression of the anisotropic effects we still retain the
  spherical symmetry for the non-relativistic amplitude field $\tPhi(r)$ but we regard now
  the anisotropic effect on the electrostatic gauge field energy $\ERe \left( \Rightarrow
    \EReg \right)$ and its mass equivalent $\MMe\crm^2 \left( \Rightarrow \tMMeg\crm^2
  \right)$. The reason is that we think those effects to be the leading ones concerning
  the expansion of all fields with respect to the magnitude of anisotropy.
\end{sloppypar}

The starting point for the calculation of the ``anisotropic'' gauge field energy $\EReg$
is (\ref{eq:IV.69}). Inserting therein the present decomposition (\ref{eq:IV.100}) of the
anisotropic potential $\pgAn\rt$ with the anisotropic part $\Aan\rt$ (\ref{eq:IV.103})
yields an analogous decomposition of the gauge field energy $\EReg$ (\ref{eq:IV.69}) into
a ``spherically symmetric'' part $\ERee$ (\ref{eq:IV.70}) and an ``anisotropic'' part
$\Eegan$, i.\,e.
\begin{equation}
\label{eq:IV.108}
\EReg = -\frac{\hbar\crm}{4\pi\as}\,\int d^3\vr\; \left|\left| \vec{\nabla}\; \pgAn\rt \right|\right|^2 = \ERee + \Eegan \;,
\end{equation}
where the ``anisotropic'' part $\Eegan$ is given in terms of the radial correction potential $\pgA^\mathsf{III}(r)$ by
\begin{equation}
\label{eq:IV.109}
\Eegan \Rightarrow \Eegiii = -\frac{\hbar\crm}{\as} \left\{ \pr e_3 \cdot \int\limits_0^\infty dr\,r^2\,\left( \frac{d\,\pgA^\mathsf{III}(r)}{dr} \right)^2 + \pr f_3 \cdot \int\limits_0^\infty dr\,\left( \pgA^\mathsf{III}(r) \right)^2 \right\} \;,
\end{equation}
since we are satisfied for the moment with the third-order approximation. Here, the coefficients $\pr e_3$ and $\pr f_3$ are determined by the angular correction potential $\pgA^\mathsf{III}(\vartheta)$ through
\begin{subequations}
\begin{align}
\label{eq:IV.110a}
\pr e_3 &\doteqdot \int\frac{d\,\Omega}{4\pi} \left( \pgA^\mathsf{III}(\vartheta) \right)^2 \\
\label{eq:IV.110b}
\pr f_3 &\doteqdot \int\frac{d\,\Omega}{4\pi} \left( \frac{d\;\pgA^\mathsf{III}(\vartheta)}{d\vartheta} \right)^2 \;.
\end{align}
\end{subequations}
Since the angular correction potential $\pgA^\mathsf{III}(\vartheta)$ is explicitly known, cf. equation (\ref{eq:IV.104c}), both coefficients are also known numerically, see the following table for $\boldsymbol{\elp = 3}$:
\newpage
{\centerline {\boldmath$\lP=3$}}
\begin{center}
\label{tablelp3}
\begin{tabular}{|c||c|c|c|c|}
\hline
$\elz$ & $\quad 0 \quad $ & $\quad \pm 1 \quad $ & $\quad \pm 2 \quad $ & $\quad \pm 3\quad $ \\
\hline\hline
$\parbox[0pt][3em][c]{0cm}{}\pr e_3$ & $\frac{1}{80}$ & $\frac{9}{1280}$ & $0$ & $\frac{5}{256}$ \\
\hline
$\parbox[0pt][3em][c]{0cm}{}\pr f_3$ & $\frac{3}{40}$ & $\frac{27}{640}$ & $0$ & $\frac{15}{128}$ \\
\hline
$\parbox[0pt][3em][c]{0cm}{}\pr m_3$ & $\frac{1}{16}$ & $\frac{9}{256}$ & $0$ & $\frac{25}{256}$ \\
\hline\hline
$\parbox[0pt][3em][c]{0cm}{}\frac{\pr f_3}{\pr e_3}$ & $6$ & $6$ & & $6$ \\
\hline
$\parbox[0pt][3em][c]{0cm}{}\frac{\pr m_3}{\pr e_3}$ & $5$ & $5$ & & $5$ \\
\hline
\end{tabular}
\end{center}
\vskip 0.5cm

Besides the gauge field energy $\EReg$ there is a further gauge field contribution to the
energy functional $\tEEePhi$ (\ref{eq:IV.92}) whose presence is needed in order that the
extremal equations do coincide with the eigenvalue system, i.\,e. the Poisson constraint
$\tNNGeg$
\begin{equation}
\label{eq:IV.111}
\tNNGeg \doteqdot \EReg - \tMMeg\crm^2 \equiv 0 \;,
\end{equation}
which is the anisotropic generalization of the spherically symmetric constraint $\tNNGee$
as shown by equation (\ref{eq:IV.90}). The ``anisotropic'' gauge field energy $\EReg$ is
given by equations (\ref{eq:IV.108})--(\ref{eq:IV.109}), and the anisotropic
generalization $\tMMeg\crm^2$ of the isotropic $\tMMee\crm^2$ (\ref{eq:IV.89}) must look
as follows:
\begin{equation}
\label{eq:IV.112}
\tMMeg\crm^2 = -\hbar\crm \int d^3\vr\;\pgAn\rt\,\pgko\rt \;.
\end{equation}
Substituting herein again the decomposition of the anisotropic potential $\pgAn\rt$
(\ref{eq:IV.100}) and the factorization (\ref{eq:IV.14}) of the charge density $\pgko\rt$
lets this mass equivalent $\tMMeg\crm^2$ (\ref{eq:IV.112}) reappear also as a sum of the
``isotropic'' part $\tMMee\crm^2$ (\ref{eq:IV.89}) and ``anisotropic'' part
$\tMMegan\crm^2$, i.e.
\begin{equation}
\label{eq:IV.113}
\tMMeg\crm^2 = \tMMee\crm^2 + \tMMegan\crm^2 \;,
\end{equation}
where the ``anisotropic'' part consists here solely of the third-order correction potential $\pgA^\mathsf{III}\rt$ $\left( \doteqdot \pgA^\mathsf{III}(\vartheta) \cdot \pgA^\mathsf{III}(r)\right)$, i.\,e.
\begin{equation}
\label{eq:IV.114}
\tMMegan\crm^2 \Rightarrow \tMMegiii\crm^2 \doteqdot -\hbar\crm\,\pr m_3 \cdot \int\limits_0^\infty dr\;r\,\pgA^\mathsf{III}(r)\,\tPhi^2(r) \;.
\end{equation}
Here, the angular part $\pgA^\mathsf{III}(\vartheta)$ of $\pgAan\rt$ (\ref{eq:IV.107})
builds up the numerical coefficient $\pr m_3$ in the following way
\begin{equation}
\label{eq:IV.115}
\pr m_3 \doteqdot \int d\Omega\;\pgA^\mathsf{III}(\vartheta) \cdot \pgko(\vartheta) = \int d\Omega\;\pgA^\mathsf{III}(\vartheta) \cdot \frac{f_R^2(\vartheta) + f_S^2(\vartheta)}{4\pi\sin\vartheta}
\end{equation}
(for the numerical values of $\pr m_3$ see the above table on p.~\pageref{tablelp3}).

With respect to the \emph{principle of minimal energy} there is now an important
circumstance; namely the splitting of the ``anisotropic'' Poisson constraint $\tNNGeg$
(\ref{eq:IV.111}) into certain subconstraints. First observe here that, due to the
splitting of $\EReg$ (\ref{eq:IV.108}) and its mass equivalent $\tMMeg\crm^2$
(\ref{eq:IV.113}), the original Poisson constraint $\tNNGeg$ splits naturally up into the
``spherically symmetric'' part $\tNNGee$ (\ref{eq:IV.90}) and an ``anisotropic'' part
$\tNNegan$:
\begin{equation}
\label{eq:IV.116}
\tNNGeg = \tNNGee + \tNNegan \;,
\end{equation}
with the ``anisotropic'' part being given in an self-evident way as
\begin{equation}
\label{eq:IV.117}
\tNNegan = \Eegan - \tMMegan\crm^2 \;,
\end{equation}
i.\,e. more concretely since we are presently satisfied with the first non-trivial approximation order $\left( \sim \pgA^\mathsf{III}(\vartheta) \right)$
\begin{equation}
\label{eq:IV.118}
\tNNegan \Rightarrow \tNNegiii \doteqdot \Eegiii - \tMMegiii\crm^2 \equiv 0 \;.
\end{equation}
This means that, because of considering both gauge field constituents $\peAn(r)$ and
$\pgA^\mathsf{III}(r)$ as independent field degrees of freedom, we can impose both Poisson
constraints (\ref{eq:IV.90}) and (\ref{eq:IV.118}) \emph{independently} from each
other. Thus, the one constraint $\tNNGeg = 0$ (\ref{eq:IV.111}) is broken up into the two
independent constraints $\tNNGee = 0$ (\ref{eq:IV.90}) and $\tNNegan = 0$
(\ref{eq:IV.118}).

Summarizing the results for the ``anisotropic'' gauge field energy (i.\,e. more precisely:
the energy of the anisotropic gauge field configurations), we find it convenient to split
up the present energy functional $\tEEePhi$ (\ref{eq:IV.92}) into a (Dirac) matter part
$\DtEEePhi$ and a gauge field part $\GtEEePhi$ which, however, enters the present
electrostatic approximation of the energy functional only via its electric component
$\etEEePhi$, i.\,e. we put
\begin{equation}
\label{eq:IV.119}
\tEEePhi = \DtEEePhi + \etEEePhi
\end{equation}
with the matter part being given by
\begin{equation}
\label{eq:IV.120}
\DtEEePhi = 2\,{}^{(p)}\Ekin_{\,[\Phi]} + 2\,\lambda_s \cdot \tNNPhi
\end{equation}
and the electrostatic gauge field part by
\begin{equation}
\label{eq:IV.121}
\etEEePhi = \EReg + \lGe \cdot \tNNGeg \;.
\end{equation}
Obviously, each of both contributions consists of a physical term and a constraint; but the proper meaning of the splitting (\ref{eq:IV.116}) refers to the \emph{principle of minimal energy}
\begin{equation}
\label{eq:IV.122}
\delta\;\tEEePhi = 0
\end{equation}
which is required to reproduce the complete system of eigenvalue equations (including the gauge field part thereof). In this respect, the observation is now that the gauge potential $\pgAn\rt$ is contained solely in the electric part $\etEEePhi$ (\ref{eq:IV.121}). Consequently, it is sufficient to consider only the latter part of the energy functional for the purpose of deducing the gauge field equations. More concretely, since the gauge field has been decomposed into the two independent field degrees of freedom $\peAn(r)$ and $\pgA^\mathsf{III}(r)$, the corresponding field equations must be deducible from (\ref{eq:IV.122}) through the two separate extremalization processes
\begin{subequations}
\begin{align}
\label{eq:IV.123a}
\delta_0\;\etEEePhi &= 0 \\
\label{eq:IV.123b}
\delta_\mathsf{III}\;\etEEePhi &= 0 \;,
\end{align}
\end{subequations}
where $\delta_0$ $\left( \delta_\mathsf{III} \right)$ denotes the extremalization with respect to $\peAn(r)$ $\left( \pgA^\mathsf{III}(r) \right)$, resp., and the matter part $\DtEEPhi$ (\ref{eq:IV.120}) plays no part at all.

\begin{center}
  \large{\textit{Non-Relativistic Gauge Field Equations}}
\end{center}

Now that all the details of the gauge field part $\etEEePhi$ (\ref{eq:IV.121}) of the energy functional have been clarified, one can proceed in determining the corresponding extremal equations. Since the spherically symmetric approximation $\peAn(r)$ is conceived to be an independent field, the associated extremalization process (\ref{eq:IV.123a}) appears in a more detailed form as
\begin{equation}
\label{eq:IV.124}
\delta_0\,\left\{ \ERee + \lGe \cdot \tNNGee \right\} = 0
\end{equation}
and yields of course the former ``spherically symmetric'' Poisson equation (\ref{eq:IV.93}) with $\lGe = -2$. Clearly, this is not a new result; but the ``anisotropic'' counterpart (\ref{eq:IV.123b}) is just what we need in order to discuss the elimination of the energetic degeneracy of the states due to different $\elz$ (but the same $\elp$).

That second extremalization process (\ref{eq:IV.123b}) reads in a more detailed form
\begin{equation}
\label{eq:IV.125}
\delta_\mathsf{III}\,\left\{ \Eegiii + \lGe \cdot \tNNegiii \right\} = 0
\end{equation}
where the ``anisotropic'' energy $\Eegiii$ is specified by equation (\ref{eq:IV.109}) and its mass equivalent $\tMMegiii\crm^2$ by equation (\ref{eq:IV.114}). Thus the extremalization process (\ref{eq:IV.125}) generates the following \emph{quadrupole equation} for $\pgA^\mathsf{III}(r)$ [Zitat]
\begin{align}
\label{eq:IV.126}
\pr e_3 \cdot \left( \frac{d^2\,\pgA^\mathsf{III}(r)}{dr^2} + \frac{2}{r}\,\frac{d\,\pgA^\mathsf{III}(r)}{dr} \right) &- \pr f_3 \cdot \frac{\pgA^\mathsf{III}(r)}{r^2} \\
&= -\pr m_3 \cdot \as\,\frac{\tPhi^2(r)}{r} \;. \nonumber
\end{align}
Amazingly enough, this equation adopts the same form for all values of angular momentum $\elz$ (besides the trivial case $\elz = \pm 2$ for $\elp = 3$):
\begin{equation}
\label{eq:IV.127}
\frac{d^2\,\pgA^\mathsf{III}(r)}{dr^2} + \frac{2}{r}\,\frac{d\,\pgA^\mathsf{III}(r)}{dr} - 6\,\frac{\pgA^\mathsf{III}(r)}{r^2} = -5\,\as\,\frac{\tPhi^2(r)}{r} \;,
\end{equation}
see the table of numerical coefficients on p.~\pageref{tablelp3}. By comparing this
quadrupole equation for the radial correction potential $\pgA^\mathsf{III}(r)$ to the
Poisson equation (\ref{eq:IV.93}) for the spherically symmetric approximation $\peAn(r)$
shows that the coupling of $\pgA^\mathsf{III}(r)$ to the spherically symmetric amplitude
field $\tPhi(r)$ is five times stronger than the corresponding coupling of
$\peAo(r)$. This, however, does not entail that the energy $\Eegiii$ (\ref{eq:IV.109})
located in the field $\pgA^\mathsf{III}(r)$ is greater than that being concentrated in the
isotropic field $\peAn(r)$. Indeed, we will readily see that the anisotropy energy due to
$\pgA^\mathsf{III}(r)$ is only a small fraction of the ``isotropic'' energy $\ERee$ due to
$\peAn(r)$.

Though all the states with the same value of $\elp$ ($=3$) but different $\elz$ ($= 0$, $\pm 1$, $\pm 3$) share the same correction potential $\pgA^\mathsf{III}(r)$, this does not entail that they all possess the same anisotropy energy $\Eegiii$ (\ref{eq:IV.109}). In order to elucidate this fact, rewrite the latter quantity in the following form
\begin{equation}
\label{eq:IV.128}
\Eegiii = - \pr e_3 \cdot \frac{\hbar\crm}{\as} \int\limits_0^\infty dr\;r^2\,\left\{ \left( \frac{d\,\pgA^\mathsf{III}(r)}{dr} \right)^2 + 6\,\left( \frac{\pgA^\mathsf{III}(r)}{r} \right)^2 \right\} \;,
\end{equation}
so that these anisotropy energies for $\elz = 0$, $\pm 1$, $\pm 3$ must obey the following interrelationship:
\begin{align}
\label{eq:IV.129}
\Eegiii\Big|_{\elz = 0}\ :\ \Eegiii\Big|_{\elz = \pm 1}\ :\ \Eegiii\Big|_{\elz = \pm 3} &= \pr e_3\Big|_{\elz = 0}\ :\ \pr e_3\Big|_{\elz = \pm 1}\ :\ \pr e_3\Big|_{\elz = \pm 3} \\
&= 1\ :\ 0{,}5625\ :\ 1{,}5625 \nonumber
\end{align}
(provided the source term $\left( \sim \tPhi^2(r) \right)$ in the quadrupole equation
(\ref{eq:IV.127}) is kept fixed). Furthermore, since the anisotropy energy $\Eegiii$
(\ref{eq:IV.109}) is always negative and the spherically symmetric approximation alone
predicts somewhat too high energies, the corrective property of the present approximation
method is optimal for $\elz = \pm 3$. Since in this case the value of $\elz$ is itself
extremal (i.\,e. $\left| \elz \right| = \elp$), such a state would correspond to circular
motion in the Bohr-Sommerfeld picture of the atoms; and thus it would be interesting to
calculate the whole spectrum (for $\elp = \left| \elz \right| = 0$, $1, 2, 3, 4, \ldots$)
within the framework of the present approximation method (see below).

\begin{center}
  \large{\textit{Generalized Eigenvalue Equation for $\tPhi(r)$}}
\end{center}

The final act of our \emph{principle of minimal energy} must of course concern the eigenvalue equation for the non-relativistic amplitude field $\tPhi(r)$. Obviously, the wanted equation will represent some generalization of the ``spherically symmetric'' eigenvalue equation (\ref{eq:IV.94}) since the present ``anisotropic'' energy-functional $\tEEePhi$ (\ref{eq:IV.119})--(\ref{eq:IV.121}) differs from its ``isotropic'' counterpart (\ref{eq:IV.92}) just by the inclusion of the anisotropy energy $\Eegiii$ (\ref{eq:IV.109}). Or in other words, the extremalization of the present ``anisotropic'' energy functional $\tEEePhi$ (\ref{eq:IV.119}) with respect to the amplitude field $\tPhi(r)$ must reproduce that ``spherically symmetric'' eigenvalue equation (\ref{eq:IV.95}) with additional inclusion of an ``anisotropic'' term $\left( \sim \pgA^\mathsf{III}(r) \right)$. Indeed, the extremalization process with respect to $\tPhi(r)$ yields now an eigenvalue equation of the expected form:
\begin{align}
\label{eq:IV.130}
-\frac{\hbar^2}{2M}\,\left( \frac{d^2}{dr^2} + \frac{1}{r}\,\frac{d}{dr} \right)\, \tPhi(r) + \frac{\hbar^2}{2Mr^2}\,\elp^2 \cdot \tPhi(r)& \\
-\hbar\crm\,\left( \peAn(r) + \pr m_3 \cdot \pgA^\mathsf{III}(r) \right)\,\tPhi(r) &= E_* \cdot \tPhi(r) \nonumber
\end{align}
where the non-relativistic energy eigenvalue $E_*$ agrees again with the Lagrangean multplier $\lambda_s$, see equation (\ref{eq:IV.96}). Here it is evident that the numerical coefficient $\pr m_3$ (\ref{eq:IV.115}) plays the role of a coupling constant for the coupling of the amplitude field $\tPhi(r)$ to the anisotropic part $\pgA^\mathsf{III}(r)$ of the gauge potential $\pgAn\rt$, see the perturbation expansion (\ref{eq:IV.103}) of $\pgAan\rt$. Concerning now the coupling strengths of the amplitude field $\tPhi(r)$ to the isotropic and anisotropic parts of the gauge potential, we see from the present eigenvalue equation (\ref{eq:IV.130}) that the ratio of couplings $1\ :\ \pr m_3$ is in the order of magnitude $1\ :\ \frac{1}{10}$, or even smaller. This says that the coupling to the isotropic part $\peAn(r)$ is dominant so that the anisotropy energy $\Eegiii$ (\ref{eq:IV.128}) can actually be expected to represent only a small correction to the result of the spherically symmetric approximation (see below).

\newpage
\begin{center}
  \large{\textit{Selecting a Plausible Trial Amplitude $\tPhi(r)$}}
\end{center}

Being motivated by the acceptable success of the precedent results~\cite{1}-\cite{2}, our choice of
an appropriate trial amplitude $\tPhi(r)$ for extremalizing the energy functional
$\tEEePhi$ is again the following:
\begin{equation}
\label{eq:IV.131}
\tPhi(r) = \Phi_*\,r^\nu \e^{-\beta r} \;,
\end{equation}
where the normalization constant must adopt the value
\begin{equation}
\label{eq:IV.132}
\Phi_*^2 = \frac{\left( 2\beta \right)^{2\nu + 2}}{\Gamma\left( 2\nu + 2 \right)}
\end{equation}
in order to satisfy the non-relativistic normalization condition
(\ref{eq:IV.91}). Consequently, the constraint term $\tNNPhi$ can be omitted for our
energy functional $\tEEePhi$ (\ref{eq:IV.92}) or $\etEEPhi$ (\ref{eq:IV.120}), resp. Thus
our trial ansatz (\ref{eq:IV.131}) contains two variational parameters (i.\,e. $\nu$ and
$\beta$) for extremalizing the corresponding energy function $\EEivbn$, see below. Surely,
such a trial amplitude (\ref{eq:IV.131}) with only two variational parameters will be too
rough in order to predict the RST energy levels with a satisfying accuracy; but it may be
sufficient for the present purpose of demonstrating the level splitting for different
values of $\elz$ (but due to the same value of $\elp$).

As mentioned above, the suggestion for a first rough estimate of the para-spectrum
consists now in taking the value of the energy functional $\tEEePhi$ (\ref{eq:IV.119}) on
the selected trial amplitude $\tPhi(r)$ (\ref{eq:IV.131}) and to look for the extremal
points of the corresponding energy function $\EEivbn$. First, the value of the kinetic
energy functional $\pEkin_{[\Phi]}$ (\ref{eq:IV.84}) on $\tPhi(r)$ (\ref{eq:IV.131})
becomes by straightforward integration
\begin{equation}
\label{eq:IV.133}
\Ekin(\nu, \beta) = \frac{e^2}{2 a_B}\,\left( 2 \beta a_B \right)^2 \cdot \ekin(\nu)
\end{equation}
with the kinetic function $\ekin(\nu)$ being given by
\begin{equation}
\label{eq:IV.134}
\ekin(\nu) = \frac{1}{2\nu + 1}\,\left( \frac{1}{4} + \frac{\elp^2}{2\nu} \right) \;.
\end{equation}
Observe here that the kinetic energy does not depend on the quantum number $\elz$ (but
exclusively $\elp$)!

Next, consider the isotropic gauge field energy $\ERee$ (\ref{eq:IV.70}) which must equal
its mass equivalent $\tMMee\crm^2$ (\ref{eq:IV.89}), provided we can use for the
calculation of these two objects the exact solution of the ``isotropic'' Poisson equation
(\ref{eq:IV.93}). In this case, we then can dispense with the ``isotropic'' Poisson
constraint term in our energy functional $\tEEePhi$ (\ref{eq:IV.119})--(\ref{eq:IV.121}),
i.\,e. $\tNNGee = 0$. However, such an exact solution $\peAn(r)$ is easily obtainable for
our selected trial amplitude $\tPhi(r)$ (\ref{eq:IV.131})--(\ref{eq:IV.132}). Here it is
convenient to pass over to dimensionless objects by putting
\begin{subequations}
\begin{align}
\label{eq:IV.135a}
\peAn(r) &= 2\,\beta\as \cdot \tanu(y) \\
\Big( y &\doteqdot 2\,\beta r\Big)\ ,
\end{align}
\end{subequations}
which recasts that ``isotropic'' Poisson equation (\ref{eq:IV.93}) in dimensionless form
\begin{equation}
\label{eq:IV.136}
\left( \frac{d^2}{dy^2} + \frac{2}{y}\,\frac{d}{dy} \right)\,\tanu(y) = -\frac{\e^{-y} \cdot y^{2\nu - 1}}{\Gamma\left( 2\nu + 2 \right)} \;.
\end{equation}
The exact solution hereof is given by~\cite{1},~\cite{4}
\begin{equation}
\label{eq:IV.137}
\tanu(y) = \frac{1}{2\nu + 1}\,\left\{ 1 - \e^{-y} \cdot \sum_{n=0}^\infty \frac{n}{\Gamma\left( 2\nu + 2 + n \right)}\,y^{2\nu + n} \right\} \;.
\end{equation}
This exact solution can now be used in order to calculate explicitly both the
``isotropic'' gauge field energy $\ERee$ (\ref{eq:IV.70}) and its mass equivalent
$\tMMee\crm^2$ (\ref{eq:IV.89}), and the result is the following:
\begin{equation}
\label{eq:IV.138}
\ERee = \tMMee\crm^2 = -\frac{e^2}{a_B}\,\left( 2\,\beta a_B \right) \cdot \epot(\nu)
\end{equation}
with the potential function $\epot(\nu)$ given by~\cite{1},~\cite{4}
\begin{equation}
\label{eq:IV.139}
\epot(\nu) = \frac{1}{2\nu + 1}\,\left( 1 - \frac{1}{2^{4\nu + 2}} \cdot \sum_{n=0}^\infty \frac{n}{2^n} \cdot \frac{\Gamma\left( 4\nu + 2 + n \right)}{\Gamma\left( 2\nu + 2 \right) \cdot \Gamma\left( 2\nu + 2 + n \right)} \right) \;.
\end{equation}

From reasons of computational simplicity it is somewhat more convenient to determine the potential function $\epot(\nu)$ (\ref{eq:IV.138}) from the mass equivalent which yields
\begin{equation}
\label{eq:IV.140}
\epot(\nu) = \frac{1}{\Gamma\left( 2\nu + 2 \right)} \int\limits_0^\infty dy\;y^{2\nu + 1}\,\e^{-y}\,\tanu(y) \;;
\end{equation}
whereas the determination from the ``isotropic'' gauge field energy $\ERee$ would necessitate to calculate a somewhat more complicated integral, i.\,e.
\begin{equation}
\label{eq:IV.141}
\epot(\nu) = \int\limits_0^\infty dy\;y^2\,\left( \frac{d\,\tanu(y)}{dy} \right)^2 \;,
\end{equation}
which then would produce a double sum in place of the simple sum (\ref{eq:IV.139}), see
the discussion of this numerical form of the Poisson identity in ref.s~\cite{1}-\cite{3}.

Finally, we have to face the problem of determining the non-relativistic anisotropy energy
$\Eegiii$ (\ref{eq:IV.109}) and its mass equivalent $\tMMegiii\crm^2$
(\ref{eq:IV.114}). The situation is here the same as for the precedent isotropic case: if
we succeed in finding the exact solution of the quadrupole equation (\ref{eq:IV.127}), the
quadrupole constraint $\tNNegiii$ (\ref{eq:IV.118}) will vanish so that we can omit it in
the wanted energy function $\EEgivbn$ which arises as the value of the functional
$\tEEePhi$ (\ref{eq:IV.119})--(\ref{eq:IV.121}) on our trial configuration $\left\{
  \tPhi(r), \peAn(r), \pgA^\mathsf{III}(r) \right\}$. But fortunately, the desired
solution can be exactly constructed by some simple mathematics which will be presented in
greater detail in \textbf{App. A}. The result is the following general form of the
anisotropy energy
\begin{align}
\label{eq:IV.142}
\Eegiii &= -\pr e_3 \cdot \frac{e^2}{a_B}\,\left(2\,\beta a_B \right) \cdot \epot^\mathsf{III}(\nu) \\
&\equiv - \pr m_3 \cdot \frac{e^2}{a_B}\,\left(2\,\beta a_B \right) \cdot \muegiii(\nu) \nonumber
\end{align}
where the ``anisotropic'' potential function $\epot^\mathsf{III}(\nu)$ is obviously the
anisotropic counterpart of the ``isotropic'' $\epot(\nu)$
(\ref{eq:IV.139})--(\ref{eq:IV.141}) and thus is a relatively complicated (but
well-determined) function of the variational parameter $\nu$, see \textbf{App. A}.

\newpage
\begin{center}
  \large{\textit{Partial Extremalization}}
\end{center}

Now that all physical contributions to the energy functional $\tEEePhi$ (\ref{eq:IV.119})
on our subset of trial configurations $\left\{ \tPhi(r), \pAn(r), \pgA^\mathsf{III}(r)
\right\}$ have been fixed, we arrive at the following result:
\begin{equation}
\label{eq:IV.143}
\EEivbn = 2\,\Ekin(\beta, \nu) + \ERee + \Eegiii \;,
\end{equation}
i.\,e.\ we get by use of the results (\ref{eq:IV.133})--(\ref{eq:IV.134}) for the kinetic energy
$\Ekin$ and the results (\ref{eq:IV.138})--(\ref{eq:IV.139}) for the ``isotropic'' gauge
field energy $\ERee$ plus (\ref{eq:IV.142}) for its anisotropic part $\Eegiii$:
\begin{align}
\label{eq:IV.144}
\EEivbn &= \frac{e^2}{a_B}\, \left( 2\,\beta a_B \right)^2 \cdot \ekin(\nu) - \frac{e^2}{a_B}\, \left( 2\,\beta a_B \right) \cdot \epot(\nu) \\
&\quad\quad\quad\quad\quad\quad\quad\quad\quad\quad\;- \pr m_3 \cdot \frac{e^2}{a_B}\,\left( 2\,\beta a_B \right) \cdot \muegiii(\nu) \nonumber \\
&\doteqdot \frac{e^2}{a_B}\,\left\{ \left( 2\,\beta a_B \right)^2 \cdot \ekin(\nu) - \left( 2\,\beta a_B \right) \cdot \etot(\nu) \right\} \;, \nonumber
\end{align}
where the total potential function $\etot(\nu)$ is evidently defined through
\begin{equation}
\label{eq:IV.145}
\etot(\nu) \doteqdot \epot(\nu) + \pr m_3 \cdot \muegiii(\nu) \;.
\end{equation}

It is important to observe here the fact that the total energy function $\EEivbn$
(\ref{eq:IV.144}) depends on the two \emph{continuous} variational parameters $\beta$ and
$\nu$ but also on the two \emph{discrete} quantum numbers $\elp$ and $\pr m_3$. The first
one $\left( \elp \right)$ is contained in the kinetic energy function $\ekin(\nu)$
(\ref{eq:IV.134}) and the second one $\left( \pr m_3 \right)$ is associated to the
discrete values of the quantum number $\elz$, cf. the table on
p.~\pageref{tablelp3}. Therefore the extremalization of the energy functional transcribes
now to the search for the extremal points of the energy function $\EEivbn$
(\ref{eq:IV.144}) with respect to the continuous variables $\beta$ and $\nu$ whereby the
discrete parameters $\elp$ and $\pr m_3$ must be kept fixed. Thus for any allowed pair of
these fixed parameters one obtains the corresponding binding energy of para-positronium,
and this provides us with the possibility of studying the question of degeneracy
(concerning the same energy for different pairs of discrete parameters), albeit only in
the present quadrupole approximation.

The wanted extremal points of the energy function $\EEivbn$ are determined now as usual by
the vanishing of its first derivatives, i.\,e.
\begin{subequations}
\begin{align}
\label{eq:IV.146a}
\frac{\partial\,\EEivbn}{\partial\beta} &= 0 \\
\label{eq:IV.146b}
\frac{\partial\,\EEivbn}{\partial \nu} &= 0 \;.
\end{align}
\end{subequations}
These two conditions are of a rather different degree of complication so that it is advantageous to first deal exactly with the simpler one (i.\,e. (\ref{eq:IV.146a})) and afterwards solve the residual condition (\ref{eq:IV.146b}) by means of an adequate numerical program. Indeed, the energy function $\EEivbn$ (\ref{eq:IV.144}) is a simple polynomial function of the first variational parameter $\beta$ so that the first extremalization condition (\ref{eq:IV.146a}) fixes $\beta$ to $\beta_*$ being given by
\begin{equation}
\label{eq:IV.147}
2\,\beta_* a_B = \frac{\etot(\nu)}{2 \cdot \ekin(\nu)} \;,
\end{equation}
and if this is substituted back in the energy function $\EEivbn$ (\ref{eq:IV.144}) one is left with the problem of determining the minmal values of the \emph{reduced} energy function ($\EEP(\nu)$, say)
\begin{equation}
\label{eq:IV.148}
\EEP(\nu) \doteqdot \EEivbn\big|_{\beta = \beta_*} = -\frac{e^2}{4\,a_B} \cdot \frac{\etot^2(\nu)}{\ekin(\nu)} \;.
\end{equation}
It is true, the denominator $\ekin(\nu)$ is here a relatively simple function, cf. \textbf{App. A}, but the numerator $\etot(\nu)$ (\ref{eq:IV.145}) is given only in form of a rather complicated power series. Therefore one will have to use some suited numerical program for determining the local minima ($\EEP^{\{n\}}$, say) of the reduced energy function $\EEP(\nu)$ (\ref{eq:IV.148}) for any possible value of the discrete parameters $\elp$ and $\elz$ ($\Leftrightarrow \pr m_3$).

\pagebreak
\begin{center}
  \emph{\textbf{4.\ Para-Spectrum}}
\end{center}

For a brief demonstration of how the energy degeneracy (with respect to $\elz$) of the
spherically symmetric approximation becomes now eliminated in the quadrupole approximation
we will restrict ourselves here to the precedently discussed quantum number $\elp = 3$
$(\Rightarrow -3 \leq \elz \leq 3)$. Clearly, similar degeneracy effects must emerge for
all quantum numbers $\elp \geq 1$ but this will be deferred to a separate paper. For the
moment, it is perhaps more instructive to consider the case $\elz = \pm \elp$ for all
values $\elp$ (i.\,e. $\elp = 1,2,3,4,\ldots$). The point here is that the preliminary
estimate of the quadrupole level splitting (\ref{eq:IV.129}) demonstrates that for $\elp =
3$ the lowest-lying energy level of the corresponding ``$\elz$-multiplet'' is that for
$\elz = \pm \elp$. Therefore the hypothesis suggests itself that this perhaps might be
true for the $\elp$-multiplet due to \emph{any} given value of $\elp$. Consequently, the
energy spectrum due to $\elz = \pm \elp$ for all values of $\elp$ is the adequate one in
order to be compared to its conventional counterpart (\ref{eq:I.2}). We will readily
plunge into this matter of question after first having glimpsed at the $\elz$-degeneracy
for $\elp = 3$.

\begin{center}
  \large{\textit{Elimination of the $\elz$-Degeneracy, $\boldsymbol{ \lP=3}  $  }}
\end{center}
For the search of the local minima of the reduced energy function~$\EEP{(\nu)}$
(\ref{eq:IV.148}) one inserts there the kinetic energy function~$\ekin{(\nu)}$
(\ref{eq:IV.134}) for~$\boldsymbol{\lP=3}$ and the total potential function~$\etot{(\nu)}$
(\ref{eq:IV.145}) which yields
\begin{equation}
  \label{eq:IV.149}
  \EEP(\nu)\big|_{l_p=3}=-\frac{e^2}{4\aB}\frac{(\epot{(\nu)}+\rklo{p}{m}_3\cdot\rogk{\mu}{e}_{|||}(\nu)
    )^2 }{\frac{ 1}{ 2\nu+1}(\frac{1}{4}+\frac{ 9}{ 2\nu} )}\ .
\end{equation}

Here, the potential function~$\epot(\nu)$ is given by equation (\ref{eq:IV.139})
and~$\rogk{\mu}{e}_{|||}$ by (\ref{eq:A.19}) or (\ref{eq:A.20}) of \textbf{App.A}. The
coefficient~$\rklo{p}{m}_3$ is linked to the possible values of~$l_z$ according to the
table on p.~\pageref{tablelp3}; and thus the four functions~$\EEP(\nu)$ for~$\lP=3$
and~$l_z=0,\pm 1,\pm 2,\pm 3$ are well-defined in order to determine the four
corresponding binding energies~$\EEP^{(4)}|_{l_z}$ as the local minima (with respect to
the variational parameter~$\nu$) of the energy function~$\EEP(\nu)$.

But observe here also that this minimalization process is based on the simple trial
function~$\tPhi(r)$ (\ref{eq:IV.131}) which has no zero. Therefore we can obtain by our
extremalization procedure only the ``groundstate'' in the effective potential which itself
appears as the sum of the centrifugal potential~$(\sim \lP^2)$ and the electrostatic
potential (due to~$\eklo{p}{A}_0(r)$ and~$\gklo{p}{A}^{|||}(r)$ in the Schr\"odinger like
equation (\ref{eq:IV.130})). This means that the principal quantum number~($\nP$, say) must
be linked here to the angular momentum quantum number~$\lP$ through
\begin{equation}
  \label{eq:IV.150}
  \nP = \lP + 1\ .
\end{equation}
Thus, the (non-relativistic) configurations under consideration are those being
characterized by the quantum numbers $\boldsymbol{ \nP=4, \lP=3, l_z=0,\pm 1, \pm 2, \pm 3}  $.
The corresponding results for the para-positronium energies~$\EEP^{(n)}$ are collected
in the following table:
\begin{center}
\label{paraE}
\begin{tabular}{|c||c|c|c|c|}
\hline
$l_z$ & $0$ & $\pm 1$ & $\pm 2$ & $\pm 3$ \\
\hline\hline
$\EEP^{(4)} \,[eV]$  & $$-0,40298\ldots & $-0,39001\ldots$ & $-0,37369\ldots$ &~$-0,42003\ldots$ \\
\hline
$\Ea{E}{4}{C}\,[eV]$ & $-0,42518\ldots$ & $-0,42518\ldots$ & $-0,42518\ldots$ &$-0,42518\ldots$ \\
\hline
$\Delta^{(4)}_\wp$ & $5,2\%$ &~$8,2\%$ &$12,1\%$ &$1,2\%$ \\
\hline
\end{tabular}
\end{center}
For an evaluation of these results one introduces the relative
deviation~$\Delta^{(4)}_\wp|_{l_z}$ from the conventional predictions through
\begin{equation}
  \label{eq:IV.151}
  \Delta^{(4)}_\wp|_{l_z} \doteqdot \frac{\Ea{E}{4}{C} -\EEP^{(4)}|_{lz} }{\Ea{E}{4}{C}}
\end{equation}
where~$\Ea{E}{4}{C}$ is the conventional prediction (\ref{eq:I.2}) for~$n=4$
\begin{equation}
  \label{eq:IV.152}
  \Ea{E}{4}{C} = -\frac{e^2}{4\aB}\cdot\frac{1}{4^2} = -0,42518\ldots [eV]\ .
\end{equation}

Obviously, the deviation from the conventional result is maximal for~$l_z=\pm 2$. Indeed
for this special value of~$l_z$ the anisotropy correction is zero,
(cf.~(\ref{eq:IV.104c})), and therefore we obtain a deviation of~$12,1\%$ which is due to
the spherically symmetric approximation, see ref.~\cite{4}. However, for all other values
of~$l_z$ one gets a smaller deviation~$\Delta^{(4)}_\wp$ which adopts its minimal value
of~$1,2\%$ just for the maximally possible value of~$l_z$, i.e. \mbox{$l_z=\pm 3\equiv \pm\lP$}.
In this way, the conventional $n^2-$fold degeneracy (\ref{eq:I.3}) is eliminated as far as
the ~$z-$component of angular momentum is concerned. Whether there is also
an~$\lP$-degeneracy for the same value of~$\nP$ with~$0\le\lP\le\nP-1$ necessitates an
extra discussion. The elimination of the~$l_z$-degeneracy is illustrated also by the
subsequent \textbf{Fig.IV.B}

\begin{center}
\epsfig{file=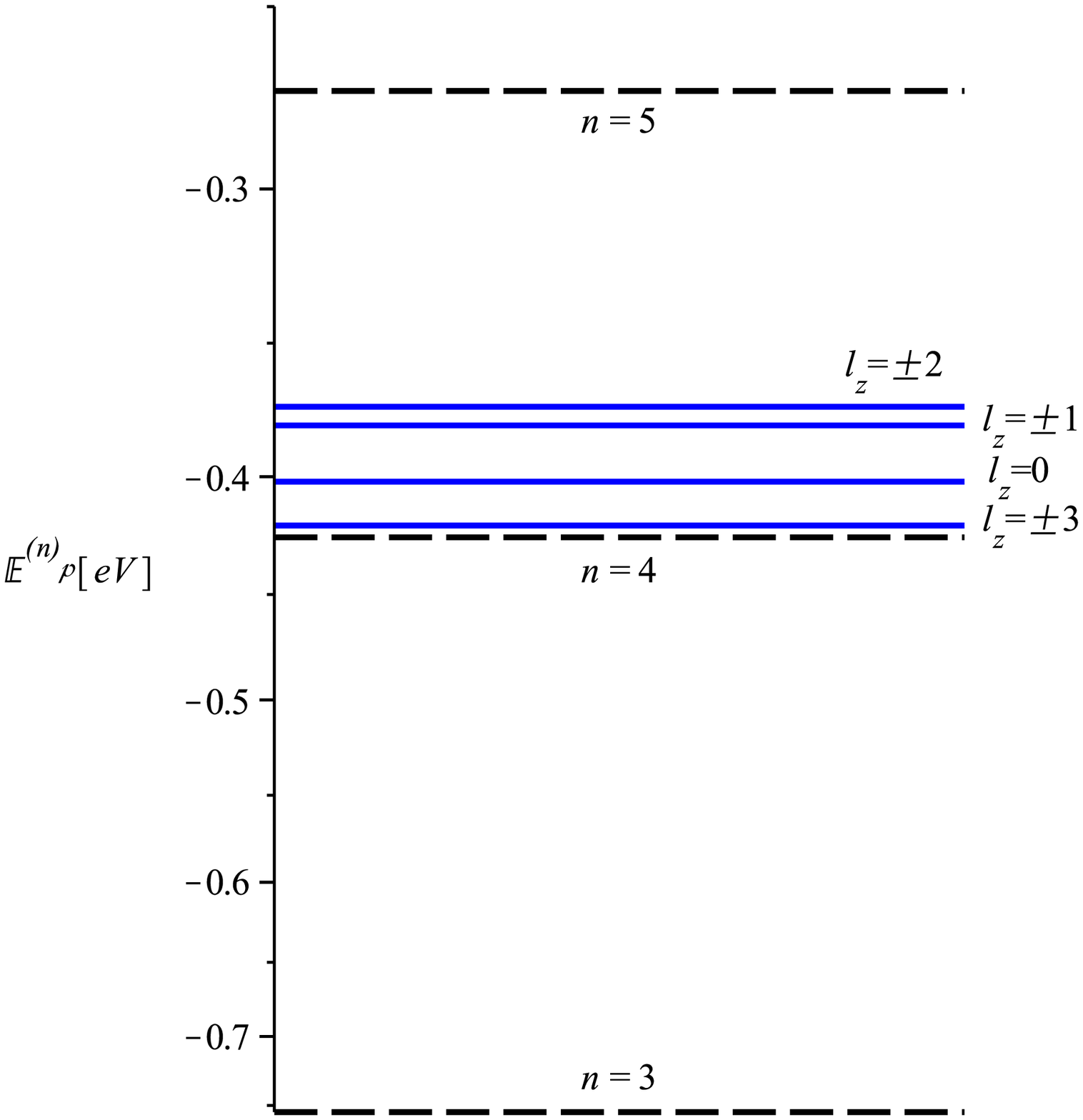,height=12cm}
\end{center}
{\textbf{Fig.~IV.B}\hspace{5mm} \emph{\large\textbf{RST Elimination of the~\boldmath$l_z$-Degeneracy
for \boldmath$\lP=3$       }   }  }
\myfigure{Fig.~IV.B: {RST Elimination of the~$l_z$-Degeneracy
\boldmath$\gkloi{p}{k}{\phi}(\vartheta) $}}
\indent\\*
\label{figIV.B}
The conventional energy levels~$\Ea{E}{n}{C}$ (\ref{eq:I.2}) (broken lines) are~$n^2-$fold
degenerated, cf.~(\ref{eq:I.3}). The RST \emph{quadrupole} approximation eliminates this
$l_z$-degeneracy for the states with principal quantum numbers~$\nP$, $\lP=\nP-1$;
here~$\nP=4,\lP=3$. Presumably, the higher-order approximations will bring the RST
predictions even closer to their conventional counterparts (\ref{eq:I.2}) (broken lines),
at least for small values of the principal quantum number~$\nP$. For the magnitude of
splitting for larger~$\nP$, see the table on~p.~\pageref{table2} and \textbf{Fig.IV.C} below.


\begin{center}
  \large{\textit{Spectrum Due to $\elz = \pm \elp$}}
\end{center}

The restriction $\elz = \pm \elp$ selects just one energy level from an $\elp$-multiplet
($-\elp \leq \elz \leq \elp$); and therefore we have to carry through all the necessary
intermediate steps for just this one multiplet term, however for all possible values of
$\elp$ ($= 1,2,3,4,\ldots$).

The first step concerns the solution of the first-order angular system
(\ref{eq:IV.8a})--(\ref{eq:IV.8b}). Here we can restrict ourselves to the case $\elz =
+\elp$ since the commutation symmetry $\elz \Rightarrow -\elz$ simply entails the
interchange $f_R(\vartheta) \Leftrightarrow f_S(\vartheta)$. Thus, for $\elz = \elp$ that
system (\ref{eq:IV.8a})--(\ref{eq:IV.8b}) becomes specialized to the following form:
\begin{subequations}
\begin{align}
\label{eq:IV.153a}
\frac{d\,f_R(\vartheta)}{d\vartheta} - \elp \cot\vartheta \cdot f_R(\vartheta) &= 2\,\elp \cdot f_S(\vartheta) \\
\label{eq:IV.153b}
\frac{d\,f_S(\vartheta)}{d\vartheta} + \elp \cot\vartheta \cdot f_S(\vartheta) &= 0 \;.
\end{align}
\end{subequations}
But here one is easily convinced that the second equation (\ref{eq:IV.153b}) does admit
exclusively the trivial solution $f_S(\vartheta) \equiv 0$ which is non-singular on the
$z$-axis $\left(\vartheta = 0,\pi\right)$. This, however, recasts the first equation
(\ref{eq:IV.153a}) to the homogeneous form
\begin{equation}
\label{eq:IV.154}
\frac{d\,f_R(\vartheta)}{d\vartheta} - \elp \cot\vartheta \cdot f_R(\vartheta) = 0 \;,
\end{equation}
and this admits the non-trivial solution (obeying of course also (\ref{eq:IV.17a}))
\begin{equation}
\label{eq:IV.155}
f_R(\vartheta) = \sqrt{\frac{2}{\pi} \cdot \frac{(2\,\elp)!!}{(2\,\elp - 1)!!}} \left(\sin\vartheta\right)^{\elp} \;,
\end{equation}
where the separate angular normalization condition (\ref{eq:IV.15a}) has already been
regarded. As a consistency test, specialize this result (\ref{eq:IV.155}) to the case
$\elp = 3$ and find just the result of the table on p.~\pageref{tablefR}, last lines.

The next step must refer to the calculation of the angular factor
$\pgA^\mathsf{III}(\vartheta)$ (\ref{eq:IV.104c}). Inserting therein the trivial function
$f_S(\vartheta) \equiv 0$ together with the result (\ref{eq:IV.155}) for $f_R(\vartheta)$
and carrying out explicitly the integration yields
\begin{equation}
\label{eq:IV.156}
\pgA^\mathsf{III}(\vartheta) = -\frac{3}{8}\,\frac{2\,\elp - 1}{\elp + 1} \cdot \left[ \cos^2 \vartheta - \frac{1}{3} \right] \;.
\end{equation}
Indeed, for $\elp = 3$ one is immediately led back to the former result
(\ref{eq:IV.104c}), last line. But once the angular pre-factor
$\pgA^\mathsf{III}(\vartheta)$ is known explicitly, one can proceed at once to the
calculation of the numerical coefficients $\pr e_3$ (\ref{eq:IV.110a}), $\pr f_3$
(\ref{eq:IV.110b}), and $\pr m_3$ (\ref{eq:IV.115}); the result is the following:
\begin{subequations}
\begin{align}
\label{eq:IV.157a}
\pr e_3 &= \frac{1}{80}\,\left( \frac{2\,\elp - 1}{\elp + 1} \right)^2 \\
\label{eq:IV.157b}
\pr f_3 &= \frac{3}{40}\,\left( \frac{2\,\elp - 1}{\elp + 1} \right)^2 \\
\label{eq:IV.157c}
\pr m_3 &= \frac{1}{16}\,\left( \frac{2\,\elp - 1}{\elp + 1} \right)^2 \;.
\end{align}
\end{subequations}
Obviously, these results do generate just those numbers of the table on
p.~\pageref{tablelp3}, last column, if the quantum number $\elp$ is again specified down
to $\elp = 3$. But more generally, we have here for any value of $\elp$ (and even for
\emph{any} allowed value of~$l_z$, see \textbf{App.C}):
\begin{subequations}
\begin{align}
\label{eq:IV.158a}
\frac{\pr f_3}{\pr e_3} &= 6 \\
\label{eq:IV.158b}
\frac{\pr m_3}{\pr e_3} &= 5 \;.
\end{align}
\end{subequations}
This ensures again that the general quadrupole equation (\ref{eq:IV.126}) adopts its specific form (\ref{eq:IV.127}); and furthermore the anisotropy energy $\Eegiii$ (\ref{eq:IV.128}) reads now
\begin{equation}
\label{eq:IV.159}
\Eegiii = - \frac{1}{80}\,\left( \frac{2\,\elp - 1}{\elp + 1} \right)^2 \cdot \frac{\hbar\crm}{\as}\,\int dr\;r^2\,\left\{ \left( \frac{d\,\pgA^\mathsf{III}(r)}{dr} \right)^2 + 6\,\left( \frac{\pgA^\mathsf{III}}{r} \right)^2 \right\} \;.
\end{equation}
The quadrupole solution $\pgA^\mathsf{III}(r)$, to be used herein, is elaborated in \textbf{App. A} and is of course the same as previously used for the special case $\elp = 3$.

Since that solution $\pgA^\mathsf{III}(r)$ is an exact one, we can evoke again the
``anisotropic'' Poisson identity in coefficient form (\ref{eq:A.18}) and thus we
ultimately obtain the wanted energy function $\EEgivbn$ in the same form as previously,
cf. (\ref{eq:IV.143})--(\ref{eq:IV.145}), with merely the coefficient of mass equivalence
$\pr m_3$ to be replaced now by its value given by equation
(\ref{eq:IV.157c}). Furthermore, because of this formal equivalence of the total energy
functions $\Egivbn$, their partial extremalization with respect to the variational parameter
works in the same way; and thus we end up with the residual problem of looking for the
minima of the reduced energy function $\EEP(\nu)$ (\ref{eq:IV.148}) where now not only the
denominator $\ekin(\nu)$ depends upon the discrete quantum number $\elp$,
cf. (\ref{eq:IV.134}), but also the denominator $\etot(\nu)$. In cosideration of the
numerical calculations, it is convenient to express this dependence on $\elp$ explicitly,
i.\,e. we finally get for the reduced energy function
\begin{equation}
\label{eq:IV.160}
\EEP(\nu) = - \frac{e^2}{4\,a_B} \cdot S_\mathcal{P}(\nu)
\end{equation}
with the spectral function $S_\mathcal{P}(\nu)$ being specified by
\begin{equation}
\label{eq:IV.161}
S_\mathcal{P}(\nu) \doteqdot \frac{\etot^2(\nu)}{\ekin(\nu)} = \frac{\left\{ \epot(\nu) + \frac{1}{16}\,\left( \frac{2\,\elp - 1}{\elp + 1} \right)^2 \cdot \muegiii(\nu) \right\}^2}{\frac{1}{2\nu + 1}\,\left( \frac{1}{4} + \frac{\elp^2}{2\nu} \right)} \;.
\end{equation}

The following table presents the energy spectrum $\EEP^{\{n\}}$
($n\Rightarrow n_\mathcal{P} = \elp + 1 = 1,2,3,\ldots$) to be determined by the local maxima
of the spectral function $S_\mathcal{P}(\nu)$ for any fixed value of $\elp$:
\begin{subequations}
  \begin{align}
\label{eq:IV.162a}
\Ea{\mathbb{E}}{n}{\wp} \doteqdot \mathbb{E}_\wp(\nu_*)\\*
\label{eq:IV.162b}
\frac{d\mathbb{E}_\wp(\nu)}{d\nu}\Big|_{\nu_*} = 0
  \end{align}
\end{subequations}
\clearpage
\begin{center}
\label{table2}
  \begin{tabular}{|c|c|c|c|c|}
  \hline
 $\nP $ & $ \EEP^{(n)} $ &  $\Ea{E}{n}{C}$\,[eV]  & $\Delta_\wp^{(n)}\,\%  $  & $\nu_*$  \\
        &  (\ref{eq:IV.162a}) &   (\ref{eq:I.2})  &  (\ref{eq:IV.151})  &  (\ref{eq:IV.162b})  \\
  \hline\hline
$ 2 $ & $ -1.58097 $ & $ -1.70072 $ & $  7.0 $ & $ 1.71724 $  \\ \hline 
$ 3 $ & $ -0.71761 $ & $ -0.75588 $ & $  5.1 $ & $ 3.86851 $  \\ \hline 
$ 4 $ & $ -0.42003 $ & $ -0.42518 $ & $  1.2 $ & $ 6.21902 $  \\ \hline 
$ 5 $ & $ -0.27878 $ & $ -0.27212 $ & $  -2.4 $ & $ 8.83210 $  \\ \hline 
$ 6 $ & $ -0.19969 $ & $ -0.18897 $ & $  -5.7 $ & $ 11.66043 $  \\ \hline 
$ 7 $ & $ -0.15062 $ & $ -0.13883 $ & $  -8.5 $ & $ 14.68005 $  \\ \hline 
$ 8 $ & $ -0.11793 $ & $ -0.10630 $ & $  -10.9 $ & $ 17.87403 $  \\ \hline 
$ 9 $ & $ -0.09499 $ & $ -0.08399 $ & $  -13.1 $ & $ 21.22077 $  \\ \hline 
$ 10 $ & $ -0.07824 $ & $ -0.06803 $ & $  -15.0 $ & $ 24.71323 $  \\ \hline 
$ 15 $ & $ -0.03690 $ & $ -0.03024 $ & $  -22.1 $ & $ 43.99860 $  \\ \hline
  \end{tabular}
\end{center}
Obviously, the RST quadrupole approximation for the lowest-lying level (i.e.~$l_z=\pm l_\wp$)
yields predictions of smaller binding energy up to~$\nP=4$ and of larger binding energy
for~$\nP>4$, as compared to the conventional levels~$\Ea{E}{n}{C}$ (\ref{eq:I.2}). The
other members of an~$l_\wp$-multiplet (i.e.~$|l_z|<l_\wp$) have smaller binding energy
(than those due to~$l_z = \pm l_\wp$) and thus do form a kind of band structure around (or
in the vicinity of) the conventional levels, see \textbf{Fig.IV.C}.
\newpage
\begin{center}
\epsfig{file=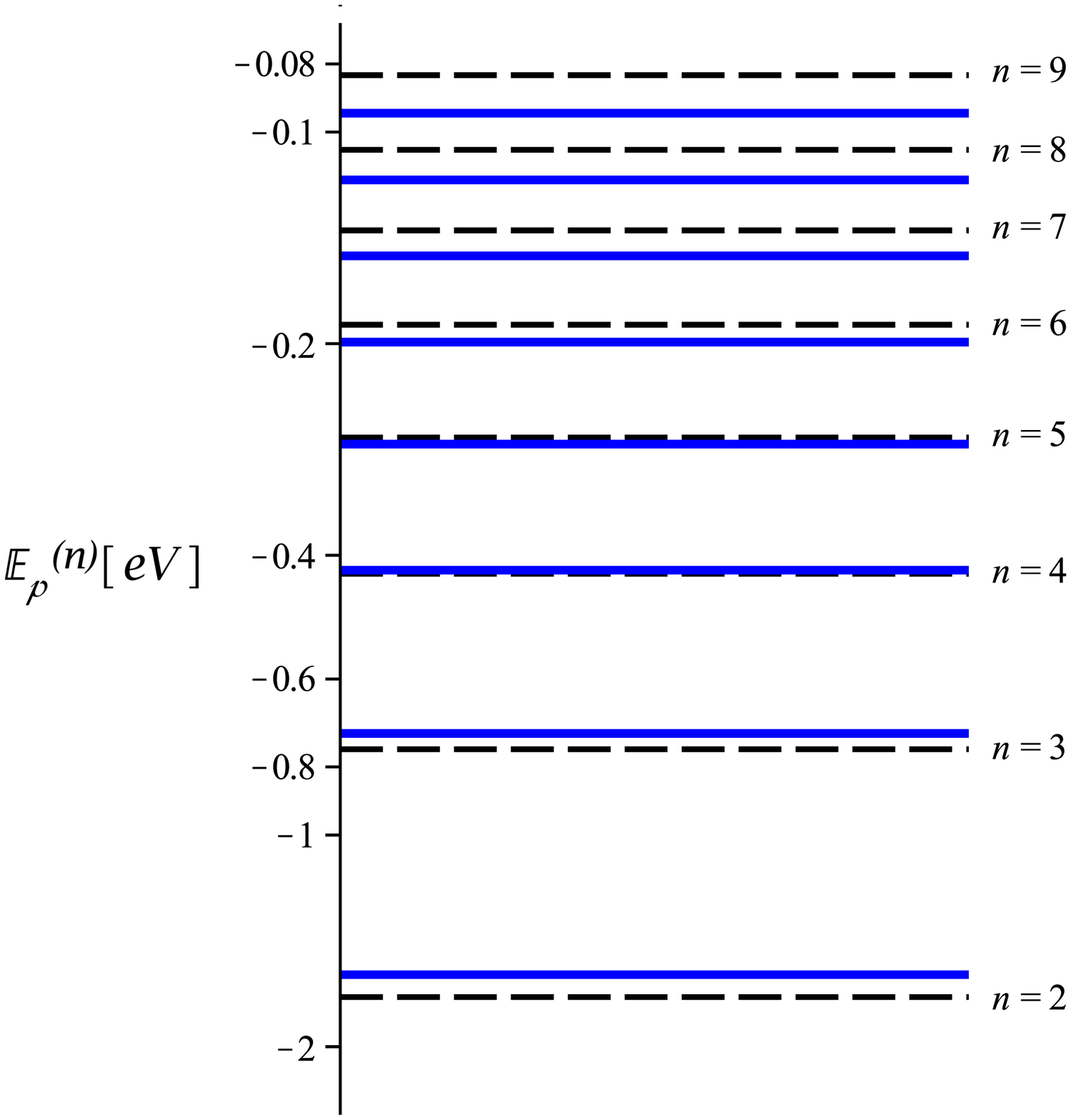,height=12cm}
\end{center}
{\textbf{Fig.~IV.C}\hspace{5mm} \emph{\large\textbf{Lowest-Lying RST Levels  {
        $\boldsymbol{  \EEP^{( n)}\Big|_{l_z=\lP} } $  }   (\ref{eq:IV.162a})     }   }  }
\myfigure{Fig.~IV.C: { }}
\indent\\*
For increasing principal quantum number~$\nP > 4$ the lowest energy levels of
an~$\lP$-multiplet $(-\lP\le l_z\le +\lP)$, with~$\lP=\nP-1$, have ever larger binding
energy in comparison to their conventional analogues~$\Ea{E}{n}{C}$
(\ref{eq:I.2}). Therefore it seems that, for sufficiently large~$\nP$, there may occur a
crossing of the conventional levels (broken lines) and RST levels (solid lines).


\newpage
\begin{center}
  \large{\textit{Magnitude of Fine-Structure Splitting}}
\end{center}

The elimination of the $\elz$-degeneracy, as being displayed by \textbf{Fig. IV.B} on
p.~\pageref{figIV.B} for $\elp = 3$, may be conceived also as a certain kind of
(electrostatic) fine-sctructure splitting of the energy levels. For such an interpretation
one would like to have an estimate for the splitting width for any $\elp$. For instance,
for the special case of {\boldmath{$\elp = 3$}} (see the table on p.~\pageref{paraE}), one observes a maximal
energy difference in the quadrupole approximation for those levels which have $\elz = 2$
and $\elz = 3$, namely an energy difference of ($0{,}42003 - 0{,}37369$)~[eV] =
0{,}04634~[eV]. This corresponds to 12{,}4\,\% of the ''spherically symmetric'' energy of
0{,}37369~[eV]. Such a magnitude of the splitting seems not to be supported by the
experimental results [6]; and this necessitates to consider the RST splitting width also
in the higher-order approximations (i.\,e. octupole, etc.). But as a first rough estimate
of that questionable splitting phenomenon it may suffice to discuss the splitting
magnitude already in the present quadrupole approximation. For this purpose, we take $\elz
= \elp$ and $\elz = \elp - 1$ for a preliminary estimate of the range of the splitting.

More concretely, we have to transcribe the former result (\ref{eq:IV.160})--(\ref{eq:IV.161})
for $\elz = \elp$ to the present choice $\elz = \elp - 1$. The first step for this is to
look for the solution of the angular system (\ref{eq:IV.8a})--(\ref{eq:IV.8b}) for our
special choice of $\elz = \elp - 1$. For this choice, the angular system appears as
\begin{subequations}
\begin{align}
\label{eq:IV.163a}
\frac{d\,f_R(\vartheta)}{d\vartheta} - (\elp - 1)\,\cot\vartheta \cdot f_R(\vartheta) &= (2\,\elp - 1) \cdot f_S(\vartheta) \\
\label{eq:IV.163b}
\frac{d\,f_S(\vartheta)}{d\vartheta} + (\elp - 1)\,\cot\vartheta \cdot f_S(\vartheta) &= - f_R(\vartheta) \;.
\end{align}
\end{subequations}
The solution of this coupled first-order system can be obtained by the method being
already described in \textbf{Subsect. IV.2} and is then found to look as follows
\begin{subequations}
\begin{align}
\label{eq:IV.164a}
f_R(\vartheta) &= \sqrt{\frac{2}{\pi}\,\left( 2\,\elp - 1 \right)\,\frac{\left( 2\,\elp - 2 \right)!!}{\left( 2\,\elp - 3 \right)!!}} \cdot \left( \sin\vartheta \right)^{\elp - 1}\,\cos\vartheta \\
\label{eq:IV.164b}
f_S(\vartheta) &= -\sqrt{\frac{2}{\pi}\,\frac{\left( 2\,\elp - 2 \right)!!}{\left( 2\,\elp - 1 \right)!!}} \cdot \left( \sin\vartheta \right)^{\elp} \;.
\end{align}
\end{subequations}
For a brief test of this result, one puts $\elp = 3$ and thus recovers both functions
$f_R(\vartheta)$ and $f_S(\vartheta)$ for $\elp = 3, \elz = 2$ as displayed by the table
on p.~\pageref{tablefR}.

Next, the angular factor $\pgAiii(\vartheta)$ (\ref{eq:IV.104c}) due to the present
solution (\ref{eq:IV.164a})-(\ref{eq:IV.164b}) must be calculated. The straight-forward
integration yields the result
\begin{equation}
\label{eq:IV.165}
\pgAiii(\vartheta) = - \frac{3}{8} \, \frac{\left( 2\,\elp - 1 \right)\left( \elp - 3 \right)}{\elp \cdot \left( \elp + 1 \right)} \cdot \left[ \cos^2\vartheta - \frac{1}{3} \right] \;.
\end{equation}
Observe here that the angular dependence of the pre-factor $\pgAiii(\vartheta)$ is always the same in the quadrupole approximation, namely [11]
\begin{equation}
\label{eq:IV.166}
\pgAiii(\vartheta) = -\frac{3}{4}\,\left( 1 - 3 \cdot \Kpgiii \right) \left[ \cos^2\vartheta - \frac{1}{3} \right] \;,
\end{equation}
with the constant $\Kpgiii$ being given by
\begin{equation}
\label{eq:IV.167}
\Kpgiii \doteqdot \int \, d\Omega \; \cos^2\vartheta \cdot \pgko(\vartheta) \;.
\end{equation}
For the present solution (\ref{eq:IV.164a})-(\ref{eq:IV.164b}) the explicit integration yields
\begin{equation}
\label{eq:IV.168}
\Kpgiii = \frac{3\,\elp - 1}{2\,\elp\,\left( \elp + 1 \right)} \;,
\end{equation}
and if this is inserted in the general expression (\ref{eq:IV.166}) one just recovers the
claimed result (\ref{eq:IV.165}). Observe also that this latter result just agrees with
the former result (\ref{eq:IV.104c}) for $\elp = 3, \elz = 2$ where $\pgAiii(\vartheta)$
must vanish!

Once the angular factor $\pgAiii(\vartheta)$ is known, cf. (\ref{eq:IV.165}), one can
calculate the corresponding quadrupole parameters $\pr e_3$ (\ref{eq:IV.110a}), $\pr f_3$
(\ref{eq:IV.110b}) and $\pr m_3$ (\ref{eq:IV.115}). The straight-forward integration
yields the following result
\begin{subequations}
\begin{align}
\label{eq:IV.169a}
\pr e_3 &= \frac{1}{80} \, \left[ \frac{\left( 2\,\elp - 1 \right) \left( \elp - 3 \right)}{\elp \, \left( \elp + 1 \right)} \right]^2 \\
\label{eq:IV.169b}
\pr f_3 &= \frac{3}{40} \, \left[ \frac{\left( 2\,\elp - 1 \right) \left( \elp - 3 \right)}{\elp \, \left( \elp + 1 \right)} \right]^2 \\
\label{eq:IV.169c}
\pr m_3 &= \frac{1}{16} \, \left[ \frac{\left( 2\,\elp - 1 \right) \left( \elp - 3 \right)}{\elp \, \left( \elp + 1 \right)} \right]^2 \;.
\end{align}
\end{subequations}
The striking feature of this result refers again to the ratios
\begin{subequations}
\begin{align}
\label{eq:IV.170a}
\frac{\pr f_3}{\pr e_3} &= 6 \\
\label{eq:IV.170b}
\frac{\pr m_3}{\pr e_3} &= 5 \;.
\end{align}
\end{subequations}
Obviously, this agrees with the corresponding ratios found also for the set of quantum
numbers $\{ \elp = 3;$ $\elz = 0, \pm 1, \pm 3 \}$, see the table on
p.~\pageref{tablelp3}; and it agrees with the ratios found for the quantum numbers $\{
\elp;\;\elz = \elp \}$ for all $\elp$. This suggests that these ratios are the same for
\emph{all} allowed conbinations of $\elp$ and $\elz$ \emph{in the quadrupole
  approximation}.

In any case, the observed ratios (\ref{eq:IV.170a})-(\ref{eq:IV.170b}) do ensure that the
general quadrupole equation (\ref{eq:IV.126}) adopts again its specific form
(\ref{eq:IV.127}) whose solution is elaborated in \textbf{App. A}. The associated
quadrupole energy $\Eegiii$ (\ref{eq:IV.128}) adopts now by use of $\pr e_3$
(\ref{eq:IV.169a}) the following form
\begin{align}
\label{eq:IV.171}
\Eegiii &= -\frac{1}{80} \left[ \frac{\left( 2\,\elp - 1 \right) \left( \elp - 3 \right)}{\elp\,\left( \elp + 1 \right)} \right]^2 \\
&\hspace{2cm}\cdot \frac{\hbar\crm}{\as} \int\limits_0^\infty dr\,r^2 \left\{ \left( \frac{d\,\pgAiii(r)}{dr} \right)^2 + 6\left( \frac{\pgAiii(r)}{r} \right)^2 \right\} \nonumber \\
&= -\frac{1}{16} \left[ \frac{\left( 2\,\elp - 1 \right) \left( \elp - 3 \right)}{\elp\,\left( \elp + 1 \right)} \right]^2 \frac{e^2}{a_B} \left( 2\,\aB \beta \right) \cdot \muegiii(\nu) \nonumber
\end{align}
where again the ``anisotropic'' Poisson identity is used in its coefficient form
(\ref{eq:A.18}). Since furthermore the ``isotropic'' gauge field energy $\ERee$
(\ref{eq:IV.138}) is the same as before, the total gauge field energy $\EReg$
(\ref{eq:IV.108}) emerges also again in the same form as before, i.\,e.
\begin{equation}
\label{eq:IV.172}
\EReg = -\frac{e^2}{a_B} \left( 2\,a_B \beta \right) \cdot \etot(\nu) \;,
\end{equation}
where $\etot(\nu)$ is now given by
\begin{equation}
\label{eq:IV.173}
\etot(\nu) = \epot(\nu) + \frac{1}{16} \left[ \frac{\left( 2\,\elp - 1 \right) \left( \elp - 3 \right)}{\elp\,\left( \elp + 1 \right)} \right]^2 \cdot \muegiii(\nu) \;.
\end{equation}
Moreover, the kinetic energy can also be taken over for the present situation so that we ultimately are left with the problem of determining the minima of the reduced energy function $\EEP(\nu)$ (\ref{eq:IV.160}) where, however, the spectral function $S_\mathcal{P}(\nu)$ is now of the following form:
\begin{equation}
\label{eq:IV.174}
S_\mathcal{P}(\nu) = \frac{\left\{ \epot(\nu) + \frac{1}{16} \left[ \frac{\left( 2\,\elp -
          1 \right) \left( \elp - 3 \right)}{\elp\,\left( \elp + 1 \right)} \right]^2
  \cdot  \muegiii(\nu) \right\}^2  }{\frac{1}{2\nu + 1} \left[ \frac{1}{4} + \frac{\elp^2}{2\nu} \right]}
\end{equation}
in place of the precedent situation (\ref{eq:IV.161}).

This formal similarity with the precedent situation admits us to determine the minima of
the reduced energy function $\EEP(\nu)$ (\ref{eq:IV.160}) by use of the same numerical
method, with merely the spectral function $S_\mathcal{P}(\nu)$ (\ref{eq:IV.161}) to be
replaced by the present one (\ref{eq:IV.174}). The corresponding results are collected in
the following table:

\begin{center}\label{table3}
  \begin{tabular}{|c|c|c|c|c|}
  \hline
 $\nP $ & $ \EEP^{(n)} $\,[eV] &  $\Ea{E}{n}{C}$\,[eV]  & $\Delta_\wp^{(n)}\,\%  $  & $\nu_*$  \\
        &  (\ref{eq:IV.162a}) &   (\ref{eq:I.2})  &  (\ref{eq:IV.151})  &  (\ref{eq:IV.162b})  \\
  \hline\hline
$ 2 $ & $ -1.65471 $ & $ -1.70072 $ & $  2.7 $ & $ 1.72555 $  \\ \hline 
$ 3 $ & $ -0.68113 $ & $ -0.75588 $ & $  9.9 $ & $ 3.76782 $  \\ \hline 
$ 4 $ & $ -0.37369 $ & $ -0.42518 $ & $  12.1 $ & $ 5.87475 $  \\ \hline 
$ 5 $ & $ -0.24146 $ & $ -0.27212 $ & $  11.3 $ & $ 8.17493 $  \\ \hline 
$ 6 $ & $ -0.17148 $ & $ -0.18897 $ & $  9.3 $ & $ 10.69620 $  \\ \hline 
$ 7 $ & $ -0.12931 $ & $ -0.13883 $ & $  6.9 $ & $ 13.43108 $  \\ \hline 
$ 8 $ & $ -0.10161 $ & $ -0.10630 $ & $  4.4 $ & $ 16.35354 $  \\ \hline 
$ 9 $ & $ -0.08228 $ & $ -0.08399 $ & $  2.0 $ & $ 19.45641 $  \\ \hline 
$ 10 $ & $ -0.06818 $ & $ -0.06803 $ & $  -0.2 $ & $ 22.71757 $  \\ \hline 
$ 15 $ & $ -0.03306 $ & $ -0.03024 $ & $  -9.3 $ & $ 41.08929 $  \\ \hline 
  \end{tabular}
\end{center}
\begin{center}
\epsfig{file=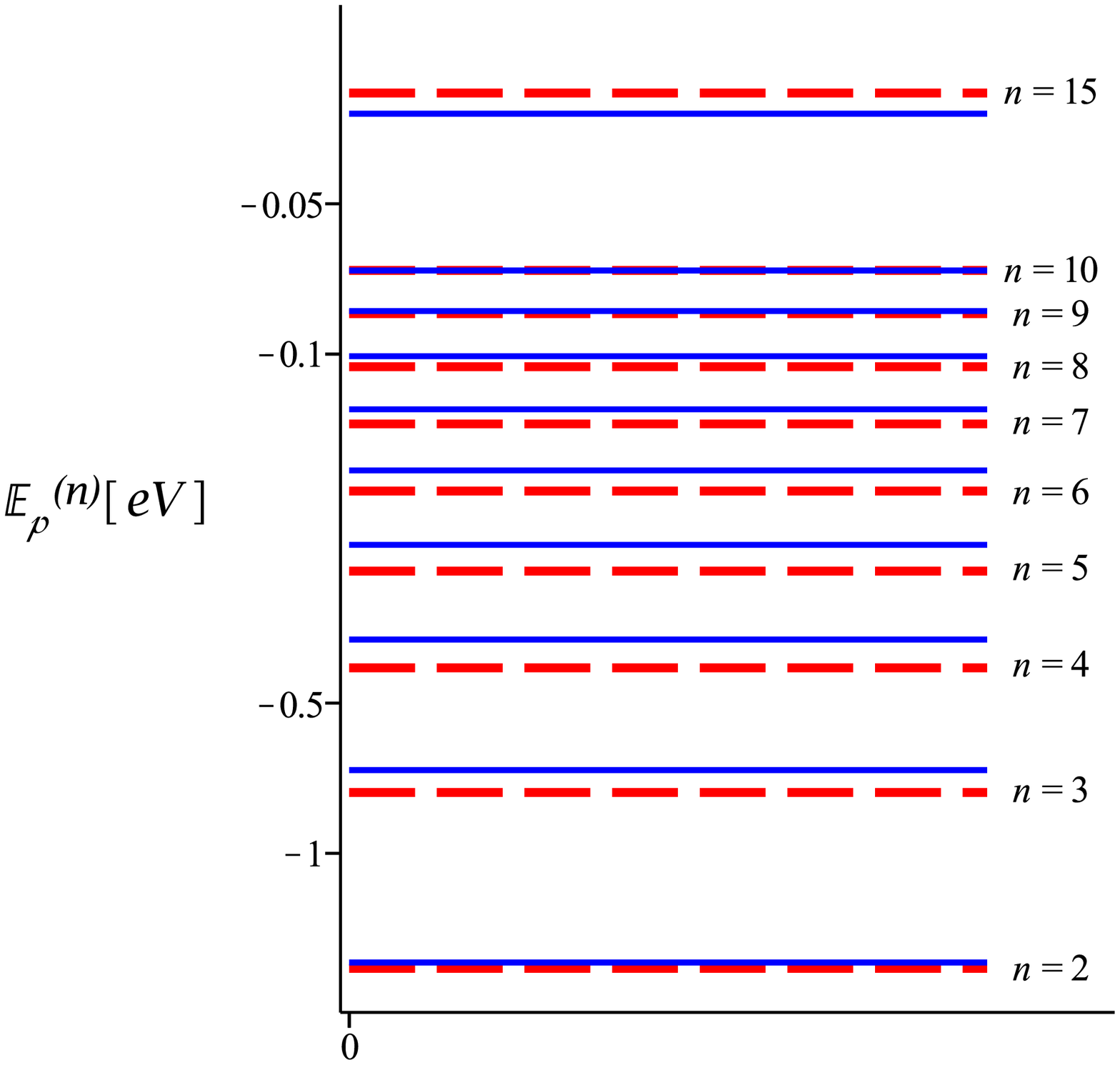,height=12cm}
\end{center}
{\textbf{Fig.~IV.D}\hspace{5mm} \emph{\large\textbf{RST Levels  {
        $\boldsymbol{  \EEP^{( n)}\Big|_{l_z=\lP-1} } $  }   (\ref{eq:IV.162a})     }   }  }
\myfigure{Fig.~IV.D: { }}
\indent\\*

For quantum numbers~$l_z=\pm (\lP-1)$ the RST levels~$\EEP^{( n)}$ (\ref{eq:IV.162a})
(solid lines) have lower bounding energy then their conventional
counterparts~$\Ea{E}{n}{C}$ (\ref{eq:I.2}) (broken lines) in the range~$2\le\lP\le 9$, see
the table on precedent p.~\pageref{table3}. However for~$\lP>9$, the RST binding energy~$\EEP^{( n)}$
becomes larger than the conventional prediction~$\Ea{E}{n}{C}$ (\ref{eq:I.2}). Thus the
RST predictions~$\EEP^{( n)}$ have lower degree of degeneracy but are placed within some
band of relatively narrow width around the conventional (highly degenerated)
levels~$\Ea{E}{n}{C}$ (\ref{eq:I.2}).

\newpage

Summarizing all the numerical results obtained up to now, it may be sufficient to collect
only the three levels for~$l_z=0,\lP-1,\lP$ (see the tables on
pp.~\pageref{table4},\pageref{table3},\pageref{table2}) into one figure in order to attain
a first impression of the RST band structure, see \textbf{Fig.IV.E} below. For low
principle quantum numbers~($n\lesssim 5$, say) the RST energy bands are neatly separated
and are situated in the neighborhood of the conventional predictions (\ref{eq:I.2})
(broken lines). But for increasing quantum numbers~($n\gtrsim 5$, say) the splitting of
the degenerated conventional levels~$\Ea{E}{n}{C}$ (\ref{eq:I.2}) becomes larger and
larger; and it seems that finally the RST spectrum may adopt a chaotic character. Whether
this is a consequence of the roughness of our approximation procedure or is a true feature
of the exact RST spectrum must be studied separately.

\epsfig{file=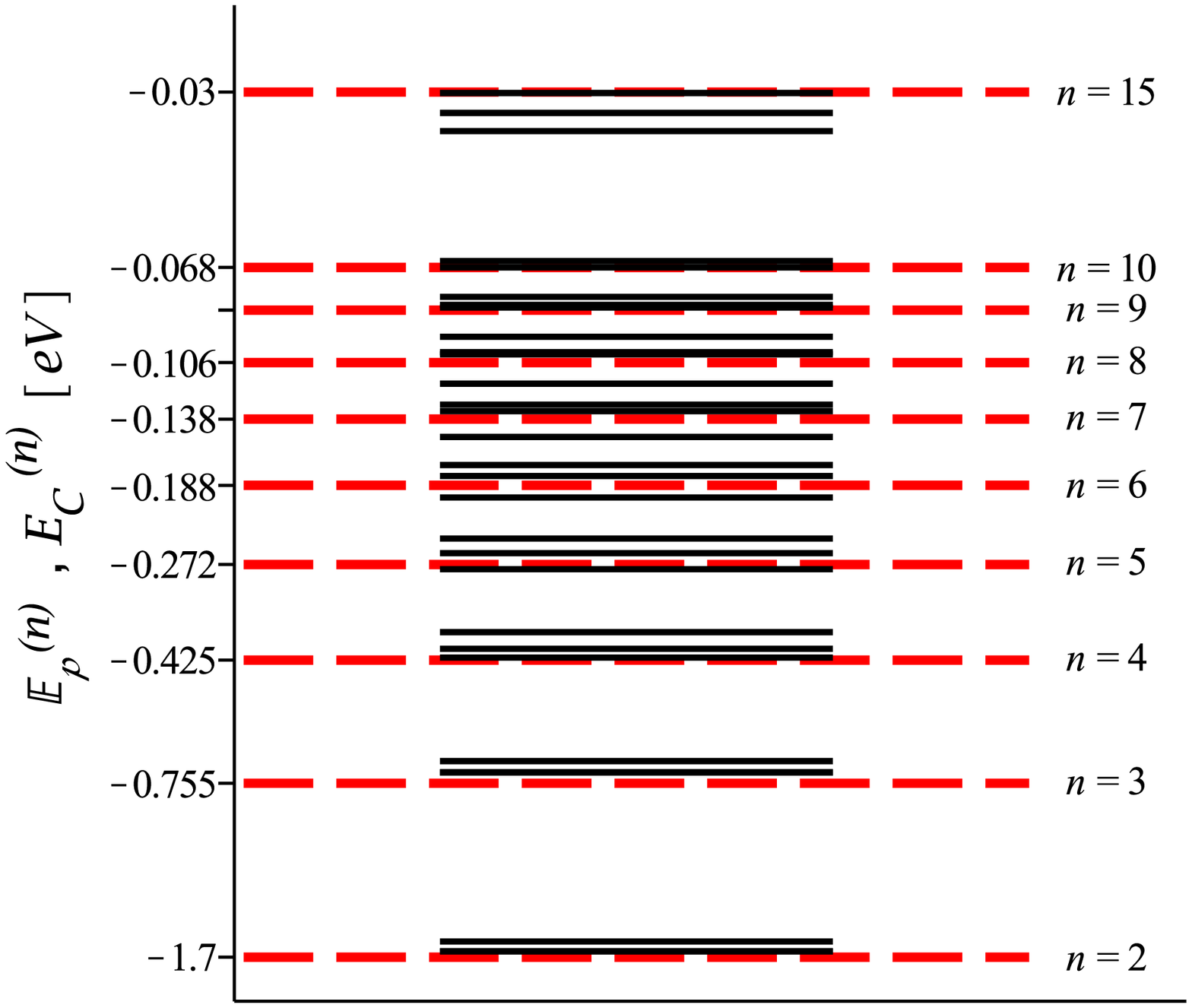,width=1.0\linewidth,height=1.2\linewidth}
\label{figIV.E}
{\textbf{Fig.~IV.E}\hspace{5mm} \emph{\large\textbf{Band Structure of the RST Levels        }   }  }
\myfigure{Fig.~IV.E: { }}
\indent\\*


\renewcommand{\theequation}{\Alph{section}.\arabic{equation}}
  \setcounter{section}{1}
  \setcounter{equation}{0}
  \begin{center}
  {\textbf{\Large Appendix A}}\\[1.5em]
  \emph{\textbf{\Large Quadrupole Energy}}
  \end{center}
  \vspace{2ex}

  In order to determine the quadrupole energy $\Eegiii$ (\ref{eq:IV.128}) we first have to
  inquire into the solution $\pgA^\mathsf{III}(r)$ of the respective quadrupole equation
  (\ref{eq:IV.127}). Passing here over again to dimensionless objects and referring to the
  selected trial amplitude $\tPhi(r)$ (\ref{eq:IV.131})--(\ref{eq:IV.132}), that
  quadrupole equation reappears now in the following form:
\begin{subequations}
\begin{gather}
\label{eq:A.1a}
\frac{d^2\,\Pnuiii(y)}{dy^2} + \frac{2}{y}\,\frac{d\,\Pnuiii(y)}{dy} - 6\,\frac{\Pnuiii(y)}{y^2} = -5\,y^{2\nu - 1} \cdot \e^{-y} \\
\label{eq:A.1b}
\left( y \doteqdot 2\,\beta r \;,\quad \Pnuiii(y) \doteqdot \frac{\Gamma(2\nu + 2)}{2\,\beta \as} \cdot \pgA^\mathsf{III}(r) \right) \;.
\end{gather}
\end{subequations}
If it were possible to find the exact solution of this equation, one could substitute it in the first-order anisotropy energy $\Eegiii$ (\ref{eq:IV.128}) which reads in terms of the dimensionless potential correction $\Pnuiii(y)$ (\ref{eq:A.1b}) as follows
\begin{equation}
\label{eq:A.2}
\Eegiii = - \pr e_3 \cdot \frac{\e^2}{a_B}\,\left( 2\,\beta a_B \right)\,\epot^\mathsf{III}(\nu) \;.
\end{equation}
Here, the potential function $\epot^\mathsf{III}(\nu)$ is the anisotropic counterpart of the ``isotropic'' $\epot(\nu)$ (\ref{eq:IV.139}) and reads in terms of the (dimensionless) anisotropic potential $\Pnuiii(y)$ (\ref{eq:A.1b})
\begin{equation}
\label{eq:A.3}
\epot^\mathsf{III}(\nu) = \frac{1}{\Gamma(2\nu + 2)^2}\,\int\limits_0^\infty dy\;y^2\,\left\{ \left( \frac{d\,\Pnuiii(y)}{dy} \right)^2 + 6\,\left( \frac{\Pnuiii(y)}{y} \right)^2 \right\} \;.
\end{equation}

Naturally, one would prefer to find the \emph{exact} solution of the (dimensionless) quadrupole equation (\ref{eq:A.1a}) in order to get the corresponding \emph{exact} value of the quadrupole energy $\Eegiii$ (\ref{eq:A.2}). Fortunately, this quadrupole problem allows us to work out the desired exact solution, albeit only in form of a power series.

\newpage
\begin{center}
  \large{\textit{The Exact Quadrupole Solution}}
\end{center}

The first observation for finding the desired solution refers to the fact that the quadrupole equation (\ref{eq:A.1a}) can be recast to the following form:
\begin{equation}
\label{eq:A.4}
\frac{1}{y^4}\,\frac{d}{dy}\,\left[ y^6 \cdot \frac{d}{dy}\,\left( \frac{\Pnuiii(y)}{y^2} \right) \right] = -5\,y^{2\nu - 1} \cdot \e^{-y} \;.
\end{equation}
Integrating this from arbitrary $y$ up to infinity ($y \rightarrow \infty$) yields for arbitrary $\nu\;\left( > \frac{1}{2} \right)$
\begin{equation}
\label{eq:A.5}
\frac{d}{dy}\,\left( \frac{\Pnuiii(y)}{y^2} \right) = -5\,\Gamma(2\nu + 4)\,\e^{-y} \cdot \sum_{n=4}^{\infty}\frac{y^{2\nu + n - 6}}{\Gamma(2\nu + 1 + n)} \;.
\end{equation}
For further integrating this intermediate result we have to observe the boundary conditions
\begin{subequations}
\begin{align}
\label{eq:A.6a}
\underset{y \rightarrow 0}{\lim} \Pnuiii(y) &= \text{const.} \cdot y^2 \\
\label{eq:A.6b}
\underset{y \rightarrow \infty}{\lim} \Pnuiii(y) &= \frac{\text{const.}}{y^3}
\end{align}
\end{subequations}
which may easily be deduced from the \emph{homogeneous} version of the quadrupole equation
(\ref{eq:A.4}). The point here is that the solutions $\Pnuiii(y)$ of the inhomogeneous
equation (\ref{eq:A.4}) (or (\ref{eq:A.1a}), resp.) must adopt the form of a solution of
the homogeneous equation in those regions of three-space where the inhomogenity ($\sim
y^{2\nu - 1} \cdot \e^{-y}$) is practically zero, i.\,e. for $y \rightarrow \infty$ and
also for $y \rightarrow 0$ for sufficiently large values of the power $\nu$.

With regard to this circumstance concerning the boundary conditions, one may further integrate (by parts) the intermediate result (\ref{eq:A.5}) in order to find by means of some simple manipulations
\begin{align}
\label{eq:A.7}
\Pnuiii(y) = \text{const.} \cdot y^2 &+ \Gamma(2\nu + 4) \cdot \frac{\e^{-y}}{y^3}\,\sum_{n=0}^\infty \frac{y^{2\nu + 4 + n}}{\Gamma(2\nu + 5 + n)} \\
&- \Gamma(2\nu - 1) \cdot y^2\,\e^{-y}\,\sum_{n=0}^\infty \frac{y^{2\nu - 1 + n}}{\Gamma(2\nu + n)} \;. \nonumber
\end{align}
Here, it remains to determine the integration constant which emerges in connection with
the solution of the homogeneous equation ($\sim y^2$). This problem can be dealt with very
conveniently by specializing that general result (\ref{eq:A.7}) to integer values of
$2\nu$. Indeed, it is easily seen by means of some simple manipulations that for this case
that general result (\ref{eq:A.7}) adopts the following form
\begin{align}
\label{eq:A.8}
\Pnuiii(y)\ \Rightarrow\ \text{const.} \cdot y^2 &+ (2\nu + 3)!\ \frac{1 - \e^{-y} \cdot \sum_{n=0}^{2\nu + 3} \frac{y^n}{n!}}{y^3} \\
&- (2\nu - 2)!\ y^2\,\left\{ 1 - \e^{-y} \cdot \sum_{n=0}^{2\nu - 2} \frac{y^n}{n!} \right\} \;. \nonumber
\end{align}

But here it is now evident that the required boundary condition (\ref{eq:A.6b}) is
violated, except the integration constant in equation (\ref{eq:A.7}) adopts the special
value const. \mbox{$\Rightarrow (2\nu - 2)!$} so that the ``integer-valued'' solution
(\ref{eq:A.8}) appears now as
\begin{align}
\label{eq:A.9}
\Pnuiii(y) = (&2\nu + 3)!\ \frac{1 - \e^{-y} \cdot \sum_{n=0}^{2\nu + 3} \frac{y^n}{n!}}{y^3} \\
+ (& 2\nu - 2)!\ y^2\,\e^{-y} \cdot \sum_{n=0}^{2\nu - 2} \frac{y^n}{n!} \;, \nonumber
\end{align}
or, by generalization of this result, the solution for general (but real-valued) $\nu$ is
\begin{align}
\label{eq:A.10}
\Pnuiii(y) = \Gamma(&2\nu + 4)\,\frac{e^{-y}}{y^3} \cdot \sum_{n=0}^\infty \frac{y^{2\nu + 4 + n}}{\Gamma(2\nu + 5 + n)} \\
+ \Gamma(&2\nu - 1)\,y^2\,\left\{ 1 - \e^{-y} \cdot \sum_{n=0}^{\infty} \frac{y^{2\nu - 1 + n}}{\Gamma(2\nu + n)} \right\} \;. \nonumber
\end{align}

Obviously, there appears to be a certain peculiarity with this result; namely concerning the behaviour of $\Pnuiii(y)$ at the origin ($y \rightarrow 0$). This can be most easily seen by inspecting a little bit more closely the ``integer-valued'' result (\ref{eq:A.9}). Indeed, this form of $\Pnuiii(y)$ can be recast to the following shape which more clearly displays the behaviour for $y \rightarrow 0$:
\begin{equation}
\label{eq:A.11}
\Pnuiii(y) = (2\nu - 2)!\,y^2 + \e^{-y} \cdot \sum_{n=0}^\infty \left[ \frac{(2\nu + 3)!}{(2\nu + 4 + n)!} - \frac{(2\nu - 2)!}{(2\nu - 1 + n)!} \right] \cdot y^{2\nu + 1 + n} \;.
\end{equation}
For instance, for $\nu = 3$ this looks as follows in the vicinity of the origin ($y \rightarrow 0$)
\begin{equation}
\label{eq:A.12}
\mathcal{P}_3^\mathsf{III}(y) \simeq 24\,y^2 - \frac{y^7}{10} + \frac{5}{66}\,y^8 + \ldots
\end{equation}
The point here is that for such a differential equation as (\ref{eq:A.4}), where the inhomogeneous term (source term) behaves as ($\sim y^{2\nu - 1}$) near the origin, one expects that the solution $\Pnuiii(y)$ should then behave there as $\sim y^{2\nu + 1}$ because the differential equation is of second order. However, this expectation is not validated by our result (\ref{eq:A.11}) since (for $\nu \geq 1$) there always dominates the ``homogeneous'' solution ($\sim y^2$) in the neighbourhood of the origin. Actually, it is the second (i.\,e. exponential) term in (\ref{eq:A.11}) which goes like $\sim y^{2\nu + 1}$ and thus meets with the original expectation.

Observe here also that one cannot simply subtract off the ``homogeneous'' solution ($\sim y^2$) since in this case one would spoil the required boundary condition (\ref{eq:A.6b}) at infinity! This is easily seen by returning to the equivalent form (\ref{eq:A.9}) which is numerically identical to the latter form (\ref{eq:A.11}) of $\Pnuiii(y)$. Finally, let us also mention that the latter integer-valued form (\ref{eq:A.11}) of the solution $\Pnuiii(y)$ admits the obvious generalization to arbitrary (but real-valued) $\nu$, i.\,e.
\begin{subequations}
\begin{align}
\label{eq:A.13a}
\Pnuiii(y) &= \Gamma(2\nu - 1) \cdot y^2 + \e^{-y} \cdot \sum_{n=0}^\infty p_n(\nu)\,y^{2\nu + 1 + n} \\
\label{eq:A.13b}
p_n(\nu) &= \frac{\Gamma(2\nu + 4)}{\Gamma(2\nu + 5 + n)} - \frac{\Gamma(2\nu - 1)}{\Gamma(2\nu + n)} \;.
\end{align}
\end{subequations}
This is a complementary form of the precedent result (\ref{eq:A.10}) or (\ref{eq:A.9}), resp. Indeed, whereas (\ref{eq:A.9}) clearly demonstrates the required boundary condition (\ref{eq:A.6b}) at infinity, the present result (\ref{eq:A.13a}) validates the condition (\ref{eq:A.6a}) near the origin.

As a check of this result, one considers the function $\Pnuiii(y)$ (\ref{eq:A.13a}) not
the result but rather an ansatz with unknown coefficients $p_n$ which depend of course on
the variational parameter $\nu$. Substituting this ansatz into the quadrupole equation
(\ref{eq:A.1a}), one gets the following recurrence formula for the determination of the
coefficients $p_n$:
\begin{equation}
\label{eq:A.14}
p_{n+2} = \frac{2\,(2\nu + 3 + n) \cdot p_{n+1} - p_n}{(2\nu + 4 + n)(2\nu + 3 + n) - 6} \;,
\end{equation}
with the two lowest-order coefficients being given by
\begin{subequations}
\begin{align}
\label{eq:A.15a}
p_0 &= - \frac{5}{(2\nu + 2)(2\nu + 1) - 6} \\
\label{eq:A.15b}
p_1 &= \frac{4\nu + 4}{(2\nu + 3)(2\nu + 2) - 6} \cdot p_0 \;.
\end{align}
\end{subequations}
From consistency reasons, the former coefficients $p_n(\nu)$ (\ref{eq:A.13b}) must then
turn out as the desired solutions of the present recurrence problem
(\ref{eq:A.14})--(\ref{eq:A.15b}). Indeed, this claim may easily be verified by a few
simple algebraic manipulations.

\begin{center}
  \large{\textit{Quadrupole Energy}}
\end{center}

For calculating now explicitly the quadrupole energy $\Eegiii$ (\ref{eq:A.2}) we simplify again the numerical computation by resorting to the quadrupole identity (\ref{eq:IV.118}). This means that we do not substitute our solution $\Pnuiii(y)$ in the ``anisotropic'' potential function $\epot^\mathsf{III}(\nu)$ (\ref{eq:A.3}) in order to obtain the anisotropy energy $\Eegiii$ (\ref{eq:A.2}), but rather we substitute the solution $\Pnuiii(y)$ into the ``anisotropic'' mass equivalent $\tMMegiii\crm^2$ (\ref{eq:IV.114}) which reads in the dimensionless notation
\begin{equation}
\label{eq:A.16}
\tMMegiii\crm^2 = - \pr m_3\,\frac{e^2}{a_B}\,\left( 2\,\beta a_B \right) \cdot \muegiii(\nu)
\end{equation}
with the mass-equivalent function $\muegiii(\nu)$ being defined through
\begin{equation}
\label{eq:A.17}
\muegiii(\nu) \doteqdot \frac{1}{\Gamma(2\nu + 2)^2} \int\limits_0^\infty dy\;y^{2\nu + 1}\,\e^{-y} \cdot \Pnuiii(y) \;.
\end{equation}
This procedure is legitimate because our general quadrupole solution (\ref{eq:A.10}) is an \emph{exact} solution of the quadrupole equation (\ref{eq:A.1a}) so that the ``anisotropic'' Poisson identity (\ref{eq:IV.118}) is actually satisfied. Of course, this is the same effect as mentioned already in connection with the spherically symmetric approximation, cf. the remarks below equation (\ref{eq:IV.139}). Consequently, substituting both results (\ref{eq:A.2}) and (\ref{eq:A.16}) into the ``anisotropic'' Poisson identity (\ref{eq:IV.118}) we obtain the numerical identity of both energy coefficients $\epot^\mathsf{III}(\nu)$ (\ref{eq:A.3}) and $\muegiii(\nu)$ (\ref{eq:A.17}):
\begin{equation}
\label{eq:A.18}
\pr e_3 \cdot \epot^\mathsf{III}(\nu) \equiv \pr m_3 \cdot \muegiii(\nu) \;,
\end{equation}
which may be used as a consistency check for the numerical correctness of the computer programs.

The explicit computation of $\muegiii$ (\ref{eq:A.17}) may be performed now by means of
either the precedent form (\ref{eq:A.10}) of the solution $\Pnuiii(y)$ or also by means of
its second form (\ref{eq:A.13a})--(\ref{eq:A.13b}). In the first case (\ref{eq:A.10}) we
find
\begin{align}
\label{eq:A.19}
\muegiii(\nu) &= \frac{\Gamma(2\nu + 4)}{\Gamma(2\nu + 2)^2} \cdot \sum_{n=0}^\infty \left[ \frac{1}{2^{4\nu + 3 + n}} \cdot \frac{\Gamma(4\nu + 3 + n)}{\Gamma(2\nu + 5 + n)} \right] \\
&+ \frac{\Gamma(2\nu - 1)}{\Gamma(2\nu + 2)^2} \cdot \Gamma(2\nu + 4) - \frac{\Gamma(2\nu-1)}{\Gamma(2\nu + 2)^2} \cdot \sum_{n=0}^\infty \left[ \frac{1}{2^{4\nu + 3 + n}} \cdot \frac{\Gamma(4\nu + 3 + n)}{\Gamma(2\nu + n)} \right] \;, \nonumber
\end{align}
and in the second case
\begin{equation}
\label{eq:A.20}
\muegiii(\nu) = \frac{\Gamma(2\nu + 4) \cdot \Gamma(2\nu - 1)}{\Gamma(2\nu + 2)^2} + \frac{1}{\Gamma(2\nu + 2)^2} \cdot \sum_{n=0}^\infty p_n(\nu)\,\frac{\Gamma(4\nu + 3 + n)}{2^{4\nu + 3 + n}} \;.
\end{equation}
Of course, both results for $\muegiii(\nu)$ must be identical (thus providing a confidence test for the corresponding computer programs). For half-integer values of $\nu$ ($\leadsto \nu = 1, \frac{3}{2}, 2, \frac{5}{2}, \ldots$) the mass-equivalent function $\muegiii(\nu)$ (\ref{eq:A.20}) can also be represented by a finite sum:
\begin{equation}
\label{eq:A.21}
\muegiii(\nu) = \frac{(2\nu + 3)(2\nu + 2)}{(2\nu + 1)!}\, \left\{  (2\nu - 2)! \cdot \LLnu - 5 \cdot \KKnu \right\}
\end{equation}
with
\begin{subequations}
\begin{align}
\label{eq:A.22a}
\KKnu &= \sum_{n=0}^{2\nu - 3} \frac{n!}{(n+6)!}\,\sum_{m=0}^{n} \frac{1}{2^{2\nu + 4 + n - m}} \cdot \frac{(2\nu + 3 + n - m)!}{(n-m)!} \\
\label{eq:A.22b}
\LLnu &= 1 - \sum_{n=0}^{4} \frac{1}{n!\,2^{2\nu - 1 + n}} \cdot \frac{(2\nu - 2 + n)!}{(2\nu - 2)!} \;,
\end{align}
\end{subequations}
see \textbf{Appendix E} of ref.~\cite{1}. Inserting now this result (\ref{eq:A.20}) via the
Poisson identity (\ref{eq:A.18}) in the expression (\ref{eq:A.2}) for the anisotropy
energy yields the anisotropic gauge field contribution to the energy functional $\tEEePhi$
(\ref{eq:IV.119}) which thus adopts the simple form (\ref{eq:IV.144}) because all
constraints are automatically satisfied.

\setcounter{section}{2}
\setcounter{equation}{0}

\begin{center}
{\textbf{\Large Appendix B}}\\[1.5em]
\emph{\textbf{\Large Special Case }} {\Large\boldmath $\elz = 0,\ \forall  \elp$}
\end{center}
\vspace{2ex}

The situation for vanishing quantum number $\elz$ and arbitrary $\elp$ $(\elp =
0,1,2,3,\ldots)$ represents a special case insofar as the relevant \emph{quadrupole
objects} $\pgko(\vartheta)$, $\pgAiii(\vartheta)$, $\pr e_3$, $\pr f_3$, $\pr m_3$,
$\EEegiii$ are all independent of the quantum number $\elp$.

In order to validate this assertion, one first turns to the angular density $\pgko(\vartheta)$
\begin{equation}
\label{eq:B.1}
\pgko(\vartheta) \doteqdot \frac{f_R^2(\vartheta) + f_S^2(\vartheta)}{4\pi\sin\vartheta}
\end{equation}
which for the product ansatz may be separately normalized to unity,
cf. (\ref{eq:IV.15a}). Now differentiate this object and find by use of the first-order
angular system (\ref{eq:IV.8a})--(\ref{eq:IV.8b})
\begin{equation}
\label{eq:B.2}
\frac{d\,\pgko(\vartheta)}{d\vartheta} + \cot\vartheta \cdot \pgko(\vartheta) = \frac{\elz}{2\pi \sin\vartheta} \, \left\{ 2 f_R(\vartheta) f_S(\vartheta) + \cot\vartheta \left[ f_R^2(\vartheta) - f_S^2(\vartheta) \right] \right\} \;.
\end{equation}
For the presently considered case $\elz = 0$, this becomes simplified to
\begin{equation}
\label{eq:B.3}
\frac{d\,\pgko(\vartheta)}{d\vartheta} + \cot\vartheta \cdot \pgko(\vartheta) = 0
\end{equation}
with the solution (normalized according to the prescription (\ref{eq:IV.15a}))
\begin{equation}
\label{eq:B.4}
\pgko(\vartheta) = \frac{1}{2\pi^2} \cdot \frac{1}{\sin\vartheta} \;.
\end{equation}
Observe here that our demand $\elz = 0$ eliminates completely the quantum number $\elp$
from the equation (\ref{eq:B.2}) so that the resulting solution (\ref{eq:B.4}) is actually
independent of $\elp$! Of course, this independency must then be transferred to all the
other objects which derive from the angular density $\pgko(\vartheta)$.

The first one of these objects is the anisotropic potential correction $\pgAiii(\vartheta)$ (\ref{eq:IV.104c}), being defined quite generally by
\begin{equation}
\label{eq:B.5}
\pgAiii(\vartheta) \doteqdot \frac{3}{2} \int d\Omega' \, \left( \hat{\vr} \sdot \hat{\vr}\,' \right)^2 \, \left[ \bgkn(\vartheta') - \frac{1}{4\pi} \right] \;,
\end{equation}
or equivalently by
\begin{equation}
\label{eq:B.6}
\bgAiii(\vartheta) = - \frac{3}{4}\,\left( 1 - 3\,K^\mathsf{III}_{\{p\}} \right) \cdot \left[ \cos^2\vartheta - \frac{1}{3} \right] \;,
\end{equation}
with the numerical constant $K^\mathsf{III}_{\{p\}}$ being defined through
\begin{equation}
\label{eq:B.7}
K^\mathsf{III}_{\{p\}} \doteqdot \int d\Omega \, \cos^2\vartheta \cdot \pgko(\vartheta) \;.
\end{equation}
This constant is easily calculated for the present angular density $\pgko(\vartheta)$ (\ref{eq:B.4}) and yields the result
\begin{equation}
\label{eq:B.8}
K^\mathsf{III}_{\{p\}} = \frac{1}{2} \;.
\end{equation}
Thus the anisotropy correction $\pgAiii(\vartheta)$ (\ref{eq:B.6}) becomes
\begin{equation}
\label{eq:B.9}
\pgAiii(\vartheta) = \frac{3}{8}\,\left[ \cos^2\vartheta - \frac{1}{3} \right] \;.
\end{equation}
The interesting point with this result is now that it does hold for any value of the
quantum number $\elp$, see for instance the case $\lP=3,\elz = 0$ in equation
(\ref{eq:IV.104c}).

The next quadrupole objects are the numerical constants $\pr e_3$ (\ref{eq:IV.110a}), $\pr f_3$ (\ref{eq:IV.110b}) and $\pr m_3$ (\ref{eq:IV.115}). Inserting in these definitions the present potential correction $\pgAiii(\vartheta)$ (\ref{eq:B.9}) and the angular density $\pgko(\vartheta)$ (\ref{eq:B.4}) yields for these constants
\begin{subequations}
\begin{align}
\label{eq:B.10a}
\pr e_3 &= \frac{1}{80} \\
\label{eq:B.10b}
\pr f_3 &= \frac{3}{40} \\
\label{eq:B.10c}
\pr m_3 &= \frac{1}{16} \;.
\end{align}
\end{subequations}
Observe here again that this result does hold for all values of the quantum number $\elp$,
not only for $\elp = 3$ being displayed by the table on p.~\pageref{tablelp3}. This
conclusion must then be true also for the anisotropy energy $\Eegiii$ (\ref{eq:IV.109}) or
(\ref{eq:IV.142}), resp.
\begin{equation}
\label{eq:B.11}
\Eegiii = -\frac{1}{16}\,\frac{e^2}{a_B}\,\left( 2\beta a_B \right)\cdot\muegiii(\nu) \;.
\end{equation}

Accordingly, the total electrostatic energy $\EReg$ (\ref{eq:IV.108}) becomes
\begin{subequations}
\begin{align}
\label{eq:B.12a}
\EReg &= -\frac{e^2}{a_B}\,\left( 2\beta a_B \right) \cdot \etot(\nu) \\
\label{eq:B.12b}
\etot(\nu) &= \epot(\nu) + \frac{1}{16} \muegiii(\nu)
\end{align}
\end{subequations}
which finally lets appear the energy function $\EEP(\nu)$ (\ref{eq:IV.148}) as
\begin{equation}
\label{eq:B.13}
\EEP(\nu) = -\frac{e^2}{4a_B} \cdot S_\mathcal{P}(\nu) \;,
\end{equation}
with the specific spectral function $S_\mathcal{P}(\nu)$ being given now by
\begin{equation}
\label{eq:B.14}
S_\mathcal{P}(\nu) = \frac{\left[ \etot(\nu) \right]^2}{\ekin(\nu)} = \frac{\left[ \epot(\nu) + \frac{1}{16} \cdot \muegiii(\nu) \right]^2}{\frac{1}{2\nu + 1}\,\left( \frac{1}{4} + \frac{\elp^2}{2\nu} \right)} \;.
\end{equation}
Here, the peculiarity due to the special case $\elz = 0$ becomes now manifest: in contrast
to the general situation ($\elz \neq 0$), e.\,g. (\ref{eq:IV.161}) or (\ref{eq:IV.174}),
the numerator is independent of the quantum number $\elp$ which thus influences
exclusively the denominator being represented by the kinetic energy $\Ekin$
(\ref{eq:IV.133})--(\ref{eq:IV.134}). Or, equivalently, the electrostatic interaction
$\EReg$ (\ref{eq:B.12a})--(\ref{eq:B.12b}) does not depend explicitly upon the quantum
number $\elp$ of angular momentum. From this one concludes that possibly the corresponding
energy spectrum, due to $\elz = 0$, may show some characteristic features missing for the
other values ($\elz \neq 0$) of angular momentum. The following table represents the
corresponding energy spectrum to be calculated again by the method of partial
extremalization described below equation (\ref{eq:IV.148}):

\newpage
\begin{center}\label{table4}
{\boldmath $\l_z=0,\ \forall \lP$}\\[5mm]
  \begin{tabular}{|c|c|c|c|c|}
  \hline
 $\nP $ & $ \EEP^{(n)} $\,[eV] &  $\Ea{E}{n}{C}$\,[eV]  & $\Delta_\wp^{(n)}\,\%  $  & $\nu_*$  \\
        &  (\ref{eq:IV.162a}) &   (\ref{eq:I.2})  &  (\ref{eq:IV.151})  &  (\ref{eq:IV.162b})  \\
  \hline\hline
$ 2 $ & $ -1.65471 $ & $ -1.70072 $ & $  2.7 $ & $ 1.72555 $  \\ \hline 
$ 3 $ & $ -0.71761 $ & $ -0.75588 $ & $  5.1 $ & $ 3.86851 $  \\ \hline 
$ 4 $ & $ -0.40299 $ & $ -0.42518 $ & $  5.2 $ & $ 6.10402 $  \\ \hline 
$ 5 $ & $ -0.25889 $ & $ -0.27212 $ & $  4.9 $ & $ 8.50058 $  \\ \hline 
$ 6 $ & $ -0.18075 $ & $ -0.18897 $ & $  4.3 $ & $ 11.03747 $  \\ \hline 
$ 7 $ & $ -0.13356 $ & $ -0.13883 $ & $  3.8 $ & $ 13.69725 $  \\ \hline 
$ 8 $ & $ -0.10282 $ & $ -0.10630 $ & $  3.3 $ & $ 16.47923 $  \\ \hline 
$ 9 $ & $ -0.08167 $ & $ -0.08399 $ & $  2.8 $ & $ 19.36094 $  \\ \hline 
$ 10 $ & $ -0.06648 $ & $ -0.06803 $ & $  2.3 $ & $ 22.34116 $  \\ \hline 
$ 15 $ & $ -0.03013 $ & $ -0.03024 $ & $  0.3 $ & $ 38.50906 $  \\ \hline 
$ 20 $ & $ -0.01719 $ & $ -0.01701 $ & $  -1.0 $ & $ 56.43818 $  \\ \hline 
  \end{tabular}
\end{center}

The numerical evidence says here that the RST energy predictions~$\EEP^{(n)}$ (second row)
are placed \emph{above} the conventional results~$\Ea{E}{n}{C}$ (third row) up to
principal quantum numbers~$\nP\eqsim 15$. For~$\nP\gtrsim 15$ the RST binding
energy~$(-\EEP^{(n)})$ becomes larger than its conventional counterpart~$(-\Ea{E}{n}{C})$.
As far as the present numerical results are concerned, the RST energies due to~{\boldmath
  $l_z=0$} do mark the upper edge of the band of non-degenerate RST levels due to the same
quantum numbers~$\nP,\lP\,(=\nP-1)$, see \textbf{Fig.s B.I} and~\textbf{IV.E}.
\begin{center}
\epsfig{file=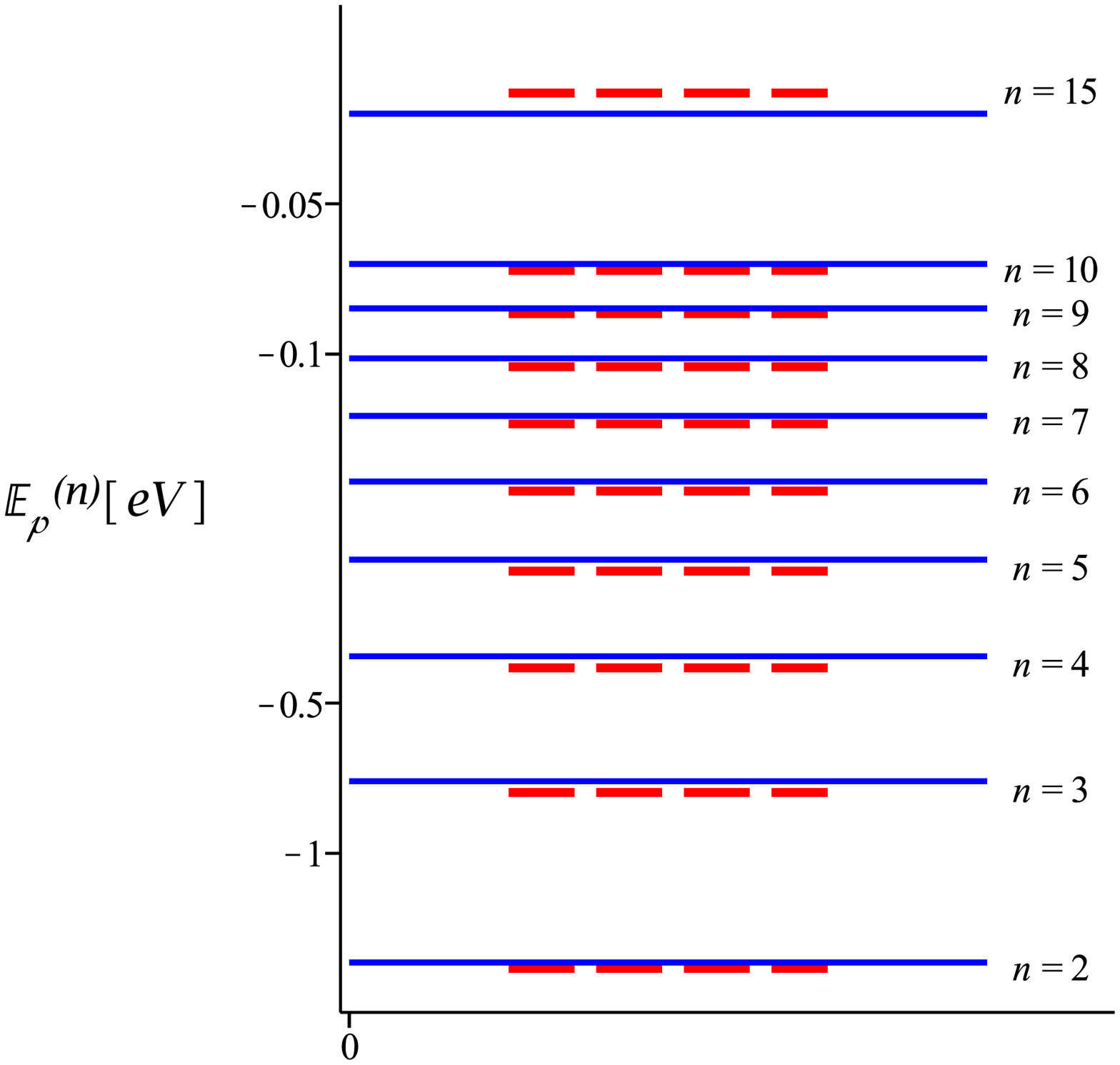,height=12cm}
\end{center}
{\textbf{Fig.~B.I}\hspace{5mm} \emph{\large\textbf{RST Levels due to~{\boldmath $l_z=0,\, \forall\lP$}
 }  }  }
\myfigure{Fig.~B.a: {
  }}
\indent

The energy predictions~$ \EEP^{(n)}$ (\ref{eq:B.13})-(\ref{eq:B.14}) due to~$l_z=0,\,
\forall\lP$ (solid lines) appear as being situated closest to their conventional
counterparts~$\Ea{E}{n}{C}$ (\ref{eq:I.2}) within the bands of non-degenerate RST levels;
and furthermore these levels (due to~$l_z=0$) do form the upper boundary of
the~$\lP$-multiplets defined by~$-\lP\le\l_z\le +\lP$. For~$\nP\gtrsim 15$, this
\emph{upper} boundary slides \emph{below} the conventional levels~$\Ea{E}{n}{C}$
(\ref{eq:I.2}).

  \setcounter{section}{3}
  \setcounter{equation}{0}

  \begin{center}
  {\textbf{\Large Appendix C}}\\[1.5em]
  \emph{\textbf{\Large Universality of the Quadrupole Ratios }} 
  {\boldmath \Large $\frac{\pr f_3}{\pr e_3}$} \emph{\textbf{ \Large and }} {\boldmath \Large $\frac{\pr m_3}{\pr e_3}$}
  \end{center}
  \vspace{2ex}

At various occasions one observes the emergence of the following ratios:
\begin{subequations}
\begin{align}
\label{eq:C.1a}
\frac{\pr f_3}{\pr e_3} &= 6 \\
\label{eq:C.1b}
\frac{\pr m_3}{\pr e_3} &= 5 \;,
\end{align}
\end{subequations}
e.\,g. for $\elp = 3$ and $|\elz| = 0,1,3$ (see the table on p.~\pageref{tablelp3}) or for
$\elz = \pm \elp$ (see equations (\ref{eq:IV.158a})-(\ref{eq:IV.158b})) or also for $\elz
= \pm \left( \elp - 1 \right)$, see equations (\ref{eq:IV.170a})-(\ref{eq:IV.170b}); and
finally for $\elz = 0, \forall\,\elp$ (see equations
(\ref{eq:B.10a})-(\ref{eq:B.10c})). Thus, one is led to the supposition that these ratios
(\ref{eq:C.1a})--(\ref{eq:C.1b}) must hold for any allowed combination of the quantum
numbers $\elz$ and $\elp$.

Indeed, it is a simple matter to convince oneself of the correctness of that
supposition. The point of departure is the general form of the anisotropic correction
potential $\pgAiii(\vartheta)$ (\ref{eq:IV.166}) which we rewrite here as
\begin{equation}
\label{eq:C.2}
\pgAiii(\vartheta) = \aiii \cdot \left[ \cos^2\vartheta - \frac{1}{3} \right]
\end{equation}
with the constant $\aiii$ being defined by
\begin{equation}
\label{eq:C.3}
\aiii \doteqdot -\frac{3}{4}\,\left( 1 - 3 \Kpgiii \right) \;.
\end{equation}
Now insert this general form of $\pgAiii(\vartheta)$ (\ref{eq:C.2}) in the definition (\ref{eq:IV.110a}) of $\pr e_3$ and find
\begin{equation}
\label{eq:C.4}
\pr e_3 = \frac{4}{45} \cdot \aiii^2 \;.
\end{equation}
Next, insert the general form (\ref{eq:C.2}) also in the definition of $\pr f_3$ (\ref{eq:IV.110b}) and find
\begin{equation}
\label{eq:C.5}
\pr f_3 = \frac{8}{15} \cdot \aiii^2 \;.
\end{equation}
Thus, both general results (\ref{eq:C.4})--(\ref{eq:C.5}) validate the general character of the first ratio (\ref{eq:C.1a}).

The case of the second ratio (\ref{eq:C.1b}) is a little bit more complicated. But inserting again the general definition of $\pgAiii(\vartheta)$, see equation (\ref{eq:IV.104c}) or (\ref{eq:B.5}), resp., in the definition of $\pr m_3$ (\ref{eq:IV.115}) yields by some simple manipulations and the normalization condition (\ref{eq:IV.15a}) for $\pgko(\vartheta)$ the following form of $\pr m_3$:
\begin{equation}
\label{eq:C.6}
\pr m_3 = \frac{3}{2} \int d\Omega\,\pgko(\vartheta) \int d\Omega'\,\left( \hat{\vr} \sdot \hat{\vr}\,' \right)^2 \pgko(\vartheta') - \frac{1}{2}
\end{equation}
After having carried out here the azimuthal integrations, this result reappears as
\begin{equation}
\label{eq:C.7}
\frac{2}{3} \left( \pr m_3 + \frac{1}{2} \right) = \left[ \int d\Omega\,\cos^2\vartheta \bgkn(\vartheta) \right]^2 + \frac{1}{2} \left[ \int d\Omega\,\sin^2\vartheta \bgkn(\vartheta) \right]^2 \;.
\end{equation}
Now observe here the definition of $\Kpgiii$ (\ref{eq:IV.167}), together with the
normalization condition (\ref{eq:IV.15a}) so that this result reads in terms of $\Kpgiii$
\begin{equation}
\label{eq:C.8}
\frac{2}{3} \left( \pr m_3 + \frac{1}{2} \right) = \left( \Kpgiii \right)^2 + \frac{1}{2} \left( 1 - \Kpgiii \right)^2 \;,
\end{equation}
i.\,e.
\begin{equation}
\label{eq:C.9}
\pr m_3 = \left( \frac{3}{2} \, \Kpgiii - \frac{1}{2} \right)^2
\end{equation}
or in terms of $\aiii$ (\ref{eq:C.3})
\begin{equation}
\label{eq:C.10}
\pr m_3 = \left( \frac{2}{3} \aiii \right)^2 \;.
\end{equation}
Now combining this with the precedent result for $\pr e_3$ (\ref{eq:C.4}) yields just the claimed ratio (\ref{eq:C.1b}).

\setcounter{section}{4}
\setcounter{equation}{0}

\begin{center}
{\textbf{\Large Appendix D}}\\[1.5em]
\emph{\textbf{\Large Angular-Momentum Algebra for Para-Positronium}}
\end{center}
\vspace{2ex}

There are certain similarities between the conventional eigenvalue problem
(\ref{eq:IV.16a})--(\ref{eq:IV.16b}) and its RST analogue
(\ref{eq:IV.8a})--(\ref{eq:IV.8b}). The quantum numbers $\elz \Leftrightarrow m$ do occur
in boath approaches in the same sense (namely as eigenvalues of $\hLz$), but the other
quantum number $\elp$ seems to emerge in a somewhat different way than is the case with
the conventional $\ell$ which is due to the square $\hvL^2$ (\ref{eq:IV.16a}). Therefore
one may raise the question whether in the RST formalism there is also an operator whose
eigenvalues are just given by the values of that second quantum numer $\elp$.

Subsequently, we will inquire into this question and we will identify the wanted operator
which has $\elp$ as its eigenvalue (see equation (\ref{eq:D.14}) below.) In the course of
this inquiry it should become clear that the RST algebra of angular momentum deserves
further elaboration beyond the present extent.

\begin{center}
  \large{\textit{Eigenvalue Problem for Angular Momentum}}
\end{center}

In order to deal with the angular momentum algebra as generally as possible, we restart
with the original eigenvalue system (\ref{eq:IV.2a})--(\ref{eq:IV.2d}). But in place of
applying the somewhat special product ansatz (\ref{eq:IV.4a})--(\ref{eq:IV.4b}), we prefer
now the following more general transformation:
\begin{subequations}
\begin{align}
\label{eq:D.1a}
\pMRpm(\vr) &= \frac{\ptMRpm(r,\vartheta, \phi)}{\sqrt{r\sin\vartheta}} \\
\label{eq:D.1b}
\pMSpm(\vr) &= \frac{\ptMSpm(r,\vartheta, \phi)}{\sqrt{r\sin\vartheta}} \;.
\end{align}
\end{subequations}
This recasts the first half (\ref{eq:IV.2a})--(\ref{eq:IV.2b}) of the eigenvalue system (\ref{eq:IV.2a})--(\ref{eq:IV.2d}) to the new shape
\begin{subequations}
\begin{align}
\label{eq:D.2a}
\frac{\partial\,\ptMRp(\vr)}{\partial r} + \frac{i}{r} \cdot \frac{\partial\,\ptMRp(\vr)}{\partial \phi} - \pAn(\vr) \cdot \ptMRm(\vr) + \frac{1}{r}\,\frac{\partial\,\ptMSp(\vr)}{\partial \vartheta} \\
- \frac{i}{r} \cot\vartheta \cdot \frac{\partial\,\ptMSp}{\partial \phi} = \frac{M + M_*}{\hbar}\crm \cdot \ptMRm(\vr) \nonumber \\[1.5em]
\label{eq:D.2b}
\frac{\partial\,\ptMSp(\vr)}{\partial r} - \frac{i}{r} \cdot \frac{\partial\,\ptMSp(\vr)}{\partial \phi} - \pAn(\vr) \cdot \ptMSm(\vr) - \frac{1}{r}\,\frac{\partial\,\ptMRp(\vr)}{\partial \vartheta} \\
- \frac{i}{r} \cot\vartheta \cdot \frac{\partial\,\ptMRp}{\partial \phi} = \frac{M + M_*}{\hbar}\crm \cdot \ptMSm(\vr) \nonumber
\end{align}
\end{subequations}
and similarly for the second half (\ref{eq:IV.2c})--(\ref{eq:IV.2d})
\begin{subequations}
\begin{align}
\label{eq:D.3a}
\frac{\partial\,\ptMRm(\vr)}{\partial r} + \frac{1}{r} \cdot \ptMRm(\vr) - \frac{i}{r}\,\frac{\partial\,\ptMRm(\vr)}{\partial\phi} + \pAn(\vr) \cdot \ptMRp(\vr) \\
- \frac{1}{r} \cdot \frac{\partial\,\ptMSm(\vr)}{\partial \vartheta} + i\, \frac{\cot\vartheta}{r} \cdot \frac{\partial\,\ptMSm}{\partial \phi} = \frac{M - M_*}{\hbar}\crm \cdot \ptMRp(\vr) \nonumber \\[1.5em]
\label{eq:D.3b}
\frac{\partial\,\ptMSm(\vr)}{\partial r} + \frac{1}{r} \cdot \ptMSm(\vr) + \frac{i}{r} \cdot \frac{\partial\,\ptMSm(\vr)}{\partial \phi} + \pAn(\vr) \cdot \ptMSp(\vr) \\
+ \frac{1}{r} \cdot \frac{\partial\,\ptMRm(\vr)}{\partial \vartheta} + i\, \frac{\cot\vartheta}{r} \cdot \frac{\partial\,\ptMRm}{\partial \phi} = \frac{M - M_*}{\hbar}\crm \cdot \ptMSp(\vr) \nonumber
\end{align}
\end{subequations}

The next substitution is again a certain generalization of the former case (\ref{eq:IV.6a})--(\ref{eq:IV.6b}), namely we try now a product ansatz for the new amplitude fields $\ptMRpm(\vr), \ptMSpm(\vr)$ of the following form
\begin{subequations}
\begin{align}
\label{eq:D.4a}
\ptMRp(\vr) &= f_R(\vartheta, \phi) \cdot \tPhip(r) \\
\label{eq:D.4b}
\ptMRm(\vr) &= f_R(\vartheta, \phi) \cdot \tPhim(r) \\
\label{eq:D.4c}
\ptMSp(\vr) &= f_S(\vartheta, \phi) \cdot \tPhip(r) \\
\label{eq:D.4d}
\ptMSm(\vr) &= f_S(\vartheta, \phi) \cdot \tPhim(r) \;.
\end{align}
\end{subequations}
When this product ansatz is inserted in the modified eigenvalue system (\ref{eq:D.2a})--(\ref{eq:D.3b}), there occurs a separation of this three-dimensional system in a two-dimensional angular problem in terms of the angular coordinates $\vartheta, \phi$ and in a one-dimensional radial problem in terms of the variable $r$. The angular subsystem looks as follows:
\begin{subequations}
\begin{align}
\label{eq:D.5a}
\left( \frac{\partial}{\partial \vartheta} + i\cot\vartheta \cdot \frac{\partial}{\partial \phi} \right)\,f_R(\vartheta, \phi) + i\,\frac{\partial\,f_S(\vartheta, \phi)}{\partial \phi} &= \elp \cdot f_S(\vartheta, \phi) \\
\label{eq:D.5b}
\left( \frac{\partial}{\partial \vartheta} - i\cot\vartheta \cdot \frac{\partial}{\partial \phi} \right)\,f_S(\vartheta, \phi) + i\,\frac{\partial\,f_R(\vartheta, \phi)}{\partial \phi} &= -\elp \cdot f_R(\vartheta, \phi) \;.
\end{align}
\end{subequations}
Here it is easy to see that this angular system represents the immediate generalization of the former, more special system (\ref{eq:IV.8a})--(\ref{eq:IV.8b}), since the latter system emerges from the present one (\ref{eq:D.5a})--(\ref{eq:D.5b}) by simply putting
\begin{subequations}
\begin{align}
\label{eq:D.6a}
f_R(\vartheta, \phi) &= \e^{i\elz\phi} \cdot f_R(\vartheta) \\
\label{eq:D.6b}
f_S(\vartheta, \phi) &= \e^{i\elz\phi} \cdot f_S(\vartheta) \;.
\end{align}
\end{subequations}
Obviously, the present more general system (\ref{eq:D.5a})--(\ref{eq:D.5b}) is nothing else than the eigenvalue problem of angular momentum for spin-zero RST systems ($s_p = 0$). This may be recast also in a more concise form, i.\,e.
\begin{subequations}
\begin{align}
\label{eq:D.7a}
e^{-i\phi} \hat{L}_+\,f_R(\vartheta, \phi) - \hat{L}_z\,f_S(\vartheta, \phi) &= \elp \cdot f_S(\vartheta, \phi) \\
\label{eq:D.7b}
e^{i\phi} \hat{L}_-\,f_S(\vartheta, \phi) + \hat{L}_z\,f_R(\vartheta, \phi) &= \elp \cdot f_R(\vartheta, \phi) \;,
\end{align}
\end{subequations}
with the usual definitions (up to $\hbar$) of the angular momentum operators
\begin{subequations}
\begin{align}
\label{eq:D.8a}
\hat{L}_+ &\doteqdot \left( \hat{L}_x + i\hat{L}_y \right) = e^{i\phi} \left\{ \frac{\partial}{\partial\vartheta} + i\cot\vartheta \frac{\partial}{\partial\phi} \right\} \\
\label{eq:D.8b}
\hat{L}_- &\doteqdot \left( \hat{L}_x - i\hat{L}_y \right) = -e^{-i\phi} \left\{ \frac{\partial}{\partial\vartheta} - i\cot\vartheta \frac{\partial}{\partial\phi} \right\}  \\
\hat{L}_z &\Rightarrow \frac{1}{i}\,\frac{\partial}{\partial\phi} \;.
\end{align}
\end{subequations}

Concerning now the remaining radial problem one finds oneself being left with (\emph{spherically symmetic approximation})
\begin{subequations}
\begin{align}
\label{eq:D.9a}
\frac{d\,\tPhip(r)}{dr} - \frac{\elp}{r} \cdot \tPhip(r) - \peAn(r) \cdot \tPhim(r) &= \frac{M + M_*}{\hbar}\crm \cdot \tPhim(r) \\
\label{eq:D.9b}
\frac{d\,\tPhim(r)}{dr} + \frac{\elp + 1}{r} \cdot \tPhim(r) + \peAn(r) \cdot \tPhip(r) &= \frac{M - M_*}{\hbar}\crm \cdot \tPhip(r) \;,
\end{align}
\end{subequations}
but this is identical to the former system (\ref{eq:IV.9a})--(\ref{eq:IV.9b}) and thus gives no new information.

\begin{center}
  \large{\textit{Second-Order Form of the Angular-Momentum Eigenvalue Problem}}
\end{center}

The special second-order differential equations (\ref{eq:IV.17a})--(\ref{eq:IV.17b}) do hint at the possibility that also the generalized form (\ref{eq:D.5a})--(\ref{eq:D.5b}) could be recast in second-order form. In order to attain this goal, it is favourable to transcribe the present eigenvalue problem of angular momentum (\ref{eq:D.5a})--(\ref{eq:D.5b}) to the two-dimensional spin space. This works as follows: First, define three operators $\hQ_x, \hQ_y, \hQ_z$ as follows
\begin{subequations}
\begin{align}
\label{eq:D.10a}
\hQ_x &\doteqdot i\cot\vartheta \, \frac{\partial}{\partial\phi} \equiv \sin\phi \cdot \hat{L}_y + \cos\phi \cdot \hat{L}_x \\
\label{eq:D.10b}
\hQ_y &\doteqdot -i \, \frac{\partial}{\partial\vartheta} \equiv \cos\phi \cdot \hat{L}_y - \sin\phi \cdot \hat{L}_x \\
\label{eq:D.10c}
\hQ_z &\doteqdot -i \, \frac{\partial}{\partial\phi} \equiv \hat{L}_z
\end{align}
\end{subequations}
and find the corresponding commutation relations as
\begin{subequations}
\begin{align}
\label{eq:D.11a}
\left[ \hQ_x, \hQ_y \right] &= \frac{i}{\sin^2\vartheta}\,\hQ_z \\
\label{eq:D.11b}
\left[ \hQ_y, \hQ_z \right] &= \left[ \hQ_z, \hQ_x \right] = 0 \;.
\end{align}
\end{subequations}

Next, define the operator $\hQQ$ to act in the suitably defined Pauli spinor bundle over the two-sphere $S^2$ by the following self-evident construction:
\begin{equation}
\label{eq:D.12}
\hQQ \doteqdot \hQ_x \sigma_x + \hQ_y \sigma_y + \hQ_z \sigma_z \equiv \hQQ_x + \hQQ_y + \hQQ_z \;,
\end{equation}
and furthermore collect both functions $f_R(\vartheta, \phi)$ and $f_S(\vartheta, \phi)$ to a Pauli two-spinor $\mathbf{f}(\vartheta, \phi)$, i.\,e. in short-hand notation
\begin{equation}
\label{eq:D.13}
\mathbf{f}(\vartheta, \phi) = \begin{pmatrix} f_R(\vartheta, \phi) \\ f_S(\vartheta, \phi) \end{pmatrix} \;.
\end{equation}
And now convince yourself that the eigenvalue system (\ref{eq:D.5a})--(\ref{eq:D.5b}) transcribes to the two-spinor form as
\begin{equation}
\label{eq:D.14}
\hQQ\,\mathbf{f}(\vartheta, \phi) = \elp \cdot \mathbf{f}(\vartheta,\phi) \;.
\end{equation}
This reveals the quantum number $\elp$ as the eigenvalue of the operator $\hQQ$. Finally multiply this eigenvalue equation (\ref{eq:D.14}) from the left by the ``quaternion'' $\hQQ$ (\ref{eq:D.12}) and thus find its desired second-order form as
\begin{equation}
\label{eq:D.15}
\hQQ^2\,\mathbf{f}(\vartheta, \phi) = \elp^2 \cdot \mathbf{f}(\vartheta, \phi) \;.
\end{equation}

Complementing finally this eigenvalue equation by
\begin{gather}
\label{eq:D.16}
\hat{\mathpzc{L}}_z\,\mathbf{f}(\vartheta, \phi) = \elz \cdot \mathbf{f}(\vartheta, \phi) \\
\left( \hat{\mathpzc{L}}_z \doteqdot \hat{L}_z \cdot \mathbf{1} \right) \nonumber
\end{gather}
it becomes evident that both eigenvalue euquations (\ref{eq:D.15}) and (\ref{eq:D.16}) are just the RST analogues of the conventional eigenvalue problem (\ref{eq:IV.16a})--(\ref{eq:IV.16b}). Of course, the simultaneous presence of two eigenvalues (i.\,e. $\elp$ and $\elz$) is a consequence of the fact that the corresponding operators do commute
\begin{equation}
\label{eq:D.17}
\left[ \hat{\mathpzc{L}}_z, \hQQ \right] = 0
\end{equation}
so that both eigenvalue equations (\ref{eq:D.14}) and (\ref{eq:D.16}) can simultaneously be valid. Observe here that the first eigenvalue equation (\ref{eq:D.14}) plays the part of a kind of ``square root'' of the equation (\ref{eq:D.15}).

The desired second-order form of the RST eigenvalue problem for angular momentum emerges now simply by explicitly writing down the squared operator $\hQQ^2$:
\begin{equation}
\label{eq:D.18}
\hQQ^2 = -\left\{ \frac{\partial^2}{\partial\vartheta^2} + \frac{1}{\sin^2\vartheta}\,\frac{\partial^2}{\partial\phi^2} \right\} \cdot \mathbf{1} + \frac{i}{\sin^2\vartheta} \, \frac{\partial}{\partial\phi}\,\sigma_z
\end{equation}
so that the equation (\ref{eq:D.15}) reads in explicit component form
\begin{subequations}
\begin{align}
\label{eq:D.19a}
- \left\{ \frac{\partial^2}{\partial\vartheta^2}+ \frac{1}{\sin^2\vartheta}\,\frac{\partial^2}{\partial\phi^2} \right\} \, f_R(\vartheta, \phi) + \frac{i}{\sin^2\vartheta}\,\frac{\partial}{\partial\phi} \, f_R(\vartheta, \phi) &= \elp^2 \cdot f_R(\vartheta, \phi) \\
\label{eq:D.19b}
- \left\{ \frac{\partial^2}{\partial\vartheta^2}+ \frac{1}{\sin^2\vartheta}\,\frac{\partial^2}{\partial\phi^2} \right\} \, f_S(\vartheta, \phi) - \frac{i}{\sin^2\vartheta}\,\frac{\partial}{\partial\phi} \, f_S(\vartheta, \phi) &= \elp^2 \cdot f_S(\vartheta, \phi) \;.
\end{align}
\end{subequations}
Observe here that the squared operator $\hQQ^2$ (\ref{eq:D.18}) is \emph{diagonal} (provided one uses the standard form of the Pauli matrices $\sigma_x, \sigma_y, \sigma_z$), and this ensures the result that the second-order equations (\ref{eq:D.19a})--(\ref{eq:D.19b}) appear \emph{decoupled}. Moreover, trying the product form (\ref{eq:D.6a})--(\ref{eq:D.6b}) we are left with the simpler system
\begin{subequations}
\begin{align}
\label{eq:D.20a}
-\frac{d^2}{d\vartheta^2}\,f_R(\vartheta) + \frac{\elz(\elz-1)}{\sin^2\vartheta}\,f_R(\vartheta) &= \elp^2 \cdot f_R(\vartheta) \\
\label{eq:D.20b}
-\frac{d^2}{d\vartheta^2}\,f_S(\vartheta) + \frac{\elz(\elz+1)}{\sin^2\vartheta}\,f_S(\vartheta) &= \elp^2 \cdot f_S(\vartheta)
\end{align}
\end{subequations}
which is nothing else than the former second-order system (\ref{eq:IV.17a})--(\ref{eq:IV.17b}).

\begin{center}
  \large{\textit{Ladder Operators}}
\end{center}

In order to find the complete set of states due to an $\elp$-multiplet (i.\,e. for $-\elp \leq \elz \leq +\elp$) it is very instructive to first collect the former results in a table (next page).

On principle, one could find all the missing members (...) of such a ``ladder'' of angular
states (due to the same value of $\elp$) by solving the coupled first-order system
(\ref{eq:D.5a})--(\ref{eq:D.5b}) separately for any member and then applying the
normalization condition (\ref{eq:IV.15a}). But it would be surely more efficient to
possess an algebraic algorithm which is able to generate the neighbouring states $\lpr
f_{R, \elz \pm 1}$, $\lpr f_{S, \elz \pm 1}$ from some point of departure $\lpr f_{R,
  \elz}$, $\lpr f_{S, \elz}$. Observe here that the symmetry $f_R(\vartheta) \Rightarrow
f_S(\vartheta)$, $f_S(\vartheta) \Rightarrow -f_R(\vartheta)$ mentioned below equation
(\ref{eq:IV.17b}) induces for $\elz = 0$ a certain peculiarity, namely the occurence of
two angular states: the upper half of the line due to $\elz = 0$ refers formally to $\elz
= 0_+$ and the lower half to $\elz = 0_-$. Indeed it is easy to see that both solutions do
obey the coupled first-order system (\ref{eq:IV.8a})--(\ref{eq:IV.8b}) for $\elz =
0$. This double solution for $\elz = 0$ is needed below when we generate the angular
states due to $\elz \neq 0$ from the states with $\elz = 0$.

\vskip 0.5cm
\begin{landscape}
\begin{center}
\label{tableladder}
\begin{tabular}{|c||c|c|}
\hline
$\elz=\,\downarrow$ & $f_{R}(\vartheta,\phi) = \e^{i\elz\phi} \cdot f_R(\vartheta)$ & $f_{S}(\vartheta,\phi) = \e^{i\elz\phi} \cdot f_S(\vartheta)$ \\
\hline\hline
$\elp$ & \multirow{2}{*}{$\sqrt{\frac{2}{\pi} \cdot \frac{(2\elp)!!}{(2\elp - 1)!!}} \cdot \e^{i\elp\phi}(\sin\vartheta)^{\elp}$} & \multirow{2}{*}{$0$} \\
(\ref{eq:IV.155}) & & \\
\hline
$\elp - 1$ & \multirow{2}{*}{$\sqrt{\frac{2}{\pi}\,(2\elp - 1) \cdot \frac{(2\elp - 2)!!}{(2\elp - 3)!!}} \cdot \e^{i(\elp-1)\phi}(\sin\vartheta)^{\elp-1}\cos\vartheta$} & \multirow{2}{*}{$-\sqrt{\frac{2}{\pi}\,\frac{(2\elp - 2)!!}{(2\elp - 1)!!}} \cdot \e^{i(\elp-1)\phi}(\sin\vartheta)^{\elp}$} \\
  (\ref{eq:IV.164a})-(\ref{eq:IV.164b}) & & \\
\hline
\vdots & \vdots & \vdots \\
\hline
$0_\pm$ & $\parbox[0pt][3em][c]{0cm}{}\sqrt{\frac{2}{\pi}} \cdot \cos(\elp\vartheta)$ & $-\sqrt{\frac{2}{\pi}} \sin(\elp\vartheta)$ \\
\cline{2-3}
(\textbf{App. B}) & $\parbox[0pt][3em][c]{0cm}{}\sqrt{\frac{2}{\pi}} \cdot \sin(\elp\vartheta)$ & $\sqrt{\frac{2}{\pi}} \cdot \cos(\elp\vartheta)$ \\
\hline
\vdots & \vdots & \vdots \\
\hline
\multirow{2}{*}{$-(\elp - 1)$} & \multirow{2}{*}{$\sqrt{\frac{2}{\pi}\,\frac{(2\elp - 2)!!}{(2\elp - 1)!!}} \cdot \e^{i(1-\elp)\phi}(\sin\vartheta)^{\elp}$} & \multirow{2}{*}{$\sqrt{\frac{2}{\pi}\,(2\elp - 1) \cdot \frac{(2\elp - 2)!!}{(2\elp - 3)!!}} \cdot \e^{i(1-\elp)\phi}(\sin\vartheta)^{\elp-1}\cos\vartheta$} \\
 & & \\
\hline
\multirow{2}{*}{$-\elp$} & \multirow{2}{*}{$0$} & \multirow{2}{*}{$-\sqrt{\frac{2}{\pi} \cdot \frac{(2\elp)!!}{(2\elp - 1)!!}} \cdot \e^{-i\elp\phi}(\sin\vartheta)^{\elp}$} \\
 & & \\
\hline
\end{tabular}
\vskip 2cm
\Large{\textbf{General Ladder Arrangement of the angular eigenfunctions}}
\end{center}
\end{landscape}

The starting point for obtaining the states with $\elz > 0$ is then the solution due to $\elz = 0_+$ (upper half) and for obtaining the states with $\elz < 0$ one starts from the angular state with $\elz = 0_-$ (lower half):
\begin{subequations}
\begin{align}
\label{eq:D.21a}
\elz = 0_+\,:\quad&f_R(\vartheta) = \sqrt{\frac{2}{\pi}} \cdot \cos(\elp\vartheta),\qquad f_S(\vartheta) = -\sqrt{\frac{2}{\pi}} \sin(\elp\vartheta) \\
\label{eq:D.21b}
\elz = 0_-\,:\quad&f_R(\vartheta) = \sqrt{\frac{2}{\pi}} \cdot \sin(\elp\vartheta),\qquad f_S(\vartheta) = -\sqrt{\frac{2}{\pi}} \cos(\elp\vartheta) \;.
\end{align}
\end{subequations}

In order to elaborate this program, it is now self-suggesting to define two kinds of \emph{ladder operators}, $\hRp$ and $\hSp$, by
\begin{subequations}
\begin{align}
\label{eq:D.22a}
\hRp &\doteqdot e^{i\phi} \left\{ \frac{\partial}{\partial\vartheta} + i\cot\vartheta \frac{\partial}{\partial\phi} \right\} \\
\label{eq:D.22b}
\hSp &\doteqdot \left\{ \frac{\partial}{\partial\vartheta} + i\cot\vartheta \frac{\partial}{\partial\phi} \right\}\,e^{i\phi} = e^{-i\phi} \hRp e^{i\phi} \;.
\end{align}
\end{subequations}
Furthermore, it is convenient to introduce an economical notation for the angular eigenfunctions $f_R(\vartheta, \phi)$ and $f_S(\vartheta, \phi)$ so that the associated quantum numbers $\elp$ and $\elz$ become immediately evident: $\lpr f_{R,\elz}(\vartheta, \phi)$ and $\lpr f_{S,\elz}(\vartheta, \phi)$. With this arrangement, we easily realize for the upper end of the above table that the highest-order state (i.\,e. $\elz = \elp$) becomes ``annihilated'' under the action of the ladder operator $\hRp$ (\ref{eq:D.22a}):
\begin{equation}
\label{eq:D.23}
\hRp\,\lpr f_{R,\elp} \simeq e^{i\phi} \left\{ \frac{\partial}{\partial\vartheta} + i\cot\vartheta\frac{\partial}{\partial\phi} \right\} \left[ e^{i\elp\phi} \cdot \left( \sin\vartheta \right)^{\elp} \right] = 0 \;.
\end{equation}
Since the highest-order state $\lpr f_{S,\elp}(\vartheta, \phi)$ of the $S$-ladder is already zero, the corresponding relation for this state $\lpr f_{S, \elp}(\vartheta, \phi)$ is trivially evident
\begin{equation}
\label{eq:D.24}
\hSp\,\lpr f_{S,\elp} = 0 \;.
\end{equation}
Thus, the combined action (\ref{eq:D.23}) plus (\ref{eq:D.24}) of both operators $\hRp$ and $\hSp$ annihilates the highest-order state of any $\elp$-multiplet.

Turning now to the next lower states (i.\,e. $\elz = \elp - 1$) of the table, i.\,e.
\begin{subequations}
\begin{align}
\label{eq:D.25a}
\lpr f_{R,\elp - 1}(\vartheta, \phi) &= \sqrt{\frac{2}{\pi}\left( 2\elp - 1 \right)
  \frac{(2\elp - 2)!!}{(2\elp - 3)!!}}\, e^{i(\lP-1)\phi}
\cdot \left( \sin\vartheta \right)^{\elp - 1}\cos\vartheta \\
\label{eq:D.25b}
\lpr f_{S,\elp - 1}(\vartheta, \phi) &= -\sqrt{\frac{2}{\pi}\frac{(2\elp - 2)!!}{(2\elp - 1)!!}} \, e^{i(\lP-1)\phi}
 \cdot \left( \sin\vartheta \right)^{\elp} \;,
\end{align}
\end{subequations}
we find again by subjecting those states to the action of the ascendive operators $\hRp$ and $\hSp$:
\begin{subequations}
\begin{align}
\label{eq:D.26a}
\hRp\,\lpr f_{R,\elp-1} &= - \frac{2\elp - 1}{\sqrt{2\elp}} \lpr f_{R,\elp} \\
\label{eq:D.26b}
\hSp\,\lpr f_{S,\elp-1} &\equiv 0 = \lpr f_{S,\elp} \;.
\end{align}
\end{subequations}
From these results one becomes tempted to suppose that the operators $\hRp$ and $\hSp$ do act as ``ladder operators'' which generate the next higher state (due to $\elz + 1$) from the precedent one (due to $\elz$) according to the recipe
\begin{subequations}
\begin{align}
\label{eq:D.27a}
\hRp\,\lpr f_{R,\elz} &= C_+(\elp, \elz) \cdot \lpr f_{R, \elz + 1} \\
\label{eq:D.27b}
\hSp\,\lpr f_{S,\elz} &= D_+(\elp, \elz) \cdot \lpr f_{S, \elz + 1}
\end{align}
\end{subequations}
with the normalization constants $C_+(\elp,\elz)$ and $D_+(\elp,\elz)$ being still to be determined. Their special values for $\elz = \elp - 1$ may be already read off from the equations (\ref{eq:D.26a})--(\ref{eq:D.26b}) as
\begin{subequations}
\begin{align}
\label{eq:D.28a}
C_+(\elp,\elp - 1) &= -\frac{2\elp - 1}{\sqrt{2\elp}} \\
\label{eq:D.28b}
D_+(\elp,\elp - 1) &= 0 \;.
\end{align}
\end{subequations}

Let us test our general construction (\ref{eq:D.27a})--(\ref{eq:D.27b}) for those special states which have $\elz = 0, \forall \elp$ (see the table on p.(?))
\begin{subequations}
\begin{align}
\label{eq:D.29a}
\lpr f_{R,0_+}(\vartheta, \phi) &= \sqrt{\frac{2}{\pi}}\cos(\elp \vartheta) \\
\label{eq:D.29b}
\lpr f_{S,0_+}(\vartheta, \phi) &= -\sqrt{\frac{2}{\pi}}\sin(\elp \vartheta) \;.
\end{align}
\end{subequations}
For this situation, one finds by straight forward calculation for the action of the operators $\hRp, \hSp$
\begin{subequations}
\begin{align}
\label{eq:D.30a}
\hRp\,\lpr f_{R,0_+}(\vartheta,\phi) &= -\sqrt{\frac{2}{\pi}} \e^{i\phi}\sin\left( \elp \vartheta \right) \\
\label{eq:D.30b}
\hSp\,\lpr f_{S,0_+}(\vartheta,\phi) &= -\sqrt{\frac{2}{\pi}} \e^{i\phi}\left\{ \elp \cdot \cos(\elp\vartheta) - \cot\vartheta\cdot\sin(\elp\vartheta) \right\} \;.
\end{align}
\end{subequations}
Comparing this to the general action of both ladder operators $\hRp$ and $\hSp$ (\ref{eq:D.27a})--(\ref{eq:D.27b}) we find that the angular functions due to $\boldsymbol{\elz = 1}$ must look as follows:
\begin{subequations}
\begin{align}
\label{eq:D.31a}
\lpr f_{R,1}(\vartheta,\phi) &= -\sqrt{\frac{2}{\pi}} \frac{1}{C_+(\elp,0)} \cdot \e^{i\phi} \sin(\elp\vartheta) \\
\label{eq:D.31b}
\lpr f_{S,1}(\vartheta,\phi) &= -\sqrt{\frac{2}{\pi}} \frac{1}{D_+(\elp,0)} \cdot \e^{i\phi} \left\{ \elp \cdot \cos(\elp\vartheta) - \cot\vartheta \cdot \sin(\elp\vartheta) \right\} \;.
\end{align}
\end{subequations}
But these two angular functions $\lpr f_{R,1}$ and $\lpr f_{S,1}$ (\ref{eq:D.31a})--(\ref{eq:D.31b}) must now obey the coupled angular system (\ref{eq:D.5a})--(\ref{eq:D.5b}) for $\elz = 1$ without producing any contradiction; and this demand can actually be satisfied provided the constants $C_+(\elp,0)$ and $D_+(\elp,0)$ are related by
\begin{equation}
\label{eq:D.32}
D_+(\elp,0_+) = \frac{\elp + 1}{\elp} \cdot C_+(\elp,0_+) \;.
\end{equation}

The symmetry operation $z \rightarrow -z$ $(\elz \Rightarrow -\elz)$ results in the substitution $f^*_R(\vartheta,\phi) \Rightarrow f_S(\vartheta,\phi)$, $f^*_S(\vartheta,\phi) \Rightarrow -f_R(\vartheta,\phi)$ (take the complex conjugate of the system (\ref{eq:D.5a})--(\ref{eq:D.5b}) or recall the remark made below equation (\ref{eq:IV.17b})); and therefore one may introduce the descendive operators $\hRm$ and $\hSm$ quite analogously to the precedent ascendive case (\ref{eq:D.22a})--(\ref{eq:D.22b}) through
\begin{subequations}
\begin{align}
\label{eq:D.33a}
\hRm &\doteqdot \left\{ \frac{\partial}{\partial\vartheta} - i\cot\vartheta\frac{\partial}{\partial\phi} \right\} \e^{-i\phi} \\
\label{eq:D.33b}
\hSm &\doteqdot -\e^{-i\phi} \left\{ \frac{\partial}{\partial\vartheta} - i\cot\vartheta\frac{\partial}{\partial\phi} \right\} = -\e^{-i\phi} \hRm \e^{i\phi} \;.
\end{align}
\end{subequations}
Here it is again obvious that these \emph{descendive} operators do act on the lowest-order states of an $\elp$-multiplet quite analogously to the \emph{ascendive} cases (\ref{eq:D.23})--(\ref{eq:D.24}) for the highest-order states, i.\,e.
\begin{subequations}
\begin{align}
\label{eq:D.34a}
\hRm\,\lpr f_{R,-\elp} &= 0 \\
\label{eq:D.34b}
\hSm\,\lpr f_{S,-\elp} &= 0 \;.
\end{align}
\end{subequations}
And the lowest-order states of an $\elp$-multiplet are again generated by
\begin{subequations}
\begin{align}
\label{eq:D.35a}
\hRm \lpr f_{R,1-\elp} &\equiv 0 = \lpr f_{R, -\elp} \\
\label{eq:D.35b}
\hSm \lpr f_{S,1-\elp} &= -\frac{2\elp - 1}{\sqrt{2\elp}} \cdot \lpr f_{S, -\elp} \;,
\end{align}
\end{subequations}
quite analogously to the ascending situation (\ref{eq:D.26a})--(\ref{eq:D.26b}).

Clearly this analogy suggests to define the action of the descendive operators $\hRm, \hSm$ quite generally through
\begin{subequations}
\begin{align}
\label{eq:D.36a}
\hRm \lpr f_{R, \elz} &= C_-(\elp,\elz) \cdot \lpr f_{R, \elz-1} \\
\label{eq:D.36b}
\hSm \lpr f_{S, \elz} &= D_-(\elp,\elz) \cdot \lpr f_{S, \elz-1} \;,
\end{align}
\end{subequations}
with the constants $C_-$ and $D_-$ to be determined later on. The special values for $\elz = 1 - \elp$ may be deduced again already from the equations (\ref{eq:D.35a})--(\ref{eq:D.35b}):
\begin{subequations}
\begin{align}
\label{eq:D.37a}
C_-(\elp, 1 - \elp) &= 0 \\
\label{eq:D.37b}
D_-(\elp, 1 - \elp) &= -\frac{2\elp - 1}{\sqrt{2\elp}} \;.
\end{align}
\end{subequations}
Recall here the analogous relations (\ref{eq:D.28a})--(\ref{eq:D.28b}) which apply to the
angular states with positive values of $\elz$ (i.\,e. the upper end of the
ladders). Furthermore, one deduces also from the survey table on p.~\pageref{tableladder} by
straightforward application of $\hRm$ and $\hSm$ the following descendive relation for the
$R$-ladder
\begin{equation}
\label{eq:D.38}
C_-(\elp,\elp) = \sqrt{2\elp}
\end{equation}
and similarly the ascendive relation for the $S$-ladder
\begin{equation}
\label{eq:D.39}
D_+(\elp,-\elp) = -\sqrt{2\elp} \;.
\end{equation}

From reasons of symmetry $(\elz \Leftrightarrow -\elz)$ one wishes again to test the consistency of the ladder ansatz (\ref{eq:D.36a})--(\ref{eq:D.36b}) for the special case $\elz = 0$, $\forall \elp$. This says that we take from the table on p. (?) the two angular functions
\begin{subequations}
\begin{align}
\label{eq:D.40a}
\lpr f_{R,0_-}(\vartheta, \phi) &= \sqrt{\frac{2}{\pi}} \cdot \sin(\elp\vartheta) \\
\label{eq:D.40b}
\lpr f_{S,0_-}(\vartheta, \phi) &= \sqrt{\frac{2}{\pi}} \cdot \cos(\elp\vartheta)
\end{align}
\end{subequations}
and let now act the descendive operators $\hRm$ and $\hSm$ (\ref{eq:D.33a})--(\ref{eq:D.33b}) on these functions:
\begin{subequations}
\begin{align}
\label{eq:D.41a}
\hRm \lpr f_{R,0_-} &= \sqrt{\frac{2}{\pi}}\e^{-i\phi} \left\{ \elp \cdot \cos(\elp\vartheta) - \cot\vartheta \cdot \sin(\elp\vartheta) \right\} \\
\label{eq:D.41b}
\hSm \lpr f_{S,0_-} &= \sqrt{\frac{2}{\pi}}\e^{-i\phi} \left\{ \elp \cdot \sin(\elp\vartheta) \right\} \;.
\end{align}
\end{subequations}
On the other hand, the ladder construction (\ref{eq:D.36a})--(\ref{eq:D.36b}) identifies here the right-hand sides with the angular eigenfunctions $\lpr f_{R,-1}$ and $\lpr f_{S, -1}$ which thereby become represented by
\begin{subequations}
\begin{align}
\label{eq:D.42a}
\lpr f_{R,-1}(\vartheta,\phi) &= \sqrt{\frac{2}{\pi}} \frac{\e^{-i\phi}}{C_-(\elp,0)}\,\left\{ \elp \cdot \cos(\elp\vartheta) - \cot\vartheta\sin(\elp\vartheta)  \right\} \\
\label{eq:D.42b}
\lpr f_{S,-1}(\vartheta,\phi) &= \sqrt{\frac{2}{\pi}} \frac{\elp \cdot \e^{-i\phi}}{D_-(\elp,0)} \cdot \sin(\elp\vartheta) \;.
\end{align}
\end{subequations}
But if our ladder hypothesis is correct, these two angular functions must obey the coupled first-order system (\ref{eq:D.5a})--(\ref{eq:D.5b}) which for the present case ($\elz = 1$) appears as
\begin{subequations}
\begin{align}
\label{eq:D.43a}
\frac{d\, \lpr f_{R,-1}}{d\vartheta} + \cot\vartheta \cdot \lpr f_{R,-1} &= (\elp - 1) \cdot \lpr f_{S,-1} \\
\label{eq:D.43b}
\frac{d\, \lpr f_{S,-1}}{d\vartheta} - \cot\vartheta \cdot \lpr f_{S,-1} &= -(\elp + 1) \cdot \lpr f_{R,-1} \;.
\end{align}
\end{subequations}
Inserting here both functions $\lpr f_{R,-1}$ and $\lpr f_{S,-1}$ (\ref{eq:D.42a})--(\ref{eq:D.42b}) leads us without any contradictions to the relation
\begin{equation}
\label{eq:D.44}
D_-(\elp,0_-) = -\frac{\elp}{1 + \elp} \cdot C_-(\elp,0_-) \;,
\end{equation}
which of course is the counterpart of (\ref{eq:D.32}) for negative values of $\elz$, i.\,e. for the lower ends of both ladders.

\begin{center}
  \large{\textit{Ladder Formalism for $\boldsymbol{\elp = 3}$}}
\end{center}

In order to become somewhat more acquainted with the peculiarities of the present algebra
of angular momentum it may be instructive to explicitly write down the action of the
ascendive and descendive operators on the members of an $\elp$-multiplet. For
$\boldsymbol{\elp = 3}$, the members are already known (see the table on p.~\pageref{tablelp3}), so that
by straightforward calculation one can determine the constants $C_+, D_+$
(\ref{eq:D.28a})--(\ref{eq:D.28b}) as well as $C_-,D_-$
(\ref{eq:D.37a})--(\ref{eq:D.37b}). Turning first to the ``$R$-ladder'', one finds the
following results (see \textbf{Fig. D.I}):

The upper edge ({\boldmath$\elp = 3$}, $\elz = 3$) is given by
\begin{equation}
\label{eq:D.45}
\threer f_{R,3}(\vartheta, \phi) = \sqrt{\frac{32}{5\pi}}\,\e^{3i\phi} \sin^3\vartheta \;,
\end{equation}
see equation (\ref{eq:IV.155}) for $\elp = \elz = 3$; and consequently the ascendive
operator $\hRp$ (\ref{eq:D.22a}) annihilates this angular state (\ref{eq:D.45}), which is
of course only a special case of the general relation (\ref{eq:D.23}). Next, consider the lower-order
state $\threer f_{R,2}(\vartheta,\phi)$
\begin{equation}
\label{eq:D.46}
\threer f_{R,2}(\vartheta, \phi) = \sqrt{\frac{80}{3\pi}}\,\e^{2i\phi} \sin^2\vartheta\cdot\cos\vartheta
\end{equation}
and find
\begin{equation}
\label{eq:D.47}
\hRp\,\threer f_{R,2} = - \frac{5}{\sqrt{6}}\,\threer f_{R,3}
\end{equation}
which then yields for the transition constant $C_+(3,2)$ (\ref{eq:D.27a})
\begin{equation}
\label{eq:D.48}
C_+(3,2) = -\frac{5}{\sqrt{6}} \;,
\end{equation}
as a specialization of the general relation (\ref{eq:D.28a}). Thus the upper edge of the
$R$-ladder for {\boldmath$\elp = 3$} is found to be in agreement with the former preliminary
remarks. This agreement does also apply to the lower edge ( {\boldmath$\elp = 3$}, $\elz =
3$). Indeed, the lowest-order non-trivial angular function is here
\begin{equation}
\label{eq:D.49}
\threer f_{R,-2}(\vartheta,\phi) = \sqrt{\frac{16}{15\pi}}\,\e^{-2i\phi} \sin^3\vartheta \;,
\end{equation}
\begin{center}
  \textbf{\large R-ladder}
\end{center}

\[\xymatrixcolsep{5pc}
\xymatrix{
  & 0  &   & \\
  & \ar@{=>}[u]_{\hat{R}_+} \rkloi{3}{f}{R,3} \ar@{-}[r]|{\ l_z=+3\ } &
 \ar@{=>}[d]_{\hat{R}_-}^{C_-(3,3)=\sqrt{6} } \rkloi{3}{f}{R,3}  & \\ 
   & \ar@{=>}[u]_{\hat{R}_+}^{C_+(3,2)=-\frac{5}{\sqrt{6} }} \rkloi{3}{f}{R,2}
 \ar@{-}[r]|{\ l_z=+2\ } & \ar@{=>}[d]_{\hat{R}_-}^{C_-(3,2)=\sqrt{10}} \rkloi{3}{f}{R,2} & \\
   & \ar@{=>}[u]_{\hat{R}_+}^{C_+(3,1)=-4 \sqrt{\frac{2}{5}} } \rkloi{3}{f}{R,1}
 \ar@{-}[r]|{\ l_z=+1\ } & \ar@{=>}[d]_{\hat{R}_-}^{C_-(3,1)=2\sqrt{3}}\rkloi{3}{f}{R,1}
  & \\
   & \ar@{=>}[u]_{\hat{R}_+}^{C_+(3,0_+)=-\frac{3}{2}\sqrt{3} } \rkloi{3}{f}{R,0_\pm}
 \ar@{=}[r]|{\ l_z=0\ } &  \ar@{=>}[d]_{\hat{R}_-}^{C_-(3,0_-)=2\sqrt{3}}  \rkloi{3}{f}{R,0_\pm}  &  \\
  & \ar@{=>}[u]_{\hat{R}_+}^{C_+(3,-1)=-\frac{4}{\sqrt{3}} } \rkloi{3}{f}{R,-1}
 \ar@{-}[r]|{\ l_z=-1\ } &  \ar@{=>}[d]_{\hat{R}_-}^{C_-(3,-1)=\sqrt{10}} \rkloi{3}{f}{R,-1}  & \\
  & \ar@{=>}[u]_{\hat{R}_+}^{C_+(3,-2)=-\sqrt{\frac{5}{2}} } \rkloi{3}{f}{R,-2}
 \ar@{-}[r]|{\ l_z=-2\ } &  \ar@{=>}[d]_{\hat{R}_-}^{C_-(3,-2)=0} \rkloi{3}{f}{R,-2} & \\
0 \ar@{-}[r]  & \rkloi{3}{f}{R,-3}
 \ar@{-}[r] | {\ l_z=-3\ } &  \ar@{=>}[d]_{\hat{R}_-} \rkloi{3}{f}{R,-3} \ar@{-}[r]  & 0 \\
 & & 0 & 
}
\]
\\[2cm]
{\textbf{Fig.~D.1}\hspace{5mm} \emph{\large\textbf{Ladder Operations for
\boldmath$\rkloi{\lP}{f}{R,l_z}(\vartheta,\phi),\ \lP=3 $        }   }  }


\pagebreak
and this angular function becomes annihilated under the action of the descendive operator
$\hRm$ (\ref{eq:D.35a}). i.e.
\begin{equation}
\label{eq:D.50}
\hRm\,\threer f_{R,-2} \equiv 0 \;,
\end{equation}
as demanded by the general relation (\ref{eq:D.34a}), cf. also (\ref{eq:D.37a}).

But now a somewhat more subtle point is to be discussed, namely the occurence of two
different angular states for $\elz = 0$, see the table on p.~\pageref{tableladder}:
\begin{subequations}
\begin{align}
\label{eq:D.51a}
\threer f_{R,0_+} &= \sqrt{\frac{2}{\pi}}\, \cos(3\vartheta) = \sqrt{\frac{2}{\pi}}\, (1 - 4\sin^2\vartheta) \cos\vartheta \\
\label{eq:D.51b}
\threer f_{R,0_-} &= \sqrt{\frac{2}{\pi}}\, \sin(3\vartheta) = 3\sqrt{\frac{2}{\pi}}\, \sin\vartheta\, (1 - \frac{4}{3} \sin^2\vartheta) \;.
\end{align}
\end{subequations}
The logical necessity of \emph{two} angular states for $\elz = 0$ will readily become
evident when we now look at the action of $\hRpm$ on the neighbouring states $\elz = \pm
1$. If we ascend the $R$-ladder ($\leadsto$ left-hand side of \textbf{Fig.D.I}) beginning on the
lowest-possible step (i.\,e. $\elz = -2$) by repeatedly applying the ascendive operator
$\hRp$ (\ref{eq:D.22a}) we first come to the step with $\elz = -1$. The corresponding
angular function is
\begin{equation}
\label{eq:D.52}
\threer f_{R,-1}(\vartheta,\phi) = -\sqrt{\frac{32}{3\pi}}\,\e^{-i\phi} \sin^2\vartheta\cos\vartheta \;,
\end{equation}
and the ascent to this step reads (see \textbf{Fig.D.I})
\begin{equation}
\label{eq:D.53}
\hRp\,\threer f_{R,-2} = -\sqrt{\frac{5}{2}} \cdot \threer f_{R,-1} \;.
\end{equation}
Now we once more apply the ascendive operator and thus obtain
\begin{equation}
\label{eq:D.54}
\hRp\,\threer f_{R,-1} = -\frac{4}{\sqrt{3}}\,\threer f_{R,0_-}
\end{equation}
with $\threer f_{R,0_-}$ being given by equation (\ref{eq:D.51b}). However, in order to
further ascend now from the step with $\elz = 0$ to the higher steps $\elz = 1,2,3$ we
\emph{cannot} apply the ascendive operator $\hRp$ to that state $\threer f_{R,0_-}$
(\ref{eq:D.51b}) just reached; but we have to start from the other state $\threer
f_{R,0_+}$ (\ref{eq:D.51a}) which is also due to $\elz = 0$:
\begin{equation}
\label{eq:D.55}
\hRp\,\threer f_{R,0_+} = -\frac{3}{2}\,\sqrt{3} \cdot \threer f_{R,1} \;.
\end{equation}
Indeed one is easily convinced by straightforward calculation that the action of $\hRp$ on the state $\threer f_{R,0_-}$ (\ref{eq:D.51b}) does not lead us to the desired step $\threer f_{R,1}$ given by
\begin{equation}
\label{eq:D.56}
\threer f_{R,1} = \sqrt{\frac{24}{\pi}}\,\e^{i\phi}\sin\vartheta\,\left( 1 - \frac{4}{3} \sin^2\vartheta \right) \;.
\end{equation}
On the other hand, once the right starting state $\threer f_{R,0_+}$ is accepted for the
further ascent, the highest state $\threer f_{R,3}$ is reached by repeated action of
$\hRp$ without any inconsistency, see \textbf{Fig. D.I}.

A similar effect is observed for the descent from the highest-order state $\threer
f_{R,3}$ to the lower states by means of the descendive operator $\hRm$ (\ref{eq:D.33a}),
see \textbf{Fig. D.I} above. Here, the first transition occurs from the highest state
$\threer f_{R,3}$ (\ref{eq:D.45}) to the next lower one $\threer f_{R,2}$ (\ref{eq:D.46})
and looks as follows
\begin{equation}
\label{eq:D.57}
\hRm\,\threer f_{R,3} = \sqrt{6} \cdot \threer f_{R,2}
\end{equation}
and this yields for the transition coefficient $C(3,3)$ (\ref{eq:D.36a})
\begin{equation}
\label{eq:D.58}
C_-(3,3) = \sqrt{6} \;,
\end{equation}
in agreement with the general relation (\ref{eq:D.38}). Then we can further descend downwards by means of $\hRm$ to the step $\threer f_{R,1}$ (\ref{eq:D.56}). But the next transition downwards to the step $\elz = 0$ requires again some attention. By straightforward calculation one finds
\begin{equation}
\label{eq:D.59}
\hRm\,\threer f_{R,1} = 2\sqrt{3} \cdot \threer f_{R,0_+} \;.
\end{equation}
This says that the descent from above ($\elz > 0$) by means of $\hRm$ terminates at that
state $\threer f_{R,0_+}$ (\ref{eq:D.51a}) on the level $\elz = 0$ from which we have to
start when we wish to generate upwards the states with $\elz > 0$, see equation
(\ref{eq:D.55}).

But for the further descent, one first has to pass over from $\threer f_{R,0_+}$
(\ref{eq:D.51a}) to the other angular state $\threer f_{R,0_-}$ (\ref{eq:D.51b}) and then
one can continue to descend, i.\,e.
\begin{equation}
\label{eq:D.60}
\hRm\,\threer f_{R,0_-} = 2\sqrt{3} \cdot \threer f_{R,-1} \;.
\end{equation}
But the next descent leads us to the last non-trivial step $\threer f_{R,-2}$ (\ref{eq:D.49})
\begin{equation}
\label{eq:D.61}
\hRm\,\threer f_{R,-1} = \sqrt{10}\,\threer f_{R,-2} \;,
\end{equation}
since this last step becomes then annihilated by the action of $\hRm$, see equation (\ref{eq:D.50}).

Concerning now the ``$S$-ladder'', the situation is quite analogous to the ``$R$-ladder'',
cf. \textbf{Fig.D.I} and \textbf{Fig.D.II}. Therefore it may be sufficient to mention
here only the most striking features. The first of these refers again to the fact that for
$\elz = 0$ one has two angular states, i.\,e.
\begin{subequations}
\begin{align}
\label{eq:D.62a}
\threer f_{S,0_+} &= -\sqrt{\frac{2}{\pi}}\,\sin(3\vartheta) = -3\sqrt{\frac{2}{\pi}}\,\sin\vartheta\,\left( 1 - \frac{4}{3}\,\sin^2\vartheta \right) \\
\label{eq:D.62b}
\threer f_{S,0_-} &= \sqrt{\frac{2}{\pi}}\,\cos(3\vartheta) = \sqrt{\frac{2}{\pi}}\,\cos\vartheta\,\left( 1 - 4\sin^2\vartheta \right) \;.
\end{align}
\end{subequations}
From here, the next higher state $\threer f_{S,1}(\vartheta, \phi)$ is obtained by applying the ascendive operator $\hSp$ (\ref{eq:D.22b}) to $\threer f_{S,0_+}$ (\ref{eq:D.62a})
\begin{gather}
\label{eq:D.63}
\hSp\,\threer f_{S,0_+} = -2\sqrt{3} \cdot \threer f_{S,1} \\
\left( \leadsto D_+(3,0_+) = -2\sqrt{3} \right) \;, \nonumber
\end{gather}
whereas for obtaining the next lower state $\threer f_{S,-1}(\vartheta,\phi)$ one has to
depart from $\threer f_{S,0_-}$ (\ref{eq:D.62b}):
\begin{gather}
\label{eq:D.64}
\hSm\,\threer f_{S,0_-} = -\frac{3}{2}\,\sqrt{3} \cdot \threer f_{S,-1} \\
\left( \leadsto D_-(3,0_-) = -\frac{3}{2}\,\sqrt{3} \right) \;. \nonumber
\end{gather}

Obviously, this is all quite analogous to the case with the $R$-ladder; and the
annihilation of the highest ($\elz = +3$) and lowest ($\elz = -3$) states by means of the
actions of $\hSp$ and $\hSm$, resp., works also quite similarly, cf. (\ref{eq:D.24}) and
(\ref{eq:D.34b}). Clearly, the ladder coefficient $D_+(3,-3)$ is nothing else than the
specialization of the general case $D_+(\elp, -\elp)$ (\ref{eq:D.39}) to the present
example $\boldsymbol{\elp = 3}$
\begin{equation}
\label{eq:D.65}
D_+(\elp,-\elp) \Rightarrow D_+(3,-3) = -\sqrt{6} \;,
\end{equation}
with the converse being given by
\begin{equation}
\label{eq:D.66}
D_-(\elp,1-\elp) \Rightarrow D_-(3,-2) = -\frac{5}{\sqrt{6}} \;,
\end{equation}
cf. equation (\ref{eq:D.37b}). Somewhat more interesting are perhaps those relations
concerning simultaneously both ladders, such as (\ref{eq:D.32}) and (\ref{eq:D.44}). From
the $R$-ladder (\textbf{Fig. D.I}) we see
\begin{equation}
\label{eq:D.67}
C_+(3,0_+) = -\frac{3}{2}\,\sqrt{3} \;,
\end{equation}
and for the $S$-ladder (\textbf{Fig. D.II})
\begin{equation}
\label{eq:D.68}
D_+(3,0_+) = -2\sqrt{3} \;,
\end{equation}
thus
\begin{equation}
\label{eq:D.69}
D_+(3,0_+) = \frac{4}{3}\,C_+(3,0_+)
\end{equation}
which validates (\ref{eq:D.32}). Similarly, from the $R$-ladder (\textbf{Fig. D.I}) we take
\begin{equation}
\label{eq:D.70}
C_-(3,0_-) = 2\sqrt{3} \;,
\end{equation}
and from the $S$-ladder (\textbf{Fig. D.II}) one takes
\begin{equation}
\label{eq:D.71}
D_-(3,0_-) = -\frac{3}{2}\,\sqrt{3}
\end{equation}
thus
\begin{equation}
\label{eq:D.72}
D_-(3,0_-) = -\frac{3}{4}\,C_-(3,0_-)
\end{equation}
which validates the general claim (\ref{eq:D.44}).

Thus, it should have become obvious that there are many interrelationships between the
ladder coeffcients; but in place of studying this for special examples it would surely be
more efficient to deduce their general shape (i.\,e. for general $\elp$ and $\elz$).
\pagebreak

\begin{center}
  \textbf{\large S-ladder}
\end{center}

\[\xymatrixcolsep{5pc}
\xymatrix{
  & 0  &   & \\
 0  \ar@{-}[r] & \ar@{=>}[u]_{\hat{S}_+} \rkloi{3}{f}{S,3} \ar@{-}[r]|{\ l_z=+3\ } &
  \rkloi{3}{f}{S,3} \ar@{-}[r]  &  0 \\ 
   & \ar@{=>}[u]_{\hat{S}_+}^{D_+(3,2)= 0 }  \rkloi{3}{f}{S,2}
 \ar@{-}[r]|{\ l_z=+2\ } & \ar@{=>}[d]_{\hat{S}_-}^{D_-(3,2)=-\frac{1}{2}\sqrt{10}} \rkloi{3}{f}{S,2} & \\
   & \ar@{=>}[u]_{\hat{S}_+}^{D_+(3,1)=-\sqrt{10} } \rkloi{3}{f}{S,1}
 \ar@{-}[r]|{\ l_z=+1\ } & \ar@{=>}[d]_{\hat{S}_-}^{D_-(3,1)=-\frac{4}{\sqrt{3}} }\rkloi{3}{f}{S,1}
  & \\
   & \ar@{=>}[u]_{\hat{S}_+}^{D_+(3,0_+)=-\sqrt{12} } \rkloi{3}{f}{S,0_\pm}
 \ar@{=}[r]|{\ l_z=0\ } &  \ar@{=>}[d]_{\hat{S}_-}^{D_-(3,0_-)=-\frac{3}{2}\sqrt{3}}  \rkloi{3}{f}{S,0_\pm}  &  \\
  & \ar@{=>}[u]_{\hat{S}_+}^{D_+(3,-1)=-\sqrt{12} } \rkloi{3}{f}{S,-1}
 \ar@{-}[r]|{\ l_z=-1\ } &  \ar@{=>}[d]_{\hat{S}_-}^{D_-(3,-1)=-\frac{4}{5}\sqrt{10}} \rkloi{3}{f}{S,-1}  & \\
  & \ar@{=>}[u]_{\hat{S}_+}^{D_+(3,-2)=-\sqrt{10} } \rkloi{3}{f}{S,-2}
 \ar@{-}[r]|{\ l_z=-2\ } &  \ar@{=>}[d]_{\hat{S}_-}^{D_-(3,-2)=-\frac{5}{\sqrt{6}}} \rkloi{3}{f}{S,-2} & \\
 & \ar@{=>}[u]_{\hat{S}_+}^{D_+(3,-3)=-\sqrt{6} } \rkloi{3}{f}{S,-3}
 \ar@{-}[r] | {\ l_z=-3\ } &  \ar@{=>}[d]_{\hat{S}_-} \rkloi{3}{f}{S,-3}   &  \\
 & & 0 & 
}
\]
\\[2cm]
{\textbf{Fig.~D.II}\hspace{5mm} \emph{\large\textbf{Ladder Operations for
\boldmath$\rkloi{\lP}{f}{S,l_z}(\vartheta,\phi),\ \lP=3 $        }   }  }


\pagebreak

\begin{center}
  \large{\textit{Determination of the Ladder Coefficients}}
\end{center}

For the determination of the general ladder coefficients $C_\pm(\elp,\elz)$ and
$D_\pm(\elp,\elz)$ it is very helpful to first determine certain products of
them. Namely, by applying in subsequent order an ascendive and/or descendive ladder
operator one must come back to the original state of departure. Thus, one obtains the
following \emph{eigenvalue equations} for the operator products, e.g. for the $R$-ladder
\begin{subequations}
\begin{align}
\label{eq:D.73a}
\left( \hRp \cdot \hRm \right)\,\lpr f_{R,\elz} &= \Big[ C_+(\elp, \elz - 1) \cdot
C_-(\elp,\elz) \Big] \cdot \lpr f_{R,\elz} \\
\label{eq:D.73b}
\left( \hRm \cdot \hRp \right)\,\lpr f_{R,\elz} &= \Big[ C_-(\elp, \elz + 1) \cdot
C_+(\elp,\elz) \Big] \cdot \lpr f_{R,\elz}
\end{align}
\end{subequations}
and analogously for the $S$-ladder
\begin{subequations}
\begin{align}
\label{eq:D.74a}
\left( \hSp \cdot \hSm \right)\,\lpr f_{S,\elz} &= \Big[ D_+(\elp, \elz - 1) \cdot
D_-(\elp,\elz) \Big] \cdot \lpr f_{S,\elz} \\
\label{eq:D.74b}
\left( \hSm \cdot \hSp \right)\,\lpr f_{S,\elz} &= \Big[ D_-(\elp, \elz + 1) \cdot
D_+(\elp,\elz) \Big] \cdot \lpr f_{S,\elz} \;.
\end{align}
\end{subequations}
The eigenvalues emerging here for the special case $\boldsymbol{\elp = 3}$ can be read off directly from both \textbf{Fig.s D.I} and \textbf{D.II} and may be collected in the following table:

\begin{center}
$\boldsymbol{\elp = 3}$
\vskip 0.5cm
\begin{tabular}{|c|c||c|c|c|c|c|c|c|}
\hline
 & $\elz \Rightarrow$ & $-3$ & $-2$ & $-1$ & $0$ & $+1$ & $+2$ & $+3$ \\
\hline\hline
$\hRp \cdot \hRm \Rightarrow$ & $C_+(3,\elz - 1) \cdot C_-(3,\elz) \Rightarrow$ & & & $-5$ & $-8$ & $-9$ & $-8$ & $-5$ \\
\hline
$\hRm \cdot \hRp \Rightarrow$ & $C_-(3,\elz + 1) \cdot C_+(3,\elz) \Rightarrow$ & & $-5$ & $-8$ & $-9$ & $-8$ & $-5$ &  \\
\hline
$\hSp \cdot \hSm \Rightarrow$ & $D_+(3,\elz - 1) \cdot D_-(3,\elz) \Rightarrow$ & & $5$ & $8$ & $9$ & $8$ & $5$ &  \\
\hline
$\hSm \cdot \hSp \Rightarrow$ & $D_-(3,\elz + 1) \cdot D_+(3,\elz) \Rightarrow$ & $5$ & $8$ & $9$ & $8$ & $5$ & & \\
\hline
\end{tabular}
\end{center}
\vskip 0.5cm

With the coefficients of both types of ladders being known, one can now test also those
``inter-ladder'' relations such as (\ref{eq:D.32}) and (\ref{eq:D.44}). For the first of
these we find for the present case $\boldsymbol{\elp = 3}$:
\begin{subequations}
\begin{align}
\label{eq:D.75a}
D_+(3,0_+) &= -2\sqrt{3} \\
\label{eq:D.75b}
C_+(3,0_+) &= -\frac{3}{2}\,\sqrt{3}
\end{align}
\end{subequations}
and therefore
\begin{equation}
\label{eq:D.76}
D_+(3,0_+) = \frac{4}{3} \cdot C_+(3,0_+)
\end{equation}
which validates the general claim (\ref{eq:D.32}). The other assertion (\ref{eq:D.44}) can be validated by an analogous conclusion.

Of course, it is not very satisfying to determine all the ladder coefficients $C_\pm(\elp,\elz)$ and $D_\pm(\elp,\elz)$ by explicit calculation; this method can only be of heuristic character in order to get some feeling of which numbers are to be expected. For the purpose of deducing the general result in form of a closed formula for all admissible $\elz$ and $\elp$, one first writes down (by explicit calculation) the product operators, i.\,e. for the $R$-ladder
\begin{subequations}
\begin{align}
\label{eq:D.77a}
\hRp \cdot \hRm &= \mathbf{1} - (1 - \cot^2\vartheta) \cdot \hLz - \cot^2\vartheta \cdot \hLz^2 + \frac{\partial^2}{\partial\vartheta^2} \\
\label{eq:D.77b}
\hRm \cdot \hRp &= \frac{1}{\sin^2\vartheta} \cdot \hLz - \cot^2\vartheta \cdot \hLz^2 + \frac{\partial^2}{\partial\vartheta^2}
\end{align}
\end{subequations}
and analogously for the $S$-ladder
\begin{subequations}
\begin{align}
\label{eq:D.78a}
\hSp \cdot \hSm &= \frac{1}{\sin^2\vartheta} \cdot \hLz + \cot^2\vartheta \cdot \hLz^2 - \frac{\partial^2}{\partial\vartheta^2} \\
\label{eq:D.78b}
\hSm \cdot \hSp &= -\mathbf{1} - (1 - \cot^2\vartheta) \cdot \hLz + \cot^2\vartheta \cdot \hLz^2 - \frac{\partial^2}{\partial\vartheta^2} \;.
\end{align}
\end{subequations}
Now observe that these operator products are required to act upon the angular functions $\lpr f_{R,\elz}$ and $\lpr f_{S,\elz}$ where then the angular-momentum operator $\hLz$ (\ref{eq:D.10c}) becomes replaced by its eigenvalue $\elz$, i.\,e. we get for the $R$-ladder
\begin{subequations}
\begin{align}
\label{eq:D.79a}
\hRp \cdot \hRm &\Rightarrow 1 - (1 - \cot^2\vartheta) \cdot \elz - \cot^2\vartheta \cdot \elz^2 + \frac{\partial^2}{\partial\vartheta^2} \\
\label{eq:D.79b}
\hRm \cdot \hRp &\Rightarrow \frac{1}{\sin^2\vartheta} \cdot \elz - \cot^2\vartheta \cdot \elz^2 + \frac{\partial^2}{\partial\vartheta^2}
\end{align}
\end{subequations}
and analogously for the $S$-ladder
\begin{subequations}
\begin{align}
\label{eq:D.80a}
\hSp \cdot \hSm &\Rightarrow \frac{1}{\sin^2\vartheta} \cdot \elz + \cot^2\vartheta \cdot \elz^2 - \frac{\partial^2}{\partial\vartheta^2} \\
\label{eq:D.80b}
\hSm \cdot \hSp &\Rightarrow -1 - (1 - \cot^2\vartheta) \cdot \elz + \cot^2\vartheta \cdot \elz^2 - \frac{\partial^2}{\partial\vartheta^2} \;.
\end{align}
\end{subequations}
Furthermore, since the angular eigenfunctions $\lpr f_{R,\elz}(\vartheta, \phi)$ and $\lpr
f_{S,\elz}(\vartheta, \phi)$ must obey the second-order eigenvalue equations
(\ref{eq:IV.17a})--(\ref{eq:IV.17b}), or (\ref{eq:D.19a})--(\ref{eq:D.19b}), resp., one
can substitute from there the second-order derivatives (with respect to $\vartheta$) into
the present operator products (\ref{eq:D.79a})--(\ref{eq:D.80b} which then finally yields
their eigenvalues (\ref{eq:D.73a})--(\ref{eq:D.74b}) in terms of $\elz$ and $\elp$,
i.\,e. for the $R$-system (\ref{eq:D.73a}) and (\ref{eq:D.73b})
\begin{equation}
\label{eq:D.81}
C_+(\elp,\elz) \cdot C_-(\elp,\elz + 1) = \elz^2 - \elp^2
\end{equation}
and analogously for the $S$-system (\ref{eq:D.74a})--(\ref{eq:D.74b})
\begin{equation}
\label{eq:D.82}
D_+(\elp,\elz-1) \cdot D_-(\elp,\elz) = \elp^2 - \elz^2 \;.
\end{equation}
From this result it is evident that the action of the operator products in reverse order
entails merely a change of $\elz$ by on unity $(\elz \Rightarrow \pm 1)$, see the table
for $\boldsymbol{\elp = 3}$ on p. (?).

It should also be self-evident that the two independent equations (\ref{eq:D.81}) and
(\ref{eq:D.82}) are not sufficient in order to fix completely the four ladder coefficients
$C_\pm(\elp,\elz)$, $D_\pm(\elp,\elz)$ as functions of the two quantum numbers $\elp$ and
$\elz$. At the most, we may conclude from these equations that the desired coefficients
must be of the general form
\begin{subequations}
\begin{align}
\label{eq:D.83a}
C_+(\elp,\elz) &= \frac{\elz + \elp}{X(\elp,\elz)} \\
\label{eq:D.83b}
C_-(\elp,\elz) &= X(\elp, \elz - 1) \cdot \left( \elz - \elp - 1 \right) \\
\label{eq:D.83c}
D_+(\elp,\elz) &= \left( \elp + \elz + 1 \right) \cdot Y(\elp, \elz + 1) \\
\label{eq:D.83d}
D_-(\elp,\elz) &= \frac{\elp - \elz}{Y(\elp,\elz)} \;,
\end{align}
\end{subequations}
with unknown functions $X(\elp,\elz)$ and $Y(\elp,\elz)$. Surely, by comparison with
\textbf{Fig.s D.I} and \textbf{D.II} one could perhaps guess how these unknown functions
$X(\elp,\elz)$ and $Y(\elp,\elz)$ should look like; e.\,g. the guess
\begin{equation}
\label{eq:D.84}
X(\elp,\elz) = -\sqrt{\frac{\elp + \elz + 1}{\elp - \elz}}
\end{equation}
would not be in conflict with the coefficients $C_\pm(\elp,\elz)$ of \textbf{Fig. D.I};
but of course it is better to rigorously work out the general form of the unknown
functions $X$ and $Y$.

For this purpose, it is very helpful to first analyze a little bit closer the action of
the ladder operators $\hRpm$ and $\hSpm$ along those arguments, which led us to the former
special results (\ref{eq:D.32}) and (\ref{eq:D.44}). Indeed, when one lets act these
operators on that product form (\ref{eq:D.6a})--(\ref{eq:D.6b}) of the angular
eigenfunctions and furthermore substitutes the derivatives with respect to the variable
$\vartheta$ from the first-order equations (\ref{eq:IV.8a})--(\ref{eq:IV.8b}) one can
represent the angular functions due to $\elz$ by those due to $\elz \pm 1$, i.\,e. for the
$R$-ladder
\begin{subequations}
\begin{align}
\label{eq:D.85a}
\lpr f_{R,\elz}(\vartheta, \phi) &= \frac{\elp + \elz - 1}{C_+(\elp,\elz - 1)}\,\e^{i\elz\phi} \cdot \lpr f_{S,\elz - 1}(\vartheta) \\
(- |&\elp - 2| \leq \elz \leq \elp) \nonumber \\[2em]
\label{eq:D.85b}
\lpr f_{R,\elz}(\vartheta, \phi) &= \frac{\e^{i\elz\phi}}{C_-(\elp,\elz + 1)}\,\big\{
(2\elz + 1) \cot\vartheta \cdot \lpr f_{R,\elz + 1}(\vartheta)\ + \\
&\hspace{4cm}+(\elp + \elz + 1) \cdot \lpr f_{S,\elz + 1}(\vartheta) \big\} \nonumber \\
(- |&\elp - 1| \leq \elz \leq \elp-1) \nonumber \;,
\end{align}
\end{subequations}
and similarly for the $S$-ladder:
\begin{subequations}
\begin{align}
\label{eq:D.86a}
\lpr f_{S,\elz}(\vartheta, \phi) &= \frac{\e^{i\elz\phi}}{D_+(\elp,\elz - 1)}\,\big\{ (1 -
2\elz) \cot\vartheta \cdot \lpr f_{S,\elz - 1}(\vartheta)\ + \\
&\hspace{4cm} + (\elz - 1 - \elp) \cdot \lpr f_{R,\elz - 1}(\vartheta) \big\} \nonumber \\
(- |&\elp - 1| \leq \elz \leq \elp-1) \nonumber \\[2em]
\label{eq:D.86b}
\lpr f_{S,\elz}(\vartheta, \phi) &= \frac{\elp - (\elz + 1)}{D_-(\elp,\elz + 1)}\,\e^{i\elz\phi} \cdot \lpr f_{R,\elz + 1}(\vartheta) \\
(- &\elp \leq \elz \leq \elp-2) \nonumber \;.
\end{align}
\end{subequations}

Obviously, these inter-ladder relations do connect the angular eigenfunctions of either
ladder to those of the other ladder, albeit for a neighbouring step. Therefore, if we wish
to generate the totality of eigenfunctions for given $\elp$, we need not repeatedly solve
the eigenvalue system (\ref{eq:IV.8a})--(\ref{eq:IV.8b}) for any allowed value of $\elz$
nor apply the ascendive and descendive ladder operators but we can obtain the remaining
unknown functions from the linear combinations of a known minimal subset of
eigenfunctions. The coefficients of these linear combinations
(\ref{eq:D.85a})-(\ref{eq:D.86b}) are essentially the ladder coefficients
$C_\pm(\elp,\elz)$ and $D_\pm(\elp,\elz)$ which thus yields additional motivation for
determining their general form. Moreover, the present results
(\ref{eq:D.85a})--(\ref{eq:D.86b}) do also clarify the fact why the same angular functions
(apart from the different pre-factors and steps) do emerge on both the $R$-ladder and the
$S$-ladder (see the table for $\boldsymbol{\elp = 3}$ on p.~\pageref{tablefR}).

The final task is now to check whether those inter-ladder relations
(\ref{eq:D.85a})--(\ref{eq:D.86b}) are compatible with the original eigenvalue equations
(\ref{eq:IV.8a})--(\ref{eq:IV.8b}). This compatibility test is positive and yields one
further inter-ladder relation, namely
\begin{equation}
\label{eq:D.87}
X(\elp,\elz) \cdot Y(\elp,\elz + 1) = 1 \;,
\end{equation}
whose consistency may be exemplified by means of the ladder coefficients being displayed
by \textbf{Fig.s D.I} and \textbf{D.II} for $\boldsymbol{\elp = 3}$. But obviously this
result is \emph{not} sufficient in order to \emph{completely} fix the ladder coefficients!
The reason is here that we did not yet use the normalization condition (\ref{eq:IV.15a})
for the angular eigenfunctions. But if this is implemented for one step of \emph{one}
ladder, then one can deduce all the other coefficients of \emph{both} ladders. Thus, one
obtains for the ascent of the $R$-ladder
\begin{equation}
\label{eq:D.88}
C_+(\elp,\elz) = -(\elp + \elz) \cdot \sqrt{\frac{\elp - \elz}{\elp + \elz + 1}}\;,\quad -(\elp - 1) \leq \elz \leq \elp - 1
\end{equation}
and similarly for the descent
\begin{subequations}
\begin{gather}
\label{eq:D.89a}
C_-(\elp,\elz) = (\elp + 1 - \elz) \cdot \sqrt{\frac{\elp + \elz}{\elp - \elz + 1}}\;,\quad \left(-|\elp - 2| \leq \elz \leq \elp \right) \\
\label{eq:D.89b}
C_-(\elp,1-\elp) = 0 \;.
\end{gather}
\end{subequations}
For the $S$-ladder, the ascent is described by the following ladder coefficients
\begin{subequations}
\begin{gather}
\label{eq:D.90a}
D_+(\elp,\elz) = -\sqrt{(\elp - \elz)(\elp + \elz + 1)} \;,\quad -\elp \leq \elz \leq (\elp - 2) \\
\label{eq:D.90b}
D_+(\elp,\elp - 1) = 0
\end{gather}
\end{subequations}
and the descent by
\begin{equation}
\label{eq:D.91}
D_-(\elp,\elz) = -(\elp - \elz)\,\sqrt{\frac{\elp + \elz}{\elp - \elz + 1}} \;,\quad -(\elp - 1) \leq \elz \leq (\elp - 1) \;.
\end{equation}
The ladder coefficients displayed by \textbf{Fig.s D.I} and \textbf{D.II} do again
exemplify these results for~$\boldsymbol{\elp=3}$. Zhe solutions
(\ref{eq:D.88})-(\ref{eq:D.91}) do satisfy both product requirements (\ref{eq:D.81}) and
(\ref{eq:D.82}) and firthermore do also obey the following symmetries for the allowed
values of~$l_z$ and~$\lP$:
\begin{subequations}
  \begin{align}
    \label{eq:D.92a}
    C_+(\lP,-l_z) &= D_-(\lP,l_z)\\*
    \label{eq:D.92b}
    C_-(\lP,-l_z) &= -D_+(\lP,l_z)
  \end{align}
\end{subequations}

\begin{center}
  \large{\textit{Compact Representation}}
\end{center}

Up to now we represented the angular-momentum algebra in component form by considering separately each component $\lpr f_{R,\elz}(\vartheta, \phi)$ and $\lpr f_{R,\elz}(\vartheta, \phi)$ of the complex two vector $\lpr \mathbf{f}_{\elz}(\vartheta, \phi)$ (\ref{eq:D.13}). But for a concise survey of the essential algebraic features it may be more advantageous to resort to a more compact abstract representation. The proper eigenvalue problem has already been reformulated in such an abstract form, cf. equations (\ref{eq:D.14})--(\ref{eq:D.18}), so that this must now merely be completed by the corresponding abstract representation of the ladder formalism.

To this end, we condense the ascendive ladder operators $\hRp$ and $\hSp$ (\ref{eq:D.22a})--(\ref{eq:D.22b}) to the two-dimensional abstract object $\mathcal{\hMTp}$
\begin{equation}
\label{eq:D.92}
\hMTp \doteqdot
\left(\begin{array}{cc}
\hRp & 0 \\
0 & \hSp
\end{array}\right)
\end{equation}
and similarly for the descendive case
\begin{equation}
\label{eq:D.93}
\hMTm \doteqdot
\left(\begin{array}{cc}
\hRm & 0 \\
0 & \hSm
\end{array}\right) \;.
\end{equation}
The ascendive and descendive properties of these newly introduced objects are now expressible by their action on the two-vector $\lpr \mathbf{f}_{\elz}$, i.\,e.
\begin{subequations}
\begin{align}
\label{eq:D.94a}
\hMTp\,\lpr \mathbf{f}_{\elz} &= \mathcal{N}_+(\elp,\elz) \cdot \lpr \mathbf{f}_{\elz + 1} \\
\label{eq:D.94b}
\hMTm\,\lpr \mathbf{f}_{\elz} &= \mathcal{N}_-(\elp,\elz) \cdot \lpr \mathbf{f}_{\elz - 1} \;,
\end{align}
\end{subequations}
where the matrix-valued normalization constants $\mathcal{N}_\pm$ are given by
\begin{subequations}
\begin{align}
\label{eq:D.95a}
\mathcal{N}_+(\elp,\elz) &= -\sqrt{\frac{\elp - \elz}{\elp + \elz + 1}} \cdot
\left(\begin{array}{cc}
\elp + \elz & 0 \\
0 & \elp + \elz + 1
\end{array}\right) \\
\label{eq:D.95b}
\mathcal{N}_-(\elp,\elz) &= -\sqrt{\frac{\elp + \elz}{\elp - \elz + 1}} \cdot
\left(\begin{array}{cc}
\elz - \elp - 1 & 0 \\
0 & \elp - \elz
\end{array}\right) \;.
\end{align}
\end{subequations}
The equations (\ref{eq:D.94a})--(\ref{eq:D.94b}) say that the action of $\hMTp$ lets \emph{increase} the quantum number $\elz$ of the eigenvector $\lpr \mathbf{f}_{\elz}$ by one unit, and $\hMTm$ lets it \emph{decrease} by one unit. Indeed it ts easy to see that this behaviour is an immediate consequence of the following relations
\begin{subequations}
\begin{align}
\label{eq:D.96a}
\left[ \hMLz, \hMTp \right] &= \hMTp \\
\label{eq:D.96b}
\left[ \hMLz, \hMTm \right] &= -\hMTm \;,
\end{align}
\end{subequations}
which themselves are easily deducible from the corresponding commutation relations for $\hRpm$ and $\hSpm$:
\begin{subequations}
\begin{align}
\label{eq:D.97a}
\left[ \hLz, \hRpm \right] &= \pm \hRpm \\
\label{eq:D.97b}
\left[ \hLz, \hSpm \right] &= \pm \hSpm \;.
\end{align}
\end{subequations}
Thus, we arrive at
\begin{subequations}
\begin{align}
\label{eq:D.98a}
\hMLz \hMTp\,\lpr \mathbf{f}_{\elz} &= (\elz + 1) \cdot \hMTp\,\lpr \mathbf{f}_{\elz} \\
\label{eq:D.98b}
\hMLz \hMTm\,\lpr \mathbf{f}_{\elz} &= (\elz - 1) \cdot \hMTm\,\lpr \mathbf{f}_{\elz}
\end{align}
\end{subequations}
which express the actions of $\hMTpm$ without any reference to the normalization constants $\mathcal{N}_\pm$.

With respect to the other quantum number $\elp$ angular momentum it is also instructive to
consider the commutator of these abstract ladder operators $\hMTpm$, i.\,e.
\begin{equation}
\label{eq:D.99}
\left[ \hMTp, \hMTm \right] = \mathbf{1} - 2\hQQ_z \;.
\end{equation}
This result for the ladder operators may be exemplified by their immediate actions
(\ref{eq:D.94a})--(\ref{eq:D.94b}) on the eigenvectors $\lpr \mathbf{f}_{\elz}$ where one
can make use of the commutative relations of the (matrix-valued) normalization constants
$\mathcal{N}_\pm(\elp,\elz)$:
\begin{equation}
\label{eq:D.100}
\mathcal{N}_-(\elp,\elz) \cdot \mathcal{N}_+(\elp,\elz-1) - \mathcal{N}_+(\elp,\elz) \cdot \mathcal{N}_-(\elp,\elz+1)  = 1 - 2\elz\sigma_z \;.
\end{equation}
But of course, the claim (\ref{eq:D.99}) does hold quite generally and independently of
the actions (\ref{eq:D.94a})--(\ref{eq:D.94b}) on the eigenstates of angular
momentum. This can easily be verified by reference to the component form
\begin{subequations}
\begin{align}
\label{eq:D.101a}
\left[ \hRp, \hRm \right] &= 1 - 2\hat{\mathrm{Q}}_z \\
\label{eq:D.101b}
\left[ \hSp, \hSm \right] &= 1 + 2\hat{\mathrm{Q}}_z \;.
\end{align}
\end{subequations}
The result (\ref{eq:D.99}) says that the sequence of ascent~$(\hat{\tau}_+)$ and
descent~$(\hat{\tau}_-)$ (or vice versa) does one not lead back to the original angular
state~$\lpr \mathbf{f}_{\elz}$ (not even up to an overall pre-factor).

\renewcommand{\refname}{{\bfseries \Large References}}

\end{document}